\documentclass[10pt,preprint2]{aastex}
\onecolumn
\usepackage{bm}
\usepackage{graphicx}
\usepackage{latexsym,color}
\usepackage{amssymb}
\usepackage{amsmath,amssymb,amsthm,mathrsfs,amsfonts,dsfont}
\usepackage{color}
\usepackage[hyperfootnotes=false]{hyperref}
\usepackage{enumitem}
\makeatletter
\def\namedlabel#1#2{\begingroup
	#2%
	\def\@currentlabel{#2}%
	\phantomsection\label{#1}\endgroup
}
\makeatother

\shorttitle{Ringed  accretion disks}
\shortauthors{D. Pugliese and Z. Stuchl\'{\i}k}

\begin{document}

\title{{Ringed  accretion disks: equilibrium configurations}}
\author{D. Pugliese and Z. Stuchl\'{\i}k}
\affil{
Institute of Physics  and Research Centre of Theoretical Physics and Astrophysics, Faculty of Philosophy \& Science,
  Silesian University in Opava,
 Bezru\v{c}ovo n\'{a}m\v{e}st\'{i} 13, CZ-74601 Opava, Czech Republic
}
\email{d.pugliese.physics@gmail.com;zdenek.stuchlik@physics.cz}
\begin{abstract}
We  investigate a model of ringed accretion disk,  made up   by several rings rotating around a  supermassive Kerr black hole    attractor. Each toroid of the ringed  disk is governed by the   General Relativity
   hydrodynamic Boyer  condition  of equilibrium configurations  of rotating perfect fluids. Properties of the tori can be then determined  by an appropriately defined  effective potential reflecting the background  Kerr geometry and the centrifugal effects. The ringed disks could be created  in various regimes during  the evolution of matter configurations  around supermassive black holes. Therefore, both corotating and  counterrotating  rings have to be considered as being a constituent of the ringed disk. We  provide  constraints on the model parameters for the existence and stability of  various ringed  configurations  and discuss  occurrence of accretion onto the Kerr black hole  and possible launching of jets from the  ringed disk. We demonstrate that  various  ringed disks can be characterized by a maximum number of rings. We present also a perturbation analysis based on evolution of  the oscillating components of the  ringed disk. The dynamics of the  unstable phases  of the ringed disk evolution seems to be promising in  relation to  high energy phenomena demonstrated in active galactic nuclei.
\end{abstract}

\newcommand{\ti}[1]{\mbox{\tiny{#1}}}
\newcommand{\im}{\mathop{\mathrm{Im}}}
\def\be{\begin{equation}}
\def\ee{\end{equation}}
\def\bea{\begin{eqnarray}}
\def\eea{\end{eqnarray}}
\newcommand{\tb}[1]{\textbf{\texttt{#1}}}
\newcommand{\ttb}[1]{\textbf{#1}}
\newcommand{\rtb}[1]{\textcolor[rgb]{1.00,0.00,0.00}{\tb{#1}}}
\newcommand{\btb}[1]{\textcolor[rgb]{0.00,0.00,1.00}{\textbf{#1}}}
\newcommand{\ntb}[1]{\textcolor[rgb]{0.00,0.00,0.00}{\textbf{#1}}}
\newcommand{\otb}[1]{\textcolor[rgb]{0.90,0.45,0.00}{\tb{#1}}}
\newcommand{\gtb}[1]{\textcolor[rgb]{0.00,0.77,0.00}{\tb{#1}}}
\newcommand{\il}{~}
\newcommand{\rc}{\rho_{\ti{C}}}
\newcommand{\dd}{\mathcal{D}}
\newcommand{\lie}{\mathcal{L}}
\newcommand{\Mie}{\mathcal{M}}
\newcommand{\Tem}{T^{\rm{em}}}
\section{Introduction}
The most energetic processes in the Universe are related to accretion disks around black holes or some alternative objects with extremely strong gravity. Such processes with extremely large radiative energy output, combined with ejection of matter associated with jet-like structures emerging from extremely small central regions, are observed in quasars and active galactic nuclei (AGN) where supermassive black holes, with mass in the interval $(10^6 - 10^{10})M_{\odot}$, are assumed in the center \citep{Ziolkowski:2005ag}. A scaled down version of these energetic processes is observed in the so called microquasars where stellar mass black holes are expected in the  center \citep{Remillard:2006fc}.
The enormous energy emitted by the accretion disks in quasars or AGN, in the form of electromagnetic radiation and jets, can be attributed to the strong gravity of the central black hole when the gravitational binding energy of accreting matter is transformed into radiation. The efficiency of conversion of rest energy of accreting matter to radiated energy is limited by the energy of the innermost stable circular geodesic \citep{Bardeen-Nature70}. In the field of non-rotating black holes, the conversion efficiency $\sim 6\%$ is large enough, enabling to explain the energy output in both quasars  and microquasars. In the field of the near-extreme Kerr black holes, the efficiency approaches $42\%$ \citep{Bardeen:1972fi}, while in the field of Kerr naked singularities the conversion efficiency can be much larger, exceeding even the $100\%$ related to annihilation processes, and approaching $157\%$ for the near-extreme Kerr naked singularities \citep{ZS-1980,SHT-CQG-2011}.

There are at least three important aspects of the accretion disk structure which in turn lead to a classification of different disk models: the geometry (the vertical thickness) for which we can distinguish geometrically thin or thick disks, the matter accretion rate (defining sub- or super-Eddington luminosity), and the optical depth (i.e., transparent or opaque disks), \citep{AbraFra}.
More specifically, geometrically thin disks are modelled as the standard Shakura-Sunayev (Keplerian) disks \citep{[S73],[SS73],Nov-Thorne73,Page-Thorne74}, the ADAF (Advection-Dominated Accretion Flow) disks \citep{Ab-Ac-Schl14,Narayan:1998ft}, and the slim disks \citep{AbraFra}. Geometrically thick disks are modelled as the Polish Doughnuts (P-D) \citep{Koz-Jar-Abr:1978:ASTRA:,Abr-Jar-Sik:1978:ASTRA:,Jaroszynski(1980),Stu-Sla-Hle:2000:ASTRA:,
Rez-Zan-Fon:2003:ASTRA:,Sla-Stu:2005:CLAQG:,
Stu:2005:MPLA,PuMonBe12,PuMon13}, or the ion tori \citep{Rees1982}. The ADAF disks and the ion tori have relatively low accretion rates (sub-Eddington), while the P-D disks have very high (super-Eddington) accretion rates. The P-D and slim disks have high optical depth being thus opaque, while the ion tori and the ADAF disks have low optical depth being thus transparent.
In the geometrically thin disks, dissipative viscosity processes are relevant for accretion, being usually attributed to the magnetorotational instability of the local magnetic fields \citep{Hawley1984,Hawley1987,Hawley1990,Hawley1991,DeVilliers}. In the toroidal disks, pressure gradients are crucial \citep{Abr-Jar-Sik:1978:ASTRA:,arXiv:0910.3184}. The magnetic pressures of toroidal magnetic fields could be also relevant \citep{Komiss,Adamek:2013dza,Hamersky:2013cza}. Notice the influence of the global internal magnetic toroidal fields can  be represented by 1D string loops \citep{Larsen94,Kolos:2013bca,Stuchlik:2012ry,Stuchlik:2014foa,Cre-Stu13}. While an off-equatorial ``levitating'' tori can be modelled by the dielectric \citep{Kovar11,Slany:2013rml,Kovar:2014tla} or kinetic \citep{Cremaschini:2013jia} models of toroidal charged structures if a large scale external magnetic field is present in vicinity of the central black hole.
{
In different situation the accretion disks could be with relatively high precision modelled by using a well defined Pseudo-Newtonian potential \citep{Pac-Wii,Abr-Cal-Nob:1980:ASTRJ2:,Stu-Kov:2008:INTJMD:,arXiv:0910.3184},  while  the structure of many geometrically thin Keplerian disks and geometrically thick toroidal disks is deeply regulated by the geodesic structure of the black hole spacetime
 \citep{Nov-Thorne73,Abr-Jar-Sik:1978:ASTRA:,Stu:2005:MPLA,AbraFra}. However, here we focus on the fully general relativistic approach.
While we  do not consider in the present paper the role of the magnetic fields, postponing this issue to future investigations.

In this article we propose a model of ringed accretion disk (or \emph{macro-configuration})  formed   by several toroidal structures, the \emph{rings} (or  \emph{sub-configurations}), orbiting a supermassive Kerr attractor and centered in its equatorial plane--\citep{Pugtot}.
The centers of all the individual tori are therefore   coplanar,  coinciding with the equatorial plane of the black hole. This assumption of coincidence between the orbital planes of the rings and the equatorial plane of the axisymmetric attractor, is adopted here as the simplest scenario for the first construction of such models. In fact, the majority of the current analytical and numerical models of accretion configurations assumes the  axial symmetry of the extended accreting matter. There is indeed  an  ``inheritance'' of the symmetries of   1D linear test particle model  in circular  motion,  for the  toroidal 2D accreting structures  orbiting  extended matter under pressure gradients, characterized by a more or less significant verticalization of the accretion disk. These symmetries are also characteristic for the model of ringed disks.

A ringed disk consists thus of several tori which  could be created due to various (succeeding) accretion regimes during the evolution of the matter configurations around the supermassive black hole.
Construction of a ringed disk in a Kerr spacetime can be realized in two orbital stability regions, corresponding to the stable corotating and counterrotating circular orbits  relative to the rotation of the black hole attractor.
 Of course, various accretion periods could create both corotating and counter-rotating toroidal configurations given by various initial conditions determining the succeeding accretion periods, e.g., by tidally destroyed stars corotating or counterrotating with respect to the attractor \citep{Bardeen:1972fi,ZS-1980}. {The individual toroidal remnant structures can be thus both corotating or counter-rotating. In fact, in modelling evolution of supermassive black holes in AGNs both corotating and counter-rotating accretion stages are mixed \citep{Volonteri:2002vz,Volonteri:2003kr}.}
On the other hand, in the galactic nuclei containing a supermassive black hole, several separated accretion regimes could  be occurred in the past, leaving some remnants in the form of toroidal structures that could be reanimated, if some additional matter is supplied into the  vicinity of the central black hole due to tidal distortion of a star, or some cloud of interstellar matter is captured by the strong gravity.

The precise arrangement of ringed disks and the relations between the rings will be here  completely characterized. The results of this work demonstrate significant constraints on different features of the  macro-configurations, in particular on the number of rings and their specific angular momentum (differential rotation of the ringed disks). Noticeably  the ringed disk model could be also seen as a unique geometrically thin accretion disk.

We trace the definition of the ringed disk model, focusing on the equilibrium configurations that ensure its physical reliability and, simultaneously, can be considered as starting points of instability phases.
In fact, an essential aspect concerns the unstable phases, associated to the (super-Eddington) accretion and formation of jets.

The ringed disks have an internal structure represented by their \emph{decomposition}, or the set of individual tori. The internal dynamics of the ringed disks is related to the dynamics of its units (individual tori). The internal dynamics of these tori may result in instabilities, associated to the accretion onto the black hole or creation of jets \citep{Meier},  leading to potentially unstable phases for the entire ringed disk.
In principle, an infinite number of individual tori centers with maximal pressure can exist. However   the tori as actually extended objects, are not 1D strings. Consequently the extension of an individual torus is constrained by its internal dynamics (for example, by the equilibrium of the pressure gradients and gravitational and inertial forces), and by the fact that the individual tori are not isolated, but actually they exist as a part of the entire structure composing the ringed disk. In such a structure, each ring must keep its identity, avoiding the overlap of material from one ring to another.
 Possible collisions of rings, feeding or accretion from a ring to another one, are considered to be unstable phases of the ringed disk, which can not be immediately described by the model here considered \citep{Boy:1965:PCPS:}.

The unstable states of a ringed disk system provide a classification of possible non-equilibrium configurations.
 The study of the equilibrium tori will be then the  starting point for a future analysis of the oscillation modes in the structure of the relativistic ringed disks  which can be related to various astrophysical phenomena.
The results  in this respects  seem to be promising in view of the phenomenology associated with both the stable and unstable  configurations. The radially oscillating tori of the ringed disk could be related to the high-frequency quasi periodic oscillations observed in non-thermal X-ray emission from compact objects \citep{TAKS05,Stuchlik:2013esa}. A strong overflow of matter from a torus could be related to a jet formation.
As a consequence of this analysis, we are able to locate the instability points (one or several) associated with accretion onto the Kerr accreitor,  of the  launching of jets inside the ringed macro-configuration, and the distribution of the specific angular momentum initially characterizing the unstable phases. Particularly interesting would be analysis of the unstable phases in which an interaction with the source is associated with consequent changes of the geometrical characteristics of the spacetime, by mutation of the mass and spin parameters of the Kerr attractor. The mutation of the geometry in turn determines a change of the dynamic properties of matter in accretion, resulting in an iterative process to be analyzed, which could be a form of runaway instability \citep{Abr-Nat-Run,Abramowicz:1997sg,Font:2002bi,Rez-Zan-Fon:2003:ASTRA:,Lot2013,Hamersky:2013cza,ergon}.

\medskip

We provide here  a precise definition and a detailed description of the ringed accretion configurations in full general relativity (GR), assuming each individual torus of the ringed disk be made of perfect fluid. The tori are governed by the equipotential surfaces that can be closed, giving stable equilibrium configurations, and open, giving unstable, jet-like structures caused by the relativistic instability due to the Paczynski mechanism.
More specifically, the individual toroidal  (thick disk) configurations (the rings or sub-configurations) are  here described by purely hydrodynamic (barotropic) models, for which the time scale of the dynamical processes (regulated by the gravitational and inertial forces) is much lower than the time scale of the thermal ones (heating and cooling processes, radiation) that is lower than the time scale of the viscous processes. Where the effects of strong gravitational fields are dominant with respect to the  dissipative ones and predominant to determine  the unstable phases of the systems\citep{F-D-02,Abramowicz:2004vi,Igumenshchev,AbraFra,Pugtot,Pac-Wii},
see also \citep{Hawley1990,Fragile:2007dk,DeVilliers,Hawley1987,Hawley1991,Hawley1984,Fon03}.
The entropy is constant along the flow and, according to the von Zeipel condition, the surfaces of constant angular velocity $\Omega$ and of constant specific angular momentum $\ell$ coincide \citep{M.A.Abramowicz,Chakrabarti0,Chakrabarti,Zanotti:2014haa}. This implies that the rotation law $\ell=\ell(\Omega)$ is independent of the equation of state \citep{Lei:2008ui,Abramowicz:2008bk}. The perfect fluid equilibrium tori can be classified in a given spacetime by the specific angular momentum distribution function $\ell(r)$, and a constant $K$ determining the matter content of the tori \citep{Abr-Jar-Sik:1978:ASTRA:}
\footnote{The Polish Doughnut model \citep{AbraFra} represents an important example of a thick, opaque and super-Eddington, radiation pressure supported accretion disks cooled by advection with low viscosity. The morphology of a P-D model in equilibrium is a fat torus centered on the attractor. In the open configurations, critical phases are characterized by funnels of material with highly super-Eddington luminosity \citep{Abr-Nat-Run}.}.
 These configurations have been often adopted as the initial conditions in the set up for simulations of the MHD (magnetohydrodynamic) accretion structures\citep{}\citep{Igumenshchev,Shafee} and \citep{Fragile:2007dk,DeVilliers,
arXiv:0910.3184}.

The model of the tori that is used in this paper is stationary and axisymmetric, being adapted to the symmetries of the Kerr spacetime. The tori are governed by ``Boyer's condition'' of the analytic theory of equilibrium configurations of rotating perfect fluids \citep{Boy:1965:PCPS:}. The toroidal structures of orbiting barotropic perfect fluid are determined by an effective potential reflecting the spacetime geometry and distribution of the specific angular momentum $\ell(r)$ of the orbiting fluid -- such an approach is well known and widely adopted  \citep{Abr-Jar-Sik:1978:ASTRA:,Pugtot,Abramowicz:1996ap,FisM76,Raine,PuMonBe12,Lei:2008ui,PuMon13}, and is applied in various contexts \citep{2011,Rez-Zan-Fon:2003:ASTRA:,Stuchlik:2012zza,Sla-Stu:2005:CLAQG:,astro-ph/0605094,Stu-Sla-Hle:2000:ASTRA:,arXiv:0910.3184,Stu-Kov:2008:INTJMD:,Stuchlik:2014jua}. Many features of the tori dynamics and morphology like their thickness, their stretching in the equatorial plane, and the location of the tori are predominantly determined by the geometric properties of spacetime via the effective potential. The gradient of the effective potential is regulating the pressure gradient of the fluid in the Euler law governing dynamics of the perfect fluid \citep{Koz-Jar-Abr:1978:ASTRA:}. The boundary of any stationary, barotropic, perfect fluid body is determined by an equipotential surface, i.e., the surface of constant pressure (the Boyer surface) that is orthogonal to the gradient of the effective potential \citep{Boy:1965:PCPS:,arXiv:0910.3184}.
The equipressure surfaces that could be closed (determining equilibrium configurations), or open (determining jets). The special case of cusped equipotential surfaces allows for the accretion onto the central black hole \citep{Koz-Jar-Abr:1978:ASTRA:,Abr-Jar-Sik:1978:ASTRA:,
Jaroszynski(1980),Pac-Wii,Abr-Cal-Nob:1980:ASTRJ2:} or excretion in spacetimes reflecting the cosmic repulsion \citep{Stu-Sla-Hle:2000:ASTRA:,Rez-Zan-Fon:2003:ASTRA:,Sla-Stu:2005:CLAQG:,Stu:2005:MPLA,arXiv:0910.3184}. The outflow of matter through the cusp occurs due to an instability in the balance of the gravitational and inertial forces and the pressure gradients in the fluid, i.e., by the so called Paczynski mechanism of violation of mechanical equilibrium of the tori \citep{Jaroszynski(1980)}. The cusp governing accretion onto the black hole has to be located near the event horizon \citep{Abr-Jar-Sik:1978:ASTRA:,cc,Pac-Wii}, while the cusp governing the excretion has to be located near the so called static radius where gravitational attraction of the black hole is just balanced by the cosmic repulsion \citep{Stu-Sla-Hle:2000:ASTRA:,Stu:2005:MPLA} independently on the rotation of the black hole \citep{Stuchlik:2003dt}.

\medskip

The possibility of constructing the ringed disk model has been mentioned in \citep{PuMon13}. To our knowledge the present article is the first attempt to define and characterize this model of accretion disk that could, we believe, to be of some significance for the high energy astrophysics related especially to accretion onto supermassive black holes, and the extremely energetic phenomena in quasars and AGN that could be observable by the planed X-ray observatory ATHENA \footnote{http://the-athena-x-ray-observatory.eu/}.

The new ringed disk model requires  the introduction of a number of new definitions, and in this article the formalism  was developed in details and extensively explained. To make more clear and simple the  description and explanation of this model, the article has been  structured in two parts: brief first descriptive part,   Secs.\il(\ref{Sec:models}) and (\ref{Sec:intro}), illustrates the model, the second part, Sec.\il(\ref{Sec:caracter}),  presents in detail the obtained results. In order to lighten the reading, we have introduced Appendices, where various aspects of the model are detailed, and parts of the analysis, dealing with some specific topics, such as an introduction to the perturbation approach, were developed.

\medskip

{In details the plan of the article is the following: in Sec.\il(\ref{Sec:models}) we briefly summarize the thick accretion disk model. The ringed disk model is introduced in Sec.\il (\ref{Sec:intro}) where the main definitions are provided. Particularly, in Sec.\il(\ref{Sec:unsta}), we define the unstable macro-configurations in the framework of ringed disk model. The definition of the major morphological features of the ringed disks is presented in Sec.\il\ref{Sec:principaldef}), closing first part of this work.
Section\il(\ref{Sec:caracter}), constituting the second part  of this article, is the essential part of our analysis providing an extensive characterization of the compositions of the ringed disks. After discussing the role of the model parameters in Sec.\il(\ref{Sec:roleofp}) and (\ref{Sec:onK}),
an effective potential of the ringed macro-structure is defined in Sec.\il(\ref{Sec:effective}). In Sec.\il(\ref{Sec:rolel}), we deepen the role of the $\ell$-parameter introducing the definition of the differential rotation and angular momentum of the ringed macro-structures in Sec.\il(\ref{Sec:diff-rot}). These definitions are used to find the results developed in Sec.\il(\ref{Sec:procedure}). In Section\il(\ref{Sec:i4cases})
we apply this analysis to characterize two main types of the ringed macro-configurations as previously introduced: each disk can be only a combination of these two types or one of them. Finally, Sec.\il(\ref{Sec:K=K}) closes the second part of this work with the study of some important limiting cases.  Discussion  and future perspectives can be found in Sec.\il(\ref{Sec:conc}). A summary with  conclusions are in Sec.\il(\ref{Sec:s-E}).
 Some Appendix sections follow: Sec.\il(\ref{Sec:App-mor-p}) presents some notes of the ringed disk morphology and the effective potential of the ringed disk.  In Sec.\il(\ref{Sec:app-maxmin}) we deepen the properties of the ring angular momentum, discussing some results of general relevance for the tori governed by the Boyer model.   We investigate some aspects of the macro-configuration with one or more critical points, and in particular the macro-configuration with the open (unstable) configurations associated with jets created by accreting matter in Sec.\il\ref{App:opem}).  General considerations on the perturbations of the ringed disks and their equilibrium can be found in Sec.\il(\ref{Sec:pertur}). Sec.\il(\ref{App:notes-index}) contains a collection of results we refer to along the analysis. Table.\il(\ref{Table:pol-cy}) provides  a list of   the main symbols and relevant notation  used throughout this article.
}

%
\begin{table*}
\centering
\resizebox{1.1\textwidth}{!}{%
\begin{tabular}{lll}
 \hline \hline
$ C$&   cross sections of the closed Boyer surfaces (equilibrium torus)&Secs.\il(\ref{Sec:models},\ref{Sec:intro})\\
$ C_x$&   cross sections of the closed cusped  Boyer surfaces (accretion torus)&Secs.\il(\ref{Sec:models},\ref{Sec:intro})\\
$ O_x$&   cross sections of the open cusped  Boyer surfaces&Secs.\il(\ref{Sec:models},\ref{Sec:intro})
\\
 $\partial C$&  the contour of the surface $C$ in the  equatorial section& Fig.\il(\ref{Fig:Quanumd})
 \\
 $\mathbf{C}^n\equiv \bigcup^n_{i=1} C_i$ & ringed  disk of order $n$ (macro-configuration)
 &Fig.\il(\ref{Fig:Quanumd})
 \\
 $(y^i_3, y^i_1)$& inner and outer edge of $C_i$ ring& Fig.\il(\ref{Fig:Quanumd})
 \\
$\overbrace{\mathbf{C}}_{{m}}$& {mixed} $\ell$counterrotating sequences&   Sec.\il(\ref{Sec:intro})  Fig.\il(\ref{Figs:RightPolstex})-top.
\\
 $\overbrace{\mathbf{C}}_{{s}}$& {isolated} $\ell$counterrotating sequences&  Sec.\il(\ref{Sec:intro})
 Fig.\il(\ref{Figs:RightPolstex})-bottom\\
 $\mathbf{C_{\odot}^n}$& configuration
 $i\in\{1,...,n-1\}:\; \partial C_i\cap\partial C_{i+1}=\{ y_1^{i}=y_3^{i+1}\}$& Fig.\il(\ref{Figs:CrystalPl})
 \\
 $\mathfrak{r}$& {rank}   of
$\mathbf{C_{\odot}^n}$& Sec.\il(\ref{Sec:unsta})
\\

$\mathbf{C^n_x}$&
 Ringed disk with at last one sub-configuration $C^i_x$&
 Fig.\il(\ref{Figs:CrystalPl})
\\
$\mathfrak{r_x}$ & {rank}  of   $\mathbf{C^n_x}$&Sec.\il(\ref{Sec:unsta})
\\
 $\mathbf{C_{\odot}^x}$& ringed  disk $\mathbf{C_{\odot}}$ with a  $C^{i}_x$ ring:  $y_1^{i-1}=y_3^{i}=y_{cusp}$
&Sec.\il(\ref{Sec:unsta})
\\
$\lambda_i\equiv ]y_1^i,y_3^i[$ $\lambda_i\equiv y_1^i-y_3^i$&  {elongation}, range and measure, of the  Boyer surface&  Fig.\il(\ref{Fig:Quanumd})
\\
$\bar{\Lambda}_{i+1,i}\equiv] y_3^{i+1},y_1^i[$ $\bar{\lambda}_{i+1,i}\equiv y_3^{i+1}-y_1^i$
 &  {spacing}, range and measure & Fig.\il(\ref{Fig:Quanumd}).
\\
 $\Lambda_{\mathbf{C}^n}$ $\lambda_{\mathbf{C}^n}$&  elongation, range and measure,  of $\mathbf{C}^n$& Eq.\il(\ref{Eq:chan-L-corby})
 \\
  $h_{{\mathbf{C}}^n}$& {height}  of $\mathbf{C}^n$& Eq.\il(\ref{Eq:max-height})
  \\
 $R_{\mathbf{C}^n}$& {thickness}   of  $\mathbf{C}^n$&
\\
$R_{\mathbf{C}_{\odot}^n}$&  {thickness} of $\mathbf{C}_{\odot}^n$& Eq.\il(\ref{Eq:wolne}), Sec.\il(\ref{Sec:unsta})
\\
 $\left.V_{eff}^{\mathbf{C}^n}\right|_{K_i}$&  effective potential of the  {decomposed} $\mathbf{C}^n$ &
Sec.(\ref{Sec:effective}) Eq.\il(\ref{Eq:def-partialeK}).
\\
$V_{eff}^{\mathbf{C}^n}$&   effective potential   of the configuration  ${\mathbf{C}^n}$
&Eq.\il(\ref{Eq:Vcomplessibo}).
\\
$\ell_{i/i+1}\equiv\ell_i/\ell_{i+1}$&
 ratio in  specific  angular momentum of $C_i$ and $C_{i+1}$&
Eq.\il(\ref{ratiolll})
\\
$\bar{\ell}_{\mathbf{C}}^n$&  area angular momentum  of  ${\mathbf{C}^n}$& Eq.\il(\ref{Eq:stand})
\\
$\bar{\bar{\ell}}_{\mathbf{C}}^n$& volume angular momentum  of  ${\mathbf{C}^n}$& Eq.\il(\ref{Eq:stand})
\\
$\bar{\ell}_{\mathbf{C}^n}^l$&  leading angular momentum  of  ${\mathbf{C}^n}$& Eq.\il(\ref{Def:leadingl})
\\
$\ell_h$&  apparent angular momentum  of  ${\mathbf{C}^n}$& Eq.\il(\ref{Def:leadingl})
\\
$r^{\pm}_{\mathcal{M}}$& maximum point    of
derivative $\partial_r(\mp \ell^{\pm})$ for $a/M$ respectively& Sec.\il(\ref{Sec:procedure},\ref{Sec:app-maxmin})
\\
$a_{\aleph_1}=0.382542M$&$a_{\aleph_1}:\ell_{\gamma}^-=-\ell_{+}(r_{mso}^-)$&
 Fig.\il(\ref{Figs:RightPolstex})
 \\
$ a_{\aleph}\approx0.508864526M$& $ a_{\aleph}:
-\ell_{mso}^+(a_{\aleph})=\ell_{\gamma}^-(a_{\aleph})$&
Eq.\il(\ref{Eq:al-1}) Fig.\il(\ref{Figs:Aslanleph1l})
\\
$a_{\aleph_2}\equiv 0.172564M$ & $a_{\aleph_2}:
-\ell_{mso}^+(a_{\aleph_2})=\ell_{mbo}^-(a_{\aleph_2})$& Sec.\il(\ref{Sec:procedure})
\\ $a_{\aleph_0}\equiv0.390781M$&$a_{\aleph_0}\in]a_{\aleph_1},a_{\aleph}[:\ell_{\gamma}^-=-\ell_{mbo}^+$
&
Eq.\il(\ref{Eq:alm-infla})\\
\hline\hline
\end{tabular}}
\caption{{Lookup table with the main symbols and relevant notation  used throughout the article. Links to associated sections, definitions and figures are also listed.}}
\label{Table:pol-cy}
\end{table*}
%
\section{Thick accretion disk model in the Kerr spacetime}\label{Sec:models}
In this work we  consider   toroidal  configurations of  perfect fluids orbiting a    Kerr black hole \textbf{(BH)} attractor. The Kerr  metric tensor can be
written in the Boyer-Lindquist (BL)  coordinates
\( \{t,r,\theta ,\phi \}\)
as follows
\bea \label{alai}&& ds^2=-dt^2+\frac{\rho^2}{\Delta}dr^2+\rho^2
d\theta^2+(r^2+a^2)\sin^2\theta
d\phi^2+\frac{2M}{\rho^2}r(dt-a\sin^2\theta d\phi)^2\ ,
\\
\nonumber
&&
 \rho^2\equiv r^2+a^2\cos\theta^2, \quad \Delta\equiv r^2-2 M r+a^2,
\eea
here $M$ is a mass parameter and the specific angular momentum is given as $a=J/M\in]0,1]$, where $J$ is the
total angular momentum of the gravitational source.
The horizons $r_-<r_+$ and the outer static limit $r_{\epsilon}^+$ are respectively given by:
\bea
r_{\pm}\equiv M\pm\sqrt{M^2-a^2};\quad r_{\epsilon}^{+}\equiv M+\sqrt{M^2- a^2 \cos\theta^2};
\eea
there is $r_+<r_{\epsilon}^+$ on   $\theta\neq0$  and  $r_{\epsilon}^+=2M$  in the equatorial plane, $\theta=\pi/2$.  The extreme Kerr black hole  has spin-mass ratio $a/M=1$, while  the non-rotating  limiting case $a=0$ is the   Schwarzschild metric.
As the line element (\ref{alai}) is independent of $\phi$ and $t$,  the covariant
components $p_{\phi}$ and $p_{t}$ of a particle four--momentum are
conserved along the   geodesics, therefore\footnote{We adopt the
geometrical  units $c=1=G$ and  the $(-,+,+,+)$ signature, Greek indices run in $\{0,1,2,3\}$.  The   four-velocity  satisfy $u^a u_a=-1$. The radius $r$ has unit of
mass $[M]$, and the angular momentum  units of $[M]^2$, the velocities  $[u^t]=[u^r]=1$
and $[u^{\varphi}]=[u^{\theta}]=[M]^{-1}$ with $[u^{\varphi}/u^{t}]=[M]^{-1}$ and
$[u_{\varphi}/u_{t}]=[M]$. For the seek of convenience, we always consider the
dimensionless  energy and effective potential $[V_{eff}]=1$ and an angular momentum per
unit of mass $[L]/[M]=[M]$.}
the quantities
\be\label{Eq:after}
{E} \equiv -g_{\alpha \beta}\xi_{t}^{\alpha} p^{\beta},\quad L \equiv
g_{\alpha \beta}\xi_{\phi}^{\alpha}p^{\beta}\ ,
\ee
are  constants of motion, where  $\xi_{t}=\partial_{t} $  is
the Killing field representing the stationarity of the Kerr geometry and  $\xi_{\phi}=\partial_{\phi} $
is the
rotational Killing field.
The constant $E$, for
timelike geodesics, may be interpreted as representing the total energy of the test particle
 coming from radial infinity, as measured  by  a static observer at infinity, and  $L$  as the axial component of the angular momentum  of the particle.
 The
Kerr metric (\ref{alai}) is also invariant under the application of any two different transformations: $x^\alpha\rightarrow-x^\alpha$
  for one of the coordinates $(t,\phi)$, or the metric parameter $a$ and   the    test particle dynamics is invariant under the mutual transformation of the parameters
$(a,L)\rightarrow(-a,-L)$. Thus  one can  limit the  analysis of the test particle circular motion to the case of  positive values of $a$
for corotating  $(L>0)$ and counterrotating   $(L<0)$ orbits with respect to the black hole.

We focus  here on the case of
 a one-species particle perfect  fluid (simple fluid),  described by  the  energy momentum tensor
\be\label{E:Tm}
T_{\alpha \beta}=(\varrho +p) u_{\alpha} u_{\beta}+\  p g_{\alpha \beta},
\ee
where $\varrho$ and $p$ are  the total energy density and
pressure, respectively, as measured by an observer moving with the fluid whose four-velocity $u^{\alpha}$  is
a timelike flow vector field.  For the
symmetries of the problem, we always assume $\partial_t \mathbf{Q}=0$ and
$\partial_{\varphi} \mathbf{Q}=0$, being $\mathbf{Q}$ a generic spacetime tensor.
The  fluid dynamics  is described by the \emph{continuity  equation} and the \emph{Euler equation} respectively:
\bea\label{E:1a0}
u^\alpha\nabla_\alpha\varrho+(p+\varrho)\nabla^\alpha u_\alpha=0\, ,\quad
(p+\varrho)u^\alpha \nabla_\alpha u^\gamma+ \ h^{\beta\gamma}\nabla_\beta p=0\, ,
\eea
where the projection tensor $h_{\alpha \beta}=g_{\alpha \beta}+ u_\alpha u_\beta$ and $\nabla_\alpha g_{\beta\gamma}=0$ \citep{MTW,Pugliese:2011aa,Pugtot}.
 We investigate  the  fluid toroidal  configurations centered on  the  plane $\theta=\pi/2$, and  defined by the constraint
$u^r=0$. No
motion is assumed in the $\theta$ angular direction, which means $u^{\theta}=0$.
We assume moreover a barotropic equation of state $p=p(\varrho)$. The  continuity equation  in Eq.\il(\ref{E:1a0}) 
 %
is  identically satisfied as consequence of the conditions.
From the Euler  equation (\ref{E:1a0}) one can obtain
\be\label{Eq:scond-d}
\frac{\partial_{\mu}p}{\varrho+p}=-{\partial_{\mu }W}+\frac{\Omega \partial_{\mu}\ell}{1-\Omega \ell},\quad \ell\equiv \frac{L}{E}
,\quad W\equiv\ln V_{eff}(\ell),\quad V_{eff}(\ell)=u_t= \pm\sqrt{\frac{g_{\phi t}^2-g_{tt} g_{\phi \phi}}{g_{\phi \phi}+2 \ell g_{\phi t} +\ell^2g_{tt}}},
\ee
%
the function $W$ is Paczynski-Wiita  (P-W) potential,  $\Omega$ is the relativistic angular frequency of the fluid relative to the distant observer, and  $V_{eff}(\ell)$ provides an effective potential for the fluid, assumed here  to be  characterized by a  conserved and constant specific angular momentum $\ell$  (see also \citep{Lei:2008ui,Abramowicz:2008bk}).
The procedure adopted  in the present article
borrows from the  Boyer theory on the equipressure surfaces applied to a  P-D  torus:
  the Boyer surfaces are given by the surfaces of constant pressure  or\footnote{{More generally $\Sigma_{\mathbf{Q}}$ is the  surface $\mathbf{Q}=$constant for any quantity or set of quantities $\mathbf{Q}$.}}  $\Sigma_{i}=$constant for \(i\in(p,\rho, \ell, \Omega) \), \citep{Boy:1965:PCPS:,Raine}, where the angular frequency  is indeed $\Omega=\Omega(\ell)$ and $\Sigma_i=\Sigma_{j}$ for \({i, j}\in(p,\rho, \ell, \Omega) \). The toroidal surfaces  are the equipotential surfaces of the effective potential  $V_{eff}(\ell) $, considered  as function of $r$,  solutions $ \ln(V_{eff})=\rm{c}=\rm{constant}$ or $V_{eff}=K=$constant.

The model is regulated by the couple of parameters $\mathbf{p}\equiv (\ell,K)$ which together uniquely identify each Boyer surface.
Similarly to the case of the test particle dynamics,
the  function  $V_{eff}(\ell)$  in Eq.\il(\ref{Eq:scond-d})  is invariant under the mutual transformation of  the parameters
$(a,\ell)\rightarrow(-a,-\ell)$, therefore we can limit the analysis to  positive values of $a>0$,
for \emph{corotating}  $(\ell>0)$ and \emph{counterrotating}   $(\ell<0)$ fluids and    we adopt the notation $(\pm)$  for  counterrotating or corotating matter  respectively.
The P-D,  could be defined as a  ``geometric''  model of  thick accretion disks as many  features   of this disk model  are mostly determined by the  geometric   properties of the spacetime background  as given by some  notable  radii  $\mathcal{R}\equiv\{r_{mso}^{\pm},r_{mbo}^{\pm},r_{\gamma}^{\pm}\}$, regulating  the particle dynamics:  the \emph{marginally circular orbit} for timelike particles  $r_{\gamma}^{\pm}$,  the \emph{marginally  bounded orbit}  is $r_{mbo}^{\pm}$, and the \emph{marginally stable circular orbit}  $r_{mso}^{\pm}$ \citep{Pu:Kerr, MTW,chandra42,Pugliese:2013zma}.  Timelike  circular orbits  can fill  the spacetime region $r>r_{\gamma}^{\pm}$, stable orbits are in $r>r_{mso}^{\pm}$ for counterrotating and corotating particles respectively, and  $E_{\pm}(r_{mbo}^{\pm})=1$.
Given $r_i\in \mathcal{R}$,  we adopt  the  notation for any function $\mathbf{Q}(r):\;\mathbf{Q}_i\equiv\mathbf{Q}(r_i)$, therefore for example $\ell_{mso}^+\equiv\ell_+(r_{mso}^+)$, and more generally given the radius  $r_{\ast}$ and the function  $\mathbf{Q}(r)$,  there is $\mathbf{Q}_{\ast}\equiv\mathbf{Q}(r_{\ast})$.

The  closed Boyer surfaces   cross the equatorial plane $\theta=\pi/2$, in  $y_i=y_i(a;\mathbf{p})$,  $i\in\{1,2,3\}$,   where $y_2<y_3<y_1$, as  shown in Figs.\il(\ref{Fig:Quanumd})\placefigure{Fig:Quanumd}.
We  consider the orbital region $\Delta r_{crit}\equiv[r_{Max}, r_{min}]$, whose boundaries correspond to the  maximum and minimum points of the effective potential respectively.
The inner edge  of the Boyer surface  must be   $y_3\in\Delta r_{crit}$,
the outer edge  is $y_1>r_{min}$.   A further  matter configuration  (with solution $y_2$) closest to the black hole is  at $r<r_{max}$. In the closed cusped  configuration $C_x$, there  is $y_3=y_2$

The centers $r_{cent}$  of the closed configurations $C_{\pm}$ are located at the minimum points  $r_{min}>r_{mso}^{\pm}$  of the effective potential, where the hydrostatic pressure is maximum. The toroidal surfaces are characterized by $K_{\pm}\in [K^{\pm}_{min}, K^{\pm}_{Max}[ \subset]K_{mso}^{\pm},1[$ and  momentum $\ell_+<\ell_{mso}^+<0$ or $\ell_->\ell_{mso}^->0$ respectively. 
The limiting case of $K_{\pm}=K_{min}^{\pm}$ corresponds to a one-dimensional ring of matter  located in  $r_{min}^{\pm}$.
The maximum points of the effective potential $r_{Max}$  are   minimum points of the hydrostatic pressure or, P-W  points  of  gravitational and hydrostatic instability.
From these points  an  accretion  overflow of matter from the  closed, cusped  configurations in   $C^{\pm}_x$ (see Fig.\il(\ref{Figs:CrystalPl})) can occur from the instability point  $r^{\pm}_x\equiv r_{Max}\in]r_{mbo}^{\pm},r_{mso}^{\pm}[$ towards the attractor, if $K_{Max}\in]K_{mso}^{\pm},1[$  with proper angular momentum $\ell\in]\ell_{mbo}^+,\ell_{mso}^+[\cup]\ell_{mso}^-,\ell_{mbo}^-[$, respectively,  for counterrotating or corotating matter. Otherwise  there can be  funnels of  material in jets from an open configuration   $O^{\pm}_x$ with   $K^{\pm}_{Max}\geq1$, launched from the point $r^{\pm}_{J}\equiv r_{Max}\in]r_{\gamma}^{\pm},r_{{mbo}}^{\pm}]$ with proper angular momentum $\ell\in ]\ell_{\gamma}^+,\ell_{mbo}^+[\cup]\ell_{mbo}^-,
  \ell_{\gamma}^-[$. Equilibrium configurations, with topology $C$, are possible for $\pm\ell_{\mp}>\pm\ell_{mso}^{\mp}$ centered in  $r>r_{mso}^{\mp}$ respectively, but no  maxima of the effective potential are for $\pm\ell_{\mp}>\ell_{\gamma}^{\pm}$, and therefore, for those angular momenta only equilibrium configurations are possible.

Each  couple $\mathbf{p}$ identifies uniquely  a Boyer surface: as discussed in \citep{Pugtot}
by a variation of the  $\mathbf{p}$ parameters, we  can consider  a matrix  $\mathfrak{B}_{\mathbf{p}}$  of corotating and  counterrotating configurations on  $\Sigma_{\mathbf{p}}\equiv\Sigma_{K}\otimes\Sigma_{\ell}$;    keeping one parameter of the couple  $\mathbf{p}\equiv(\ell,K)$ constant and   changing the other one in the set of values  for the formation of a Boyer surface,
 we could consider the ``projection''  $\mathfrak{B}_{\mathbf{p}_\mathfrak{j}}\equiv\mathfrak{B}_{\mathbf{p}}/\Sigma_{\mathbf{p}_\mathfrak{i}}$ of $\mathfrak{B}_{\mathbf{p}}$  on the constant surface $\Sigma_{\mathbf{p}_\mathfrak{i}}$, i.e. the  sequence, array or column of elements of  $\mathfrak{B}_{\mathbf{p}_\mathfrak{j}}$ on  the  constant $\mathbf{p}_\mathfrak{i}$ surfaces, and assigning     $\mathbf{p}_\mathfrak{j}$ ($\mathbf{p}_\mathfrak{i}$ can be either $\ell$ or $K$) as a family  parameter.
In this work we focus on the elements of $\mathfrak{B}_{\mathbf{p}}$  as defined on a  $\Sigma_t$ surface,  and we investigate   the possible multiple-configurations    on the  fixed $\Sigma_t$.
Since the  toroidal configuration  can be corotating  $\ell a>0 $ or counterrotating   $\ell a<0$, with respect to the black hole  $a>0$, then assuming    several  toroidal configurations,  say the couple $(C_a, C_b)$, with proper angular momentum $(\ell_a, \ell_b)$  orbiting  in   the equatorial plane of a given Kerr \textbf{BH},   we need to introduce   the concept  of
 \emph{$\ell$corotating} disks,  defined by  the condition $\ell_{a}\ell_{b}>0$, and \emph{$\ell$counterrotating} disks  by the relations   $\ell_{a}\ell_{b}<0$,  the two $\ell$corotating tori  can be both corotating $\ell a>0$ or counterrotating  $\ell a<0$ with respect to the central attractor.

\section{Introduction of the  ringed disk model  and definitions}\label{Sec:intro}
We first briefly  introduce  the ringed accretion disk model (or macro-configuration of tori), providing  some  fundamental definitions and explicit  formalism that will be used in  the  following discussion. It is convenient  to introduce the notation of  $C$ more specifically for the cross sections of the closed Boyer surfaces  and generally {for boundaries  $\partial C$}, the contour of the surface in the  equatorial section, as shown in  Fig.\il(\ref{Fig:Quanumd}). Therefore  the cross section of each  torus has boundary  $\partial C$  (barotropic surfaces) and the fluid fills  $C$.  For simplicity of notation, $C_i$ will refer generally to the Boyer surfaces  meaning that it could be   possibly in its unstable phase, we could consider then  the cusped topology $ C^x_i $. In any case, when it will be relevant, we will specify this possibility. The critical phase of an $ O_x $ topology, will be considered a part, as a special case.

We question the existence of  ``structured'' toroidal accretion disks constituted by several tori orbiting and centered around a given  Kerr attractor.   The aim is  to provide a  model for the  matter \emph{structure} of the multiple  thick configurations defined as  the union\footnote{We  intend the set formed by a number $n$ of tori orbiting one Kerr \textbf{BH}  attractor, the failure of rigor of this definition is hereby replaced by the intuitive immediacy, as  it is shown in Figs.\il(\ref{Fig:Torc}). The  Latin subscript in $ C_i$  usually indicates, when otherwise specified the $ i- $ {sub-configuration} of decomposition, ordered according to the scheme described in the text, see also Figs.\il(\ref{Fig:Quanumd}
).}  $\mathbf{C}^n\equiv \bigcup^n_{i=1} C_i$  of   closed, {cusped or not,  torus  \emph{separated} \emph{sub-configurations} $C_i\bigcap C_j=0$ and $\partial C_i\bigcap \partial C_j=\{\emptyset,y_1, y_3\}$} of P-D thick accretion disks, whose intersection is the null set or at last the inner or outer edge of the configurations as specified below (this analysis  does not take into account the lobe disk associate to the solution $y_2$), see Fig.\il(\ref{Fig:Quanumd}) and Fig.\il(\ref{Figs:CrystalPl}). The lower-case Latin indices   $i, j,...a, b...$, either superscript or subscript, are indices  of {sub-configurations} and indicate  (uniquely in this model) each toroidal component (\emph{ring}) of the ringed structure. For a double configuration, formed by a couple of rings, in general with $\mathbf{\mathbf{(o)}}$ we refer here to  any quantity related  to  the external (outer) configurations  of the couple and $\mathbf{(i)}$ to the more internal, the inner  one with respect to the attractor.

We define \emph{order} of the macro-configuration $\mathbf{C}^n$ the number $n\in[1,\infty]$ of the sub-configurations of $\mathbf{C}^n$, and the  set of $\{C_i\}_{i=1}^{n}$  as  \emph{decomposition} of the $\mathbf{C}^n$ macro-configuration  of the $n$-order, the decomposition will be  considered an \emph{ordered  sequence} of configurations in the sense specified below. In Fig.\il(\ref{Fig:Torc})  a pictorial representation of a  $\mathbf{C}^3$ ringed disk of order  $n=3$ is shown. Then, even if
each sub-configuration is characterized  here by a well defined and constant specific angular momentum $\ell_i=$constant, as discussed in \citep{Lei:2008ui,Abramowicz:2008bk},   this assumption can be relaxed in  more general P-D models.
The macro-structure inherits many of the symmetries of each toroidal sub-configuration of its decomposition, for example, the symmetry for reflection on the equatorial plane:
we  specify  that  we are  considering $r^i_ {min}\in\Sigma_{\theta=\pi/2}$ where  $r^i_{min}=r_{cent}^i$ coincides with the center of the $C_i$ sub-configuration   located in the equatorial plane of the disk,   which always coincides with the symmetry plane of the attractor. All the rings are thus assumed coplanar.  Therefore we limit  the study to the  Boyer configuration sections  on the equatorial plane.

 The generalization of this set up to other situations in which the rings are not  coplanar  or with  unique   plane symmetry  (the symmetry plane of the disk)  different from  the  symmetry plane  of the attractor is also possible.{
In this article we proceed  to  characterize  possible  decompositions of a $n$-order ringed disk    fixing a certain set of characteristics $\mathbf{C}^n$.
The decomposition is  in fact uniquely determined by the set $\mathcal{S}\equiv \{n, \mathbf{p}_{\mathbf{C}^n}\} $, where $\mathbf{p}_{\mathbf{C}^n}\equiv\{\mathbf{p}_i\}_{i=1}^n$, $n$ is the decomposition order and $i$ is as usual the index of sub-configuration, i.e. each $\mathbf{p}_i$ uniquely identifies one element, ring, of a ringed disk, its morphology as  its location on the equatorial plane, as discussed  in Sec.\il(\ref{Sec:models})\footnote{{This is indeed,  accordingly, the maximum available information as it is extractable from the model of the single ring   introduced  in Sec.\il(\ref{Sec:models}), i.e., the maximum number of characteristics of each individual ring, element of the decomposition, which  constitutes a ringed disk. A macro-configuration $\mathbf{C}^n$  at a   fixed   order  $n$ \emph{only}, in general   has not a unique  decomposition. In fact  there can be more then one set of different rings, $\{C_i\}_{i=1}^n$ and $\{\tilde{C}_j\}_j$ with different sequences of parameters $\mathbf{p}_{\mathbf{C}^n}\equiv\{\mathbf{p}_i\}_{i=1}^n \neq\{\mathbf{\tilde{p}}_i\}_{i=1}^n\equiv\mathbf{\tilde{p}}_{\mathbf{C}^n}$.
One could ask how many  decompositions are possible  for a disk (or  in how many ways one can decompose a ring accretion disk) with a proper fixed set of  characteristics. Viceversa what is, if it exists, the minimum number of morphological characteristics (such as the torus thickness, the disk elongation etc) or dynamics (the angular momentum distribution in its decomposition) to be  specified, in order  to uniquely determine the decomposition $\mathbf{C}^n$ (or equivalently $\mathbf{p}_{\mathbf{C}^n}$)  of the  disk.}}.}

Given a decomposition of $\mathbf{C}^n$, a \emph{sequence} or also equivalently {\emph{sub-sequence}}   of rings, is a set  $\{C_i\}$ of $n_*\leq n$ elements of its decomposition. In particular  it is important to consider the  $\ell$corotating sequences, formed by all the $\ell$corotating rings    of the decompositions  $\ell_i\ell_j>0$,  and  we  call $\ell$counterrotating  sequences  of the decomposition of  $\mathbf{C}^n$ the two $\ell$corotating sequences of this, formed by the rings $\{C_{i_-}\}_{1_-}^{n_-}$ and  $\{C_{i_+}\}_{1_+}^{n_+}$ with    $\ell_{i_{-}}>0$ and $ \ell_{i_{+}}<0$  respectively. Then it is $n=n_++n_-$.

 In Fig.\il(\ref{Figs:CrystalPl}) we   show  different  ringed  disks of  the order  $n=4$ made by  only   $\ell$corotating sequence of counterrotating rings, while  Fig.\il(\ref{Figs:RightPolstex}) shows different ringed disks  of the  order  $n=4$  with  $\ell$counterrotating sub-sequences.

We discuss  the conditions  for the existence    of a    $\mathbf{C}^n$ configuration in terms of the  $\mathcal{S}$ set.  We prove that the set $\mathcal{S}$ consists of mutually dependent elements, in particular  we will provide    constraints (specifically an  upper limit) on the configuration order $n$ depending on  the values  of $\mathbf{p}_{\mathbf{C}^n}$. The relative rotation of the fluids belonging to different rings of the macro-structure has a decisive role in determining the ringed disk. In fact we will show that   the conditions imposed on the $\ell$counterrotating sequences   of the decomposition  are particularly restrictive.
On the other side, it is well  known that  the effective  potential (\ref{Eq:scond-d}) for a Kerr-\textbf{BH} background  does  not have any   double  minimum or maximum.
As a consequence of this  fact, the set of  sub-configurations $C_i\subset \mathbf{C}^n$, cannot be generated by an unique fixed   constant $\mathbf{p}_i=\mathbf{p}_j$, from the same effective potential $V_{eff}(\ell,a,r)$ inferred  by the Euler equation. We   consider in fact the disjoint union of the sub-configurations, or at last  the  union with a couple of  double  points only for each contour of one ring (i.e. the conditions $y^i_3=y^{i-1}_1$ and $y^{i+1}_3=y^{i}_1$--see below).

 Each configuration  $C_i $ is thus a solution of an Euler equation for a  unique constant  $\mathbf{p}_i$. The tori of the decomposition  are linked  by some  boundary conditions, raised from the fact that they are not isolated  elements but  part of a macro-structure $\mathbf{C}^n$--we will discuss in details this point in the next Sections.  These boundary conditions  determine the macro-structure  as one (structured) body  along with  a proper  internal dynamics. In other words, although the tori are  not tied by the same  potential $V_{eff}(\mathbf{p})$,  their  presence as sub-configuration  in $\mathbf{C}^n$ imposes some limits,  constraining their angular momentum or thickness, their locations, number  and their equilibrium.
 They  may even lead to  some instability phenomena for $\mathbf{C}^n$, or  feeding of  one disk from another.

\medskip

More precisely, in a more \emph{general} situation, one could generalize the definition of macro-configuration  considering a toroidal configuration made up by  a set of  $n$ tori, obtained by the  Euler equation and  associated to   $n$ generic parameters $\mathbf{p}_{\mathbf{C}^n}=\{\mathbf{p}_i\}_{i=1}^n$, that   can then  be intertwined, looped or separated at less than a double point  or
\begin{description}
\item[\namedlabel{Intertwined}{\emph{Intertwined}}] tori are
${C}_{i+1}\cap {C}_{i}\neq\emptyset$ \emph{and} $\partial {C}_{i+1}\cap \partial {C}_{i}\neq\emptyset$.
  We say   that configurations made up  by  intertwined   tori  have an \emph{ ordered decomposition} meaning that it is possible to order the sequence of disks  $C_i$ or   ${C}_{i}<{C}_{i+1}$  {if  and only if } ${r}_{cent}^i<r_{cent}^{i+1}$ for  $i\in\{1,...,n\}$.
\item[\namedlabel{Ringed}{\emph{Looped}}] tori
are defined by the condition  ${C}_{i}\subset {C}_{i+1}$ in the equatorial plane,  i.e a no-crossing disk loop occurs. Turning unstable, even if formed,  they could possibly turn into  a single, energetically more favorable  configuration, for  the fluid fills the entire
contour  $(\partial C_i)$ at equal $\ell$ to ensure the existence and stability
of the  orbiting matter. In other words, a loop of  tori is defined as the macro-structure $\mathbf{C}^n\equiv\bigcup_{i=1}^{n} C_i:\;C_{i}\subset C_{i+1}$ and $\partial C_{i}\bigcap \partial C_{i+1}=\emptyset$ $\forall i\in\{1,n-1\}$, where $n$  is  the  order of the loop. These configurations  may have   some relevance in  models with a non-constant  angular momentum distribution along  each disk  \citep{Lei:2008ui}.  We say that the macro-configuration $\mathbf{C}^n$  made up by    looped tori   admits an \emph{ordered decomposition}, if the sequence of sub-configurations $C_i$ satisfies  ${C}_{i}\subset{C}_{i+1}$ and $\partial{C}_{i}\subset{C}_{i+1}$ for  $i\in\{1,...,n\}$.
\begin{figure}[h!]
\centering
\begin{tabular}{cc}
\includegraphics[scale=.36]{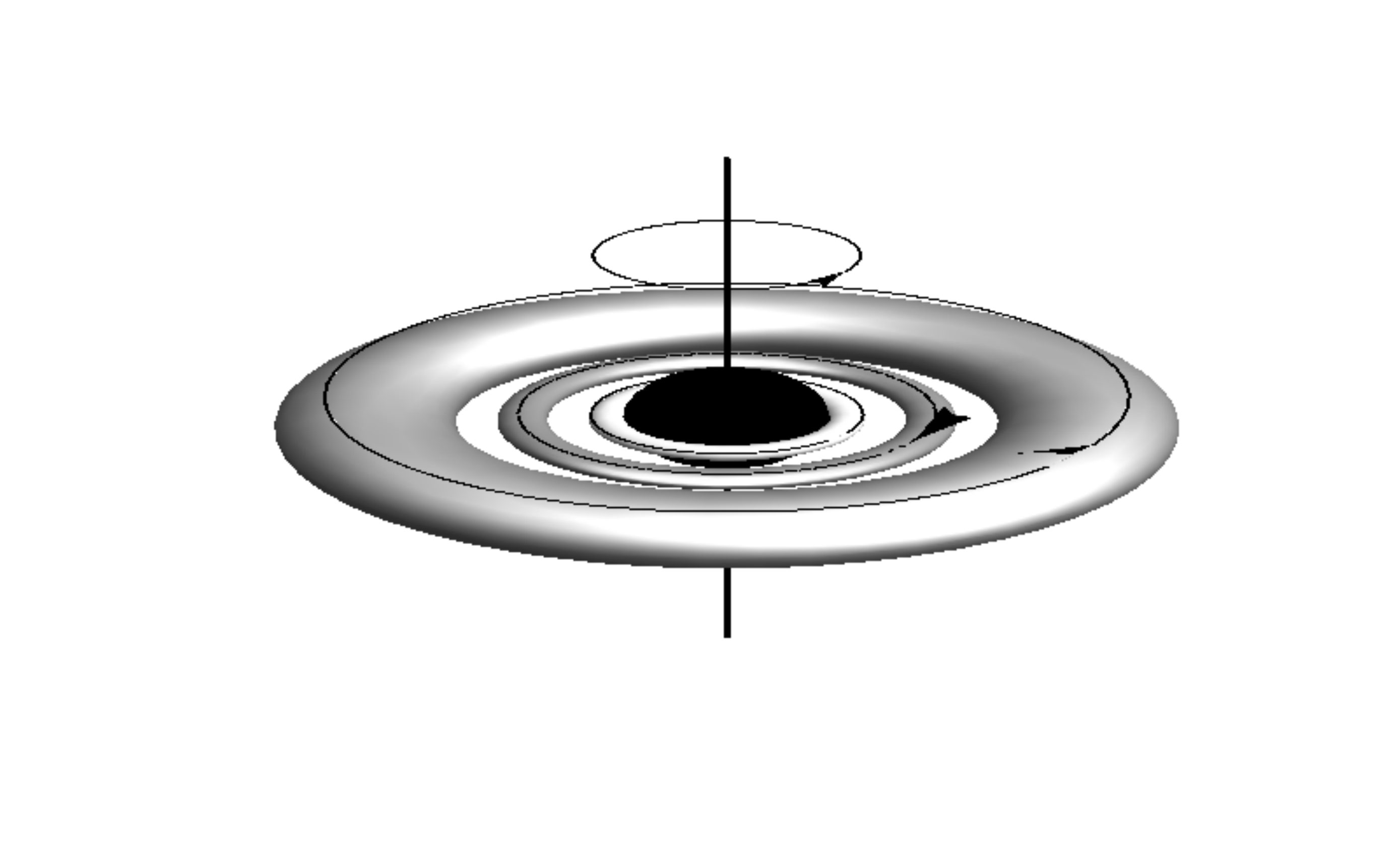}
\includegraphics[scale=.3]{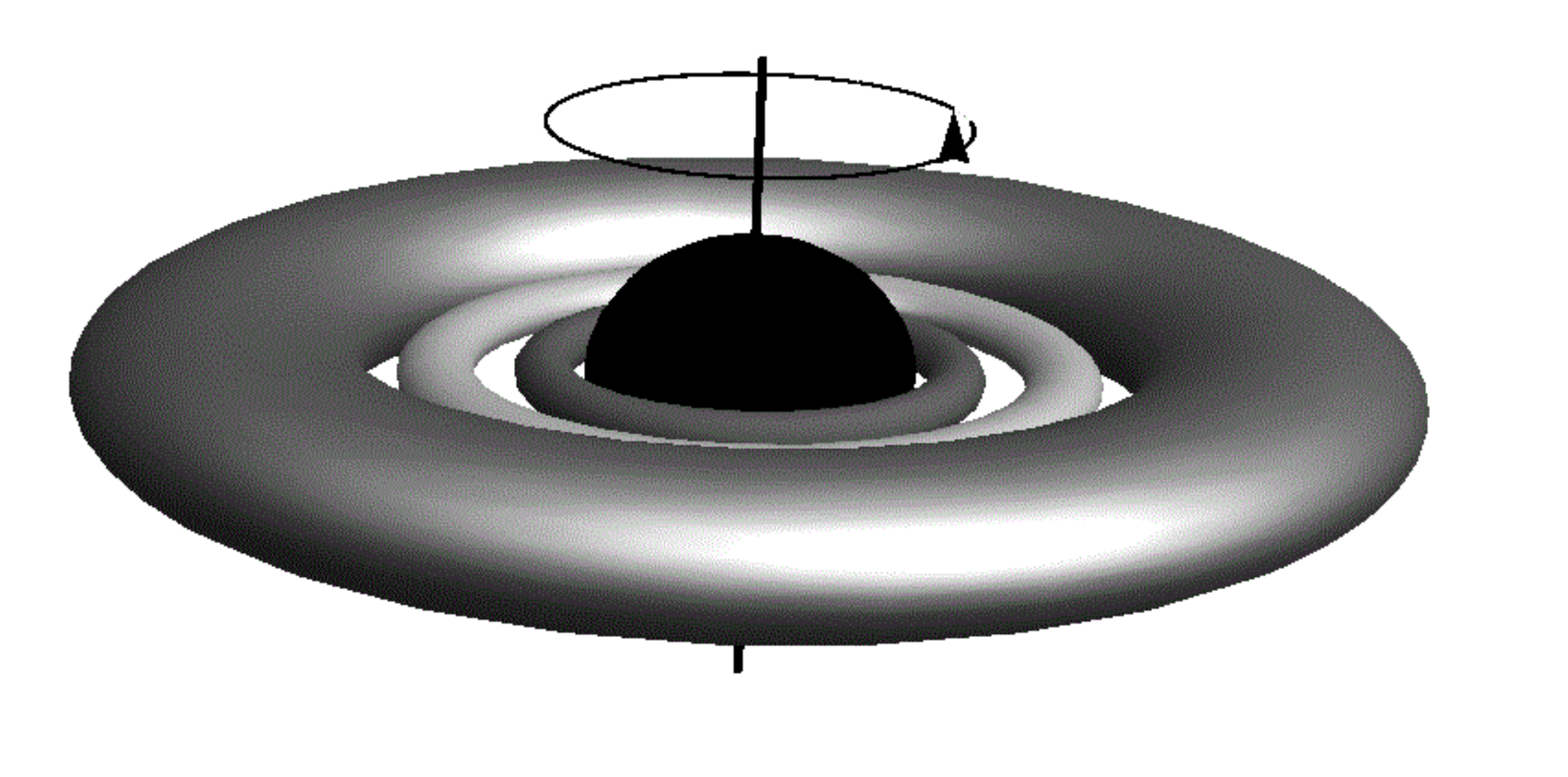}
\end{tabular}
\caption[font={footnotesize,it}]{\footnotesize{ Pictorial representation of a ringed accretion disk $\mathbf{C}^3$ of  order  $n=3$. The black region  is   $r<r_{+}$ where $r_+$ sets the outer horizon for a black hole attractor.}}\label{Fig:Torc}
\end{figure}
\begin{figure}[h!]
\centering
\begin{tabular}{c}
\includegraphics[scale=.4]{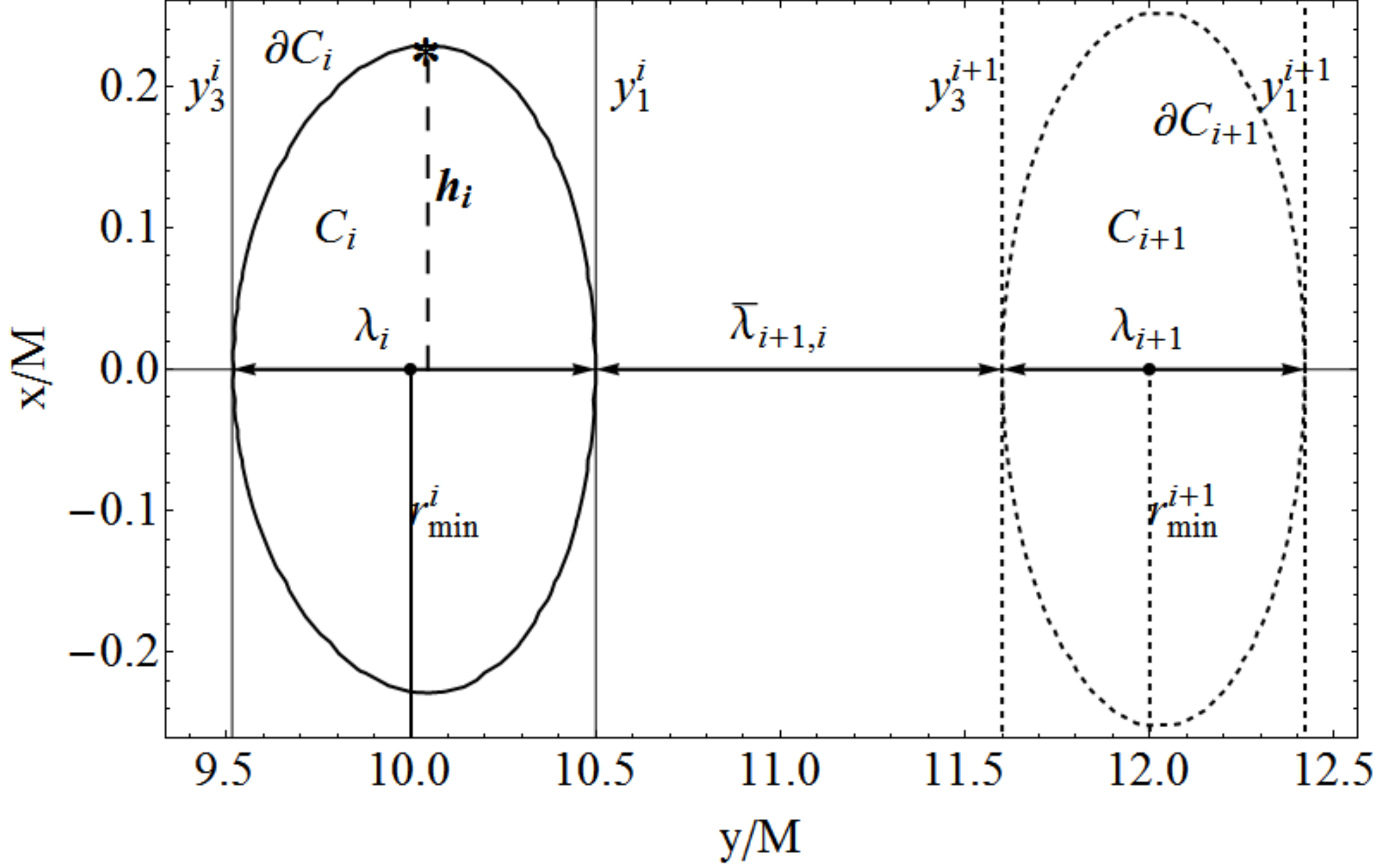}\\
\includegraphics[scale=.4]{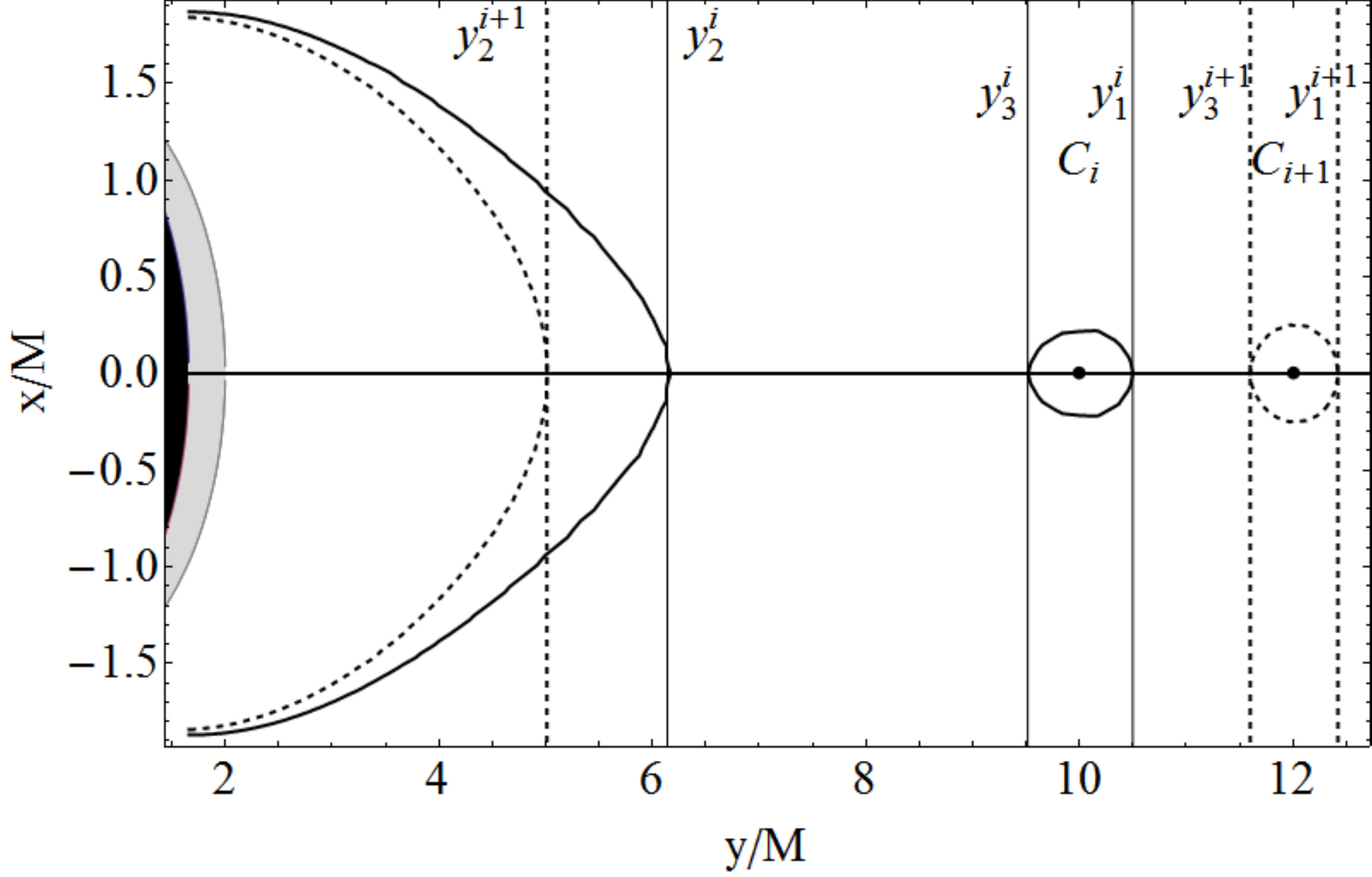}
\end{tabular}
\caption[font={footnotesize,it}]{\footnotesize{Cross sections on the equatorial plane of the consecutive   Boyer surfaces of two separated  rings $C_i<C_{i+1}$, with boundaries $\partial C_i$ and $\partial C_{i+1}$, the rings centers  $r_{min}=r_{cent}$ are signed with points and the lines $r_{min}=$constant. With $r/M=\sqrt{x^2+y^2}$ and  $(x,y)$ are Cartesian coordinates. The inner $y_3$ and outer $y_1$ edge are also signed, $\lambda_a=y_1^a-y_3^a$ for $a\in\{i,i+1\}$ are the rings elongation and $\lambda_{i+1,i}=y_3^{i+1}-y_1^i$ is the spacing among the rings. { $h_i\equiv x_{Max}^i$  is the  height associated to  the sub-configuration $C_i$, or   the maximum point  of the surface $\partial C_i$ } The $\ell$corotating  rings, rotate around a black hole  attractor with spin
$a=0.75M$ and specific angular momentum
$\ell_i=-4.2897$ and
$\ell_{i+1}=-4.41883$. Black region in $r<r_+$, and gray region in $r<r_{\epsilon}^+$, where $r_+$ is the spacetime outer horizon  while $r_{\epsilon}^+$ is the static limit.}}\label{Fig:Quanumd}
\end{figure}
 A particular kind of looped tori are the
\textbf{\emph{centered}}
tori defined as loops  of tori where $ \forall \{i,j\}\in\{1,...n\}\; r^{i}_{min}=r^{j}_{min}$, where $ r^{i}_{min}$ is  the disk center and  also the \emph{loop  center}.  We note that sub-configurations with   $\ell_{\mathbf{(i)}}=\ell_{\mathbf{(o)}}$  are (necessarily) looped (in particular we stress it has to be $\partial C_{i}\bigcap \partial C_{i+1}=\emptyset$ that means they cannot be interwined) \emph{and}  they are (necessarily) {centered} (i.e.  $ r^{\mathbf{(i)}}_{min}=r^{\mathbf{(o)}}_{min}$, $r_{min}$ being the disk center). 
\item[\namedlabel{Separated}{\emph{Separated}}]   tori  are defined,   for a $n$-order macro-structure $\mathbf{C}^n=\bigcup_1^n C_{i}$, according to the following condition
\be\label{Def:separated}
C_i\bigcap C_j=0\quad\mbox{ and}\quad \partial C_i\bigcap \partial C_j=\{\emptyset,y_1^i=y_3^j\}\quad\mbox{ where }\quad i<j.
\ee
For $n=2$, a double configuration, ${C}_{\mathbf{(i)}}\cap {C}_{\mathbf{(o)}}=\emptyset$ or those with  $y_1^{\mathbf{(i)}}=y_3^{\mathbf{(o)}} $ where  the outer edge of the inner rings coincides with the inner  edge of the outer ring. In other words for macro-configurations made by  separated tori, the penetration  of  a ring within another ring  is  not possible. However, as the condition $y_1^{\mathbf{(i)}}=y_3^{\mathbf{(o)}} $  can hold,   in a limit situation  the  collision of matter between the two surfaces at contact point $y_1^i=y_3^j$ could be possible. Ringed disks   $\mathbf{C}^n$, made by separated  tori,  have an \emph{ordered decomposition}, if the sequence of rings  $C_i$ satisfies   {${C}_{i}<{C}_j$  if and only if $r_{cent}^i\equiv r_{min}^{i}<r_{min}^j\equiv r_{cent}^j$ for  $i<j$, ${C}_i$ being the inner one, closest to attractor, with respect to ${C}_j$}. Within these definitions, the rings  $(C_i, C_{i+1})$ and $(C_{i-1}, C_{i})$ are \emph{consecutive} if  $C_{i-1}<C_i<C_{i+1}$. Figs.\il(\ref{Fig:Quanumd},\ref{Figs:CrystalPl}). Then one can clarify condition  (\ref{Def:separated}) by saying that  $\partial C_i\bigcap \partial C_j=\emptyset $ or it is $\partial C_i\bigcap \partial C_j=\{\emptyset,y_1^i=y_3^j\}$ if $j=i+1$.
\end{description}
see also \citep{Pugtot}.

In this work  we consider  only macro-configurations constituted by separated tori. We introduce also the following  definitions for the $\ell$counterrotating sub-sequences of a  decomposition of the  order $n=n_++n_-$, {of     \emph{isolated} $\ell$counterrotating sequences   if  $C_{n_-}<C_{1_+}$ or $C_{n_+}<C_{1_-}$  and \emph{mixed} $\ell$counterrotating sequences if $\exists\; {i_+}\in[1_+, n_+]:\; C_{1_-}<C_{i_+}<C_{n_-}$ or viceversa  $\exists\; {i_-}\in[1_-, n_-]:\; C_{1_+}<C_{i_-}<C_{n_+}$. Alternately,}
\begin{description}
\item[\namedlabel{(-)}{(i-a)}] \emph{isolated} $\ell$counterrotating sequences $\overbrace{\mathbf{C}}_{{s}}$: $[r_{min}^{1_-},r_{min}^{n_-}]\cap[r_{min}^{1_+},r_{min}^{n_+}]=0$, then the couple of rings $C_{n_{\pm}}$  are consecutive and we say   the $\ell$counterrotating  isolated sub-sequences  of order $n_{\pm}$ respectively are consecutive, Fig.\il(\ref{Figs:RightPolstex})-top.
\begin{figure}[h!]
\begin{center}
\begin{tabular}{cc}
 \includegraphics[scale=0.3]{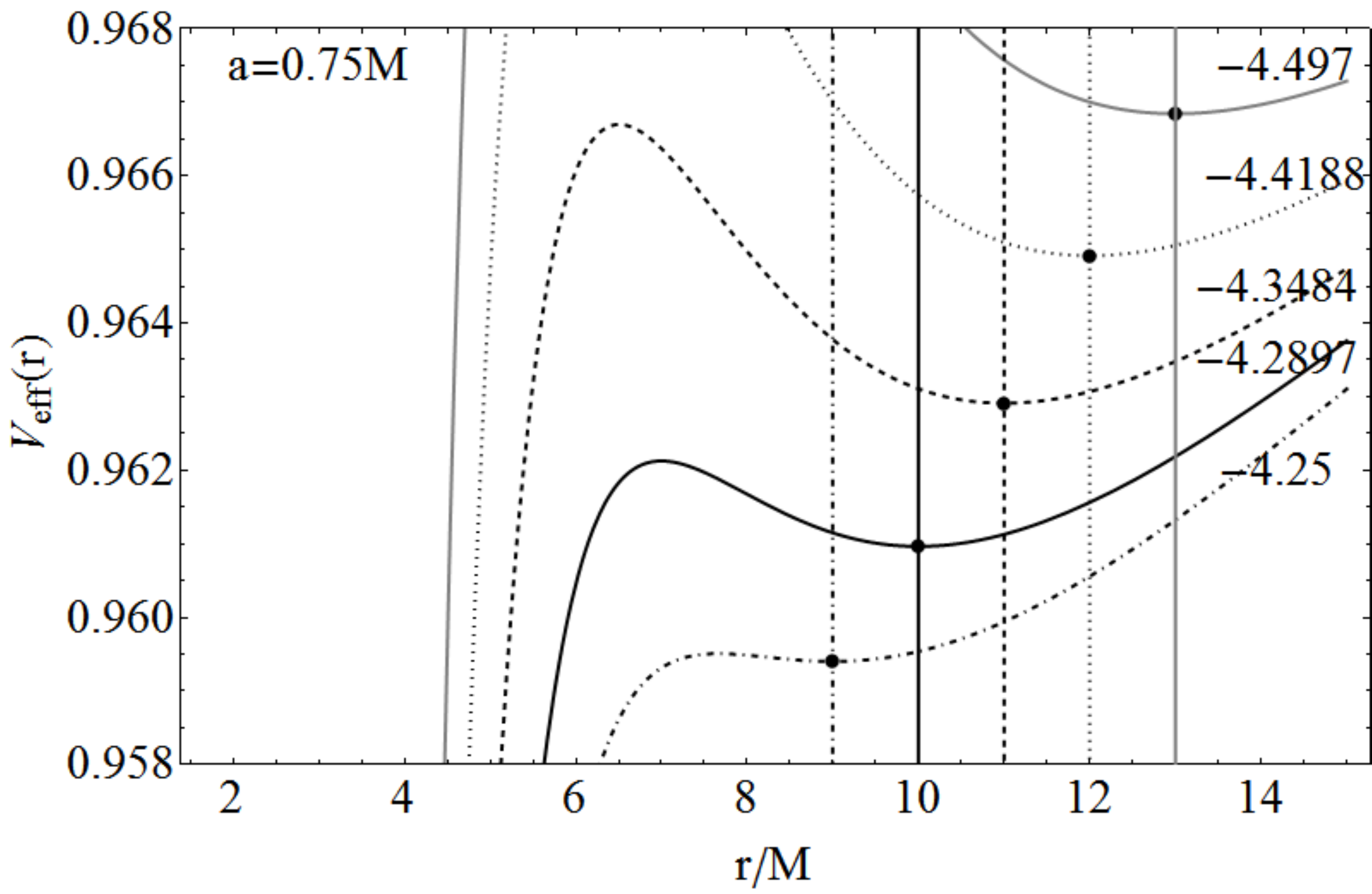}
 \includegraphics[scale=0.3]{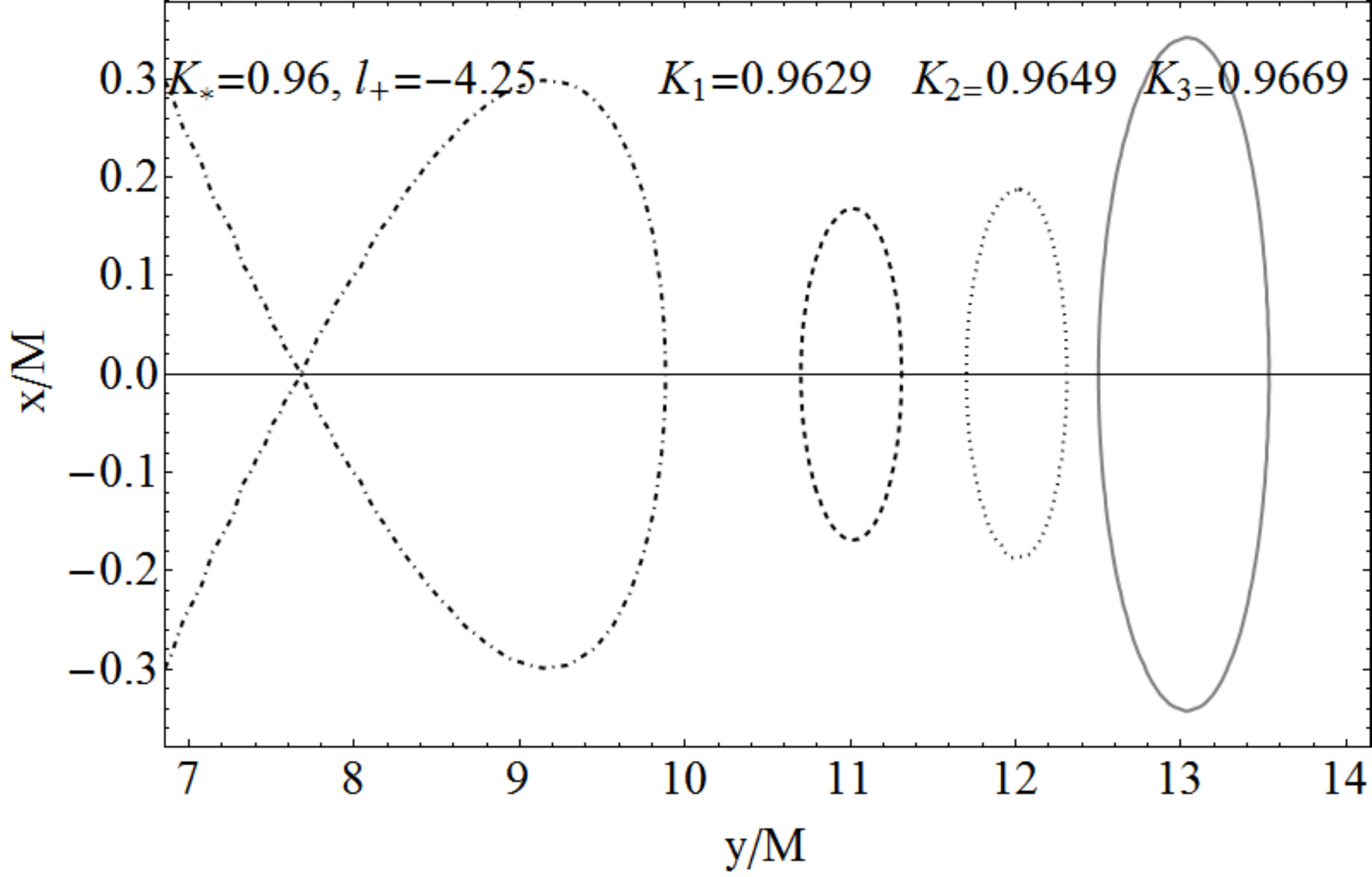}
 \\
 \includegraphics[scale=0.3]{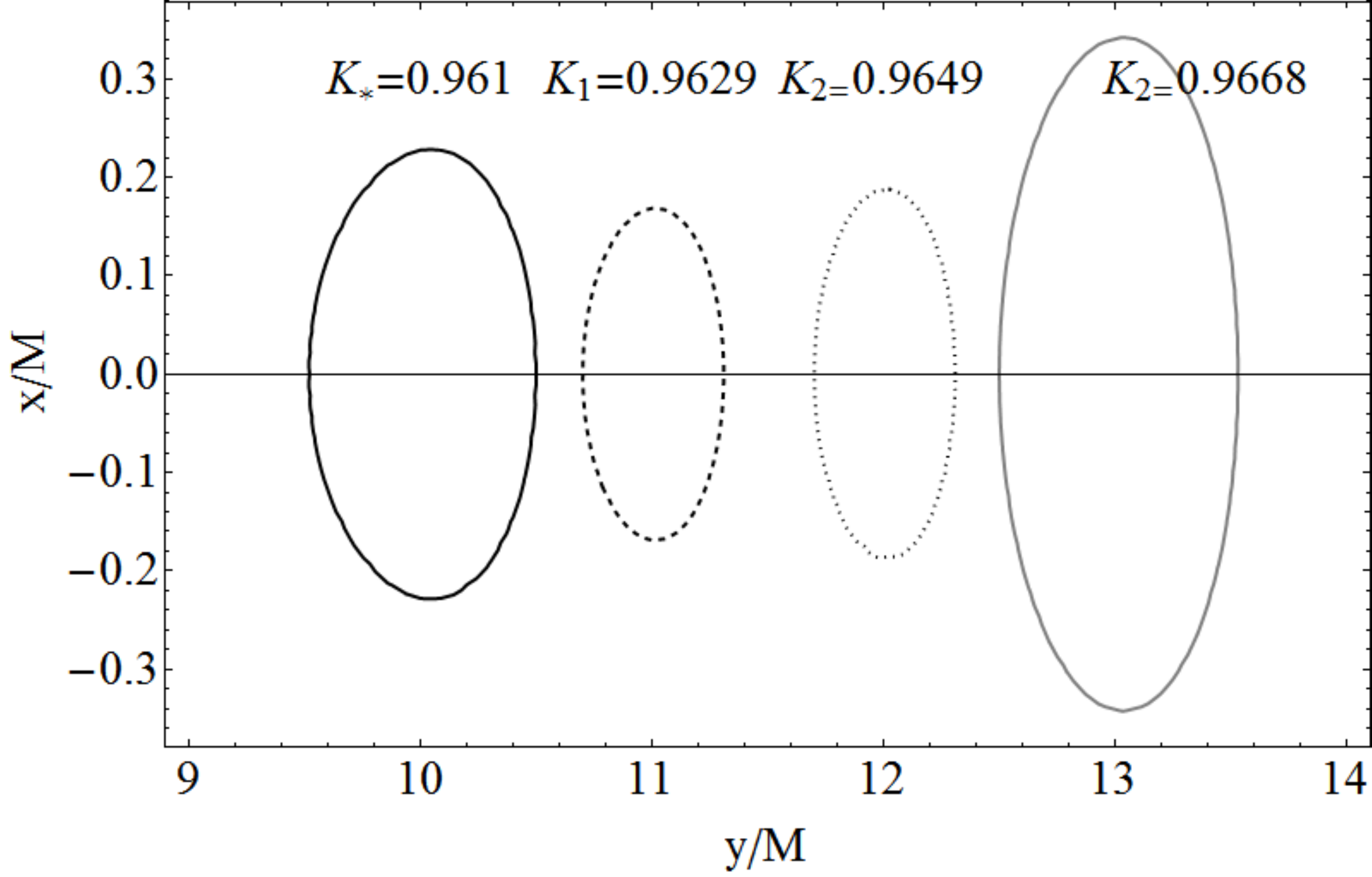}
 \includegraphics[scale=0.3]{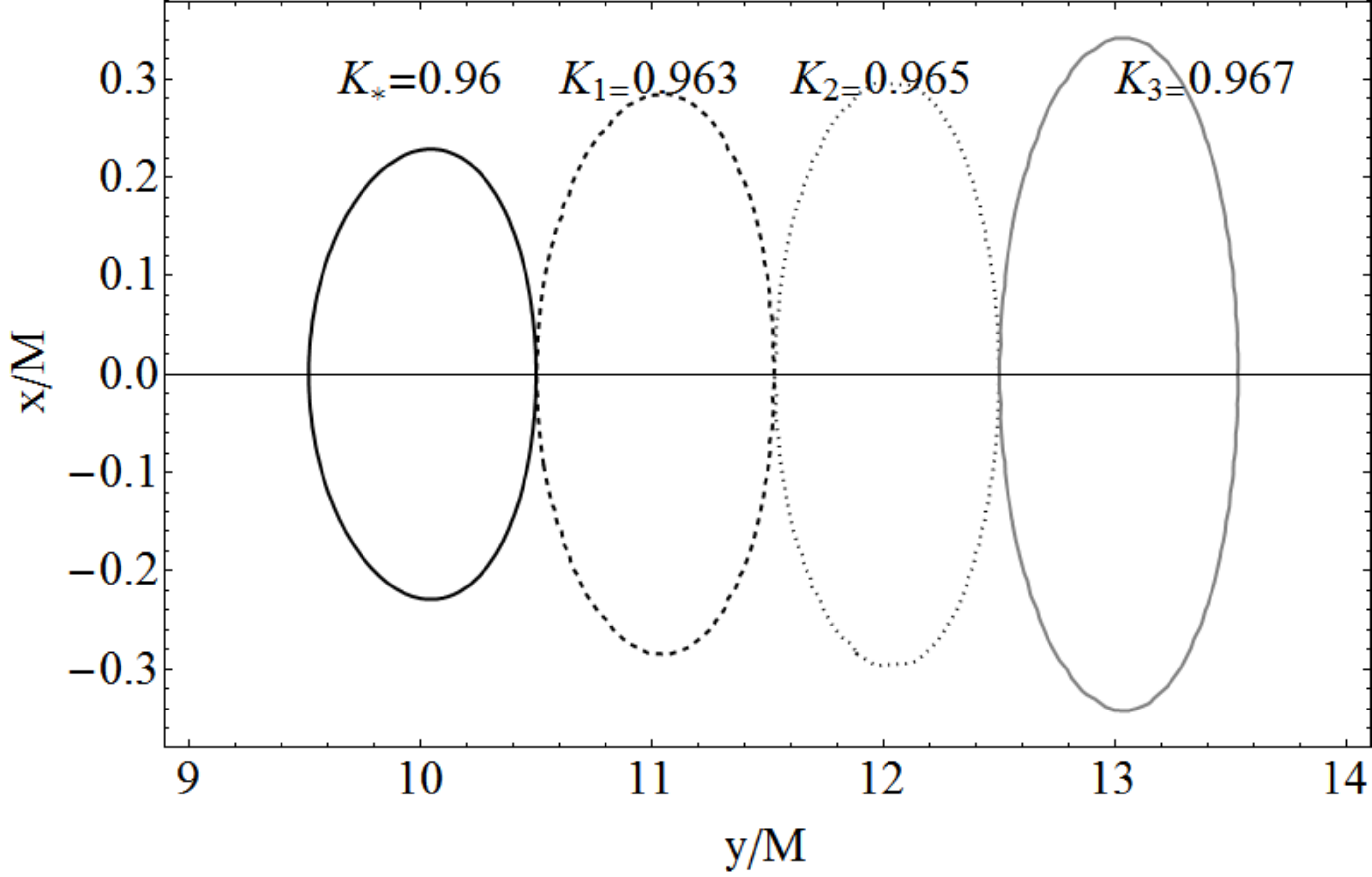}
\end{tabular}
\caption{Spacetime spin $a=0.75M$: $\ell$corotating sequences $\bar{\mathfrak{C}}_{1b}$-configurations ($\ell_i\ell_j>0$) of counterrotating disks  ($\ell_i a<0$ $\forall i j$). It follows the discussion in Sec.\il(\ref{Sec:rolel}),  Eq.\il(\ref{Eq:cononK}). Upper left panel shows the effective potential as a function of $r/M$
for different values of $\ell_+$  indicated close to the curves. The  minimum points  of the  potential are indicated by points and the lines  $r_{min}=$constant. The  other panels show  sections of  the decompositions associated to each specific angular momentum,
the values of $K_i$ for each Boyer configuration, are indicated in the  figures. Upper left panel: Ringed disk $\mathbf{C}_{x}^4$ in accretion.  Bottom left panel:  separated  ringed  $\mathbf{C}_{\odot}^4$ disk. The specific angular momenta have been set so as to have a distance between the centers $ \delta_{min}^{i+1,i} =M $ constant with the index configuration, then the
algorithm  for the choice of the sequence $\{K_i\}_{i=1}^n$ was fixed assuming $K_*=K_0= V_{eff}(\ell_0,r_{min}^0+1/2)$, and
$
K_{i}=V_{eff}(\ell_i, r_{min}^{i-1}+0.7)$.   Right-bottom panel: saturated ringed $\mathbf{C}_{\odot}^4$ disk of maximum rank $\mathfrak{r}_{Max}=3$. In the  sequences  $\{K_i\}_{i=1}^n$ the difference  $K^i_{min}-K_{i}\approx0$ are of the order of  $10^4$ c.a  but the elongation  $\lambda_i$ are not neglible.}\label{Figs:CrystalPl}
\end{center}
\end{figure}
\begin{figure}[h!]
\begin{center}
\begin{tabular}{cc}
 \includegraphics[scale=0.3]{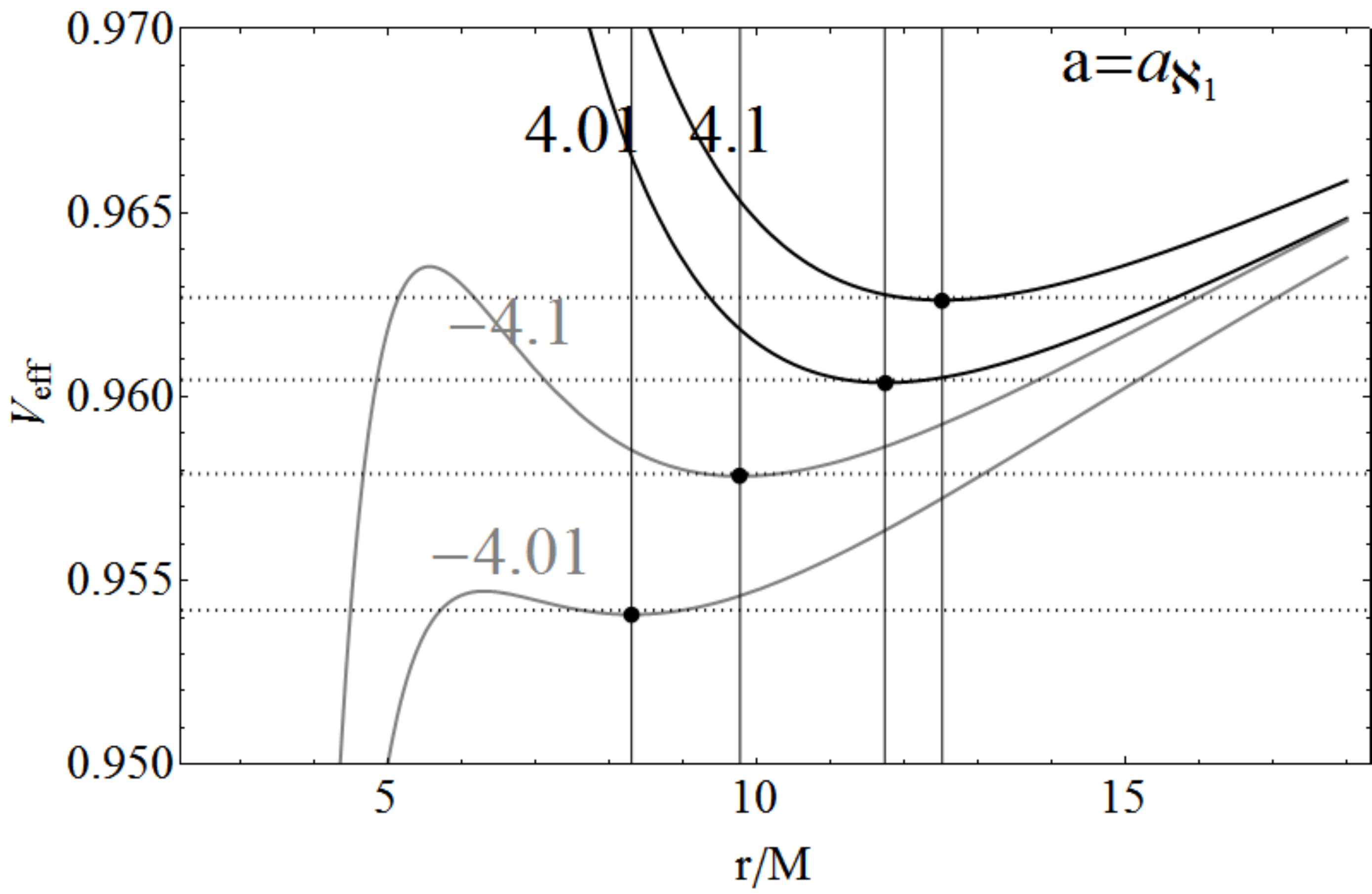}
 \includegraphics[scale=0.3]{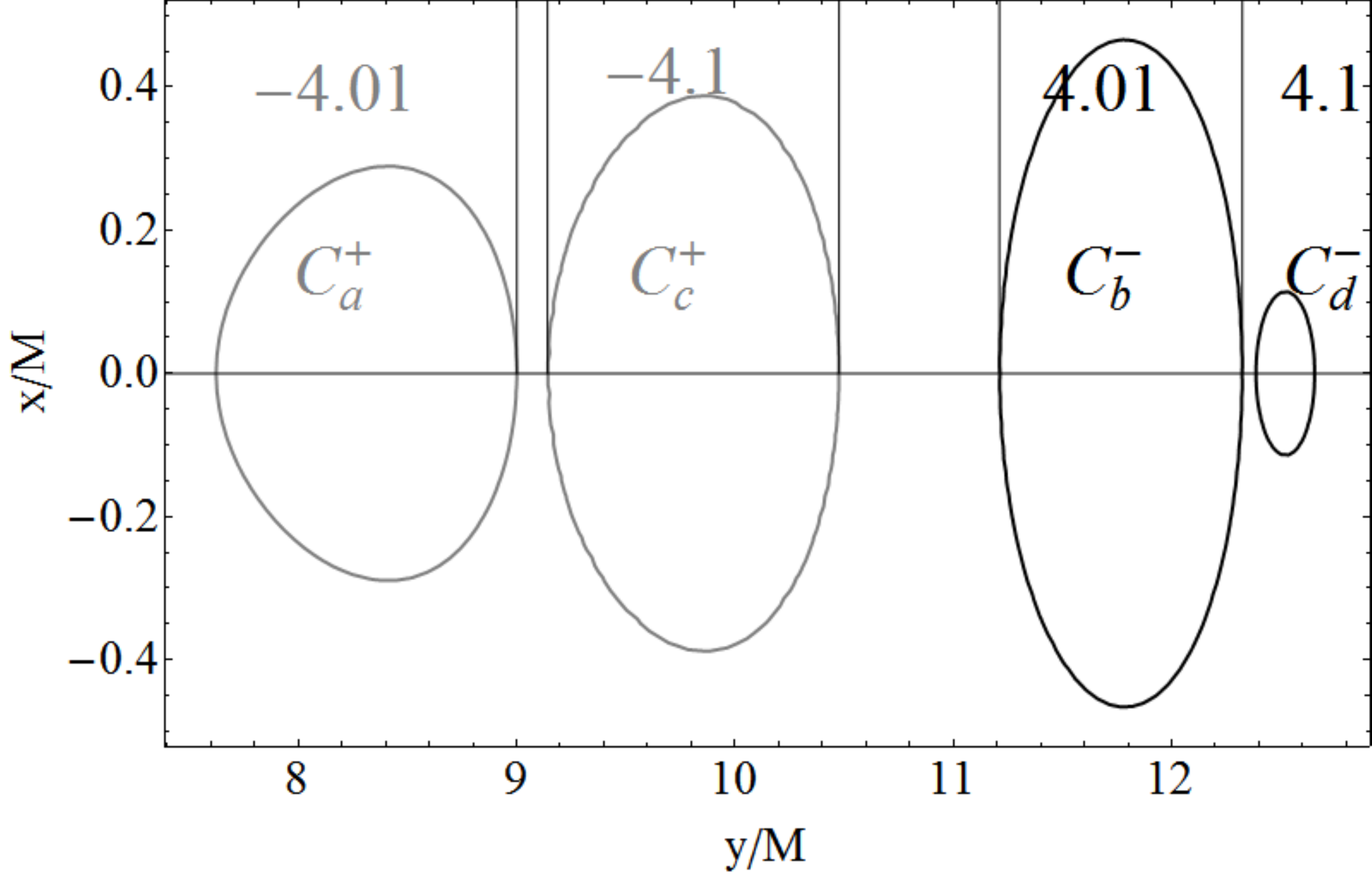}
 \\
 \includegraphics[scale=0.3]{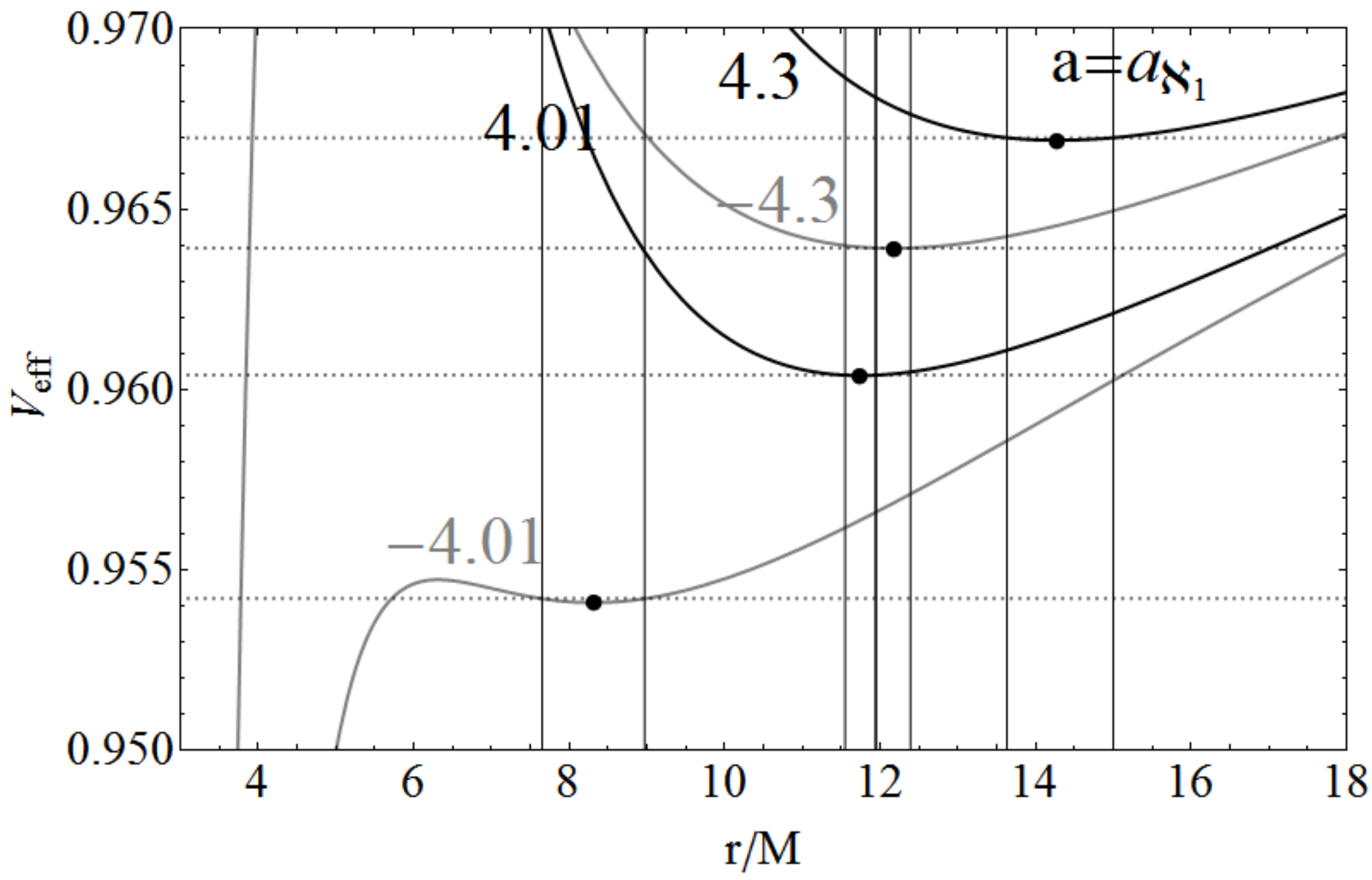}
 \includegraphics[scale=0.3]{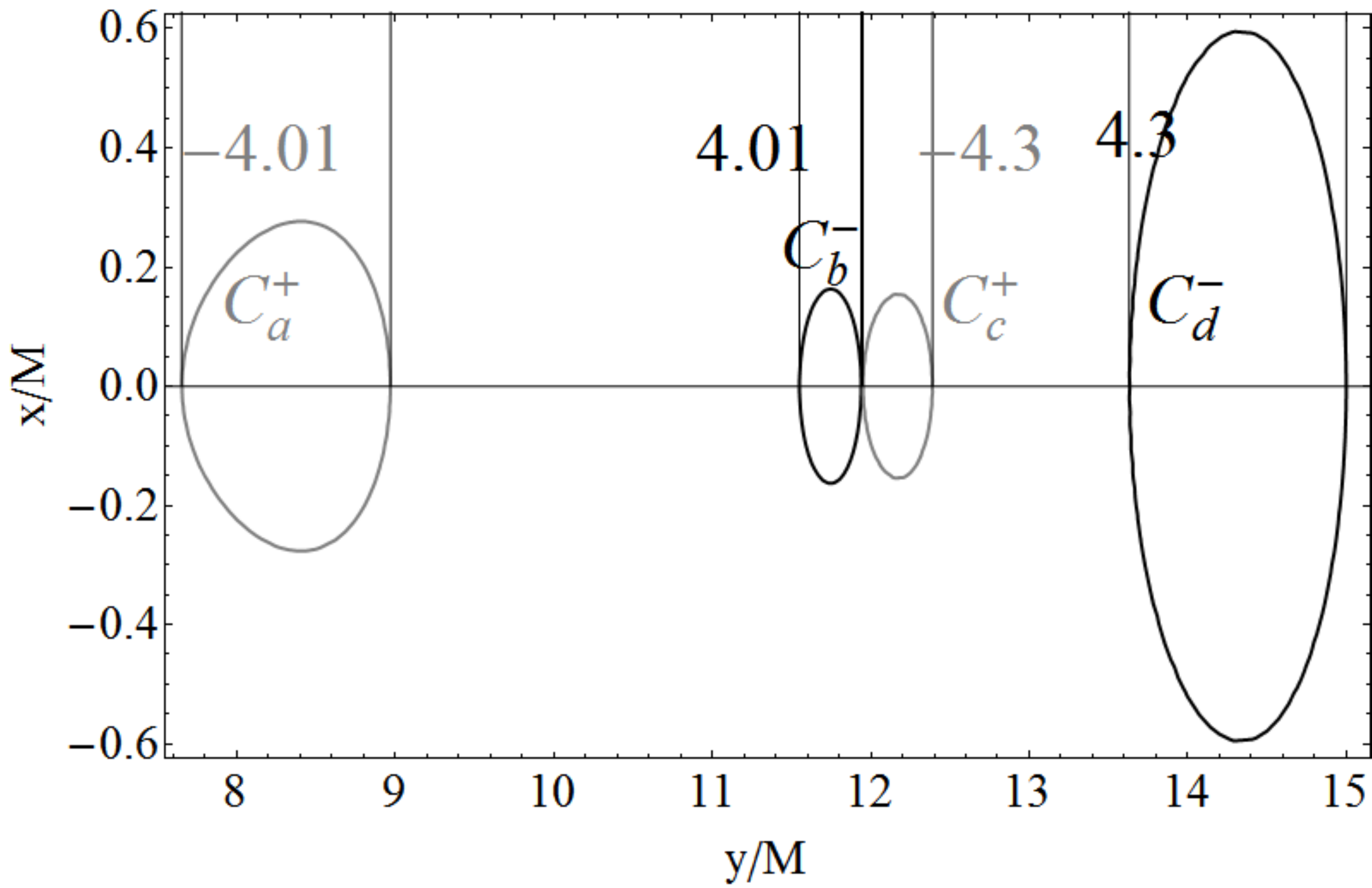}
 \end{tabular}
\caption{$\ell$counterrotating sequences. Spacetime $ a_{\aleph_1}=0.382542M$
The outer horizon is at
$r_+=1.92394M$ the marginally bounded orbits are located in
$r_{mbo}^+=4.73417M$ and
$r_{mbo}^-=3.18903M$, the marginally circular orbits are in
$r_{\gamma}^+=3.41407M$,
$r_{\gamma}^-=2.51784M$.
Left panels: effective potentials as functions of $r/M$. The specific angular momenta are  close to the curves in gray  for counterrotating  matter and black for corotating fluids. Points sign the minimum of the functions. Vertical lines are the ring edges  as determined by the sequences of $\{ K_i \}_{i=1}^n$.
Right panels:  Boyer surfaces  of the decomposition, the curves  associated with each specific  angular momentum of the effective potential are shown with curves of the same color of its potential.
It follows the discussion  in Sec.\il(\ref{Sec:procedure})
  and in particular Eq.\il(\ref{Eq:cononKlcontro}).
Upper panels show isolated sequences, 
bottom panels show mixed sequences. 
}\label{Figs:RightPolstex}
\end{center}
\end{figure}
    \\
\item[\namedlabel{(--)}{(i-b)}] \emph{mixed} $\ell$counterrotating sequences $\overbrace{\mathbf{C}}_{{m}}$:  $[r_{min}^{1_-},r_{min}^{n_-}]\cap[r_{min}^{1_+},r_{min}^{n_+}]\neq0$ Fig.\il(\ref{Figs:RightPolstex})-bottom.
 \end{description}
We shall neglect the inner  lobe  of  a ring $C_i$ in equilibrium and associated with  $y_2$ as in Fig.\il(\ref{Fig:Quanumd}). In fact, we assume that the existence  of this innermost surface  does not influence the outer lobe of the  Boyer  surface, when this is    far from its critical cusped phase $C^i_x$. The basis of this   assumption  relies in the fact that for each closed toroidal Boyer  $C$-configuration
the two lobes are separated, as long as the  $\mathbf{p}$ parameter of the  closed $C$-ring remains far from the critical values, where the ring  morphology changes for $C$   to $C_x$. The cross is then a contact point between the two surfaces, and a  P-W instability cusp, in other words  $y^i_2=y^i_3$.

Therefore,  at the equilibrium the two lobes of a $C_i$ ring,  both regulated by Eq.\il(\ref{Eq:scond-d}) at equal $\mathbf{p}_i$,  without considering any other  further  interaction due  for example to possible  magnetic fields of the disk environment,  are  geometrically separated and  dynamically independent.

{In Sec.\il(\ref{Sec:unsta}) we introduce three  special  macro-configurations associated with the instability phases of the ringed  disk. In Sec.\il(\ref{Sec:pertur})   some general considerations on the perturbations of the ringed disks and their equilibrium follow.
This section closes in Sec.\il(\ref{Sec:principaldef})   with a review of the  major morphological features of the ringed disk $\textbf{C}^n$, while the height and thickness of  $\textbf{C}^n$ are finally  detailed in  Sec.\il(\ref{Sec:App-mor-p}).}
\subsection{Unstable macro-configurations}\label{Sec:unsta}
We focus on  the possible unstable ringed configurations $\mathbf{C}^n$.
 There are  two main instabilities which lead to (global) instability for a $\mathbf{C}^n=\bigcup^nC_i$ disk  and  correspondingly  two distinct models (with degenerate topology) of unstable configurations.

 Firstly, the instability arising inside   the decomposition of the macro-structure, and  induced by  the action of $C_i$  ring in equilibrium  on the other (consecutive) $C_{i\pm 1}$ ring, for overlapping of materials with consequent collisions. Secondly  the  P-W instability of one  or more  rings of the decomposition, in its unstable phase.
 With the consequent    destabilization of  the entire decomposition,   resulting eventually  in a different  topology,  where the rings are   no more separated.
The unstable ringed disks   can be in one of the following macro-configurations
 \begin{description}
 \item[The macro-structure
$\mathbf{C_{\odot}^n}$]  In this model the disk could lead to possible unstable decompositions  with feeding of matter from  one ring to  the other consecutive one, or collision of the orbiting material of the two disks. For these disks there is  at last  {$i\in\{1,...,n-1\}:\; C_i\cap C_{i+1}=\emptyset$ and $\partial C_i\cap\partial C_{i+1}=\{ y_1^{i}=y_3^{i+1}\}$}, as clarified in  Fig.\il(\ref{Figs:CrystalPl});  we could define $y_1^i$ as a \emph{double} point of the decomposition. We define  \emph{rank} $\mathfrak{r}$  of a
$\mathbf{C_{\odot}^n}$ ringed disk  the number of double points of its decomposition. The  maximum possible rank for an $n$-order configuration $\mathbf{C_{\odot}^n}$  is $\mathfrak{r}_{Max}=n-1$ where $\forall i\in\{1,...,n-1\}\; C_i\cap C_{i+1}=\{y_1^i=y^{i+1}_3\} $. If the ringed disk  $\mathbf{C_{\odot}^n}$ has the maximum rank, we call it   \emph{saturated} ringed disk. Fig.\il(\ref{Figs:CrystalPl}) shows an example of saturated ringed disk of the order $n=4$ and maximum rank $\mathfrak{r}_{Max}=3$.
\item[The macro-structure $\mathbf{C^n_x}$] This ringed disk   has at last one sub-configuration $C^i_x$ in its decomposition, that is  a ringed torus  with a  self-cusped topology, the cross  being a  P-W accretion point, with $\pm\ell_{\mp}\in]\pm\ell_{mso}^{\mp},\pm\ell_{\gamma}^{\mp}[$, see Fig.\il(\ref{Figs:CrystalPl}).  We define \emph{rank} $\mathfrak{r_x}$  of the ringed disk  $\mathbf{C^n_x}$ the number of P-W instability points in its decomposition. For a ringed disk of order $n$, the maximum rank could be in general $\mathfrak{r_x}_{Max}=n$.We characterize these particular decompositions in details in the following Sections, especially  the location of $C^i_x$ in the mixed or isolated decompositions. A $\mathbf{C}_x^n$ topology could be  also a $\mathbf{C}_{\odot}^n$ leading  to the following  third macro-structure:
     \item[The macro-structure $\mathbf{C_{\odot}^x}^n$] This  $\mathbf{C_{\odot}^x}$ disk is characterized   at last by a $C^{i}_x$ ring where  $y_1^{i-1}=y_3^{i}=y_{cusp}$. Clearly  a $\mathbf{C_{\odot}^n}$   can be   also  a  $ \mathbf{C_x^n}$ but  not a   $\mathbf{C_{\odot}^x}^n$ one. In any case ringed disk of   $\mathbf{C_{\odot}^x}^n$ topology violates the condition of separation  or  non penetration of matter that  will  be discussed in more details later.
\end{description}
%
%
\subsection{Definition of the major morphological features of the ringed disk $\textbf{C}^n$}\label{Sec:principaldef}
There  we focus on the properties of  separated ringed disks $\mathbf{C}^n$.
We character    the   disk  morphology      introducing the definition   of  the  thickness  and     elongation of the ringed disk.
The macro-structure $\mathbf{ C}^n$, as made  by an  ordered set or  sequence of separated rings $C_i$ of its decomposition, has a  disconnected topology. The  symmetry for reflection on the equatorial plane   is inherited by the  macro-structure $\mathbf{C}^n$,  and the  symmetry plane of each $C_i$  and $\mathbf{C}^n$ is also the symmetry plane of the attractor.
It is useful to introduce   a  toroidal surface $\partial\mathbf{ C}^n$ the for macro-structure $\mathbf{ C}^n$, plotted as the  regular toroidal  surface    generated by the circular cross-sections,  obtained as the envelope surface $\partial\mathbf{ C}$ of the (closed)   $\partial C_i$,  and the associated solid torus including  the spacings (the  vacuum regions among the sub-configurations as formally defined below), that we identify with $\mathbf{ C}^n$. The introduction of this  surface,  could be useful in the definition of some morphological properties of the ringed accretion disk, such as the thickness, and the height defined as the respective quantities of its  envelope.  One may estimate  the total (global) dimensions of the ringed disks, by an envelope surface.
The curve  $\partial\mathbf{ C}$ is tangent  the  sub-configurations at least  in the maximum points: $y_{Max}^i\in\partial\mathbf{ C}$. The surface $\partial\mathbf{ C}$  has yet  a regular constant generatrix  as generated from tori with regular generatrix, and the symmetry for reflection on the equatorial plane and the symmetry relative to  the symmetry axis  are inherited by the macro-configuration, but the envelope surface may have more maximum points   $y_{Max}^i\in \partial C_i\cap\partial{\mathbf{C}}$.  However, this surface can be better defined by introducing an  effective potential for configuration\footnote{
Actually the equation for the envelope could be
$\mathcal{V}_{eff}(i,x,y)=V_{eff}(i,x,y)-K_i=0$, and $\partial_i\mathcal{V}_{eff}(i,x,y)=0$ where the envelope index $i$  is the (ordered) decomposition index attached to each sub-configuration, then  as $\partial_i\equiv \partial_{r_{min}}$.}  $\mathbf{C}_n$ what will be discussed in  Sec.\il(\ref{Sec:effective}).

 In the following part  we  give the definitions of some main features of the ringed torus on  the basis of morphological characteristics of the single ring of its decomposition.
\begin{description}
\item[]
The \emph{inner edge}, $\partial\mathbf{C}_-^n$, of any macro-configuration $\mathbf{C}^n$  of order $n$ is  located in the equatorial plane and associated to the solution $y_3^1$  that is $\partial\mathbf{C}_-^n=\partial{C_1}$ for the inner edge of the inner ring of its decomposition and the \emph{outer edge}  $\partial\mathbf{C}_+^n$ to  the solution $y_1^n$  of the outer ring of the decomposition see Fig.\il(\ref{Fig:Quanumd}).
\\
\item[]
For any sub-configuration  $C_i\subset\textbf{C}^n$ we introduce the range  $\lambda_i\equiv ]y_1^i,y_3^i[$ whose measure  $\lambda_i\equiv y_1^i-y_3^i$ is the \emph{elongation} of the  Boyer surface cross section  $\partial C_i$  on the equatorial plane as shown in  Fig.\il(\ref{Fig:Quanumd}).
 \\
 \item[]
For any $i\in\{1,...n-1\}$  the range  $\bar{\Lambda}_{i+1,i}\equiv] y_3^{i+1},y_1^i[$ whose measure $\bar{\lambda}_{i+1,i}\equiv y_3^{i+1}-y_1^i$
 defines the \emph{spacing} between two subsequent sub-configurations indicated   $C_i$ and $C_{i+1}$ respectively,  see  Fig.\il(\ref{Fig:Quanumd}).
 \end{description}
It is clear that in the case of macro-configuration   $\mathbf{C}_{\odot}^n$ of rank $\mathfrak{r}$,  there are $\mathfrak{r}$ rings of  its decomposition where $C_i:\;\bar{\lambda}_{i+1,i}=0$  see Fig.\il(\ref{Figs:CrystalPl}). In the following we will use indifferently, when not necessary to specify, the term elongation and spacing for $\lambda_i$ e $\bar{\Lambda}_i$  and their measures respectively.

Thus the elongation  $\Lambda_{\mathbf{C}^n}$ of $\mathbf{C}^n$  is:
 \be\label{Eq:chan-L-corby}
  \Lambda_{\mathbf{C}^n}\equiv[y_3^1- y_1^n]=
\left(\bigcup_{i=1}^{n}{\Lambda}_{i}\right)\bigcup\left(\bigcup_{i=1}^{n-1}\bar{\Lambda}_{i+1,i}
\right)=
{\Lambda}_{n}\cup\bigcup_{i=1}^{n-1}\left(\bar{\Lambda}_{i+1,i}\cup{\Lambda}_{i}\right),
 \ee
 whose measure is  the  elongation $\lambda_{\mathbf{C}^n}$ of the  (separated) configuration  $\mathbf{C}^n$, that is explicitly:
\be\label{Eq:lambdac1}
\lambda_{\mathbf{C}^n}\equiv y_1^n-y_3^1=\sum_i^n \lambda_i+\sum_{i=1}^{n-1}\bar{\lambda}_{i+1,i}\geq \lambda_{\mathbf{C}^n}^{inf}\equiv\left.\sum_i^n \lambda_i\right|_{\sum_i\lambda_i}.
 \ee
Equation (\ref{Eq:lambdac1}) shows that the minimum value $\lambda_{\mathbf{C}^n}^{inf}$ of the elongation $\lambda_{\mathbf{C}^n}$ is achieved, \emph{at fixed $\sum_i^n \lambda_i$}, when $\bar{\lambda}_{i+1,i}=0;\forall i$ that is for a $\mathbf{C}^n_{\odot}$ configuration of rank $\mathfrak{r}=\mathfrak{r}_{Max}$.

{We should note that the total elongation of the ringed disk  $\lambda_{\mathbf{C}^n}$ can be regarded, as shown by the second equality of Eq.\il(\ref{Eq:lambdac1}), as the addition of two components: the sum  $\mathcal{S}_{\lambda}>0$ of the elongations of each of its sub configurations, which is a  strictly positive quantity, and the sum $\mathcal{S}_{\bar{\lambda}}\geq0$ of its spacings,   where this   quantity may be positive or zero.
There is   $\mathcal{S}_{\bar{\lambda}}=0$   for the saturated configuration $\mathbf{C}^n_{\odot}$,  realizing simultaneously \textbf{(a.)} the ringed  configuration of minimum elongation $\left.\lambda_{\mathbf{C}^n}\right|_{ \mathcal{S}_{\lambda} }$ \emph{at fixed sum  $\mathcal{S}_{\lambda}$ }  and, \textbf{(b.)} consistently  the maximum elongation $\lambda_{\mathbf{C}^n}$, when the sum  $\mathcal{S}_{\lambda}$ is, at fixed  order  $n$, a variable.}

{The property \textbf{(a.)} comes straightforwardly, once one considers the key aspect that the sum $\mathcal{S}_{\lambda}$  has been  kept fixed,  independently from  $\mathcal{S}_{\bar{\lambda}}$. This  is immediately obtained from Eq.\il(\ref{Eq:lambdac1}). To be more explicit if, for example,  we consider  $\mathcal{S}_{\lambda}$  as a known data, by canceling the second contribution to the total elongation  derived   from the displacement, then $\lambda_{\mathbf{C}^n}$ has the sole, invariant, $\mathcal{S}_{\lambda}$ component. This immediately implies that all the ringed disks with equal $\mathcal{S}_{\lambda}$ have the same minimum elongation when they are saturated, i.e., the  part on the equatorial plane occupied by matter is the same \emph{independently}  of the   distribution of angular momentum, or the  order $n$, which can be $n=1$ or also arbitrarily large.}

{
The property  \textbf{(b.)} stands as  $\mathcal{S}_{\lambda}$ is  considered no longer as a fixed quantity.
We should note in fact that,  if we evaluate  the infimum
  $\inf{\lambda_{\mathbf{C}^n}}$   of the total  elongation, by minimizing  $\sum_i^n \lambda_{min}^i$, (no longer regarded as fixed) i.e. rearranging the decomposition 
  to minimize each term $ \lambda_i $ (or rather to minimize  the total sum $\mathcal{S}_{\lambda}$), one would  disregard the fact  that the  spacings  $\bar{\lambda}_{i+1,i}$ are indeed related to the   functions   $\lambda_i$.  As a consequence of this fact the minimization of $\lambda_i$ could lead to $\bar{\lambda}_{i+1,i}\neq0$.
 Viceversa, the sub-configurations  satisfying  $\bar{\lambda}_{i+1,i}\neq0$ could lead  to a greater  $\sum_i^n \lambda_i$. While, at fixed  $\sum_i^n \lambda_i$, the  inequality  in (\ref{Eq:lambdac1}) is certainly  verified as it  is  $\bar{\lambda}_{i+1,i}\geq0$, and thus the definition  $\lambda_{\mathbf{C}^n}^{inf}$ is not inconsistent.
It can be shown that the  $n$-order decomposition   of the (fixed) macro-structure  realizing  the condition  $\bar{\lambda}_{i+1,i}=0\;\forall i$ maximizes the total sum $S_{\lambda}=\sum_i^n \lambda_i$ of the elongations. Thus this decomposition    provides the maximum elongation for the ringed accretion disk.
  In other words in the former case \textbf{(a.)} the  elongation  $\lambda_{\mathbf{C}^n}$ is minimized  as function of $\mathcal{S}_{\bar{\lambda}}$ with  $\mathcal{S}_{\lambda}$ being a fixed parameter. Therefore, by minimizing  $\mathcal{S}_{\bar{\lambda}}$ at fixed  $\mathcal{S}_{\lambda}$, in the  \textbf{(b.)} case we maximize  $\lambda_{\mathbf{C}^n}$ considering  $\mathcal{S}_{\lambda}$ as a function of $\mathcal{S}_{\bar{\lambda}}$, showing that  the first is maximized when the second is minimized\footnote{{ The  issue  if any ringed disk, at fixed  $\{\ell_i\}_{i=1}^n$, could be  set in a  saturated  configuration (i.e. $\mathcal{S}_{\bar{\lambda}}=0$) is faced in Sec.\il(\ref{Sec:procedure}) where we will discuss more thoroughly this problem, also considering the constraints that the corresponding sequence of parameters $\{K_i\}_{i=1}^n$ has to satisfy in order to be realized as saturated configuration.}
}.}

It is convenient to  show  the dependence of the elongation ${\lambda}_{\mathbf{C}^n} $ from the  spacing $\bar{\lambda}_{\mathbf{C}^n}$ defined below, by explicit the sum in (\ref{Eq:lambdac1}):
\bea\label{Eq:lambdac1s}
\lambda_{\mathbf{C}^n}&=&y_1^n-y_3^1=
\sum_{i=1}^n (y_1^{i}-y_3^i)+\sum_{i=1}^{n-1}(y_3^{i+1}-y_1^i)=\sum_{i=1}^{n-1} (y_3^{i+1}-y_3^i) +\lambda_n
\\\label{spaziamenti} \bar{\Lambda}_{\mathbf{C}^n}&\equiv&\bigcup_{i=1}^{n-1}\bar{\Lambda}_{i+1,i},\quad
\bar{\lambda}_{\mathbf{C}^n}\equiv\sum_{i=1}^{n-1}\bar{\lambda}_{i+1,i}=
+(y_3^n-y_1^1)-	\left.\sum_{i=2}^{n-1}\right|_{n>2}\lambda_i\geq0.
 \eea
Equation (\ref{Eq:lambdac1s}) relates the elongation $\lambda_{\mathbf{C}^n}$ of the $n$-order configuration $\mathbf{C}^n$ to the elongation $\lambda_n$ of the $n$th-sub-configuration  $C_n$ and to the  relative position of the inner edges of the remaining  sub-configurations of the decomposition, so that one need to know only   $\lambda_n$ and the location of the inner edges on the interior rings $y_3^i$, to establish  the elongation $\lambda_{\mathbf{C}^n}$. Because the sum  in Eq.\il(\ref{Eq:lambdac1s}) does not include any term related to the $n$th-configuration, the minimization of the two distances   at the right side of   Eq.\il(\ref{Eq:lambdac1s}), is   independent of each other.

Equation (\ref{spaziamenti}) is the  sum  of the spacings $\bar{\lambda}_i$ {(we note that this is precisely $S_{\bar{\lambda}}$)},  it indicates  how large portion of the macro-structure $\mathbf{C}^n$ is not filled with the fluid.

Moreover, it explicits
the dependence of $\bar{\lambda}_{\mathbf{C}^n}$, from the elongations $\lambda_i$ of the rings $C_i\in\{2,n-1\}$.
 For
a  configuration of the order $n=2$, there  is   $\bar{\lambda}_{\mathbf{C}^2}\equiv\bar{\lambda}_{2,1}=
+(y_3^2-y_1^1)$. However,  since
  the ring elongations contribute negatively to the total spacing,
we infer from the inequality in (\ref{spaziamenti}) that
$y_3^n-y_1^1$ (the distance between the inner edge of the external configuration  $\partial C_n^-$  and the  outer one $\partial C_1^+$ of the inner one) is greater that the sum of the elongations of the $ 2,...,n-1$ rings.  The sum does  not include any term related to the $C_1$ and $C_n$ sub-configurations, as such  minimization of the two terms in (\ref{spaziamenti}) can be considered a priori independently.

Then the smaller  is the elongation of the interior configurations   the larger is the
  spacing $\bar{\lambda}_{\mathbf{C}^n}$,  where
the minimum  ${\bar{\lambda}}_{\mathbf{C}^n}=\bar{\lambda}_{\mathbf{C}^n}^{inf}=0$ is realized  in Eq.\il(\ref{spaziamenti}) when $\left.\sum_{i=2}^{n-1}\right|_{n>2}\lambda_i=(y_3^n-y_1^1)$, which corresponds to a saturated $\mathbf{C}_{\odot}^n$ configuration. {A definition of \emph{height} and \emph{thickness} of the ringed disk  is discussed in Sec.\il(\ref{Sec:App-mor-p}), this allows to characterize morphologically the disk as a single body.}
\section{Decompositions and analysis  of the model parameters}\label{Sec:caracter}
Any sub-configuration  $C_i$  in equilibrium\footnote{For the considerations of this section and in the most of this work we will assume the existence of $\Delta r_{crit}^i$, in particular  the existence of a  maximum point of the effective potential.
But we should specify here, as already mentioned earlier, that the minimum, center of the  configuration, always exists for  $\pm\ell^{\mp}>\pm\ell_{mso}^{\mp}$,
while the minimum point, corresponding to instability exists only if $\pm\ell^{\mp}\in]\pm\ell_{mso}^{\mp},\pm\ell_{\gamma}^{\mp}[$ with the consequence that  if  $\pm\ell^{\mp}\geq\pm\ell_{\gamma}^{\mp}$, only equilibrium configurations may exist.
For   $\pm\ell^{\mp}>\ell_{\gamma}^{\mp}$,  it is possible generalize these definitions as done in \citep{coop}} satisfies the following property:
\be\label{Eq:ulter}
\Delta r_{crit}^i\cap\lambda_i=\{y_3^i\},\quad \forall i.
\ee
or, we can write  the condition (\ref{Eq:disbase}) explicitly as
\be\label{Eq:disbase}
r_{mbo}^i<r_{Max}^i<y_3^i<r_{min}^i<y_1^i\quad \mbox{and}\quad  r_{min}^i>r_{mso}^i.
\ee
%
For a $n$-order configuration $\mathbf{C}^n$  the separation condition:
\be\label{Eq:condi-sss}
 \mathbf{C}^n:\;
\lambda_i\cap\Lambda_j= \emptyset\; \forall i,j\in \{1,...,n\}\quad\mbox{and}\quad \mathbf{C}_{\odot}^n:\;
\lambda_i\cap\Lambda_j=\{y_1^i=y_3^j\} \; \forall i<j\in \{1,...,n\}
\ee
 must be satisfied,
where $\lambda_i$ are the elongation  ranges as introduced in Sec.\il(\ref{Sec:principaldef}). Conditions  (\ref{Eq:condi-sss}) are in fact equivalent to the condition of separation or no penetration of the material in Eq.\il(\ref{Def:separated}).
To fix the ideas we can consider  the case of
the  macro-structures $\mathbf{C}^2$,
then we  use explicitly the condition (\ref{Eq:condi-sss}) for the case of $n=2$.
Considering the role of the  maximum  and minimum points, we obtain the following conditions:
{
\bea\nonumber\label{Condizioni}
&&
r_{Max}^{\mathbf{(i)}}\leq y_3^{\mathbf{(i)}}<r_{min}^{\mathbf{(i)}}<y_1^{\mathbf{(i)}}\leq y_3^{\mathbf{(o)}}<r_{min}^{\mathbf{(o)}}<y_1^{\mathbf{(o)}},
\\\label{Condizionibar}
&&
r_{mso}^{\mathbf{(i)}}<r_{min}^{\mathbf{(i)}}<r_{min}^{\mathbf{(o)}}\quad\mbox{and}\quad r_{mso}^{\mathbf{(o)}}<r_{min}^{\mathbf{(o)}}.
\eea}
On the other hand, the generalization  of Eqs.\il(\ref{Condizioni},\ref{Condizionibar}) to an $n$-order configuration is straightforward.

Relation Eq.\il(\ref{Condizioni}) is a very  general condition  for  existence of the separated configurations. This relation  is  necessary for the existence of disjoint configurations for that  it can be considered as a  criterion to establish   the \emph{inner} $\mathbf{(i)}$ and \emph{outer} ${\mathbf{(o)}}$ configurations,  Fig.\il(\ref{Fig:Quanumd}).
However Eq.\il(\ref{Condizioni}) do not indicate the relative position of the marginally stable orbits, which obviously determines the order of the corotating and counterrotating rings, neither the  relative location  of the maximum points which, as   shown in the next Sections, could be able by itself to  constraint  in some cases  the order $n$ of  the decomposition.
It is thus necessary to fix the couple $\Delta_{mso}^{\mathbf{\mathbf{(i,o)}}}\equiv(r_{mso}^{\mathbf{(i)}},r_{mso}^{\mathbf{(o)}})$ and $\Delta_{Max}^{\mathbf{\mathbf{(i,o)}}}\equiv(r_{Max}^{\mathbf{(i)}},r_{Max}^{\mathbf{(o)}})$ as related to each other. If the measure  $\delta_{mso}^{\mathbf{\mathbf{(i,o)}}}=r_{mso}^{\mathbf{(o)}}-r_{mso}^{\mathbf{(i)}}=0$, the configurations are $\ell$corotating, while  for $\delta_{mso}^{\mathbf{(i,o)}}>0$, the outer  ring is counterotating while the inner one  is corotating,  and viceversa for $\delta_{mso}^{\mathbf{\mathbf{(i,o)}}}<0$.

Within the  {condition in Eq.\il(\ref{Condizionibar})}  for a couple of rings,  two cases can occur, according  to the location of the maximum points $\Delta_{Max}^{\mathbf{(i,o)}}$  and  to the permutation  in the ordered  couple
$(r_{min}^{\mathbf{(i)}},r_{Max}^{\mathbf{(o)}})$; the case $r_{min}^{\mathbf{(i)}}\equiv r_{Max}^{\mathbf{(o)}}$  will be discussed in Sec.\il(\ref{Sec:rmin=rMx})). More generally for a   $\mathbf{C}^n$ configuration, there are the  possibilities
\begin{description}
\item[\namedlabel{C0}{$\bar{\mathfrak{C}}_0$}:] corresponding  to $ \Delta_{cri}^{i}\cap\Delta_{cri}^{i+1}=\emptyset$, or explicitly
 $r_{Max}^{i}<r_{min}^{i}<r_{Max}^{i+1}<r_{min}^{i+1}$;
\item[]
{Then there can be
$r_{Max}^{i+1}<r_{min}^{i}$.}

 Considering the role of $r_{Max}^{i}$, it can be either:
\begin{description}
 \item[\namedlabel{C1a}{$\bar{\mathfrak{C}}_{1a}$}:] that is $r_{Max}^{i+1}\in\Delta_{crit}^{i}$, explicitly
 $r_{Max}^{i}<r_{Max}^{i+1}<r_{min}^{i}<r_{min}^{i+1}$   or
 \item[\namedlabel{C1b}{$\bar{\mathfrak{C}}_{1b}$}:]  that is $ \Delta_{crit}^{i}\subset\Delta_{crit}^{i+1}$ or explicitly $r_{Max}^{i+1}<r_{Max}^{i}<r_{min}^{i}<r_{min}^{i+1}$ Fig.\il(\ref{Figs:CrystalPl}).
 \end{description}
 \end{description}
These cases, detailed in Sec.\il(\ref{Sec:i4cases}),  and their generic combination in the decomposition of the order $n$ exhaust all the possibilities: the ringed disk can be $\bar{\mathfrak{C}}_0$, {$\bar{\mathfrak{C}}_{1a}$ and $\bar{\mathfrak{C}}_{1b}$}, or could  have also a decomposition with consecutive rings belonging to one or the other cases.
{
In the remainder of this section we characterize
the role of the parameter $\mathbf{p}=(\ell, K)$ in Sec.\il(\ref{Sec:roleofp}), by focusing on
 the $K$-parameter in the ringed disks, in Sec.\il(\ref{Sec:onK}), and
$\ell$-parameter and the relation between these in Sec.\il(\ref{Sec:rolel}).
Existence and structure of the ringed configurations is a topic of Sec.\il(\ref{Sec:i4cases})
which will continue in Sec.\il(\ref{Sec:K=K}),  with the analysis of the
limiting cases on the $K$ parameters for a $n$-order decomposition.
 Some appendix sections complete the discussion  of this section and  particularly Sec.\il(\ref{Sec:app-maxmin}), (\ref{Sec:effective}) and (\ref{Sec:pertur}).}
\subsection{The role of the parameter $\mathbf{p}=(\ell, K)$}\label{Sec:roleofp}
The parameter $\mathbf{p}_{\mathbf{C}^n}$ associated to the macro-configuration of order $n$ is a sequence of ordered  $n$-tuplet  of couples $\{\mathbf{p}_i\}_{i=1}^{n}$ where $\mathbf{{p}}_i\equiv(\ell_i, K_i)$. Setting the ordered multi-parameter  $\mathbf{p}_{\mathbf{C}^n}$ of the configuration means fixing univocally the macro-configuration $\mathbf{C}^n$ and its decomposition.  According to the condition of separation (\ref{Condizioni}), not all the choices of  $\mathbf{p}_{\mathbf{C}^n}$ are possible, then
a double (closed) configuration ($n=2$)  may exist if there is a couple of parameters
\be
(\mathbf{{p}}_{\mathbf{(i)}},\mathbf{{p}}_{\mathbf{(o)}}):\;\quad \mathbf{{p}}_{\mathbf{(i)}}\neq \mathbf{{p}}_{\mathbf{(o)}}\quad\mbox{and}\quad y_1^{\mathbf{(i)}}\leq y_3^{\mathbf{(o)}}.
\ee
Then, in analogy with the case of the single torus (ring), we can define, at fixed order $n$,
 the matrix $\mathcal{B}_{\mathbf{p}}^n$ generated by all the possible variation of the  couple  $\mathbf{p}$, and the projections  $(\mathcal{B}_{\ell}^n,  \mathcal{B}_{K}^n)$ are  generalized straightforwardly.
For an $n$-order disk with   $\mathbf{p}_i\neq \mathbf{p}_{j}$, three cases can occur:
\begin{description}
\item[
 \textbf{1.} $K_{i}=K_{j}$] This case give rise of $\mathcal{B_\ell}$  sequence.  The stability conditions    are discussed in Sec.\il(\ref{Sec:onK}) and Sec.\il(\ref{Sec:K=K}).
\item[ \textbf{2.} $\ell_{i}=\ell_{j}$]
The second case with  sequences  $\mathcal{B}_K$ of equal specific angular momentum is immediately ruled out, as these are evidently centered configurations (see also Fig.\il(\ref{Figs:Aslanleph1l})). At constant $\ell$ the effective potential is uniquely defined  as function of $r/M$, and  there is only one minimum, if $\pm\ell^{\mp}>\pm\ell_{mso}^{\mp}$, and maximum, if $\pm\ell^{\mp}\in]\pm\ell_{mso}^{\mp},\pm\ell_{\gamma}^{\mp}[$
.
The separated sub-configurations must have different specific angular momentum $\ell_i\neq\ell_j$, no separated rings with equal specific angular momentum can rotate one \textbf{BH} attractor. However, in the case of a $\ell$corotating couple $(C, O_x)$ with launching of jets from the P-W instability point $O_x$, the situation $\ell_{i}/\ell_{j}=1$ would be possible,  under a relaxation of the condition in  (\ref{Condizioni})  and  $K=K_{Max}>1$ for the open configuration governing the jets, and  $K\in]K_{min},1[$ for the concentric configuration  $C$ in  equilibrium.
\item[\textbf{3. } $K_{i}\neq K_{j}$ and $\ell_{i}\neq\ell_{j}\quad \forall (i,j)$]. This is the most general case the couple parameters are both different.
\end{description}
\subsection{On the individual role of the $K$-parameter in the ringed disks}\label{Sec:onK}
The choice of the  $ K $ parameter for the  decomposition of a  ringed disk is very strict as each $K_i$  fixes the elongation $\lambda_i$ of the  sub-configuration $C_i$.
At fixed specific angular momentum $\ell_i$ one and only one $K_i$  can be set, although  typically in an infinite but bounded range of values, constrained  only by the condition of separations between the rings in equilibrium or in accretion.  Conversely, for a fixed $K_i$,  we can set in general a number $n$  of different specific angular momenta defining  the ordered decomposition. We will use this last property to set in Sec.\il(\ref{Sec:effective}),  definition of effective potential for the ringed configuration.
The case \textbf{1.} of Sec.\il(\ref{Sec:roleofp}) $K_{i}=K_{j}$, with sequences  $\mathcal{B}_{\ell}$,  is solved for any couple of fluid specific angular momenta  $\ell_{i}\neq\ell_{j}: \; \exists!K=K_{j}=K_{i}$ where $K_{i}$ is a value of the effective potential function associated to the  ${C}_i\subset\mathbf{C}_n$.
If $K_{i}=K_{j}=K\; \forall K$ then $K$ plays the role of the  \emph{$K$-parameter or $K_{\mathbf{C}}^n$-for the macro-configuration  $\mathbf{C}^n$} entirely decomposed by a sequence  $\mathcal{B}_{\ell}$ .
This condition has to be completed then  by  condition  given in Eq.\il(\ref{Condizioni}),  
but in general the condition for  existence of a common $K$  can be stated  as follows
\be
\Delta_{\mathbf{K}}\equiv\bigcap_{i=1}^n\mathbf{K}_{crit}^i\neq\emptyset\quad \delta_{\mathbf{K}}\equiv\inf_{i=1}^n{K_{Max}^i } -\sup_{j=1}^n{ K_{min}^j}>0\quad\mbox{ where}\quad \mathbf{K}_{crit}^i\equiv [K_{min}^i,K_{Max}^i]\quad\mbox{ and}\quad \beth{(\Delta_{\mathbf{K}})}\geq1,
\ee
{where $\beth{(\Delta_{\mathbf{K}})}$ is the number of elements of the  intersection set $\Delta_{\mathbf{K}}$ or, equivalently, the number of  possible values of the   $K$ parameter as they came from the cross set. Therefore  the crossing set has  at least one element\footnote{{We assume that the couple  $(K_{min}^i,K_{Max}^i)$ is well defined for any $i$. In this case, it is clear, in fact, that the necessary condition for the existence of  any two rings of a common $K=K_i=K_j$ is  $K\in \Delta_{\mathbf{K}}$ (as it has to be, for each $i$,  $K_i\in [K_{min}^i,K_{Max}^i]$). On the other hand, there could be many more common possible values of $K$ belonging to the set $ \Delta_{\mathbf{K}}$, actually an entire range  of  values $K\in  \Delta_{\mathbf{K}}$ as it is clear, for example, from the effective potential of $C_2$ $(\ell_2=-4.4188)$ and $C_1$ $(\ell_1=-4.3484)$ in Fig.\il(\ref{Figs:CrystalPl})-upper.}  }$i$,}
  $\delta_{\mathbf{K}}$, is then the measure  of $\Delta_{\mathbf{K}}$ and the $\inf$ and $\sup$ will refer to one or more sub-configurations.  Then $K_{Max}^i= K_{Max}^j$, and/or $K_{min}^i= K_{min}^j$; these particular cases will be discussed extensively in Sec.\il(\ref{Sec:minimacoinc}).

In general,  any   $n$-order  decomposition
at fixed $\{K_i\}_{i=1}^n$ (thus also $K_i\neq K_j$) can be realized by the  sequences  $\{\ell_i\}_{i=1}^n$ with $K_i\in \Delta_{\mathbf{K}}^i$.
It is clear that  the condition $\delta_{\mathbf{K}}=0$  {could} mean  $K_{Max}^i=K_{min}^j\; \forall \{i,j\}$ \emph{or}  $\exists(i,j):K_{Max}^i=K_{min}^j$, but in this last case  the order of the configuration \emph{must} be $n_{Max}=2$, see also Fig.\il(\ref{Fig:bCKGO}).
\begin{figure}[h!]
\centering
\begin{tabular}{c}
\includegraphics[scale=.34]{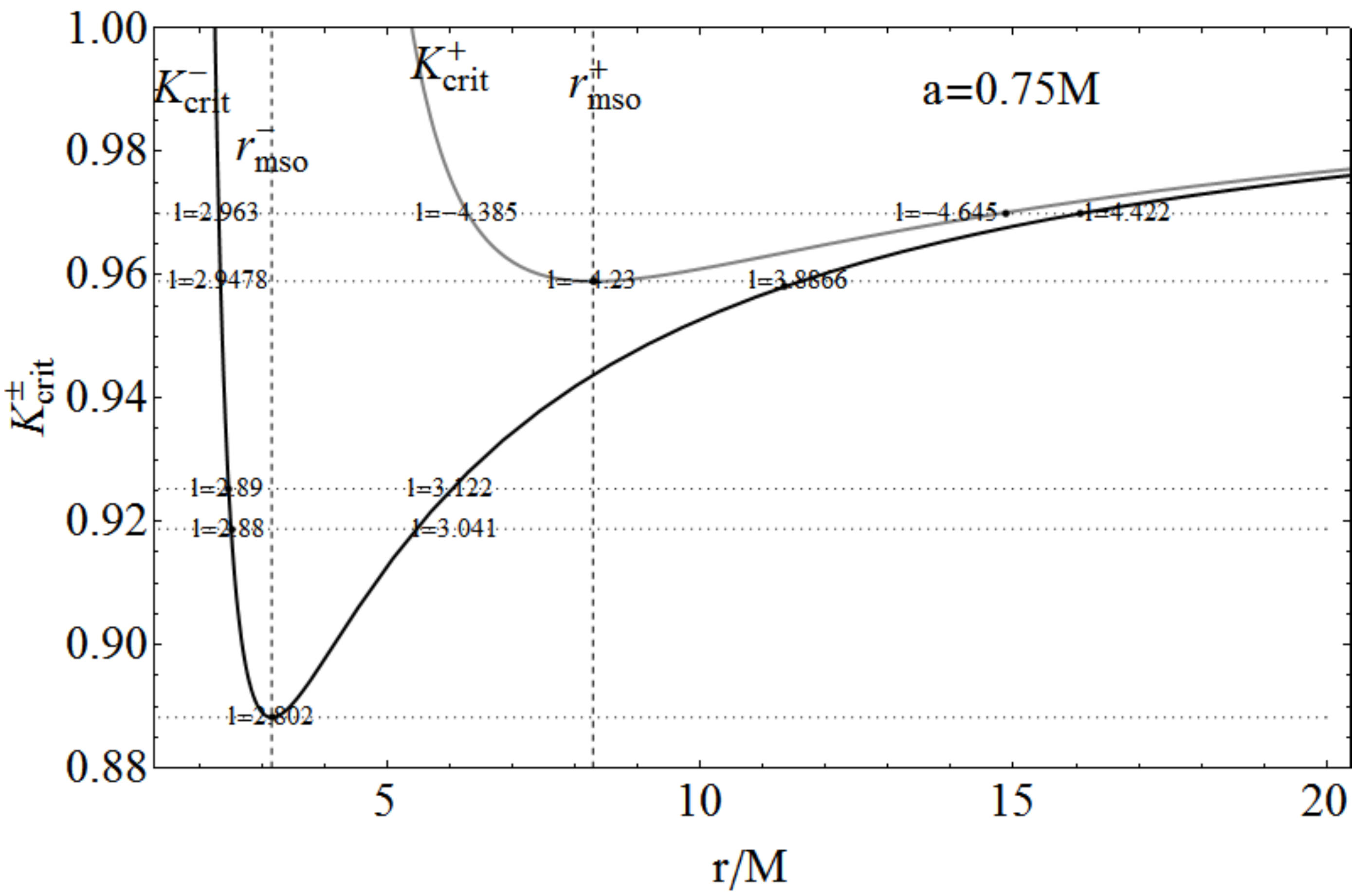}
\end{tabular}
\caption{Fluids orbiting in a  \textbf{BH}-spacetime  with $a=0.75M$. The  $K^{\pm}_{crit}=V_{eff}(\ell_{\pm})$ curves as functions of $r/M$, for corotating $K_{crit}^-$ (black curve) and counterrotating $K_{crit}^+$ matter (gray curve).
Marginally stable circular orbits $r_{mso}^{\pm}$ are signed  with dashed vertical lines.
Points on the curves set the critical points and the corresponding values of the specific angular momentum. The values $\ell=$constant are signed with dotted (horizontal) lines.
}\label{Fig:bCKGO}
\end{figure}
 Further details on the rings  $K_{Max}^i=K_{min}^j$ are provided in Sec.\il(\ref{Sec:Kamx=Mmin}).
As  $K_{mso}^-<K_{mso}^+$, in general  $\delta_{\mathbf{K}}\in]0,1-K_{mso}^{\pm}[$, where  $^{(+)}$ holds  at last one ring is counterrotating, and  $^{(-)}$ holds if all the rings are corotating\footnote{ It is a  wide margin: we imposed a cut at  $K=1$ as supremum of the range,  being in any case  $K<1$,  or a configuration of order $n$ could have  $n$ jets. The infimum of   the $K$ for each ring is clearly  $K_{mso}^{\pm}$ respectively for counterrotating and corotating rings, therefore the supremum of the range of a common $K$  is  $\sup K_{crit} -\inf K_{crit}=1-K_{mso}^{\pm}$.}.  In  Fig.\il(\ref{Figs:CrystalPl}) an example of  procedure adopted for the selection of the sequence  $\{K_i\}_{i=1}^n$ at $\{\ell_i\}_{i=1}^n$ fixed,  is demonstrated. {In Sec.\il(\ref{Sec:effective}) we introduce  the important notion of effective potential  of the ringed disk, which also enables  to  clarify further the role of the sequence of the parameters  $\{K_i\}_{i=1}^n$}.
%
%
%
\subsection{On the role of the $\ell$-parameter:  differential rotation and angular momentum of the macro-structure}\label{Sec:rolel}
The specific angular momentum of the sub-configuration $C_i$  determines the center $r_{cent}^i=r_{min}^i$, and the instability point  $r_{cusp}^i=r_{Max}^i$, determining also if,  in its unstable phase, the ring $C_i$  can generate jets of matter   \emph{or}  it will be accreting  onto the black hole.
 So far we have taken into account particularly the  reciprocal sign of the specific  angular momentum for the  two consecutive rings, but not their  magnitude. An  important feature of the ringed disk structure  is the ratio in  specific  angular momentum between two consecutive rings $C_i$ and $C_{i+1}$,
\be\label{ratiolll}
\ell_{i/i+1}\equiv\ell_i/\ell_{i+1}\neq 1,
\ee
 whose sign defines  the  $\ell$corotating or  $\ell$counterrotatings rings.
Definition (\ref{ratiolll}) can be extended for any pair of rings $C_i<C_j$ of the $\mathbf{C}^n$ decomposition and also to values $\ell_{i/i+1}=-1$, see also discussion in \citep{Pugtot}. Configurations $\ell_{i/i+1}=1$ could perhaps be permissible  for a couple  $(C,O_x)$, but providing to a weakening of the separation condition  (\ref{Condizioni}) when the decomposition includes open critical surfaces.
We can   characterize  the decompositions  of a ringed disk establishing a  relationship between the specific angular momenta of the sub-configurations in terms  of the   ratios $\ell_{i/j}$. 
  An analogue   analysis will be similarly  done in terms of the parameter  $K$, in Sec.\il(\ref{Sec:onK}) and in Sec.\il(\ref{Sec:K=K}).
 {In Sec.\il(\ref{Sec:diff-rot}) we introduce the concept of differential rotation of the macro-structure, providing also a definition for the ringed disk angular momentum. The analysis of the  role of the ring specific angular momentum $\ell_i$ in the formation of the ringed disk  continues  in Sec.\il(\ref{Sec:procedure}) where, considering the $\ell$corotating and $\ell$counterrotating sequences separately,  we shall explore  the mutual influence of the set of  parameters  $\{\ell_i\}_{i=1}^n$  and   $\{K_i\}_{i=1}^n$, showing also this relation in the special case of a saturated disk}.
\subsubsection{Differential rotation and the disk angular momentum}\label{Sec:diff-rot}
The macro-structure $\mathbf{C}^n$  is  not characterized by an univocal  specific angular momentum equal for any rings,  and  one can say  it has a \emph{differential rotation}:
meaning the ordered sequence of parameters $\{\ell_i\}_{i=1}^n:\;\ell_i\neq\ell_j$ 
and,  as discussed in  Sec.\il(\ref{Sec:effective}),
 unequivocally defining  the potential in Eq.\il(\ref{Eq:Vcomplessibo}) of the macro-structure, the order of the decomposition, but not its decomposition.
Further details on the differential rotation  will be provided also  in  Sec.\il(\ref{Sec:app-maxmin}) and Sec.\il(\ref{Sec:pertur}) where the perturbation of  the $\mathbf{C}^n$ disk will be addressed.
In this Section we introduce a  definition of (specific) angular momentum associated to the  ringed disk $ \mathbf{C}^n$,  taking into account  its  differential rotation.
\begin{description}
\item[The area angular momentum  $\bar{\ell}_{\mathbf{C}}^n$   and  the volume angular momentum $\bar{\bar{\ell}}_{\mathbf{C}}^n$ ]
   of the macro-configuration of the order $n$ are defined respectively as the algebraic sums:
\be\label{Eq:stand}
\bar{\ell}_{\mathbf{C}}^n\equiv\frac{\sum_{i=1}^n \ell_i \mathcal{A}_i}{
\sum_{i=1}^n\mathcal{A}_i},\quad \bar{\bar{\ell}}_{\mathbf{C}}^n\equiv\frac{\sum_{i=1}^n \ell_i \mathrm{V}_i}{\sum_{i=1}^n\mathrm{V}_i}\quad\mbox{and}\quad\bar{\bar{\ell}}_{\mathbf{C}}^n\approx\frac{\sum_{i=1}^n \ell_i \mathcal{A}_ir_{min}^i}{\sum^n_{i=1}\mathcal{A}_i r_{min}^i}
\ee
where  $\mathcal{A}_i$   is the surface area of the $C_i$ cross section in the equatorial plane\footnote{
 That is $\mathcal{A}_i=
 2\int_{y_3^i}^{y_1^i}\partial C_+^i d y$ where $d\lambda\equiv dy$  (the  boundary $\partial C_+$  is a properly oriented curve in $y>0$)},  and $\mathrm{V}_i$ is proportional to the total mass of the system is the interior volume of the $C_i$-ring, which is approximately evaluated in the equality  in (\ref{Eq:stand}) considering the Guldinus theorem (or Pappus'centroid theorem-see for example \citep{Guld-ref}) as the system has a constant\footnote{Standard approaches of rotating bodies  considers  rigid ($\Omega(r)=$const) and differential ($\Omega(r)\neq$const) rotation. There we shift the meaning of differential rotation of ringed disk to those ringed tori having $\ell_i(r)=$const, but $\ell_i(r)=\mbox{const}\neq \ell_j(r)=\mbox{const}$.} centroid, taking
the centroid function and its geometric centroid corresponding to the disk center and maximum pressure $r_{min}$. The first definition   $\bar{\ell}_{\mathbf{C}}^n$ in Eq.\il(\ref{Eq:stand}) relates the total angular momentum to the section of the Boyer surface  in the equatorial plane, the second $\bar{\bar{\ell}}_{\mathbf{C}}^n$, consider the total volume of each ring.
 \\
 \item[The leading $\bar{\ell}_{\mathbf{C}^n}^l$ and the apparent $\ell_h$ angular momentum]

The specific angular momentum  $\ell_1$, associated to the inner  ring $C_1$, and the momentum $\ell_n$ of the outer ring $C_n$, play a major role in the equilibrium of the decomposition and also in  determining the morphological characteristics of the ringed structure.
However, one might consider, in analogy to the definition of the thickness and height of the ringed  disk provided   respectively in Eq.\il(\ref{Eq:max-height}) and Eq.\il(\ref{Eq:wolne}). by defining
the  \emph{leading} specific angular momenta
\be\label{Def:leadingl}
\bar{\ell}_{\mathbf{C}^n}^l\equiv \sup |\ell_i|\mathcal{A}_i,\quad\mbox{or}\quad \bar{\bar{\ell}}_{\mathbf{C}^n}^l\equiv \sup |\ell_i|\mathrm{V}^i.
 \ee
 Henceforth we indicate equally these two  and the \emph{apparent} $\ell_h$ specific angular momentum of the sub-configurations with maximum heights $h$. %
\end{description}
In the following Sections,  characterizing   decompositions with sequences  $\ell$counterrotating,
we face the problem if the existence of (stable) ringed configurations with $\bar{\ell}_{\mathbf{C}}^n=0$ (or $\bar{\bar{\ell}}_{\mathbf{C}}^n=0$) for a  $\mathbf{C}^2$ disk is possible;
for example  
to be zero $\bar{\ell}_{\mathbf{C}}^2=0$ implies
 $\ell_{1/2}=-  {\mathcal{A}_2}/{\mathcal{A}_1}$ and $\bar{\bar}{\ell}_{\mathbf{C}}^2=0$ implies
 $\ell_{1/2}=-  ({\mathcal{A}_2}/{\mathcal{A}_1}) (r_{min}^2/r_{min}^1)$.

The analysis considers then particular  ratios $\ell_{i/j}$ between the specific angular momenta of the rings, but one can  consider the   relation:
\be\label{Eq:Marix-ex-spo}
|\ell_i|=|\ell_j|+\epsilon_{ji}\quad\mbox{with}\quad \epsilon_{ij} \approx0: \; \ell_i\neq\ell_k\quad\mbox{and more generally}\quad \quad r_{min}^i\neq r_{min}^k \; \forall i\neq k,
\ee
where the (trace null) totally antisymmetric  matrix of  \emph{displacements} $\Large{\mathbf{\epsilon}}:\;\epsilon_{ij}=-\epsilon_{ji}\approx 0$ would  have a leading role in the perturbation of the decomposition and therefore of  the $V_{eff}^{\mathbf{C}^n}$ potential
in (\ref{Eq:Vcomplessibo}), see  also Sec.\il(\ref{Sec:pertur}).
As  $\epsilon_{ij}\approx0$, we can  use the relation (\ref{Eq:Marix-ex-spo})  as a procedure  to build up a model of a (geometrically thin) disk
consisting  of a very large number  $n$ of possibly   $\ell
$corotating  rings,  to avoid collisions in this  peculiar disk model,   or with  $\ell$counterrotating isolated $ \overbrace{\mathbf{C}_s}$  bends \ref{(-)}) with a
given differential rotation. As proved in Secs.\il(\ref{Sec:procedure},\ref{Sec:app-maxmin}), the procedure  (\ref{Eq:Marix-ex-spo})  is particularly suitable for isolated  decompositions, or decomposition constituted by an $\ell$corotating sequence only  as in this case there is
$\partial_i |\ell|>0$, where $\partial_i$  stands for a variation with respect to the configuration index as detailed below. It would be a discrete one-dimensional  model
of geometrically thin disks, where  the only relevant dimension  is the radial  one. The hydrostatic pressure  shall have $n$ maximum  points in the equatorial plane, and would be    constant  on each ring surface, being  approximately coincident with a portion of the envelope surface of the macro-structure. However, it remains   to   define the stability of the ringed  disk in this model that could be also unstable-
see also   Sec.\il(\ref{Sec:pertur}) on the perturbations of the decomposition.
The difference with a dust disk (bundles of one-dimensional geodesics) is clearly that each ring in this model is a thick disk with a  pronounced verticalization  ($R_i\approx1$) but obedient to a similar dynamic equation: one could write $V_{eff}^i=V_{eff}^j+f(\epsilon_{ij})$,  considering  that if the two rings are $\ell$counterrotating,  $f(\epsilon_{ij})$ is not in general a small correction term  for $V_{eff}^j$.

As we have already mentioned, the potential  (\ref{Eq:Vcomplessibo}) is uniquely determined by the differential rotation,
but it is undetermined by the choice of the sequence $\{K_i\}_{i=1}^n $. Nevertheless, one
can always choose an appropriate $K_i$ for a $\ell_i$ such that the spacing $\lambda_{i\pm1}$ is small
enough, see also Secs.\ref{Sec:onK} and \ref{Sec:K=K}. In Sec.\il(\ref{Sec:procedure})  we propose a procedure  to set the suitable sequences $\{K_i\}_{i=1}^n$, and we  analyze also some limiting cases  where $\epsilon_{ij}=0$.

Concerning the gradient of the differential rotation or the derivative $\partial_i \ell$, with respect to the configuration index, we note that, if $\epsilon_{ij}$ is the distance between the two specific  angular momenta,  the distance between $(r_{min}^i, r_{min}^j)$  is fixed and it will be a function of  $\epsilon_{ij}$.
As any specific angular momentum is attached to  a  ring center, we could characterize the differential rotation in a more general sense as a discrete function   of the sub-configuration index $(i)$
and then evaluate the derivative $\partial_i$ as   $\partial_r\ell_{crit}$ for $r $ being a discrete variable with values in  $\{r_{min}^i\}_{i=1}^n$, see also Secs.\il(\ref{Sec:app-maxmin}) and (\ref{Sec:pertur}). We will show in Sec.\il(\ref{Sec:procedure}) and Sec.\il(\ref{Sec:app-maxmin}) that the differential rotation for any  $\ell$corotating sub-sequence   is always increasing in magnitude  with the configuration index (that is with the distance from the source) but not monotonically with the index (for equidistant centers).

We   note finally that, if  we start from two generating $\ell$counterrotating  rings, the \emph{seeds}  $C_1^{\pm}$, as the inner rings of the two $\ell$counterrotating sub-sequences, then,   the  displacement matrices in Eq.\il(\ref{Eq:Marix-ex-spo})  might automatically generate an isolated decomposition  with a shift  between the two subsequences    function of  $\epsilon_{1_+ 1_-}$, according to the initial data  on the generating couple
(a similar situation would be obtained by perturbing a suitable preexisting decomposition).
In other words,  a separation in the decomposition by the action of a displacement matrix  between the two $\ell$counterrotating   sub-sequences of the disk could  be obtained,  depending on the initial configuration. Assuming for simplicity to divide the process of the translational displacements generated by  the two matrices   $\epsilon_{ij}^{\pm}$,  on the  $\ell$counterrotating  sequences with  $n=n_++n_-$  with a general initial (seed) couple  $C_{\pm}$, the spacing between the two $\ell$counterrotaing bends will be
\bea\label{Sec:sum-choc-uk-Gre}
\bar{\lambda}_{\mp}=y_3^{1_-}-y_1^{n_+}\quad\mbox{if}\quad C_->C_+, \quad \bar{\lambda}_{\pm}=y_3^{1_+}-y_1^{n_-}\quad\mbox{if}\quad C_+>C_-, \quad \ell_{n_{\pm}}^{\pm}=\ell_{1_{\pm}}+\sum_{i=1_{\pm}}^{n_{\pm}-1}\epsilon^{\pm}_{i,i+1},
\eea
with the elements $\epsilon_{ij}^{\pm}$ small enough to lead to a $\overbrace{\mathbf{C}}_s$ sequence.
However,  for each $\ell$corotating sub-sequence the sum in  Eq.\il(\ref{Sec:sum-choc-uk-Gre})
increases in magnitude with the sub-sequence order  ($n_{\pm}$ respectively).
As we have already noted, the ringed disk has specific angular momenta increasing in magnitude outwardly but one can properly set the  initial rings  such that
$|\ell_-|-|\ell_+|\approx0$, and if we can choice $K_{\pm}$ such that $\mathcal{A}_-\approx \mathcal{A}_+$, then the angular momentum of the disk $\bar{\ell}^{n}_{\mathbf{C}}\approx |\ell_{\pm}| |n_+-n_-|$.
\subsubsection{Some notes on the ring specific angular momentum and the differential rotation of the disk}\label{Sec:procedure}
The curves $\ell_{\pm}$  plotted  in Fig.\il(\ref{Figs:Aslanleph1l},\ref{Figs:Aslanleph1lb}),
locate, for  $\ell_i=$constant, the critical points $r^i_{crit}=\{r^i_{Max},r^i_{min}\}$ of the $C_i$ ring, that is  the  center on  $C_i$ and the inner  edge of its unstable phase.
The investigation of the decomposition  of a $n$-order disk  with  the  unstable rings  is simplified by noting that the curves  $K^i_{crit}=$constant, plotted  in Fig.\il(\ref{Fig:bCKGO}), locate    exactly  the    inner edge  and the \emph{outer} edge of the $C_x^i$ rings but not,  directly, the ring center $r_{min}^i$  and therefore the value $K_{min}^i$.
Moreover, as the curves $K^i_{crit}=$constant  provide $K_{Max}^i$ for the unstable phases (but not $K_{min}^i$),  the information on the ring topology in its unstable phase  is  provided, when it  can accrete  on the source or create jet.
On the other hand, the location of  $r_{min}^i$ is easily  identifiable, from the plot of the specific  angular momenta, knowing $ r_{Max}^i $ from the curve $ K_{crit} $=constant:
\be\label{Eq:24}
\partial_{r_{Max}^i}r_{min}^i<0\quad\mbox{and}\quad\partial_{|\ell_i|}r_{min}^i>0,\quad
\partial_{|\ell_i|}\lambda_x^i>0,
\ee
where $\lambda^i_x$ is the elongation of $C_i$ ring in its accretion phase (if this exists).
Here and in the following  $\partial_{\mathbf{B}}\mathbf{Q}>0$ we mean  the quantity $\mathbf{Q}$ increases as the quantity $\mathbf{B}$ increases and viceversa--see also
\ref{NoteIII}.
Furthermore, the elongation  $\lambda$ of the equilibrium disk is smaller than the elongation of its accretion configuration $\lambda_x$:
\be\label{Eq:board-point}
\lambda_i<\lambda^i_x, \quad \Lambda_i\subset \Lambda^i_x, \quad  K\in]K^i_{min},K^i_{Max}[\subset]K_{mso},K^i_{Max}[,
\ee
$ \Lambda^i_x\equiv [y_3^i,y_1^i]_x$ is the elongation range of the accreting  ring.
However the  inner edge of the disk in accretion must be $\left.y_3^i\right|_x=r_{Max}^i\in ]r_b,r_{mso}[$, therefore $r_{mso}\in C^x_i$,  but for a disk in equilibrium  $y_3^i\in]r_{Max}^i,r_{mso}[$, therefore $r_{mso}\in C_i$ or
$y_3^i\in]r_{mso},r_{min}[$, where $r_{mso}{\large{\not}}{\in}C_i$ does not belong to the $C_i$ disk.
The elongation of the disk in the first configuration is larger then the elongation of the second and therefore it requires a $K_i\in]K_{mso},K_{Max}^i[$  greater then  in the second configuration.  In Fig.\il(\ref{Figs:CrystalPl})   an example of how stringent can be the choice of the  $K$ parameter in the decomposition with a ring in accretion is shown.

\textbf{$\ell$corotating rings}

We focus first on the case of the $\ell$corotating   rings as shown in  Fig.\il(\ref{Figs:CrystalPl}) and Fig.\il(\ref{Figs:Aslanleph1l},\ref{Figs:Aslanleph1lb}) where there are  the curves  $\mp\ell_{\pm}$ for different  attractors in
$r>r_{\gamma}^{\pm}$ respectively.
The inner rings of each  $\ell$corotating sequence of { not necessarily} consecutive rings, have  their specific  angular momentum  magnitude lower than the outer ones,  $|\ell_a|<|\ell_b|$  $\forall\;C_a<C_b$.
To ensure the separation between the outer lobes of the $\ell$corotating  disks in accretion as in  Eq.\il(\ref{Condizioni}),
only the inner ring of each $\ell$corotating  sub-sequences of { not necessarily } consecutive rings can be in accretion,  i.e. in $C_x$ topology, implying that no ``feeding'' is possible by a P-W instability mechanism among the $\ell$corotating rings of a ringed $\mathbf{C}_x^n$ configuration. 
The configurations $ C^i_x $  must be the  inner one of the $\ell$corotating (non necessarily consecutive) sequences. We can thus deduce  the following  result:
given a ringed disk  of order $n$ and any its decomposition with specific   angular momenta $\{ \ell_i\}_{i=1}^n$,
the subset of  not \emph{necessarily consecutive}  $\ell$corotating rings with specific angular momenta  $\{\ell_j\}_{j=1}^k$, $k\leq n$,
allows \emph{only} one configuration $C_x$--only  the  inner  ring of the disk can accrete onto the attractor  and  the  inner one has the  lowest specific angular momentum  magnitude.
Thus the outer configuration must be  in  $ C$ topology,
see  Fig.\il(\ref{Figs:CrystalPl}).
The highest possible rank   of  a $\mathbf{C}_x^n$ configuration is then $\mathfrak{r_x}_{Max}=1$, and accretion onto the black hole can occur only in the inner corotating or the inner counterrotating ring, see Fig.\il(\ref{Figs:CrystalPl}).
An unstable $\ell$corotating  couple of rings $(C, O_x)$  could be characterized by $\ell_{i/j}=1$, the couple having an  unique center of maximum pressure  provided that  $\ell_i\in]\ell_{\gamma},\ell_{mbo}[$, that is  an $r_{Max}\in]r_{\gamma},r_{mbo}[:\; K_{Max}>1$. One could have two centered configurations at equal (algebraically) specific angular momentum, a  closed one in equilibrium, and the other being open and governing a  jet with launching point placed in $r_{Max}$. In any case  also
 the couple   $(C, O_x)$  would violate the condition in Eq.\il(\ref{Condizioni}), this case will be considered in more details in Sec.\il(\ref{App:opem}).

It follows then that the rings of the  $\ell$corotating sub-sequences  of an $n$-order decomposition satisfy  the property $\bar{\mathfrak{C}}_{1b}$ as in (\ref{C1b}).
A specific characterization of the decompositions $\bar{\mathfrak{C}}_{1b}$ is  in  Sec.\il(\ref{Sec:i4cases}). It is therefore necessary to briefly discuss the condition on maximum in
$\bar{\mathfrak{C}}_1$ in Eq.\il(\ref{Condizioni}) for the  $\ell$corotating sequences:
with  reference to Fig.\il(\ref{Figs:Aslanleph1l},\ref{Figs:Aslanleph1lb}), we consider a couple of constant specific  angular momenta $\ell=\ell_i,\; i\in\{a, b\}$  with  $\ell_a<\ell_b$. Then
$ \Delta_{crit}^a\subset\Delta_{crit}^b$,
in other words the ringed $2$nd  order configuration $\mathbf{C}^2=C_a\cup C_b$ is a $\bar{\mathfrak{C}}_{1b}$ one. Clearly, this property generalizes straightforwardly to any order $n$.
However,
 it is immediate  to prove the condition {in Eq.\il(\ref{Condizionibar})}.
Increasing the spacing between the rings centers, or by adding a ring  to a $\ell$corotating ringed configuration,  it is necessary  to supply additional angular momentum to the outer ring.
\footnote{We note that  the critical  $\ell$corotating-rings and  the correspondent effective potentials are  centered in $r_{mso}$.} Considering the definition (\ref{ratiolll}), there is always $|\ell_{i/i+1}|\in]0,1[$,
 the rings do not need to be consecutive  and then  one can generalize  this relation as
\bea
0<|\ell_{i/j}|<1, \;\forall i<j \quad C_i<C_j\quad \ell_i\ell_j>0.
\eea
Another way to express this result is the following: the distance  between the two minima  $(r_{min}^a,r_{min}^b)$ is increasing with increasing   the difference between the two specific angular momenta  magnitude.  This can be seen from the slope of the curve $\ell(r)$ at $r>r_{mso}$, or from the radial first derivative of the specific  angular momentum in $r>r_{mso}$  which is always positive. The ring elongation and  thickness is  determined, as in Eqs.\il(\ref{Eq:24},\ref{Eq:board-point}), by the gap in the specific angular momentum in condition (\ref{Condizioni}).
However, it is important to note that  the derivative $\partial_r(\mp \ell^{\pm})$ does not  grow monotonically\footnote{The distance between two the centers of the  two $\ell$corotating  configurations  $|\ell_i|<|\ell_j|$ is
$\delta_{min}^{i, j}\approx\left(\nabla \ell^{\pm}\right)^{-1} (\ell_j-\ell_i)$ where  $\nabla \ell^{\pm} $ is the incremental ratio of the specific angular momentum function,  but if the configuration indices are  $j=i+\epsilon$ with $\epsilon\gtrapprox0$, that is, the two rings have very close values of the  specicic angular momentum (and  therefore very close  effective potentials) the  spacing among them is very small. The elongation is determined trivially  from the radial  derivative of the  specific  angular momentum.} but it has a maximum point  $r^{\pm}_{\mathcal{M}}$ respectively, different for each attractor, as  we detailed in  Sec.\il(\ref{Sec:app-maxmin}).
Increasing  of  the specific angular momentum   magnitude  with the radial distance  from the attractor is not constant  but it rises up to a limiting radius,
  different for the $\ell$counterrotating sequences. Radii $r^{\pm}_{\mathcal{M}}$ are functions of the attractor spin-mass ratio,  Figs.\il(\ref{Figs:PlotLKKdisol},\ref{Figs:PlotLKKdisolb}).
As proved in Sec.\il(\ref{Sec:app-maxmin}),  this is a relativistic effects that disappears in the Newtonian limit  when the orbital distance is large enough with respect to    $r^{\pm}_{\mathcal{M}}$, and $r^{+}_{\mathcal{M}}=r^{-}_{\mathcal{M}}$ disappears in the case of static spacetimes; clearly these  effects occur  due to the presence of the  intrinsic spin of the attractor.
This behavior of the specific angular momentum   distinguishes the two $\ell$corotating  sub-sequences   of corotating and    counterrotating rings   of a  generic decomposition: the \textbf{BH} spin has a stabilizing effect for the corotating  matter  and  destabilizing for the counterrotating one, as  $\partial_{|a|}r_{\mathcal{M}}^\mp\lessgtr0$.
We prove these results  in Sec.\il(\ref{Sec:app-maxmin}).

\medskip

We consider first the case of a $\ell$corotating  $ \mathbf{C}_{\odot}^3 $ configuration of rank $\mathfrak{r}\leq2$.  There must be at least a couple $y_1^a=y_3^b$:
\be
r_{Max}^a<r_{Max}^b<r_{Max}^c<r_{mso}<y_3^a<r_{min}
^c<y_1^a=y_3^b<r_{min}^b<y_1^b<y_3^a<r_{min}^a<y_1^a,\quad\mbox{where}\quad
|\ell_a|>|\ell_b|>|\ell_c|.
\ee
%
We note that the specific angular momentum of the fluid and the associated effective potential   grow in magnitude simultaneously. 
This property holds  in particular  in the critical points $K_{crit}$, as it is clear from Figs.\il(\ref{Figs:Aslanleph1l},\ref{Figs:Aslanleph1lb})  \ref{NoteI} and \ref{NoteIII}). Then, increasing the magnitude of the fluid momentum, at fixed $r$ and therefore at fixed $y_1^a=y_3^b$,  there are increasing values of $V_{eff}$ and viceversa,   it is:
\be\label{Eq:21}
V_{eff}(\ell_a,y_1^a)=K_a>K_b=V_{eff}(\ell_b,y_3^b)\quad\mbox{for}\quad y_3^b=y_1^a,\quad\ell_a\ell_b>0,\quad |\ell_{ab}|>1,
\ee
see \ref{NoteI}, \ref{NoteII} and \ref{NoteIII}.
On the other hand, the maximum in $r_{Max}\in]r_{\gamma},r_{mso}[$ decreases with the magnitude of the angular momentum  at fixed $K$, that is increasing  $K_{min}$, potentially the disk elongation  tends to increase because  the distance between the point of maximum and minimum increases; as for fixed $\ell_a:\; K_{Max}^a>K_{min}^b$--see Fig.\il(\ref{Figs:Aslanleph1l},\ref{Figs:Aslanleph1lb}).
 This relation  can be iterated for any order $n$, see also (\ref{a.})) and (\ref{b.}).

We consider now  the parameters $K$ associated with two separated sub-configurations.
For $
|\ell_a|<|\ell_b|$, it has to be
$K_{mso}<K_{min}^a<K_{min}^b<K^b$
as shown in Fig.\il(\ref{Figs:Aslanleph1l},\ref{Figs:Aslanleph1lb}),
but we can have $K_a=K_b$  or $ K_a\neq K_b$. Therefore, the relation $ (K_a,K_b) $ and the  elongation of the rings is not yet settled.
It is worth noting that there may be two different specific angular momenta   such that $K_{Max}^a=K_{min}^b$; this is the case  illustrated in \citep{Pugtot} and considered in Sec.\il(\ref{Sec:Kamx=Mmin}).

For a  $ \mathbf{C}_{\odot}^n $ ringed disk
we deduce  from Fig.\il(\ref{Fig:bCKGO},\ref{Figs:Aslanleph1l},\ref{Figs:Aslanleph1lb}),
that  fixing, say, the inner  margin
$y_3^1$ of the the inner sub-configuration $C_1$ (or equivalently $K_1$ or  $y_1^1$  fixing one of the three parameters $(y_3^i,y_1^i, K_i)$ automatically identifies the other two quantities), then the entire ringed  $ \mathbf{C}_{\odot}^n $ structure is fixed.  In fact,  considering the potential
$V_{eff}^{\mathbf{C}^n}$ in  Eq.\il(\ref{Eq:Vcomplessibo}), if it exists, then there  is  \emph{one and only one}   $\ell$corotating sub-sequence of  the ringed decomposition of  $ \mathbf{C}_{\odot}^n $
 with $y_1^{i+1} =y_3^i$.
Then we say $y_3^1$  to be  a ``starting point''  \emph{uniquely} fixing the decomposition of the  $\ell$corotating sub-sequence of  $\mathbf{C}_{\odot}^n $.

 If the ringed $\mathbf{C}_{\odot}^n $ with $n>2$ is made  by $\ell$counterrotating  rings, we can say $y_3^{1_{\mp}}$  to be  a ``starting point'' to fix the  \emph{two} sequences of $\ell$corotating disks of the decomposition of $\mathbf{C}_{\odot}^n$ of  corotating $(y_3^{1_{-}})$ and counterrotating $(y_3^{1_{+}})$ disks separately,  maintaining   then the   constraints imposed by  separation of the rings.
 So  the couple $y_3^{1_{\mp}}$  constraints the decomposition uniquely but not entirely. As  we are not considering here the $\ell$counterrotating tori together:
given $y_3^{1_{\mp}}$, we fix the two sequences separately to be combined  to  form of a  $\mathbf{C}_{\odot}^n$ disk. Then it could be in need to make a reduction of the order $n=n_++n_-$. The number of  rings in fact depends   on the differential rotation, but mostly by the choice of the ``starting point'': the greater the elongation in the equatorial plane of the first ring, then smaller in general is the second $\ell$corotating disk. So one  could say that the elongation of two consecutive  $\ell$corotating disks is inversely proportional--$\partial_{\lambda_{i}}\lambda_{i+1}<0$  exact law  could be found  that depends on the specific angular momentum, irrespectively of the  number of consecutive\footnote{The hypothesis of consecutiveness between the rings is adopted here only to explicit the exclusion of a possible  $\ell$counterrotating configurations $C^+_{i+1}$ between  two $\ell$corotating rings $C^-_i<C^+_{i+1}<C^-_{i+2}$ that would clearly impose further restrictions  on the starting point. This case is discussed  later for  the $\ell$counterrottaing rings.}  $\ell$corotating disks. Some general notes on the perturbations of $\mathbf{C}^n$  and the relation with the   elongation and spacings have been addressed in  Sec.\il(\ref{Sec:pertur}) and Sec.\il(\ref{Sec:app-maxmin}).
More generally, regarding  the relation between the order of the decomposition and the distances $\delta_{min}^{i,j}$ between two minima of an $\ell$corotating sequence, one can deduce from Fig.\il(\ref{Figs:CrystalPl})  that the elongation of the ordered $\ell$corotating disks,  although of not consecutive rings, depends on $\ell$ and the starting point of the iteration. We can further clarify these statements as follows, assuming for convenience the iterative parameter $K$ for a generic $\mathbf{C}^{n+1}$ configuration, and fixing the inner edge with $K_0$.
As discussed   in Sec.\il(\ref{Sec:onK})
then $K$  should  obey  a
necessary condition  verified for each $\ell$corotating sub-sequence. Considering the  property $\partial_{|\ell|}r_{min}>0$, it has to be
\bea\label{Eq:cononK}
&&K_0\in]K_{min}^0,{V_{eff}(\ell_0, r_{min}^1)}],
\\\nonumber
&&K_1\in ]K_{min}^1,\inf\{V_{eff}({\ell_1},r_{min}^2),V_{eff}(\ell_1, y_1^0)\}],
\\\nonumber
&&\hspace{2cm}...
\\\nonumber
&&K_i \in ]K_{min}^i,\inf\{V_{eff}(\ell_i,r_{min}^{i+1}),V_{eff}(\ell_i, y_1^{i-1})\}],\quad \forall i\in[1,n-1]
\\\nonumber
&&\hspace{2cm}...
\\\nonumber
&&K_n\in]K_{min}^n, V_{eff}(\ell_n, y_1^{n-1})].
\eea
This  is a necessary but not sufficient condition  for the existence of a $\ell$corotating ringed disk $\mathbf{C}^{n+1}$.
If it is a  $\mathbf{C}_{\odot}^{n+1}$ configuration  with rank $\mathfrak{r}=\mathfrak{r}_{Max}=n$,  then  $K_i= V_{eff}(\ell_i,y_1^{i-1})$, and it is univocally determined by the appropriate choice of $K_0$ with the constraint $K_0\in]K_{min}^0,V_{eff}(\ell_0,r_{min}^1)]$. The condition could be written as  $y_3^o:\; y_1^0\in]r_{min}^0,r_{min}^1[\cup...y_1^i\in]r_{min}^i,r_{min}^{i+1}[$.
The construction of this ringed disk is carried out in an iterative process where the starting point is not necessarily $K_0$. In the case of $\mathbf{C}_{\odot}^{n+1}$ topology,   the choice of  $K_0$ uniquely fixes the sequence $\{K_i\}_{i=0}^{n}$.

For the  case of  $\mathbf{C}^n$ ringed disk,
considering  the distance $\delta_{min}^{j,i}=r_{min}^j-r_{min}^i$ between the minima of the $C_i<C_j$, the maximum elongation $\lambda_i^{Max}$ as the upper bound realized at zero spacings between consecutive rings of the configurations $\mathbf{C}_{\odot}$ and  finally the maximum possible spacing $\bar{\lambda}^{Max}_{i+1,i}$ between the  $C_i$ and $C_{i+1}$ rings, the following relations hold:
\bea\label{Eq:neut-w}
\frac{\partial\delta_{min}^{j,i}}{\partial \ell_{j/i}}>0, \quad \frac{\partial\lambda_i^{Max}}{\partial \ell_{i/(i-1)}}>0,\quad  \frac{\partial\bar{\lambda}_{i+1,i}^{Max}}{\partial \ell_{(i+1)/i}}>0, \quad
\frac{\partial\lambda^{Max}_{i}}{\partial {\lambda}_{i-(2k+1)}}<0, \quad\frac{\partial\lambda^{Max}_{i}}{\partial {\lambda}_{i-2k}}>0,\quad 
\forall i<j\quad k\in\mathds{N}.
\eea
see also \ref{NoteIII}, \ref{NoteIV}, these variations could  indeed be evaluated exactly, for example  in the treatment of the perturbations as discussed in  Sec.\il(\ref{Sec:pertur}).
However, for a  fixed  $\ell$corotating sequence $\{\ell_i\}_1^n$,
for the $\ell$corotating couple $C_a> C_b$ within the  \ref{NoteIII} and ((\ref{b.})), 
 we can have  $K_a=K_b$ or $K_a
\neq K_b\in]K_{min}^b,K_{Max}^b[\subset]K_{min}^a,K_{Max}^a[$.
In the latter case, however,  any inequality  can be satisfied because none of the two ring edges  is a critical point. There are no crossing  points of the two functions $V_{eff}^a> V_{eff}^b$ (see Fig.\il(\ref{Figs:CrystalPl}) in the stability region, or for any $a\neq b$  $\nexists r$:
$\left.V_{eff}(\ell_a)\right|_{r}=\left.V_{eff}(\ell_b)\right|_{r}$, with  $\ell_a
\ell_b>0$ (with the exclusion of  $\ell_a=\ell_a$, in that case clearly it is only one effective potential), and  for a $\mathbf{C}_{\odot}^n$ configuration Eq.\il(\ref{Eq:21})  holds.

A overview of the results for the $\ell$corotating sequences is in Table\il(\ref{Table:sum-BH})

\medskip

\textbf{$\ell$counterrotating rings}

We now study the   $\ell$counterrotating sequences of a $n$-decomposition,  considering  simultaneously  the couple of curves $\ell_{\pm}$ shown  in Figs.\il(\ref{Figs:Aslanleph1l},\ref{Figs:Aslanleph1lb}).
As discussed in \citep{Pugtot}, many  morphological characteristics of a toroidal  accretion disks show a symmetry under the transformation $\ell\rightarrow-\ell$, then we expect  a similar  symmetry to be  conserved in the macro-structure $\mathbf{C}^n$, and so find  analogies  between the $\ell$corotating  and  $\ell $counterrotating   sub-sequences.
A  property that we will  often use is the following:
\bea\label{Eq:sto-bru-cioc}
\forall\; \bar{r}>r_+, \;\; \ell>0\quad &&V_{eff}(-\ell,\bar{r})<V_{eff}(\ell,\bar{r}),
\\
&&K_{crit}(\bar{r},\ell_+)>K_{crit}(\ell_-)\quad \mbox{and }\quad -\ell_+(\bar{r})>\ell_-(\bar{r}),
\eea
and an equality in Eqs.\il(\ref{Eq:sto-bru-cioc})  occurs  only for $a=0$.
It is convenient to start our consideration with the analysis of the limiting case
$\ell^{-}=-\ell^+$, that is $\ell_{i/j}=-1$ for some $C_i$ and $C_j$. It is simple to prove that if $C_i<C_j$, then it follows that  $C_i$ is counterrotating and $C_j$ is corotating with respect to the black hole, therefore
\be\label{Eq:rmsorminrmas}
r_{Max}^-<r_{Max}^+<r_{mso}^+<r_{min}^+<r_{min}^-, \quad K_{mso}^-<K_{mso}^+<K_{min}^+<K_{min}^-,\quad K_{Max}^+<K_{Max}^-,\quad \quad\mbox{for}\quad \ell_{i/j}=-1,
\ee
see Fig.\il(\ref{Figs:Aslanleph1l},\ref{Figs:Aslanleph1lb}).
The rings with $\ell_{i/j}=-1$ are possible only in $r>r_{mso}^+$ and $\ell>-\ell_{mso}^+$,
we note that Eq.\il(\ref{Eq:rmsorminrmas}) does not fully constraint the couple $K_{\pm}$, while the relation  between the critical points  follows from the following considerations:
\be\label{Eq:hi-nu-lla-te}
\mathbf{K_{min}^+<K_{min}^-:} \quad V_{eff}(-\ell,r_{min}^+)=K_{min}^+< V_{eff}(-\ell,r_{min}^-)<V_{eff}(\ell,r_{min}^-)=K_{min}^-<V_{eff}(\ell,r_{min}^+).
\ee
The first inequality in Eq.\il(\ref{Eq:hi-nu-lla-te}) is given as  definition of minimum of $V_{eff}(-\ell)$;  using  the first of (\ref{Eq:rmsorminrmas}), the second  of  Eq.\il(\ref{Eq:hi-nu-lla-te}) is an application of Eq.\il(\ref{Eq:sto-bru-cioc}),  the third  follows from the definition of minimum point of the function $V_{eff}(\ell)$. Finally,  in the fourth inequality closing Eq.\il(\ref{Eq:hi-nu-lla-te}), the property (\ref{Eq:sto-bru-cioc}) has been applied again.   Similarly for the maxima $K_{Max}$ one has
\be\label{Eq:Schr-webN}
\mathbf{K_{Max}^+<K_{Max}^-:}\quad
K_{mso}^+<V_{eff}(-\ell,r_{Max}^+)=K_{Max}^+<V_{eff}(-\ell, r_{Max}^-)<V_{eff}(\ell,r_{Max}^-)=K_{Max}^-,
\ee
where we used   the definition of maximum point assuming that each function is well defined, i.e., in this case it means to consider the location of  $r_{\gamma}^{\pm}$,
 see Fig.\il(\ref{Fig:bCKGO}). On the other hand, the limiting case $V_{eff}(-\ell^+)=V_{eff}(\ell^-)$ is verified  only in the static geometry  with $a=0$.  Each (consecutive) couple  $\ell_{i/i+1}=-1$ is $\bar{\mathfrak{C}}_{1b}: \; \Delta_{crit}^+\in\Delta_{crit}^-$ as defined in  \ref{C1b} of Sec.\il(\ref{Sec:caracter}).    More generally, for any couple of rings, even not consecutive at  $\ell_{a/b}=-1$, the condition  $\bar{\mathfrak{C}}_{1b}$ holds, but the relationship between the critical points associated with elements of different couples, i.e., for a general ratio $\ell_{a/b}<0$, can be different.  The maximum order of a macro-configuration $\mathbf{C}^n$ where all rings have $\ell_{i/j}=-1$ is $n_{Max}=2$:  in other words, there cannot be more then two rings with $\ell_{i/j}=-1$.
However there can be possibly an infinite number of couples $\ell_{a/b}=-1$ and $\ell_{c/d}=-1$, and we consider the case of two couples  closing  this section.
This analysis, being clearly important in  investigation of   the  unstable configurations, considers the maximum points of the effective potential $V^i_{eff}$ associated with each ring (i.e. $|\ell_{\pm}|\in]|\ell_{mso}^{\pm}|,|\ell_{\gamma}^{\pm}|[$).
The inner counterrotating configuration  can be  then in accretion or even opened for jets.

 As we have already discussed, the presence of an outer ring in its unstable phase  would imply a violation of the condition (\ref{Def:separated}) on the separation of material of different rings. However, especially in the case of couples with open external configurations  in jets, this condition could be relaxed and a thorough study of this case could be done. From the  analysis of the instability of  corotating ring  of the couple  $ C_i^+ <C_j^-$ with  $ \ell_{i/j} = - 1 $, we can draw useful  general conditions of instability in the ringed disks.
It is not necessary to consider the maximum for the corotating ring, but if it exists then, for any couple of $\ell$counterrotating rings with  $\ell^{-}=-\ell^+$, one of the two conditions has to be satisfied:    
\be\label{Eq:casesCChact}
\breve{\mathrm{C}}_{I}:\;r_{Max}^+\in]r_{Max}^-,r_{mso}^-[\quad\mbox{or} \quad \breve{\mathrm{C}}_{II}:\;r_{mso}^-\in]r_{Max}^-,r_{Max}^+[,
\ee
see Figs.\il(\ref{Figs:Aslanleph1l},\ref{Figs:Aslanleph1lb}).
In general,
the  two cases $(\breve{\mathrm{C}}_{I}, \; \breve{\mathrm{C}}_{II})$ are regulated  primarily by  the parameter $K_{crit}$  and the properties of the \textbf{BH} attractor  through its spin-mass ratio $a/M$. Indeed, for $a/M$ sufficiently low and specific angular momentum sufficiently large
\footnote{With reference to Fig.\il(\ref{Figs:Aslanleph1l}) for  $C_x^-$ ring, requiring \(K_-=K_{Max}^-<1\), and \(r_{Max}^-\in]r_{mbo}^-,r_{mso}^-[ \), it has to be
\(\ell_-\in ]\ell_{mso}^-,\ell_{mbo}^-[\), for the    $O_x^-$ rings, requiring $K_-=K_{Max}^->1$, and \(r_{Max}^-\in]r_{\gamma}^-,r_{mbo}^-[\),
it has to be  \(\ell_-\in]\ell_{mbo}^-,\ell_{\gamma}^-[\), and  similar relations can be found for the counterrotating case.},
 there is always a specific angular momentum  $\ell=\mp\ell_{\pm}$ with $r_{Max}^-$; these  conditions are discussed in the following points.
\begin{description}
\item[Existence of $r_{Max}^-$]
The first necessary  condition    for the existence of $r_{Max}^-$ with   $\ell_{i/j}=-1$   is $\ell_{\gamma}^-\geq-\ell_{mso}^+$, this is the case  only of ringed disks orbiting a sufficiently low spin of the  attractor
\be\label{Eq:al-1}
a<a_{\aleph}\approx0.508864526M:
-\ell_{mso}^+(a_{\aleph})=\ell_{\gamma}^-(a_{\aleph})
\ee
 see Figs.\il(\ref{Figs:Aslanleph1l},\ref{Figs:Aslanleph1lb},\ref{Figs:PlotLKKdisol},\ref{Figs:PlotLKKdisolb})).
\begin{figure}[h!]
\begin{center}
\begin{tabular}{cc}
 \includegraphics[scale=0.3]{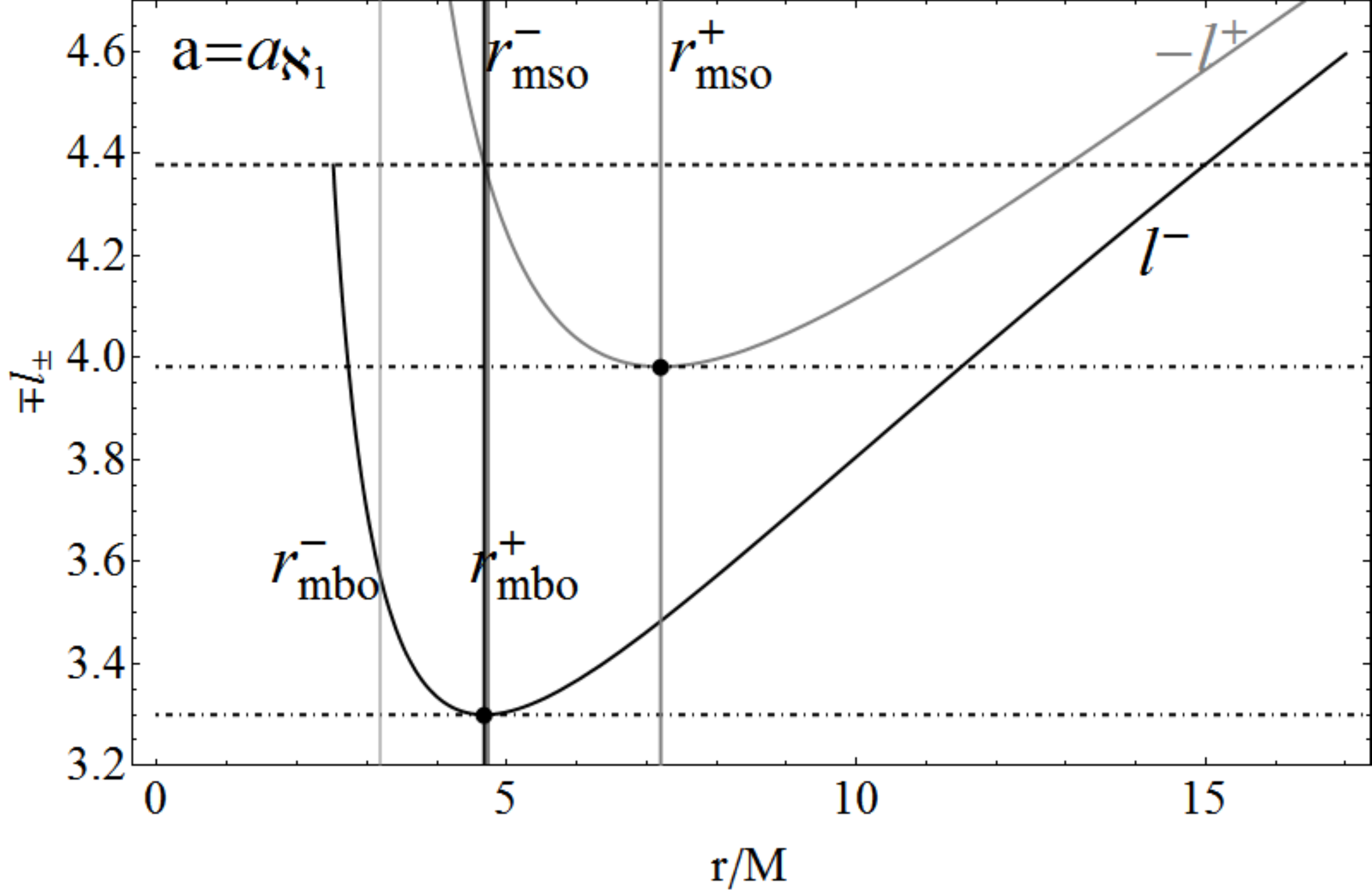}
 \includegraphics[scale=0.3]{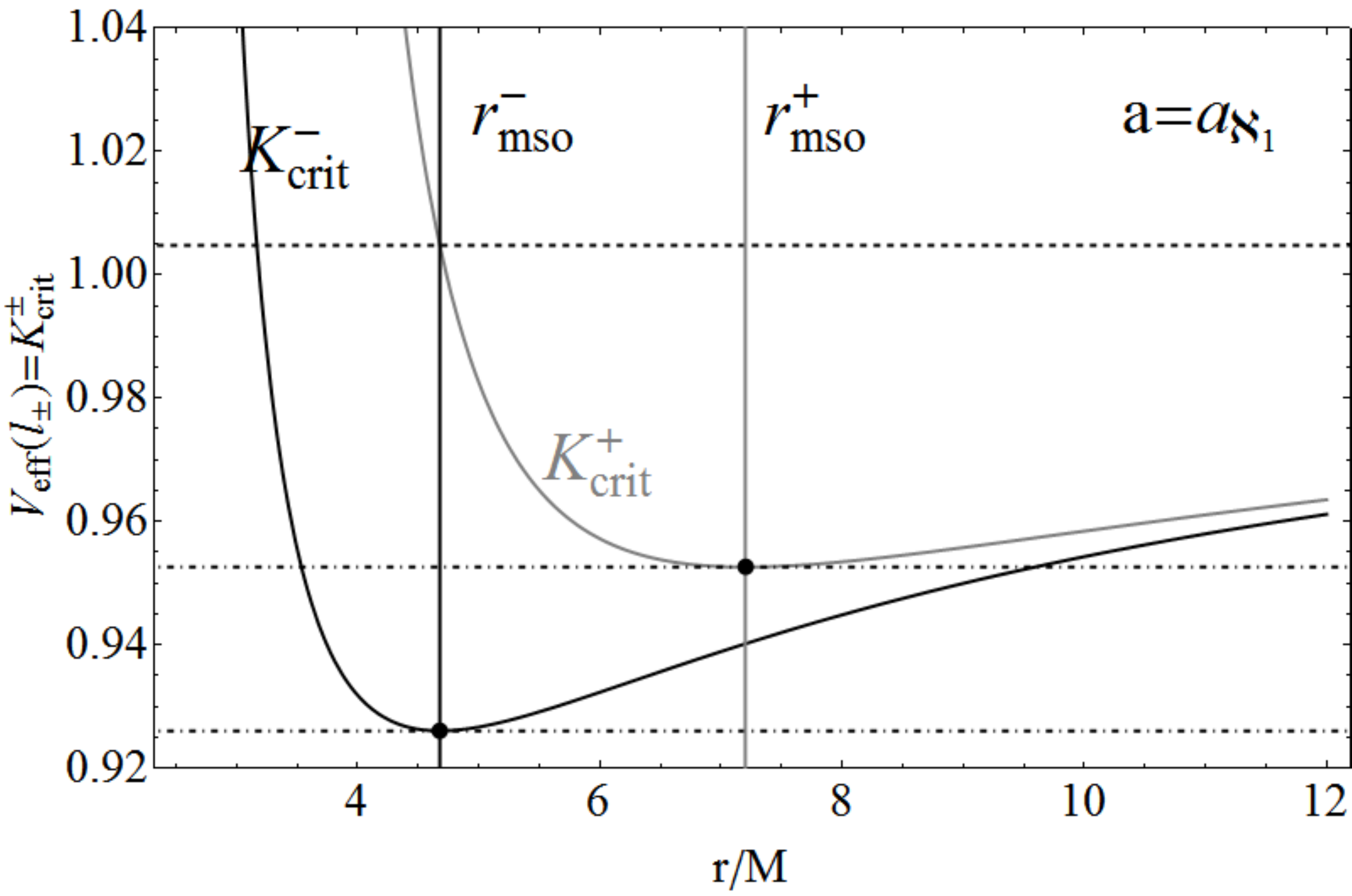}
 \\
 \includegraphics[scale=0.3]{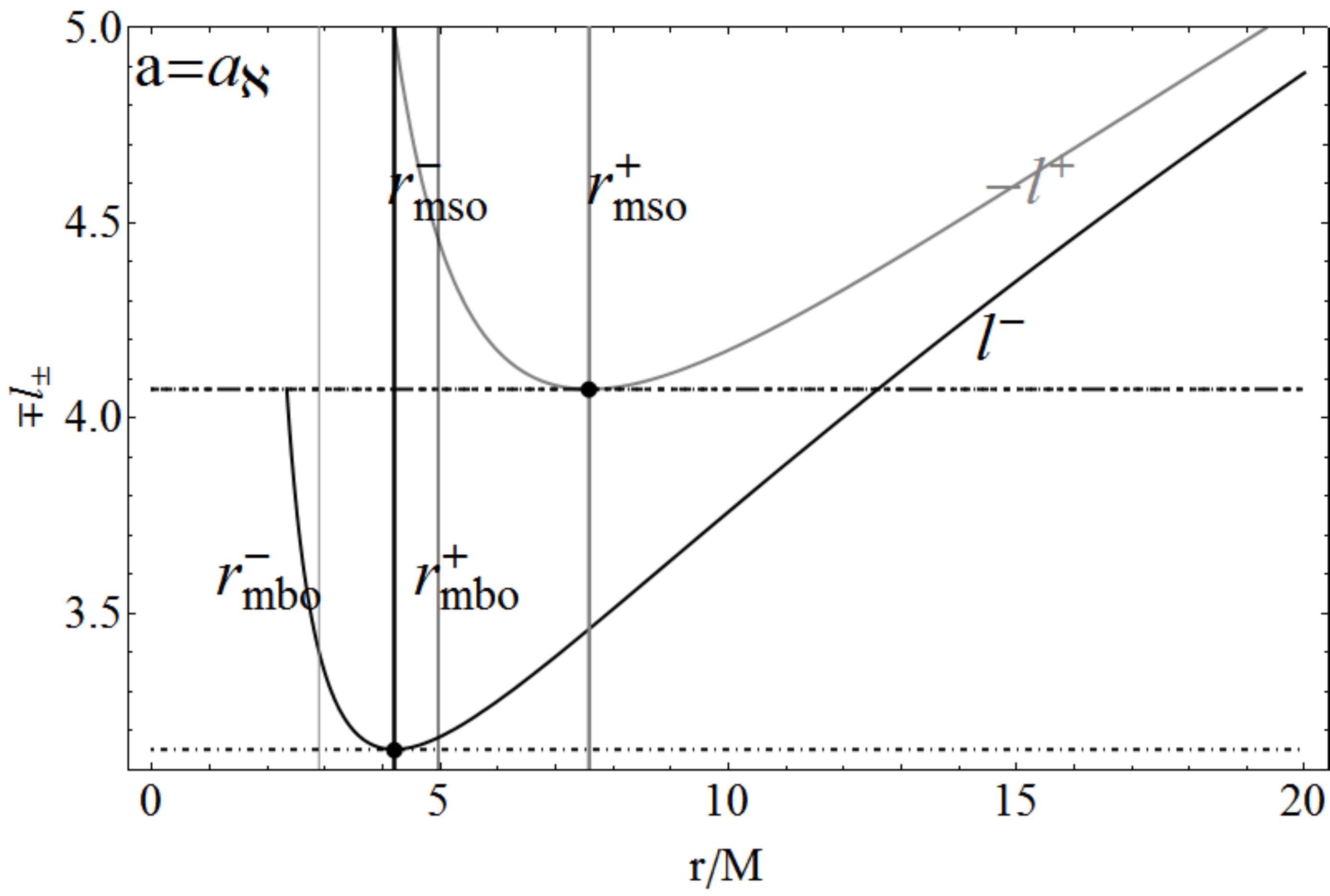}
 \includegraphics[scale=0.3]{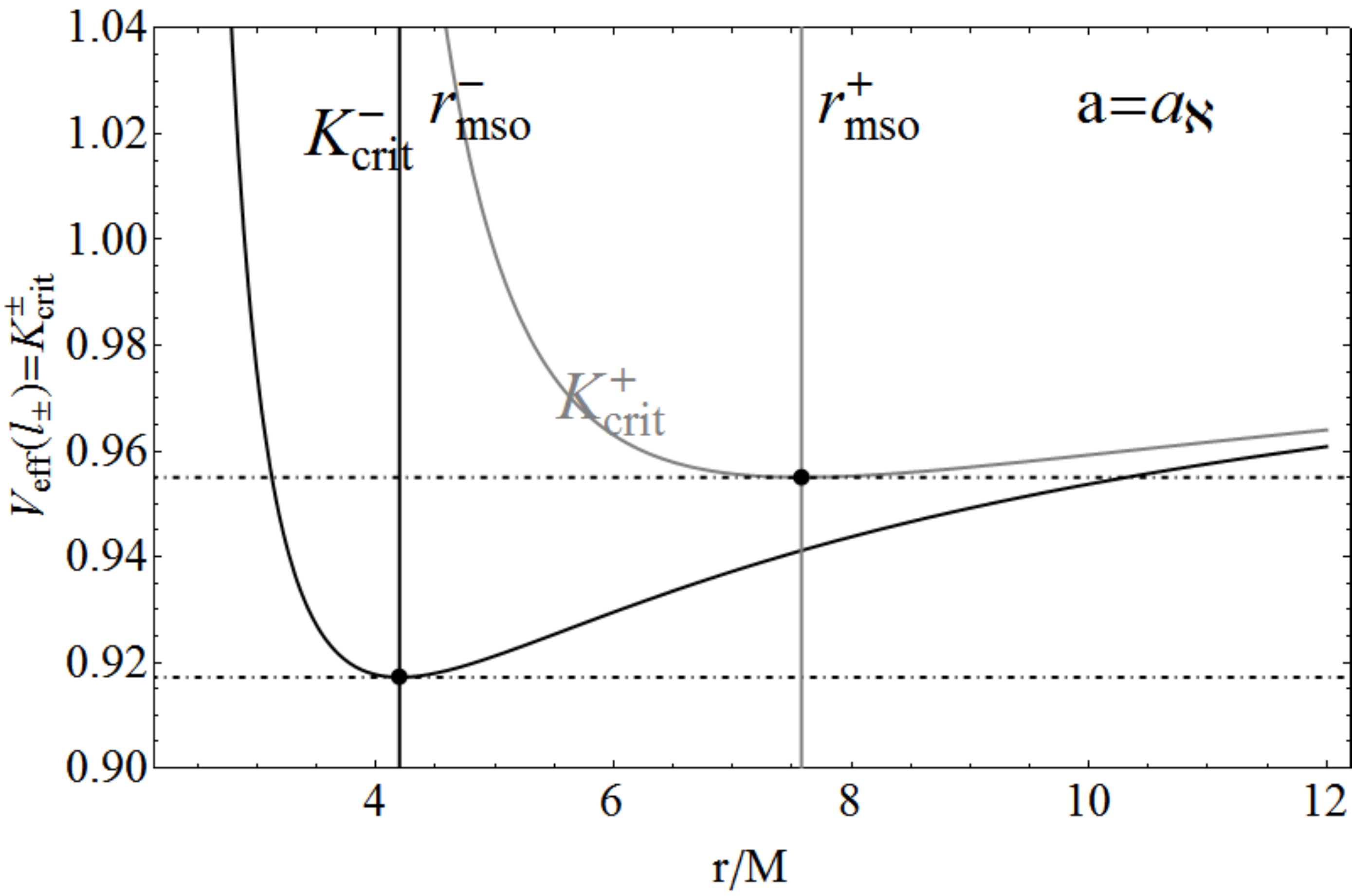}
\end{tabular}
\caption{
Left panels show  the fluid specific angular momentum for corotating $\ell_-$ (black curves)
  and counterrotating $\ell_+$  (gray curves) as functions of the $r/M$. Right panels are the  $K^{\pm}_{crit}=V_{eff}(\ell_{\pm})$ as functions of $r/M$.
Marginally stable circular orbits $r_{mso}^{\pm}$ are signed  with points  on the curves and black/gray lines $r_{mso}^{\pm}=$constant respectively, see also Fig.\il(\ref{Figs:Aslanleph1lb}). On $a_{\aleph_1}=0.382542M:\;\ell_{\gamma}^-=-\ell_{+}(r_{mso}^-)$ and
$a_{\aleph}=0.508865M:\;\ell_{\gamma}^-=-\ell_{mso}^+$, see also Fig.\il(\ref{Figs:Aslanleph1lb}). For the spacetime $a= a_{\aleph_1}$
it is
$r_+=1.92394, $
$r_{mbo}^+=4.73417M$,
$r_{mbo}^-=3.18903M$,
$r_{\gamma}^+=3.41407M$,
$r_{\gamma}^-=2.51784M$. Spacetime  $a= a_{\aleph} =0.508865M$
it is
$r_+=1.86085M$,
$r_{mbo}^+=4.96558M$,
$r_{mbo}^-=2.89276M$,
$r_{\gamma}^+=3.54085M$,
$r_{\gamma}^-=2.33381M$. Where $r_+$ is the outer horizon, $r_{{mbo}}^{\pm}$ are the marginally bounded orbits, $r_{\gamma}^{\pm}$ the marginally circular orbits.
}\label{Figs:Aslanleph1l}
\end{center}
\end{figure}
\begin{figure}[h!]
\begin{center}
  \includegraphics[scale=0.27]{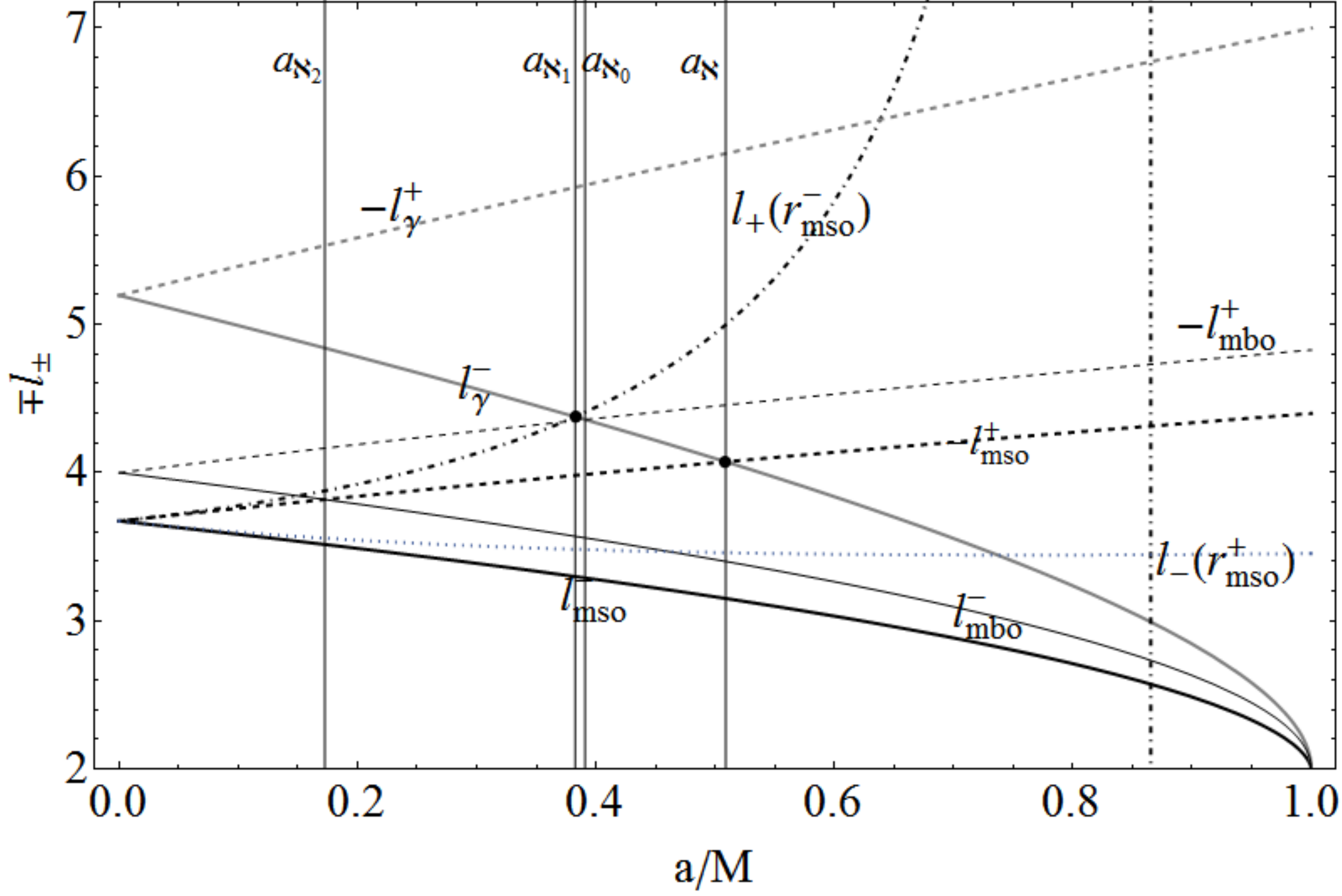}
\caption{Angular momenta for corotating $\ell_->0$ and counterrotating  $\ell_+<0$ fluids, on the last stable circular orbits $r_{mso}^{\pm}$ as functions of the spin-mass ratio of the attractor $a/M$.  $r_{\gamma}^{\pm}$  are the marginally circular orbits,  marginally bounded orbits are $r_{mbo}^{\pm}$ .}\label{Figs:Aslanleph1lb}
\end{center}
\end{figure}
There are  no circular geodesic  orbits at $r<r_{\gamma}^-$, thus if
$\ell=\mp\ell_{\pm}$
such that  $\ell>-\ell_{mso}^+>\ell_{\gamma}^-$,
there is no maximum point for the effective potential of the corotating matter, formally then there is no P-W instability point.
For $a=a_{\aleph}$, there is   $\mp\ell_{\pm}=-\ell_{mso}^+$, and the limiting case $\ell^-=-\ell_{mso}^+$  gives rise to a limiting  unstable line configuration of counterrotating dust material   with the  unstable point at $r_{mso}^+$.
The constraint  $a<a_{\aleph}$ is necessary for the corotating ring of the couple to be unstable. In general this condition applies as well for a  general ratio $\ell_{a/b}<0$ where the inner, corotating ring is unstable  and the distance among the centers, $\delta_{min}^{ba}<\delta_{mso}^{\pm}$.
 However, this condition is only necessary for the existence of  configurations with  equal magnitude of specific  angular momentum and unstable points.  To  distinguish the two cases in Eq.\il(\ref{Eq:casesCChact}), we require a  further constraint on the spacetime spin. These conditions only consider the role of the specific  angular momentum of the couple, once assumed that the maximum should be located in  $r>r_{\gamma}^-$, but they do not set the topology of the unstable configuration. Following the same reasoning, and as $\ell_{mso}^-<\ell_{mbo}^-<\ell_{\gamma}^-$ for  a $C^-_x $ inner ring,  it has to be    $-\ell_{mso}^+<\ell_-\in]\ell_{mso}^-,\ell_{mbo}^-[\subset]\ell_{mso}^-,\ell_{\gamma}^-[$;  in particular,  $\ell_{mbo}^->-\ell_{mso}^+$, see Fig.\il(\ref{Figs:Aslanleph1lb}).  This situation occurs only for $\ell$counterrotating  couple  orbiting attractors with dimensionless spin $a<a_{\aleph_2}\equiv 0.172564M$  where there $
-\ell_{mso}^+(a_{\aleph_2})=\ell_{mbo}^-(a_{\aleph_2})$. For higher spacetime spin  $a>a_{\aleph_2}$, there  is
$\ell_{mbo}^-\in]\ell_{mso}^-,-\ell_{mso}^+[$   and $C_-$ cannot be in accretion phase. Similarly for open configuration       $O_x^-$, it has to be $-\ell_{mso}^-<\ell_-\in]\ell_{mbo}^-,\ell_{\gamma}^-[$, but for this to happen, the condition   $-\ell_{mso}^+<\ell_{\gamma}^-$ has to be satisfied, {i.e. for } $a<a_{\aleph}$. The location of the points of instability can be inferred  from  Fig.\il(\ref{Figs:PlotLKKdisol}). This discussion will be summarized and  explained later.
 \\
 \item[Conditions $\breve{\mathrm{C}}_I$ and $\breve{\mathrm{C}}_{II}$]

 \mbox{}

  For $\breve{\mathrm{C}}_I$ with the     condition $r_{Max}^+<r_{mso}^- $ it should be \footnote{ \label{fnote1} We remind that in terms of the fluid specific  angular momentum only ${C}_{x}$ configurations are at $K<1$ and $\ell\in]\ell_{mbo}^+,\ell_{mso}^+[\cup]\ell_{mso}^-,\ell_{mbo}^-[$,
 ${C}$ rings are for $K\in]K_{mso}^{\pm},1[$,  $\ell<\ell_{mso}^+<0\cup\ell>\ell_{mso}^->0$
and $  r>r_{mso}^{\pm}$,
  $ {O}_{x}$ open configurations  at $K>1$ $\ell\in ]\ell_{\gamma}^+,\ell_{mbo}^+[\cup]\ell_{mbo}^-,
  \ell_{\gamma}^-[$.}
$\mp\ell^{\pm}\in]-\ell_{mso}^+,\ell_{\gamma}^-[$.
Then $ \ell_\gamma^-> -\ell_{mso}^+$
but also  $\ell_{\gamma}^->-\ell^+(r_{mso}^-)$,
see Figs.\il(\ref{Figs:Aslanleph1l},\ref{Figs:Aslanleph1lb},\ref{Figs:PlotLKKdisol},\ref{Figs:PlotLKKdisolb}).
This can be  satisfied only for \textbf{BH} attractors with  $a<a_{{\aleph}_1}\equiv0.38254M$, where for $a=a_{{\aleph}_1}$ there  is $\ell_{\gamma}^-=-\ell^+(r_{mso}^-)$. On the other side, double configurations with
$\breve{\mathrm{C}}_{II}$ are possible\footnote{ In terms of the fluid  specific angular momentum only ${C}_{x}$ configurations have $\ell\in]\ell_{mbo}^+,\ell_{mso}^+[\cup]\ell_{mso}^-,\ell_{mbo}^-[$.
For equilibrium rings ${C}$,  it has to be    $\ell<\ell_{mso}^+\cup\ell>\ell_{mso}^-$
. As pointed out in   \citep{Pugtot}, one can write these results in terms of a rationalized angular momentum $\bar{\ell}\equiv\ell/a$.} in the spacetimes $ a\in]a_{{\aleph}_1},a_{{\aleph}}[$.
\item[Instability of the inner counterrotating ring $C_+$]

\mbox{}

 The role of the radius $ r_{mbo}^+$  is to  distinguish the two cases-- accretion with
 $r_{mbo}^+<y_3^+=r_{Max}^+<r_{mso}^+$,  and launching of jets with  $r_{\gamma}^+<y_3^+=r_{Max}^+\leq r_{mbo}^+<r_{mso}^+$ respectively.
Then it is clear that if  $K_{Max}^+\geq1$, for configurations with open jet from the point  $r_J=r_{Max}^+$,  the corresponding corotating inner ring with   $\ell=-\ell^+$   can not exist, because  it would violate the condition   of no matter penetration. In reality  this condition could be relaxed and, although in  this work we neglect the analysis of open sub-configurations, we can trace some conclusions here, discussing   this case in Sec.\il(\ref{App:opem}).  In some cases, double configurations with $\ell_{i/j}=-1$ can not give rise to any instability with accretion or launching of jets (by P-W mechanism).
Accretion  starts from the point:
\be\label{Eq:firstkeycons}
C_+<C_-\quad r_{Max}^+\in]r_{{mbo}}^+, r_{mso}^+[,\quad\mbox{with}\quad K_{Max}^+<1,\quad |\ell|\equiv \ell_-=-\ell_+>-\ell_{mbo}^+>\ell_{\gamma}^-,
\ee
 the last equality is to ensure the potential associated to the  corotating ring of the pair has a maximum point.
However, for the accretion,  where $r_{Max}^+>r_{{mbo}}^+>r_{{mbo}}^-$, it can be  either
 \bea\label{Eq:CaCb-rinorma}
&&\breve{\mathrm{C}}_{\alpha}:\;r_{Max}^-<r_{mbo}^-<r_{mbo}^+<r_{Max}^+\quad\mbox{with}\quad K_{Max}^+<K_{Max}^-<1 \quad\mbox{or}\quad
\\
&& \breve{\mathrm{C}}_{\beta}:\;r_{mbo}^-<r_{Max}^-<r_{mbo}^+<r_{Max}^+\quad\mbox{with}\quad  K_{Max}^+<1\leq K_{Max}^-,
\eea
where we used Eq.\il(\ref{Eq:Schr-webN}). The conditions above distinguish only  inner  accreting configurations.
Note that, if  the  outer ring $C_-$ cannot generate jets, for  $K_{Max}^-<1$ (that is the case $\breve{\mathrm{C}}_{\alpha}$), then also the inner ring $C_+$ does not fulfil the necessary condition for the generation of the jets.
On the other hand,  if  $K_{Max}^->1$,   then
 it could be that $K_{Max}^+>1$, implying  possibly a double jet from the inner point $r^-_J=r_{Max}^-$ (with matter corotating with the attractor)  and  the outer point of launch $r_J^+=r_{Max}^+$  (with matter counterrotating  with respect to  the central object).  For  the case $\breve{\mathrm{C}}_{\beta}$, we can obtain a combined  instability with  the \emph{inner} (corotating) funnel of matter and the \emph{outer} (counterrotating) accretion point, see for further details  discussion in Sec.\il(\ref{App:opem}).

The two cases in Eq.\il(\ref{Eq:CaCb-rinorma}) are discriminated by the spin requiring a value of:
\be \label{Eq:alm-infla} a_{\aleph_0}\equiv0.390781M\in]a_{\aleph_1},a_{\aleph}[:\ell_{\gamma}^-=-\ell_{mbo}^+.
\ee
 For $a >a_{\aleph_0}$ there is $ \ell_{\gamma}^-<-\ell_{mbo}^+$, when  $K_{Max}^+<1$, and we obtain the cases  $\breve{\mathrm{C}}_{\alpha}$ or $\breve{\mathrm{C}}_{\beta}$.
Clearly, to satisfy  $\ell_{mbo}^->-\ell_{mso}^+$, with open corotating ring, it has to be
$a<a_{\aleph_2}\equiv 0.172564M:
-\ell_{mso}^+=\ell_{mbo}^-$. For higher spacetime spin  $a>a_{\aleph_2}$, it is
$\ell_{mbo}^-<-\ell_{mso}^+$,   and there cannot be
open corotating configurations in jets (within the condition $\ell_{i/j}=-1$).
To have  an open configuration in the spacetimes  with $a>a_{\aleph_2}$,  one has to ensure
a ratio  $|\ell_{i/j}|$ sufficiently high and, therefore, the necessary separation between the points of maximum pressure  (the minimum $r_{min}^{\pm}$ of the effective potential function see also \ref{NoteIII}). This distance
can be easily evaluated and it increases  with the attractor spin, as can be seen in Figs.\il(\ref{Figs:Aslanleph1l},\ref{Figs:Aslanleph1lb}),
evaluating
$\partial_{a_*}(\ell_{mbo}^-+\ell_{mso}^+)<0$
and correspondingly
$\partial_{a_{*}}(r_{mso}^+-r_{mbo}^-)>0$.
\end{description}
We focus now on the constraints of the parameter $K$
for the existence  of  the $\ell$counterrotating sequences  of separated  rings (or with  double points as in $\mathbf{C}_{\odot}$) with couples $\ell_{a/b}=-1$.
Then similarly to condition (\ref{Eq:cononK})  for the $\ell$corotating sub-sequences, we can set the necessary condition
\bea\label{Eq:cononKlcontro}
&&K_+\in]K_{min}^+,{V_{eff}(-\ell, r_{min}^-)}],
\quad
K_-\in ]K_{min}^-,\inf\{V_{eff}({\ell},r_{min}^2),V_{eff}(\ell, y_1^+)\}],
\eea
where we  considered the existence of a third ring centered in  $r_{min}^2$.

If there is only a couple,   then
\be\label{Eq:some-al}
K_+\in]K_{min}^+,{V_{eff}(-\ell, r_{min}^-)}] \quad\mbox{and}\quad K_-\in ]K_{min}^-,V_{eff}(\ell, y_1^+)].
\ee
We note that  consistency   of the constraint   Eq.\il(\ref{Eq:cononKlcontro})  on the $K_+$ and
Eq.\il(\ref{Eq:some-al})  is ensured by Eq.\il(\ref{Eq:hi-nu-lla-te}).
The  constraint on $K_-$ in both  Eqs.\il(\ref{Eq:cononKlcontro},\ref{Eq:some-al})  is guaranteed  by the constraint on $K_+$ and the definition of minimum point.
It deserves to say that the range
 $]K_{mso}^+,V_{eff}(-\ell,r_{min}^-)[$,
is not null since  it is
$K_{mso}^+=V_{eff}(-\ell, r_{mso}^+)<V_{eff}(-\ell, r_{min}^-)$,
however the range  for $K$ increases as the difference
$(r_{min}^--r_{mso}^+)
 $ increases
 and this clearly increases with the magnitude of the specific angular momentum.
 In any case, the distance $\delta_{min}^{-,+}=(r_{min}^--r_{min}^+)$, and possibly also the distance between the maxima $\delta_{Max}^{+-}=(r_{Max}^+-r_{Max}^-)$, decreases  with increasing of the specific angular momentum magnitude, see also \ref{NoteIII} and \ref{NoteIV}.
In analogy with Eq.\il(\ref{Eq:neut-w}), we arrive to the relations
\bea\label{Eq:lastdon+-}
\frac{\partial\delta_{min}^{-,+}}{\partial |\ell|}<0
\quad \frac{\delta_{min}^{+-}}{\partial |\ell|}<0\quad  \left.\frac{\partial\bar{\lambda}^{-,+}_{Max}}{\partial |\ell|}\right|_{{\lambda}_{+}}<0 \quad
\left.\frac{\partial{\lambda}_{+}^{Max}}{\partial |\ell|}\right|_{\bar{\lambda}_{-,+}}<0,
\eea
where  $\bar{\lambda}_{-,+}^{Max}$ is the maximum spacing  at fixed elongation  ${\lambda}_{+}$,  and  ${\lambda}_{+}^{Max}$ is  the maximum elongation,
see also \footnote{This  could be exactly evaluated, considering that
$(\ell^-+\ell^+)=0$ for  large  $r$,
as $(-6a/r)$.
Close to the static case  (for very slow attractors) there is
$\ell^-+\ell^+\approx2(4M - 3 r) a/(r-2 M)^2$, for    $r>6M$.} Figs\il(\ref{Figs:Aslanleph1l},\ref{Figs:Aslanleph1lb},\ref{Figs:RightPolstex}). Finally, we note that relations (\ref{Eq:lastdon+-})  show the existence of possible   constraints in the number of rings  with respect to a range of specific angular momenta--see  \ref{NoteIV} for the elongation $\lambda_{\pm}\neq0$.

\medskip

\textbf{{Decomposition of  order $n$:  $\ell$counterrotating rings $\ell_a \ell_{b}<0$}}

We can trace some considerations on the  $\ell$contourrotating   couples at $|\ell_{(i+1)/i}|>1$ from the analysis of the multiple configuration  with couples $|\ell_{(i+1)/i}|=1$.

We  start our discussion considering the  $\ell$counterrotating sequence in \emph{accretion}  i.e.
\be
\exists\; i:\; \ell_i \ell_{i+1}<0\quad\mbox{ with}\quad C^{i+1}_x.
\ee
 Then
the outer configuration { has to  be} counterrotating as  $C^{i+1}_x$ clearly marks the infimum of the counterrotating sequence  whose centers can only be placed at  $r_{min}^j>r_{min}^{i+1}$. This does not imply that its sequences are isolated in the disk, see definition \ref{(-)}.
Indeed we can easily prove  this  by noting that
 for two configurations $C_{i}< C_x^{o}$,  it is
\be
r_{Max}^i\leq y_{3}^i<r_{mso}^i<r_{min}^i<y_1^i\leq y_3^o=r_{Max}^o<r_{mso}^o<r_{min}^o<y_1^o \quad \mbox{with}\quad  r_{Max}^o>r_{mbo}^o.
\ee
In particular, $r_{mso}^i<r_{mso}^o$ that proves $C_i=C_i^-$ and $C_x^o=C_x^{o_+}$. Moreover,
 the necessary condition for the closed configuration is :
\be
r_{Max}^+>r_{mbo}^+>r_{mbo}^-,\quad K_{min}^-<K_{min}^+<K_+=K_{Max}^+<1,\quad \ell_-<-\ell_+,
\ee
see \ref{III.}  for the relation between $K_{min}$ and Figs\il(\ref{Figs:Aslanleph1l},\ref{Figs:Aslanleph1lb},\ref{Fig:bCKGO}).
If  $r_{Max}^+\in]r_{\gamma}^+, r_{{mbo}}^+]$ then there can be open rings, allowing  for funnels of counterrotating  material  from a point $r_{Max}^+$ of the disk--see Fig.\il(\ref{Figs:openFig}). Further considerations are in Sec.\il(\ref{App:opem}). It is worth noting that the point of launching of jets $r^+_{J}\equiv r_{Max}^+$ is located at $r^+_{J}>r_{min}^-$, considering  conditions in Eq.\il(\ref{Eq:board-point}).

However, we are mostly interested in the  situation where  the inner ring of the decomposition has to be unstable, therefore
\be
\exists i\;:\; \ell_i \ell_{i+1}<0\quad\mbox{ with}\quad C^{i}_x.
\ee
This condition alone   is not sufficient to establish the  distribution of the specific angular momentum.  
 We can set some conclusions by  considering together the curves of $K_{crit}$ as in Fig.\il(\ref{Fig:bCKGO}), the curves $\ell_{\pm}$ as in Figs.\il(\ref{Figs:Aslanleph1l},\ref{Figs:Aslanleph1lb}) and Figs.\il(\ref{Figs:PlotLKKdisol},\ref{Figs:PlotLKKdisolb}) for the radii of the set $\mathcal{R}$:
\begin{description}
\item[From Fig.\il(\ref{Figs:PlotLKKdisol},\ref{Figs:PlotLKKdisolb}):]
As  $r_{mso}^-
<r_{mso}^+$,  the inner unstable configuration, in accretion or jets, is $C_i^x=C_-^x$. The  accretion point    or  jet launching of corotating matter  approaches the source as  its  spin-mass ratio increases, viceversa for  the counterrotating matter.
The  measure $\delta_J$ of orbital region $\Delta_J$  where the launch of jets,   or the accretion $(\delta_x,\Delta_x)$, occurs  reduces  for the corotating matter.
This fact
emphasizes a preference for the equilibrium configurations  of corotating matter for
 too large spin-mass ratio of the black hole, while the situation is just the opposite for the  counterrotating matter. We can summarize the situation for  corotating and counterrotating  fluids as follows:
\be\label{Eq:ban-appl-cho}
\Delta_J^{\pm}<\Delta_x^{\pm}.\quad \delta_J^{\pm}<\delta_x^{\pm},\quad \partial_{a_*}\delta_J^{\pm}\gtrless0,\quad \partial_{a_*} \delta_x^{\pm}\gtrless0, \quad a_*\equiv a/M.
\ee
The \textbf{BH} spin-mass ratio distinguishes the morphology and topology of the  $\ell$counterrotating
rings as  in general the orbital instability regions  for  corotating matter are   smaller then   the     counterrotating ones, or $\delta_c^{-}<\delta_c^{+}$ for $c\in\{J, x\}$,  for any spin except $ a = 0 $, where there is no point to distinguish the two $\ell$counterrotating subsequences. We note that a similar relation between the orbital range as for their measures  $\Delta_c^{-}<\Delta_c^{+}$ for $c\in\{J, x\}$,  is only possible for the rings in rotation around attractors with spin-mass ratios sufficiently high. As explained below, this behavior  occurs  also for  the range of the  specific angular momentum of the unstable configurations, further discussion can be found also in  Sec.\il(\ref{App:opem}) where  we deepen the case of a $\ell$counterrotating critical  couple  of  sub-configurations.
For the couple  $C_x^-<C^{+}$, it is in general:
\be\label{Eq:le-no-ji}
\partial_{\ell_-}K_{Max}^->0,\quad\partial_{K_-}\lambda_->0,\quad\ell_-\in[\ell^-_{mso},\ell^-_{mbo}].
\ee
For a $C_-$ configuration, we can have  also  $\ell_->\ell^-_{mbo}$, for $K_-<1$ (analogously for  a counterrotating ring).
Therefore, with increasing attractor spin one can say on the basis of the orbital regions  allowed for the points of instability that  the unstable corotating configurations   are significantly disadvantaged with respect to the counterrotating ones.
In other words, according also  to the considerations in \citep{Pugtot}, the rotation of the black hole favors the stability of the
corotating configurations, and has a destabilizing effect on the counterrotating ones, separating therefore the $\ell$counterotating rings of the decomposition\footnote{ This situation  could be described using the analogy with  the electromagnetic interactions in the dynamics of electric charges in the Kerr-Newman spacetimes,  suggesting the spin-orbital momentum couple  in $\ell a\gtrless0 $, as coupling between spin charges, as the electromagnetic couple  $ Qq \gtrless0 $ for the attractor and particle respectively in the electrical charges case, see for example \citep{ergon}}.
 Furthermore, at high values of attractor  spin, the specific angular momentum required for the corotating ring to be in accretion is closest to the minimum $\ell_{mso}^-$. Then $\ell_{\gamma}^-$   close to $\ell_{{mbo}}^-$  is  required for a jet--this analysis should be completed with the study of the momentum   ranges  and the range of the $K$ parameters.
\\
\item[
From Fig.\il(\ref{Figs:Aslanleph1l}) and Fig.\il(\ref{Figs:Aslanleph1lb})]
Consider, in analogy to  Eq.\il(\ref{Eq:ban-appl-cho}), the ranges for the specific angular momentum:
\be\label{Eq:ban-appl-choell}
\delta\ell_J^{\pm}>\delta\ell_x^{\pm},\quad \partial_{a_*}\delta\ell_J^{\pm}\gtrless0,\quad \partial_{a_*} \delta\ell_x^{\pm}\gtrless0,\quad \delta\ell_s^{-}<\delta\ell_s^{+}\quad\mbox\quad s\in\{J, x\},\quad a_*\equiv a/M.
\ee
For the unstable corotating configurations, $\Delta\ell^-_s$ and their measures $\delta\ell_s^-$  decrease with  increasing  spin, as the   ratio $\ell/a$ increases. The opposite happens for the counterrotating rings of  the decomposition.
In other words, we need more specific angular momentum for the  accretion of
a counterrotating ring (or to launch a counterrotating-jet of matter) thus for the
corotating  ones. This
situation, is less and less pronounced as the spin of the attractor decreases  or  even inverted for  spin-mass ratios small enough, namely   $a<a_{\aleph_1}$ for the open sub-configuration   and  $a<a_{\aleph_2}$   for those in accretion. Indeed, we note that similarly to what observed for the range of orbital instability and their measurements, the condition $\Delta\ell_s^{-}<\Delta\ell_s^{+}$ $ s\in\{J, x\}$ only applies to rings around attractors fast enough, that is with  high spin-mass ratio with respect to the limits in $\{a_{\aleph_0}, a_{\aleph},a_{\aleph_2}\}$.
However, although this analysis would suggest that the unstable modes for the counterrotating  rings are disadvantaged at large spin, we note from Eq.\il(\ref{Eq:ban-appl-choell}) that
the gap between the $\ell$counterrotating  momentum of the  accretion phases  increases  with  the spin,
distinguishing clearly and unequivocally the two configurations, but at small spin
  this gap tends to zero or changes sign.
Moreover, the ratio $\ell/a$  for corotating  rings decreases with the spin while for the counterrotating increases in magnitude with the \textbf{BH} spin.
 The range of possible specific  angular momenta for corotating rings in
accretion or jets decreases with the spin-mass ratio to a smaller, bounded  set of possible values, while it increases for the counterrotating ones. For corotating rings   it would be   therefore  increasingly  difficult to select the  specific angular momentum for the emergence of the  unstable mode, particularly  for the  jets.
Large specific  angular  momentum corotating jets  would then  be favorite although $\delta_J^-$ (and $\Delta_J^-$)  decreases with the spin, confirming the preference for the stable corotating rings for fast attractors. Then it is also more  easy to move from one critical topological class to the  other  at very high  spin. In the region very close to the source, the  possibility, for a small change in  the specific  angular momentum, of a  transition  from the accretion  to jet launch  becomes more  significant then for  the lower spin. The opposite is
 the situation for the counterrotating  matter.
\\
\item[
From Fig.\il(\ref{Fig:bCKGO})]
The specific angular momentum and the $K$  parameter required for the accretion and the opening up in jet in the  corotating  sub-configurations orbiting   slow attractor
is greater than for faster  sources, instead  the inverse holds  for the counterrotating ones, for which at  high-spin
   greater specific angular momentum in magnitude and greater parameter $K$ are  required.
This situation is clearly less evident for small spin-mass ratios
 where the distinction between $\ell$counterrotating rings  becomes less
dominant.
As the Eq.\il(\ref{Eq:board-point}) holds, we can draw some
qualitative considerations on the spacing and elongations of the rings.
Being:
\be
\partial_{a_*} (1-K^{\pm}_{mso})\lessgtr0, \quad \partial_{a_*}|\Delta K_{mso}^{\pm}|>0,\quad a_*\equiv a/M,
\ee
for the two kinds of rings   the values of $K_{\pm}\in]K_{mso}^{\pm},1[ $, and therefore the range of supremum elongations $\lambda_x$ at the accretion   in the two kinds of rings are smaller for the counterrotating rings at higher spin  with similar elongations, $K_+>K_{mso}^+$. The  rings (in equilibrium or in accretion)  are   morphologically similar around faster attractors if they are counterrotating, the inverse is for  corotating rings.
As just noted, the stability of corotating rings for high attractor  spins is favoured and the elongation of the corotating rings can be  very diversified.

If $C_x^+<C_-$, then the couple is located at $r>r_{{mbo}}^+$,
see also \ref{NoteIII}.
This arrangement does not allow the instability in the outer disk confined to distant regions.
If $C_x^-<C_+$ and therefore the couple is whole confined at $r>r_{mbo}^-$,  the outer ring can be in accretion and in some spacetimes  it could open up for jets, see end of  Sec.\il(\ref{App:opem}). Then one can refer to the relations (\ref{Eq:ban-appl-cho}) and  (\ref{Eq:ban-appl-choell}). 
We conclude this part by noting
that if the outer configurations allows   jets, the point of launch  will be outer then the accretion point  if orbiting sufficiently fast attractors. For slow attractors it can happen that the inner accreting disk overlapS the  couterrotating jets and even be  $r_{Max}^-\lessapprox r_{Max}^+: \; K_{Max}^\mp\lesseqgtr1$; further details can be found in Sec.\il(\ref{App:opem}).
\end{description}
In conclusion, an  increase of the attractor spin-mass ratio acts we could say ``centrifugally'', separating   spatially and morphologically the $\ell$counterrotating rings of the ringed  disk decomposition.
If the ringed disk is unstable, it  is interacting with the source.
Such  an interaction can result in a slow change in the spin and mass
of the attractor  with a consequent backreaction on the disk itself.  This effect might have relevant consequences  as  a runway instability  of relativistic thick disks \citep{Abr-Nat-Run,Abramowicz:1997sg,Font:2002bi,Lot2013,Hamersky:2013cza,ergon}.
Then  there could occur a the spin up or down of the attractor
and a consequent change of  the
ring disk stability and equilibrium, finally  leading to a positive or negative feedback process with
alternate phases, periodic or quasi periodic, up to a
asymptotic adjustment towards equilibrium  or conversely a
further evolutionary phase of instability.  This clearly depends on several
factors including the initial decomposition of the ringed disk, and in particular  its differential rotation and  topology in accretion.
Indeed, the attractor-ringed disk  interaction takes place from the inner  ring $ C_1 $, but a perturbation induced by this interaction has
consequences on the entire stability of the ringed disk, mainly as consequence of the
change of the spin-mass ratio   and  a change of the parameters of the
 first ring,  see also Sec.\il(\ref{Sec:pertur}).

\medskip

We close this Section  considering  two couples  of rings in equilibrium at equal magnitude of  momenta,  suppose
\be
|\ell_{mso}^+|<|\ell_a|=|\ell_b|<|\ell_c|=|\ell_d|\quad \mbox{and} \quad C_a<C_b\quad\mbox{ and}\quad C_c<C_d.
\ee
Then
$
-\ell_{mso}^+<-\ell_a^+=\ell_b^-<-\ell_c^+=\ell_d^-
$,
%
and considering the  couples $\ell_{a/b}=-1$, the particular Eq.\il(\ref{Eq:rmsorminrmas}), and
using the properties of the  $\ell$corotaing sequences summarized  in (\ref{a.}) and (\ref{b.}), for  the couples  $(C_a^{+},C_c^+)$ and $(C_b^-,C_d^-)$, 
 we finally obtain the following relations:
\bea\nonumber
&&C_a^+<C_b^-\qquad r_{Max}^{b_-}<r_{Max}^{a_+}<r_{min}^{a_+}<r_{min}^{b_-}\qquad K_{min}^{a_+}<K_{min}^{b_-}\qquad K_{Max}^{a_+}<K_{Max}^{b_-}
\\\nonumber
&&{\Large{\mathbf{\wedge}}} \qquad\wedge\qquad\qquad{\Large{\vee}} \quad\quad\vee\quad\quad{\Large{\wedge}} \qquad\wedge\qquad\qquad{\Large{\mathbf{\wedge}}} \qquad \wedge\qquad\qquad{\Large{\mathbf{\wedge}}} \quad\qquad\wedge
\\\label{Fig:schema}
&&C_c^+<C_d^-\qquad r_{Max}^{d_-}<r_{Max}^{c_+}<r_{min}^{c_+}<r_{min}^{d_-}\qquad K_{min}^{c_+}<K_{min}^{d_-}\qquad K_{Max}^{c_+}<K_{Max}^{d_-}.
\eea
By using the  conditions on $K_{min}$  and   $K_{Max}$, one cannot obtain any  upper boundary on  $K_{Max}$ and so   a constraint on the existence of the open configurations for jets or closed for accretion. However, the couple $(C^+_a,C^-_d)$, inner and outer rings of the macro-structure, plays  an essential role in the  determination of  the precise decomposition and instability of this $\mathbf{C}^4$ macro-structure.
From the relations (\ref{Fig:schema}), it is clear that if  $C_d^-$  does not satisfy the necessary condition for the jet formation, then no jets can be launched at  any point of the entire configuration.
 On the other hand,  to have  the inner  $C_a^+$  related to jets it  the necessary condition for  all the  rings of the decomposition has to be guaranteed.
On the basis of these considerations only, one could say that the outer corotating  ring  $C^d_-$ of the decomposition is  the most  favoured    for a jet,  consistently  the  launch  takes place in the   point $r_J^{d_-}$ closest to the  attractor, as seen from the second square  of (\ref{Fig:schema}).  The same argument can be applied to  the couple $(C_b^-,C_c^+)$  when the launch of jet would happen in the middle of the decomposition. The  determination of  the location for  the couple $(C_b^-,C_c^+)$ requires  further discussions on  the specific angular momentum of the pair.
This case has to  be further restricted when considered together with the properties (\ref{Eq:cononKlcontro}) derived essentially  from the analysis of the specific angular momenta.

This ringed disk of the order $n=4$  represents an interesting case for   investigation of the  equilibrium  and   the instability with accretion or launching of jets.
Here, as in the  following,  we are considering  limiting  configurations  as  $ | \ell_{a/b} |=1 $.  The physical meaning,  and therefore the interest in these peculiar cases, lies  in the fact that, as it follows from the continuity of the  functions, the results  apply, or  the most part can be straightforwardly  extended, to the non exact case (as  $| \ell_{a/b} |\neq1 $ or to be more precisely, one could say ${|1-|\ell_{a/b}||}/(|\ell_{c/a}|-1)\in]0,1[$).

However, not all of the \textbf{BH} attractors allow indifferently one or the  settings for the jet launching. Indeed, one has to consider  the strict constraints provided by the specific angular momentum
where
 $r_{Max}^{\pm}\in]r_{\gamma}^{\pm},r_{{mbo}}^{\pm}[$ and \textsuperscript{\ref{fnote1}} $\ell\in] \pm\ell_{mbo}^{\mp},\pm \ell_{\gamma}^{\mp}[$,  Figs\il(\ref{Figs:Aslanleph1l},\ref{Figs:Aslanleph1lb}). In the cross of these regions,  multiple jets are   possible only for    sufficiently slow \textbf{BH} attractors.
Considering   the results from Eqs.\il(\ref{Eq:firstkeycons})traced for the instability when   $\ell_{a/b}=-1$, we infer that there could be four points of jets launching if  $a<a_{\aleph_1}$,
 or four accretion  points say  for
$a \lessapprox a_{\aleph_2}$.
We point out that  the accretion phase  for   the corotating couple
$(C_b^-,C_d^-)$ is prevented in this scheme by the condition (\ref{Def:separated}), assumed  to avoid the  penetration of matter with the inner second couple.

Relations (\ref{Fig:schema}) are not sufficient  to set uniquely the decomposition. One can say in  some way that the first square  (\ref{Fig:schema})   has two vertices  in $(C_d^-,C_a^+)$, providing  respectively  the inner  and outer edge of $\mathbf{C}^4$ crucially  constraining  the stability of the entire  system. Starting from a vertex of the square and considering  the segment  $]C_a^+,C_d^-[$ with extremes on the  two vertices, one  has to establish the relative position of $C_b^-$ and  $C_c^+$. According to the commutation in  the couple $(C_b^-,C_c^+)$, one can obtain two different sequences depending on the specific  angular momentum of the inner counterrotating configuration. Precisely,
it can be either isolated or mixed  $\ell$counterrotating sub-sequence:
\bea\label{Eq:conms}
&&
\overbrace{\mathbf{C}}_{{s}}:\;-\ell^+_c\in]-\ell_a^+,-\ell^+(r_{min}^{b_-})[,\quad C^+_a<C_c^+<C_b^-<C_d^-,\\
&&
\overbrace{\mathbf{C}}_{{m}}:\;-\ell^+_c>-\ell^+(r_{min}^{b_-}),\quad C^+_a<C_b^-<C_c^+<C_d^-;
\eea
see \ref{(-)} and \ref{(--)},
the two sequences are  in Fig.\il(\ref{Figs:RightPolstex}).
Notice that the same consideration can be done for a   generic relation $\ell_{a/b}<-1$, the main difference here is the further simplification as we assumed to consider any specific  angular momenta  with  $|\ell|>|\ell_{mso}^+|$.
Let us focus first on the isolated   case $\overbrace{\mathbf{C}}_{{s}}$,  where the center of the counterrotating ring $C_b^+\in]C_a^+,C_c^-[\subset]C_a^+,C_d^-[$. Its distance is easily assessable from the  solutions of the equation for the specific  angular momentum and,  as shown in Fig.\il(\ref{Figs:RightPolstex}), it decreases with  increasing distance from the source, but not with the  spin of the attractor. Thus one can say that, at great distances,  it is less  favoured for separate decompositions with couples  of  equal (or close)  specific angular momentum magnitude, which should therefore be a wild feature of the accretion disks very close to the source  or associated with very large spins, that is for $R=r/a>r_{mso}^+/a$ small enough.
In general,
by increasing the spin of the attractor, the distance between the curves $\mp\ell^{\pm}$ increases, consistently   the distinction between corotating and counterrotating matter becomes more relevant. In other words, at fixed $r$  the difference
$(-\ell^+-\ell_-)$ increases, but also $\mp\ell_{mso}^{\pm}$ increases, and it is  evident from the limit $a<a_{\aleph}$ imposed to get the maximum points for the corotating ring
that
\be
\left.\partial_{a_*}(-\ell^+-\ell_-)\right|_r>0\quad \mbox{and}\quad \left.\partial_{a_*}(\delta_{min}^{-+})\right|_{\mp\ell^{\pm}}>0, \quad a_*\equiv a/M.
\ee
In terms of the dimensionless radius  $R\equiv r/a$,  one can say that  for   $R$ sufficiently large,   the distance between the curves is very small and the $\overbrace{\mathbf{C}}_{{s}}$  decomposition is possible for very close specific  angular momenta magnitudes, i.e. $\ell_{a/c}\approx-1$. Brief analysis   shows that  such rings are  thinner compared with the overall size of the ringed disks with  very small spacings. This is due to the small gap imposed on the parameter  $K$. In this sense,   for large values of $R$, the sequences  or  sub-sequences $\overbrace{\mathbf{C}}_{{m}}$
 are favoured,  while for small $R$,  configurations $\overbrace{\mathbf{C}}_{{s}}$ 

One can  rewrite Eq.\il(\ref{Eq:conms}) by saying that if there are two pairs  $C_a^+<C_b^-$ where
$\ell_b^-=-\ell_a^+ +\epsilon_{ab}$, to add a third counterrotating ring $C_c$, the following conditions should be satisfied:
\bea
\overbrace{\mathbf{C}}_{{s}}:\;C_c^+\in]C_a^+,C_b^-[\quad \mbox{then}\quad-\ell^+_c\in]-\ell_a^+, -\ell^+(r_{min}^b)[,\quad
\overbrace{\mathbf{C}}_{{m}}:\;C_c^-\in]C_a^+,C_b^-[\quad \mbox{then}\quad\ell^-_c\in]\ell_b^-(r_{min}^a), \ell_b^-[.
\eea
As can be seen from Fig.\il(\ref{Figs:RightPolstex}),
there is a region of possible common values of the specific angular momentum magnitude where the possibility to insert a couple of equal  magnitude of specific angular momentum is confined to $ \ell_c^-=-\ell_c^+\in[-\ell_{a}^+,\ell_b^-[$.
On the other hand, there are forbidden specific angular momenta for one kind of ring and the other respectively, more precisely
for a corotating sub-configuration with $\ell_c^-\Large{\not}{\in}\;]\ell_b^-,-\ell^+(r_{min}^b)[$,  with high specific  angular momenta possible on the contrary for the counterrotating rings, and  viceversa
for low momenta $-\ell_c^+\Large{\not}{\in}\;]\ell^-(r_{min})^a,-\ell_{a}^+[$, possible instead for the corotating matter. Additional constraints can be  provided by  the analysis of  the  $K$ parameter.
It is  clear  that in the first case, the distance between the configurations is  reduced as compared to the inner range  of specific angular momenta, in which the ringed disks are more spaced.

Finally in Table\il(\ref{Table:sum-BH}) is an overview of the results for the $\ell$counterrotating sequences.
\subsection{On the existence and structure of the ringed configurations}\label{Sec:i4cases}
In this Section we briefly discuss  the  ringed configurations  according to the classifications \ref{C1}, \ref{C1a} and \ref{C1b}   introduced in Sec.\il(\ref{Sec:caracter}).
The results presented here derive from the considerations outlined in Sec.\il(\ref{Sec:procedure}) and we show  in particular that some decompositions  are possible only for limiting number of rings--see also Table\il(\ref{Table:sum-BH}).
\begin{description}
\item[{Configuration $\bar{\mathfrak{C}}_0$}]   
%

%
Consider a couple of rings with   $ \Delta_{cri}^{\mathbf{(i)}}\cap\Delta_{cri}^{\mathbf{(o)}}=\emptyset$. Then separated configurations can \emph{certainly} exist as it follows from the definition of minimum point, and the inner  ring can be in accretion.
Condition  $\bar{\mathfrak{C}}_0$,
 clearly \emph{excludes} {\textbf{\large{$ {\ell}$}}}corotating matter configurations.
 Multiple configurations of this kind are necessarily {\textbf{\large{$ {\ell}$}}}counterrotating, and   $r_{mso}^{\mathbf{(i)}}=r_{mso}^-$ and $r_{mso}^{\mathbf{(o)}}=r_{mso}^+$. That  is, the inner ring must be corotating with  respect to the black hole,  $C_-<C_+$, and  the region $[r_{min}^{-},r_{Max}^{+}]\subset\Delta_{mso}^{\mp}$--see also Sec.\il(\ref{Sec:procedure}).
Furthermore,  from  analogue considerations  it follows  that the maximum order of the decomposition
\footnote{We use the condition $\bar{\mathfrak{C}}_0:\;\Delta_{cri}^{i}\cap\Delta_{cri}^{j}=\emptyset\; \forall i,\;j\in\{1,...,n\}$ where in particular
$\Delta_{cri}^{i}\cap\Delta_{cri}^{i+1}=\emptyset\; \forall i,\in\{1,...,n\}$  combined with Eq.\il(\ref{Eq:ulter}),  and clearly $r_{mso}\in\Delta_{cri}^{\mathbf{(i})}$. Indeed, consider a  sequences of   $n=3$  closed configurations,  then   explicitly: $r_{Max}^{(1)}<r_{mso}^{(1)}<r_{min}^{(1)}<r_{Max}^{(2)}<r_{mso}^{(2)}<r_{min}^{(2)}<r_{Max}^{(3)}
<r_{mso}^{(3)}$  which is contradictory for any kind  of rotating matter.}
 is  $n_{Max}(\bar{\mathfrak{C}}_0)=2$.
However, we  note that
 the hypothesis of  {\textbf{\large{$ {\ell}$}}}counterrotating disks
 is essential but not sufficient: one can always have {\textbf{\large{$ {\ell}$}}}counterrotating disks in the Schwarzschild
spacetime where $a=0$, but  $\mathbf{C}^2$ satisfying $\bar{\mathfrak{C}}_0$ are not allowed for\footnote{In the static spacetime, we do not distinguish   corotating or counterrotating configurations ($\ell a\lessgtr0$) and \emph{therefore} $\ell_i \ell_j\lessgtr0$. Clearly, $\ell$counterrotating  fluids  can orbit around one \textbf{BH} attractor but from the point of orbital stability, and hence the toroidal configurations in equilibrium, it is  the rotation of the attractor to give meaning to concept of  $\ell$counterrotating  or $\ell$cororotating.
 Then  the rings in the Schwarzschild geometry can be considered  always $\ell$corotating.
} $a=0$.
As such  one can say that $\bar{\mathfrak{C}}_0$  configurations are  features proper of the general relativistic effects induced by   the rotation of the   attractor, with no equivalent in the static geometry (meaning here $R^{-1}=a/r\approx0$). Possibly the outer ring could open up for jets and this is the case of double  critical sub-configurations considered  in Sec.\il(\ref{App:opem})  and particularly  in  Sec.\il(\ref{Sec:procedure}).
The  measure   $\delta r_{mso}^{\pm}=\sup{\bar{\lambda}_{o,i}}$  increases with increasing $a/M$, from the Schwarzschild  to the extreme Kerr spacetime,
 and the couple of rings can be    more spaced with  the  innermost one approaching the black hole.

 However we need to distinguish  between the configurations orbiting \textbf{BH} attractors at $a\in]0,a_{\aleph_1}[$ where $r_{mbo}^+<r_{mso}^-$. In these spacetimes, the outer counterrotating  ring $C_o^+$ cannot open for jets. This case is possible  instead for  attractors  with $a\in]a_{\aleph_1},M[$ where $r_{mbo}^+>r_{mso}^-$  and the outer ring could open in jets. In any case,  the  specific angular momenta must satisfy the relations $-\ell^o_+\in]-\ell_{mso}^+,-\ell_{+}(r_{mso}^-)[$ and
$\ell_i^-\in]\ell_{mso}^-,\ell_-(r_{mso}^+)[$.
%
%
%
%
%
%
%
The unstable configurations for this couple are  $(\mathbf{C_{\odot}^x}^2,\mathbf{C^2_{\odot}},\mathbf{C^2_x})$.
The configurations $\mathbf{C^2_{\odot}}$, characterized by zero spacing   $\bar{\lambda}_{2, 1}=0$ at fixed specific angular momentum  can  exist with a proper choice of $K_{\pm}$; see also  discussion in Sec.\il(\ref{Sec:procedure}).
\\
\item[Configurations {$\bar{\mathfrak{C}}_{1a}$-$\bar{\mathfrak{C}}_{1b}$}]
The cases {$\bar{\mathfrak{C}}_{1a}$-$\bar{\mathfrak{C}}_{1b}$} were introduced in  Sec.\il(\ref{Sec:caracter})  and they are is  characterized by the condition  $r_{Max}^{i+1}<r_{min}^{i}$.  In order to trace some conclusions on  the {$\bar{\mathfrak{C}}_{1a}$-$\bar{\mathfrak{C}}_{1b}$} decompositions, we should focus on the role of the maximum points.

 As   a closed surface centered in the minimum always exists (for definition of  critical point), one can always check for  two separated {$\bar{\mathfrak{C}}_{1a}$-$\bar{\mathfrak{C}}_{1b}$ } surfaces. But the location of the maximum $r_{Max}^{i+1}$, when it exists, selects the kind  of rotating matter, if it is corotating or counterrotating.  The   configurations are  therefore divided in  the two cases discussed below:
\begin{description}
\item[
{ $\bar{\mathfrak{C}}_{1a}:\;r_{Max}^{i+1}\in\Delta_{crit}^{i}$}]

\mbox{}

In  $\bar{\mathfrak{C}}_{1a}$ case,      $r_{Max}^{i}<r_{Max}^{i+1}<r_{min}^{i}<r_{min}^{i+1}\;\forall i\in\{1,...,n-1\}$, and the rings have to be $\ell$counterrotating  (see for example (\ref{a.})).

If
$
C_-<C_+:\;
r_{Max}^-< r_{Max}^+<r_{min}^-<r_{min}^+$. Assuming  proper conditions on the  specific angular momenta  magnitude  (see for example Fig.\il(\ref{Figs:Aslanleph1l},\ref{Figs:Aslanleph1lb})),  it is clear that the maximum  order of the decomposition is\footnote{It is clear that one cannot add a third (outer) ring to the $\bar{\mathfrak{C}}_{1a}$ couple $
C_-<C_+$ for  (\ref{a.}) it should be $\ell$counterrotating with respect to the outer of the first couple, that is corotating, but if it is then for (\ref{a.}) with respect to the inner ring $C_-$ it cannot satisfies the property  $\bar{\mathfrak{C}}_{1a}$.} $n_{Max}(\bar{\mathfrak{C}}_{1a})=2$; moreover, we  note that    the location of
$r_{mso}^{\pm}$ should be clarified and therefore the ratio of the specific angular momenta. This follows the discussion in the end of Sec.\il(\ref{Sec:procedure}).

If
$
C_+<C_-:\;
r_{Max}^+< r_{Max}^-<r_{mso}^-<r_{mso}^+<r_{min}^+<r_{min}^-$
 could not be possible, but around very slow attractors with $a\approx0$, see Figs.\il(\ref{Figs:Aslanleph1l},\ref{Figs:Aslanleph1lb}),
it would be at a fixed  $\bar{r}$, $\ell_{-}(\bar{r})+\ell_+(\bar{r})\approx0$, and
one could consider the   couple as  $\ell$corotating and apply (\ref{a.}).
\\
\item[ $\bar{\mathfrak{C}}_{1b}:\;\Delta_{crit}^{i}\subset\Delta_{crit}^{i+1}$]

\mbox{}

This is the case defined in \ref{C1b} and as it follows from discussion in Sec.\il(\ref{Sec:rolel}), it can be a $\ell$corotating sequence of rings with $n_{Max}(\bar{\mathfrak{C}}_{1b})=\infty$, or  an $\ell$counterrotating one.
It can be either the couple $\ell_{(i+1)/i}=-1$, with the outer corotating ring,
$\ell_->-\ell_+$, or also $-\ell_+>\ell_-$ within proper constraints on the specific angular momentum ratios.
In general, as condition $\bar{\mathfrak{C}}_{1b}$ does not fix the  differential rotation of the ringed disk, in principle there could be  an infinite number of rings and
further constraints on the order of the decomposition may come from  the  additional requirements provided on the  macro-structure effective potential by means of the boundary conditions in terms of the sequence $\{K_i\}_{i=1}^n$.
\end{description}
\end{description}
\subsection{Limiting cases on the $K$ parameters for an $n$-order decomposition}\label{Sec:K=K}
In this Section we focus on the sequences  $\{K_i\}_{i=1}^n$.  Similarly to what seen for the case of constrained differential rotation of the ringed disk  realized adopting specific and  known relations between the elements of  $\{\ell_i\}_{i=1}^n$, we here prove that, considering particular relations for the elements in the  sequences $\{K_{crit}^i\}_{i=1}^n$,    in some cases  the decomposition order $n$ turns to be  upper  bounded, implying various constraints on the morphology of the entire ringed structure.   The $K_i$ parameter   sets the elongation of each sub-configurations univocally, and then  it is able to provide a limit on the  number of rings  for a macro-configuration,   that  is intrinsically linked to the effective potential affecting  the fluid and  containing  information on the gravitational field and the centrifugal force each ring is subjected to. Some basic considerations on the role of this parameter has been provided in Sec.\il(\ref{Sec:onK}).
In the following, we focus  on some limiting cases for the values of  $K_{crit}^i$  as plotted in Fig.\il(\ref{Fig:bCKGO}):
for the couple of rings $(C_a,C_b)$ we investigate the case of $K^a_{Max}=K^b_{min}$ in  Sec.\il(\ref{Sec:Kamx=Mmin}) and
$K^a_{Max}=K^b_{Max}$ or $K_{min}^a=K_{min}^b$  in Sec.\il(\ref{Sec:minimacoinc}), while in Sec.\il(\ref{Sec:order4}) we explore the four order ringed disk $\mathbf{C}^4$ characterized by a unique $K_{crit}$. This Section closes  in Sec.\il(\ref{Sec:rmin=rMx}) with the analysis of the couple of rings at $r_{min}^a=r_{Max}^b$.
The examination of these limiting cases, characterized by a  known and exact  relation  between $r_{crit}$ or $K_{crit}$, is interesting because it  makes evident  possibility to  provide a strict    upper bound   for the decomposition  order and therefore to constraint the number of rings of a ringed disk; the maximum order provided in these cases is a number of the  set $n_{Max}\in\{2,4\}$.  As consequence of  the continuity of the effective potential function, many of the considerations traced for these cases are  valid,  or can be easily extended  for a slight change of the relation  between the  parameters--see also Sec.\il(\ref{Sec:rolel}),. This indeed turns to be suitable for the perturbation analysis as sketched in Sec.\il(\ref{Sec:pertur}))  and the investigation of more reliable situations where a fine tuning of the critical values imposed with the limiting cases  is  not required. An overview of the main results of this  analysis is also in  Table\il(\ref{Table:sum-BH})).

Taking into account the  considerations of the   \ref{NoteIV}   one could show that the  number of rings  can be largely constant, for a little change in the model parameters,  having indeed to guarantee the ring separation\footnote{On the other hand, the accurate assessment of the ``slight  change''   to maintain constant  the configuration order can be easily assessable as one consider specific cases fixing the  $a/M$ parameter.}.
\subsubsection{Rings $(C_{a},C_{b})$: $K_{Max}^{a}=K_{min}^{b}$}\label{Sec:Kamx=Mmin}
Firstly we analyze the case  of  two rings $(C_a,C_b)$   with $K_{Max}^{a}=K_{min}^{b}$, therefore we consider the   curves $K_{crit}=$constant in  Fig.\il(\ref{Fig:bCKGO}).
In general it is:
\be\label{Eq:tiny}
K_{mso}^a<K_{min}^a<K_a<K_{Max}^a=K_{min}^b<K_b<K_{Max}^b, \quad K_{Max}^a<1,
\ee
meaning that the $C_a$ ring  cannot be open.
The maximum number of separated configurations with  $K_{Max}^{a}=K_{min}^{b}$  is $n_{Max}=4$, two of them having to be  {\large{$\ell$}}counterrotating, $\ell_{\mathbf{(i)}}\ell_{\mathbf{(o)}}<0$, and the others {\large{$\ell$}}corotating, $\ell_{\mathbf{(i)}}\ell_{\mathbf{(o)}}>0$, where $\mathbf{(i)}$ and $\mathbf{(o)}$ are for inner and outer ring of the couple.

Indeed at   $K_{min}=K_{min}^a>K_{mso}^+$ there can be two $\ell$counterrotating  rings,  say $C_a^-$ and  $C_a^+$, with $K_{min}^{a_-}=K_{min}^{a_+}\equiv K_{min}^a$,
 and, respectively, given  $K_{Max}\equiv K_{Max}^b$ with  $K_{min}^a=K_{Max}^b$ or $K_{min}^a\neq K_{Max}^b>K_{mso}^+$. Thus  there is a $\ell$counterrotating couple:
 \be\label{Eq:four-gre-uck}
(C_b^-,C_b^+):\;K_{mso}^-< K_{mso}^+<K_{Max}^{b_-}=K_{Max}^{b_+}\equiv K_{Max}^b=K_{min}^{a_-}=K_{min}^{a_+}\equiv K_{min}^a,\quad n_{Max}=4.
\ee
This case will be discussed in Sec.\il(\ref{Sec:minimacoinc}).
Here we detail  the situation for the couple $C_a^{\pm}$ and $C_b^{\pm}$ considering separately the $\ell$corotating pair and the $\ell$counterrotating ones.
\begin{description}
\item[
\textbf{{\large{$\ell$}}corotating rings}]

\mbox{}

For a $\ell$corotating  couple we consider the curves $K_{crit}^{\pm}$ shown in Fig.\il(\ref{Fig:bCKGO}). Then Eq.\il(\ref{Eq:tiny})  holds  and  to fix the ideas, we suppose that      it is  $K_{min}^a<K_{min}^b $,    and
\be
\ell_a\ell_b>0,\quad\ell_a<\ell_b,\quad C_a<C_b,\quad r_{Max}^b<r_{Max}^a\leq y_3^a<r_{min}^a<y_1^a<y_3^b<r_{min}^b<y_1^b;
\ee
see also \ref{NoteI} and  Figs\il(\ref{Figs:Aslanleph1l},\ref{Figs:Aslanleph1lb}) where:
\be
K_{mso}^a=K_{mso}^b<K_{min}^a<K_a=K_{Max}^a=K_{min}^b<K_b<K_{Max}^b, \quad K_{Max}^a<1,
\ee
 so that  the outer ring $C_b$ cannot be unstable,
 and the inner  $C_a$ cannot be in accretion. This is because of Eq.\il(\ref{Eq:board-point}) implying that  if $\lambda_a=\lambda_a^x$, then $y_1^a=y_{min}^b$. In conclusion, considering that these results apply for the two  $\ell$corotating sequence of the decomposition, the ringed disk $\mathbf{C^4_x}$, within condition (\ref{Eq:tiny})  cannot accrete  into the black hole  and more generally cannot be unstable according to a P-W instability, while the possibility that the disk can be in $ \mathbf{C}_{\odot}^n $ topology will be briefly discussed  in Sec.\il(\ref{Sec:order4}). %
\\
\item[
\textbf{$\ell$counterrotating rings}]

\mbox{}

For two  $\ell$counterrotating rings,  when Eq.\il(\ref{Eq:tiny})  holds,
suppose to fix the ideas that it is   $K_{Max}^a=K_{min}^b$
with    $\ell_a\ell_b<0$. Then we can realize  one of the following two possibilities
\be
\mathbf{(C_a^-,C_b^+)}:\;K_{mso}^+<K_a^-<K_{Max}^{a_-}=K_{min}^{b_+}<K_b^+<1,\quad
\mathbf{(C_a^+,C_b^-)}: \; K_{mso}^+<K_a^+<K_{Max}^{a_+}=K_{min}^{b_-}<K_b^-<1.
\ee
and, respectively, the two rings $C_a^{\pm}$  can be in equilibrium or in accretion but cannot give rise to jets.
\begin{description}
\item[]
In the first case
\be
\mathbf{(C_a^-,C_b^+)}:\quad r_{Max}^{a_-}<r_{mso}^-<r_{mso}^+<r_{min}^{b_+}\quad \mbox{and}\quad K_{min}^{a_-}<K_{a}^-<K_{Max}^{a_-}=K_{min}^{b_+}<K_b^+<K_{Max}^{b_+},
\ee
in other words
it is
$
K_a^-<K_b^+$
and
$K_{min}^{a_-}<K_{min}^{b_+}$.
For the last condition  to be satisfied, it can be either
\be
 r_{Max}^{b_+}<r_{Max}^{a_-}<r_{min}^{a_-}< r_{min}^{b_+}\quad (\mbox{i.e.} \quad C_a^-<C_b^+\quad\mbox{ and}\quad \ell_a^-<-\ell_{b}^+),
\ee
a stable ringed disk, i.e. non accreiting if $C_1=C_a^-$, or
\be
r_{Max}<r_{min}^{b_+}<r_{min}^{a_-}<\bar{r}_-:\;V(\ell_a^-,\bar{r}_-)=K_{min}^{b_+}\quad (\mbox{i.e.}\quad C_b^+<C_a^-).
\ee
The corotating configuration $\bar{C}_-$ with minimum in $\bar{r}_-$ has clearly $\ell_->\ell_a^-$,
and the couple $(C_b^+,\bar{C}_-)$  constitutes  the case of $K_{min}^+=K_{min}^-$ analyzed below.
To characterize the couple  in  the two cases $C_a^-<C_b^+$ or $C_b^+<C_a^-$, we need to discuss the location of $r_{min}^{a_-}$, and the situation is different for different spacetime classes, (see also results in \ref{(i)} \ref{(ii)} (\ref{a.}) and \ref{b.})\footnote{However, considering \ref{NoteII} and the relations presented  in Eqs.\il(\ref{Eq:sto-bru-cioc}-\ref{Eq:Schr-webN}), we remind that
in  $r>r_{mso}^+$, the  $\ell$counterrotating rings have both
minimum points; then   $K_{min}^-(\ell_-,r_{min})<K_{min}^+(\ell_+,r_{min})$ at fixed $r_{min}$
where
$\ell_-<-\ell_+$, see Figs.\il(\ref{Figs:Aslanleph1l},\ref{Figs:Aslanleph1lb}). If, on the other hand,
 $K_{min}^-(\ell_-,r_{min})< K_{min}^-(\ell_1^-,r_{min}^1)$ for some $(\ell_1^-,r_{min}^1)$,
then   $\ell_-<\ell_1^-$  for  $r_{min}<r_{{min}_1}$, see Fig.\il(\ref{Figs:Aslanleph1l},\ref{Figs:Aslanleph1lb}). Suppose that
$K_{min}^-<K_{min}^+$ and
$r_{min}^-=r_{min}^+$
then  $\ell_-<-\ell_+$
and
$r_{Max}^-<r_{mso}^-<r_{mso}^+<r_{min}^-=r_{min}^+$.
 It is clear that the position of maximum $r_{Max}^+$ depends on
$r_{\gamma}^+<r_{mbo}^+<r_{mso}^-<r_{mso}^+<r_{Max}^-<r_{Max}^+$.}.
\\
\item[]
In the second case
\be
\mathbf{(C_a^+,C_b^-)}:\quad K_{mso}^-<K_{mso}^+<K_{min}^{a_+}<K_a^+<K_{Max}^{a_+}=K_{min}^{b_-}<K_b^-<K_{Max}^{b_-},
\ee
in particular
$K_a^+<K_b^-$
and
$K_{min}^{a_+}<K_{min}^{b_-}$; from this it follows that
\be\label{Eq:fo-d-n-w}
r_{min}^{a_+}<r_{min}^{b_-}\quad\mbox{therefore}\quad C_a^+<C_b^-,
\ee
see Eqs.\il(\ref{Eq:sto-bru-cioc}-\ref{Eq:Schr-webN}) and also
 Fig.\il(\ref{Fig:bCKGO}). Equation (\ref{Eq:fo-d-n-w})  is not sufficient to fix the relative location of $r_{Max}$,  neither the relation between  the specific angular momentum,  but the two cases  listed  in  \ref{III.} should be discussed.
 The inner ring is the counterrotating one of the couple and it cannot be open, but it could be in accretion. This can happen if the ringed disk has order $n=2$, with one  only counterrotating couple. If a third or fourth rings are added within (\ref{Eq:condi-sss}), then results for the $\ell$corotating sequence apply, and P-W unstable mode can be {formed} for the ringed disk.

 However, from  the consideration of Eq.\il(\ref{Eq:board-point}), for the inner ring $C_a^+\subset \mathbf{C}^2$  to be in accretion it  has to be:
\be
\lambda_a^+=\lambda_x^+ \quad\mbox{ and therefore}\quad  K_a=K_{Max}^{a_+}=K_{min}^{b_-}\quad \mbox{with}\quad y_3^{a_+}=y_{Max}^{a_+}<r_{mso}^+<y_{1}^{a_+}< y_{min}^{b_-}.
\ee
From the  relation between the minima only, we are not able to fix uniquely  the sign of  $\ell_-+\ell_+$, in order to do this,  we need the information on $K_{Max}$, (consider \ref{I.} \ref{II.} \ref{III.}).

In any case, for the inner torus $C_a^+$ in accretion one finds
 \be\label{Eq:area-BH-inter}
\quad\mbox{if}\quad\delta r^{\mp}_{min,1}\equiv(r_{min}^{a_-}-y_1^{a_+})\quad\mbox{then}\quad  \partial_{r_{min}^{a_-}}\delta r_{min,1}^{\mp}<0, \quad
  \partial_{y_3^{a_+}}\delta r_{min,1}^{\mp}>0, \quad \partial_{K^+_a}\delta r_{min,1}^{\mp}<0,
  \ee
 with obvious implications on the spacings and then in the density of the rings for  more complicated decompositions with  one couple satisfying the condition (\ref{Eq:tiny}). The ranges of variation in (\ref{Eq:area-BH-inter}) must be related to the existence of appropriate values for the counterrotating configurations   in accretion, and the  corotating in equilibrium,  considered  due to Eq.\il(\ref{Eq:tiny}).
However, as there is    $y_3^{a_+}\in]r_{mbo}^+,r_{mso}^+[$, then   one has to consider differently the attractors with spins
\bea
a<a_{\aleph_1}:\;r_{mbo}^-<r_{mbo}^+<r_{mso}^-<r_{mso}^+,\quad\mbox{or}\quad
a>a_{\aleph_1}:\;r_{mbo}^-<r_{mso}^-<r_{mbo}^+<r_{mso}^+,
\eea
distinguished by the relative position of $(r_{{mbo}}^+, r_{mso}^-)$.
\end{description}
\end{description}
\subsubsection{Rings $(C_a,C_b)$: $K_{min}=K_{min}^a=K_{min}^b$ and  $K_{Max}=K_{Max}^c=K_{Max}^d$}\label{Sec:minimacoinc}
The rings  $(C_a,C_b)$  with
$K_{min}^a=K_{min}^b$
 \emph{must} be $\ell$counterrotating or
\be
\mbox{if}\; K_{min}^a=K_{min}^b\;\mbox{then}\quad \ell^+_a\ell^-_b<0, \quad  C_a^+<C_b^-\quad\mbox{and}\quad n_{Max}=2,
 \ee
 %
see also Sec.\il(\ref{Sec:onK}).
Analogously.  rings   with equal   $K_{Max}$ must be  $\ell$counterrotatings  and  $n_{Max}=2$.
However,  as   mentioned in Sec.\il(\ref{Sec:Kamx=Mmin})- Eq.\il(\ref{Eq:four-gre-uck}), if
$K_{min}^a=K_{min}^b$,  then  it is possible to find  two $\ell$counterrotating sub-configurations
\be
(C_c, C_d):\;
K_{Max}^c= K_{Max}^d=K_{min}^a=K_{min}^b,\quad\mbox{ then}\quad
r_{Max}^{b_-}<r_{Max}^{a_+}<r_{min}^{a_+}<r_{min}^{b_-},\quad n_{Max}=4.
\ee
We would get exactly the same situation from
 fixing first $K_{Max}^c= K_{Max}^d$ for two $\ell$counterrotating  rings,  and  it is  $r_{Max}^{d_-}<r_{Max}^{c_+}$.
Then  we could set in this way a ringed disk
by setting an unique  $K_{crit}$. We note finally that the condition $K_{mso}^+<K_{Max}^a=K_{Max}^b$ is not sufficient to locate the disks.  This procedure   requires discussion of the existence condition on the maximum points provided here for example in Sec.\il(\ref{Sec:procedure}).

The couples at equal $K_{min}$
can exist only for
\be
K\in]K_{mso}^+,1[\quad\mbox{and}\quad r>r_{mso}^+\quad\mbox{with}\quad r_{min}^{a_+}<r_{min}^{b_-}
 \ee
 see \ref{III.}.
There is
 \be
 \partial_{K_{min}}\bar{\lambda}_{ab}<0, \quad\mbox{  and}\quad\partial_{K}\bar{\lambda}_{ab}<0\quad\mbox{ where }\quad K=K_a=K_b,
 \ee
and we will prove below that the case $K=K_a=K_b$ is always possible. The limiting unstable case
$K=K_{mso}^+$, saddle point for the  potential of $C_+$,  corresponds to a maximum point of the pressure for the corotating disk
 $C_-$. The distance $(r_{min}^--r_{mso}^+)$,  determined by the condition  $K_{mso}^+=K_{min}^-$, provides the maximum distance  among the ``centers'', this distance decreases with
$R\equiv r/a$; in analogy with  the curves of specific angular momenta $\ell_{\pm}=$constant  considered in Sec.\il(\ref{Sec:procedure})   in Figs\il(\ref{Figs:Aslanleph1l},\ref{Figs:Aslanleph1lb})  it is
\be
\partial_{a}\delta_{min}^{ba}>0,\quad \partial_{r}\delta_{min}^{ba}<0,
\ee
where $\delta_{min}^{ba}$ sets the separation between the centers and,
 for a fixed  $K_{min}$, it has to be   $r_{min}^+<r_{min}^-$.
The difference of the magnitude of the specific angular momentum can be easily evaluated, it  decreases with $R$
 and, according to the results in Sec.\il(\ref{Sec:app-maxmin}),
the magnitude of the specific  angular momentum increases with $r$.

We focus  now on the choice of  $K_a$ and  $K_b$:  it is easy to prove that
it can be  $ K=K_a=K_b$.
We start our consideration by noting that there always exists a radius:
\be
r_C\in]r_{min}^{a_+} r_{min}^{a_-}[\;:\; V_{eff}(\ell_a^+,r_C)=V_{eff}(\ell_b^-,r_C)\equiv K_C\in]K_{min},1[
\ee
and the pair $(K^+_a,K^-_b)$ can be set so that
$K\in]K_{min},K_C[$.
The radius $r_C$ always exists, as can be proved by
the  Bolzano's theorem on the  zeros of a continuous  function\footnote{
Indeed  the function $f\equiv V_{eff}^{a_+}-V_{eff}^{b_-}$ is  continuous   as sum
of continuous  functions,  and it takes  opposite signs in $r^{a_+}_{min}$ and
$r_{min}^{b_-}$ respectively and $r^{a_+}_{min}<r_{min}^{b_-}$, as   $f(r^{a_+}_{min})<0 $ and  $f(r^{b_-}_{min})>0$. This is because for definition $K_{min}^{a_+}=K_{min}^{b_-}$,
 there exists at last a point $r_C\in]r^{a_+}_{min},r^{b_-}_{min}[$
such that $f(r_C)=0$ and $ V_{eff}^{a_+}(r_C)=V_{eff}^{b_-}(r_C)$.}.
Then each $K$ should be set according to
$K\in]K_{min},K_C[$, and this means we can have  $K_a=K_b$, see also \citep{Pugtot}. However, clearly it is $\partial_K\lambda_i>0$ at $K=K_a=K_b$ , $\lambda_i$ being as usually the elongation of the $C_i$ ring, for $i\in\{a,b\}$ and   $\partial_K\bar{\lambda}_{ba}<0$, the maximum for the elongation $\lambda_i^{Max}$ is at $ K_C$ but then the configuration has $K_a=K_b=K_C$, and  constitute a  $\mathbf{C}_{\odot}^2$.
\subsubsection{macro-configurations $\mathbf{C}^4$: $K=K_{crit}$}\label{Sec:order4}
We focus here on the configuration $\mathbf{C}^4$ with a unique   $K=K_{crit}=K_{min}=K_{Max}$.
To fix the ideas we can set:
\be\label{Eq:notcrit}
K_{mso}^+<K_{crit}=K_{Max}^{a_-}=K_{Max}^{b_+}=K_{min}^{c_+}=K_{min}^{d_-}<1,
\quad \mbox{then}\quad
r_{Max}^{a_-}<r_{Max}^{c_+}<r_{mso}^+<r_{min}^{c_+}<r_{min}^{d_-},
\ee
see  \ref{III.} $(C_a^-,C_b^+)$ cannot be open up in jets and  the rings $(C_d^-, C_a^-)$ have  to be in equilibrium. However  the location of  $(C_a^-,C_b^+)$  and  the critical values $(r_{Max}^{b_+},r_{Max}^{d_-})$ remain undetermined.
This task however is facilitated by noting that
 $C_{a}^-$ and $C_b^+$ cannot be open and
\be\label{Eq:boueb}
r_{Max}^{i}<r_{{mbo}}^{i}\quad\forall i\in\{a,b\}\quad\mbox{then}\quad  r_{mbo}^-<r_{Max}^{a_-}<r_{mso}^-<r_{Max}^{b_+}<r_{mso}^+<r_{min}^{c_+}<r_{min}^{d_-}
\ee
with $r_{mbo}^+<r_{Max}^+$. We should locate $r_{Max}^+$ with respect to  $r_{mso}^-$.
For  condition (\ref{Eq:boueb})  to be satisfied, it has to be
$\ell_a^-\in[\ell_{mso}^-,\ell_{mbo}^-]$ and
$-\ell_b^+\in[-\ell_{mso}^+,-\ell_{mbo}^+]$. 
But it is
$[\ell_{mso}^-,\ell_{mbo}^-[<[-\ell_{mso}^+,-\ell_{mbo}^+]$
 for sufficiently fast attractors  or
$a>a_{\aleph_2}:\; \ell_{mbo}^-=-\ell_{mso}^+$, in these spacetimes   $\ell_a^-<-\ell_b^+$ and we can refer point \ref{(ii)}.

Considering   $a>a_{\aleph_2}$, the location of the  minimum points $(r_{min}^{a_-}, r_{min}^{b_+})$ is  also fixed by noting that, due to
 Eq.\il(\ref{Eq:boueb})
we can always  choice an appropriate  $K_{min}^{a_-}$ and $K_{min}^{b_+}$ to be lower than $K_{min}=K_{min}^{c_+}=K_{min}^{d_-}$,
but as
\be\label{Eq:resc-fiel}
\ell_-<-\ell_+ \quad\mbox{ then }\quad r_{min}^{a_-}<r_{min}^{b_+} \quad \mbox{and}\quad  K_{min}^{a_-}<K_{min}^{b_+}.
\ee
However,  if
\be
K_{min}^{a_-}<K_{min}^{b_+}<K_{min}\quad\mbox{
 then
}\quad
r_{min}^{a_-}<r_{min}^{b_+}<r_{min}^{c_+}<r_{min}^{d_-}.
\ee
Thus in  ringed  disks around these attractors  only mixed configurations $\overbrace{\mathbf{C}}_{m}$ \ref{(--)} exist.

We focus now on the second class of attractors with $a\leq a_{\aleph_2}$.
This class is split by the spin $a_{\aleph_0}:\;-\ell^+(r_{mso}^-)=\ell_{mbo}^-$, see Eq.\il(\ref{Eq:alm-infla}).
But in the whole class, there is  a region
where the specific angular momentum could also be set in:
\be
[-\ell_{mso}^+,\ell_{mbo}^-] =[-\ell_{mso}^+,-\ell_{{mbo}}^+]\cap[\ell_{mso}^-,\ell_{mbo}^-].
\ee
%
In this  common region, the  relation between the specific angular momenta can be inverted with respect to Eq.\il(\ref{Eq:resc-fiel}).
Considering  the minima of the $\ell$counterrotating  $(C_a^-,C_b^+)$ and the maxima of the couple $C_c^+<C_d^-$, it is convenient to focus first on the two $\ell$corotating rings separately.
We consider  then $(C_a^-,C_d^-)$,  
as if
\be
\ell_a^-\ell_d^->0\quad
K_{min}(\ell_a^-)<K_{Max}(l_a^-)=K_{min}(\ell_{d}^-)<K_{Max}(\ell_{d}^-),\quad\mbox{it follows}\quad C_a^-<C_d^-;
\ee
similarly for the  counterrotating rings $C_b^+<C_c^+$. Considering  then these relations together we obtain
\be
r_{min}^{a_-}<r_{min}^{d_-}\quad\mbox{and}\quad r_{min}^{b_+}<r_{min}^{c_+}\quad\mbox{with}\quad  r_{min}^{c_+}<r_{min}^{d_-}\quad \mbox{implying}\quad r_{min}^{b_+}<r_{min}^{c_+}<r_{min}^{d_-}.
\ee
The location of
$r_{min}^{a_-}<r_{min}^{d_-}$  remains undetermined , for this couple one can apply analogue discussion to those for  Eq.\il(\ref{Fig:schema})\footnote{ In this case however one can consider that
$r_{Max}^{a_-}<r_{Max}^{b_+}<r_{mso}^+$ and then
$\ell_a^->-\ell_{b}^+$
or
$r_{min}^{a_-}>r_{min}^{b_+}$
and analogously, if
$ K_{Max}(\ell_a^-)=K_{min}(\ell_c^+)$ then
$K_{min}(\ell_a^-)<K_{Max}(\ell_a^-)=K_{min}(\ell_c^+)<K_{Max}(\ell_c^+)$
and
$K_{min}(\ell_a^-)<K_{min}(\ell_c^+)$
and
$K_{Max}(\ell_a^-)<K_{Max}(\ell_c^+)$ with $
K_{Max}(\ell_b^+)<K_{Max}(\ell_{c}^+)$ and then $K_{min}(\ell_b^+)<K_{Max}(\ell_b^+)<K_{Max}(\ell_{c}^+)$.}.
\subsubsection{Rings $(C_a,C_b)$: $r_{min}^a=r_{Max}^b$}\label{Sec:rmin=rMx}
We finally consider the case $r_{min}^a=r_{Max}^b$, where it  has to be:
\bea\nonumber
n_{Max}=2, \quad \ell_a^-\ell_b^+<0,\quad C_a^-<C_b^+\quad r_{Max}^{a_-}<r_{mso}^-<r_{min}^{a_-}=r_{Max}^{b_+}<r_{mso}^+<r_{min}^{b_+}<y_1^{b_+},
\\ \label{Eq:ban-pan-lot}
\quad\mbox{and}\quad\ell_{mso}^-<\ell_{a}^-<-\ell_{mso}^+<-\ell_{b}^+.
\eea
The outer ring $C_b^+$ has to be in equilibrium, and
it is always possible to find  a couple $(K_a^-,K_b^+)$ for the two separated sub-configurations  $[y_1^{a_-},y_3^{b_+}]\subset[r_{mso}^-,r_{mso}^+]$;   then
%
\be\label{Eq:ban-pan-lot-1}
K_{mso}^-<K_{min}^{a_-}<K_{mso}^+<K_{min}^{b_+}<K_b^+<K_{Max}^{b_+}\quad\mbox{and}\quad
 K_{mso}^-<K_{min}^{a_-}<K_a<K_{Max}^{a_-}.
\ee
Eqs \il(\ref{Eq:ban-pan-lot},\ref{Eq:ban-pan-lot-1}) do not set  completely $K_{Max}^{a_-}$, particularly with respect to $K^{b_-}_{min}$ or $K_{Max}^{b_-}$. This is important to characterize the couple in its unstable phase, as the inner, corotating  ring $C_a^-$ is accreting or, if possible, opened for jets. In the remaining part of this subsection  we will prove, at least in a certain orbital region,   that  $K_{min}^{b_+}>K_{Max}^{a_-}$, immediately ruling  out the possibility of a corotating jet for this couple.

To prove this result we shall consider  \ref{NoteIII} and \ref{NoteII}.

We should consider that
the maximum possible value of  $K_{max}^{a_-}$   is for
$r_{min}^{a_-}\in\Delta r_{mso}^{\pm}$ approaching $r_{mso}^+$,
 also $K_{m
in}^{a_-}$ approaches  $K_{Max}^{b_+}$, decreasing
  $r_{Max}^{a_-}$, while  the difference of  the two specific angular momenta increases correspondingly, or
it is:
\be
r_{Max}^{b_-}=r_{min}^{a_-}\in]r_{mso}^-,r_{mso}^+[,\quad \partial_{r_{min}^{a_-}}K_{Max}^{b_+}<0,\quad\partial_{r_{min}^{a_-}}K_{Max}^{a_-} >0,\quad\partial_{r_{min}^{a_-}}(K_{Max}^{b_+}-K_{Max}^{a_-})<0,
\ee
 see Figs.\il(\ref{Fig:bCKGO},\ref{Figs:Aslanleph1l},\ref{Figs:Aslanleph1lb}).
 Therefore  the difference  $(K_{Max}^{b_+}-K_{Max}^{a_-})$ reaches its minimum at  the upper extreme  of the  radial range,  i.e., at $r_{mso}^+$,
where $K_{min}^{b_+}<K_{Max}^{b_+}$.
Conversely, if
$\partial_r(K_{Max}^{b_+}-K_{min}^{a_-})<0$ in the range  $]r_{mso}^-, r_{mso}^+[$,
then the maximum range $(K_{Max}^{b_+}-K_{min}^{a_-})$  is reached at $r=r_{mso}^-$.  Then we evaluate the difference   $(K_{Max}^{b_+}-K_{min}^{a_-})$ close to the extreme of the orbital boundaries that is at some  $\bar{r}=r_{mso}^+-\epsilon$ and
 $\bar{r}=r_{mso}^-\pm\epsilon$ for $\epsilon\gtrapprox0$.  Summarizing, the
supremum of $K_{Max}^{a_-}$:
\be
\sup{K_{Max}^{a_-}}:\;  r_{min}^{a_-}=r_{mso}^+-\epsilon\quad \epsilon\geqslant0\quad \mbox{gives}\quad K_{min}^{a_-}\approx K_{min}^{a_-}(r_{mso}^+)<K_{mso}^+,
\ee
more precisely
\bea\label{Eq:con-sup-fin}
\sup({K_{Max}^{b_+}-K_{min}^{a_-}})>0:\quad r_{min}^{a_-}=r_{Max}^{b_+}=r_{mso}^-+\epsilon,
\\\label{Eq:con-inf-fin}
\inf({K_{Max}^{b_+}-K_{min}^{a_-}})>0:\quad r_{min}^{a_-}=r_{Max}^{b_+}=r_{mso}^+-\epsilon.
\eea
Now consider a couple  of rings which satisfies (\ref{Eq:con-sup-fin}),  having the center for the corotating ring   $r_{min}^{a_-}=r_{mso}^-+\epsilon$, then with reference to Fig.\il(\ref{Figs:Aslanleph1l}), we have
$r_{Max}^{a_-}=r_{mso}^--\tilde{\epsilon}(\epsilon)$ where  $\tilde{\epsilon}(\epsilon)\gtrapprox 0$ is a function of $\epsilon:\tilde{\epsilon}(\epsilon)<\epsilon $. It  has to be   $K_{min}^{a_-}<K_{Max}^{a_-}\approx K_{mso}^-\ll K_{Max}^{b_+}$, but $r_{min}^{a_-}=r_{Max}^{b_+}$, from which it follows that $r_{min}^{b_+}\gg r_{min}^{a_-}$ that implies  $K_{min}^{b_+}>K_{min}^{a_-}$. On the other hand,  $r_{Max}^{a_-}=r_{mso}^{-}-\tilde{\epsilon}$ also $K_{Max}^{a_-}=K_{mso}+\tilde{\tilde{\epsilon}}<K_{min}^{b_+}$, being $K_{min}^{b_+}>K_{mso}^+$.
 By  considering these relations together, we finally obtain:
 $ K_{min}^{a_-}<K_{Max}^{a_-}<K_{min}^{b_+}<K_{Max}^{b_+}$ which constitutes  the proof for points $r_{Max}^{b_+}=r_{min}^{a_-}$ close to the supremum in  (\ref{Eq:con-sup-fin}).
 In this discussion  we are clearly assuming that $r_{\gamma}^+<r_{mso}^+$  for the existence of the maximum for the  counterrotating ring.  In fact it could be that one of the elements $ r_i\in\{r_{\gamma}^+,r_{mbo}^+\}$  be  $r_{i}\in\{r_{mso}^-,r_{mso}^+\}$. The first case occurs, if the spin-mass ratio of the attractor
$a/M>0.61$ c.a.,  and $ r_{mbo}^+\in \Delta r_{mso}^{\pm}$ if  $a >a_{\aleph_1}$.

We focus now on an orbital range near the infimum  in (\ref{Eq:con-inf-fin}).
The situation in this case can not be treated as in the previous case for the orbital upper boundary. In the first place, the maximum for the corotating matter can also not exist, in fact similar conditions   to those in Sec.\il(\ref{Sec:procedure}) should be valid.
In particular  it has to be $\ell_-(r_{mso}^+)<\ell_{\gamma}^-$ which is the case only for spacetimes sufficiently slow,  say  $a/M \ll 0.7$ c.a.--see Fig.\il(\ref{Figs:Aslanleph1lb}).
 For attractors
$a/M\in]0.43, 0.7[ $ c.a. it is  $
K_{Max}^{a_-}>1>K_{Max}^{b_+}$
 and only for slower attractors, there is  $K_{Max}^{a_-}<1$. This can be easily seen
by  investigation of  the  couples orbiting attractors with center of maximum pressure near the  upper boundary $r_{mso}^+-\epsilon$: for the corotating ring the specific angular momentum is  $\ell_-(r_{mso}^+-\epsilon)$, function of $a/M$ and defines uniquely  $r_{Max}$, if this exists.  The counterrottaing ring  cannot give rise to a jet  while, for sufficiently fast attractors ($a/M\in]0.43, 0.7[ $ c.a.)  the inner  corotating  one can result  in its  unstable phase in open funnels of jets and the reverse occurs in the vicinity of lower orbital boundary $r_{mso}^-$.

\section{Discussion  and Future Perspectives}\label{Sec:conc}
In this work we  propose   a model of ringed  disk made up by   several  toroidal perfect fluid configurations, a ringed disk, orbiting  an   axially symmetric attractor, namely a supermassive Kerr    black hole. Such  ringed  disks could be created     during the evolution of matter configurations around supermassive black holes.
To model each  ring of the disk,   an hydrodynamic  model based on the Boyer theory of equipressure surfaces in General Relativity is considered.
We define and  characterize  the morphology and equilibrium of the ringed macro-configurations.  The    ringed disk, has been assumed to  posses a  well defined   equatorial plane of symmetry, coincident with the equatorial planes of each ring.  The (reflection) symmetry plane of the rotating attractor   is also the ringed disk equatorial plane.
Each ring is governed by the  Euler equation for  fixed values of the parameter couple $\textbf{p}=(\ell,K)$, and coupled to the next-neighbor  (consecutive) rings of the decomposition by    some   boundary conditions, specifical constraints on the inner and outer edges of  each ring, or equivalently by the sequence of parameters  $\{K_i\}_{i=1}^n$ of the decomposition,  by the  model set up assumed  a priori. The configurations are  characterized providing  constraints and limits on the number of rings and their dynamics (if in equilibrium or unstable) considering both the corotating or counterrotating fluids.
The unstable ringed configurations are also considered. The (structural) perturbation and unstable modes of the ringed  disks, as defined  in Sec.\il(\ref{Sec:pertur}), are also discussed.
 Perturbations of the ringed accretion structure  are namely perturbation \emph{of} and \emph{in} its decomposition into rings.
The model adopted for  each ring of the decomposition,   determined according to the Boyer condition,  is essentially governed by the geometric properties of the gravitational background  and the constants of the fluid motion associated with the symmetry of the system.
Essentially,  we based our analysis   taking advantage of these symmetries and basing our results according to various   geometric considerations, emphasizing the role of the curvature effects  and especially the rotation of the central attractor.

We can recognize   two fundamental aspects of this investigation:

 \textbf{1.}  We  considered mainly separated sub-configurations, as defined in Sec.\il(\ref{Sec:intro}), assuming  a
non-overlapping (non penetration) of
matter between the rings of the decomposition.
 This hypothesis is indeed suitable and necessary for  rings in equilibrium.  Where a penetration or intersection of sub-configurations  occurs, an instability arises,  in closed $C$ configurations these correspond to the   $\mathbf{C}^n_{\odot}$ topology. They are  interpreted as  the initial stages of the unstable modes
involving  matter collisions, or  feeding, or accretion, of a
sub-configuration to one another. In each case these are
argued to be    unstable phases of the ringed disk, not
immediately described by the  Boyer theory. These configurations, introduced and briefly discussed here, are   investigated  in more details in  future studies of the unstable modes of the ringed disks \citep{coop}.

In fact, the assumption of non-overlapping  can be relaxed for  the decompositions with at least one unstable open point,
 considering therefore the possibility of a decomposition with   launching points of the jets   as discussed in  Sec.\il(\ref{App:opem}).

\textbf{2.} A second remarkable aspect that reveals (surprisingly) as a result of this analysis is the distinction  between  $\ell$counterrotating tori, in other words the role of the relative rotation of the fluids belonging to different sub-configurations. We should note that this distinction is induced by the rotation of the central attractor and has less relevance to determine the macro-structure decomposition as the dimensionless radius $R\equiv r/a$ increases according to the analysis detailed in particular in Sec.\il(\ref{Sec:app-maxmin}). This aspect is  much more surprising since, as   emphasized  in different points, the rings are not related by one  dynamic law regulating the entire decomposition, but the considerations of the gravitational background are sufficient to deduce  very stringent constraints providing the boundary conditions in the (ad hoc) definition of the  effective potential of the macro-structure,  provided in Sec.\il(\ref{Sec:effective}).
Characterization of the equilibrium morphology of  such objects represents a starting point for  characterization of the  unstable phases  of the ringed disks.
Especially in the unstable modes, this model could prove to be a  promising approach for the   analysis of the  emergence of high energy  phenomena   as   jets. In fact, a possibly interesting perspective of this work is the study of global instability inside the ringed disk induced by the  gravitational instability or breaking  of the  geometric constraints as described in the Sec.\il(\ref{Sec:unsta}), leading  to the collapse of a ring on the Kerr accretor  or  on the  consecutive ring, or feeding of matter between the rings.
 The  unstable phases are  essentially associated with  the (super Eddington)  accretion   and jet emission. From the  description of these states we determine   the instability  points  (considering also the possibility of   several points) associated with accretion and  launching of jets from the ringed  disk  as well as
  the specific angular momentum   initially characterizing the   unstable phases.

The present study of the ringed disks can be seen as a starting  model of more complex systems. The majority of the results traced here hold or can be simply extended  to more general situations with different equation of state, for example.
The extension of  this analysis to  different  fluids, including for example the dissipative effects, or the electromagnetic contributions  an  a   GRMHD  scenario, is also possible.
This study could be easily generalized to the case of non-constant angular momentum \citep{Lei:2008ui,Abramowicz:2008bk}, or to the case of non-coplanar rings    rotating  the same attractor (see for the off-equatorial tori around compact objects \citep{Cremaschini:2013jia}).
 However we have shown in Sec.\il(\ref{Sec:effective}) that the ringed  disk  turned to be  a  geometrically thin model of the disk with negligible spacings.
The role of the second lobe, here completely neglected except in accreting configurations, should be then   deepened.
The generalization to other models for the sub-configurations could involve essentially two types of difficulties:  first one (perhaps  less relevant), is the algebraic complication in extended frameworks,   for example in  the  GRMHD thick disks,   whose  analytical treatment  today is practically  limited to the study of solutions with   magnetic field with exclusively toroidal component\citep{Komiss,AbraFra,PuMon13,Zanotti:2014haa,AEA,Adamek:2013dza,Cio-Re,Hamersky:2013cza}.
The second significant aspect is constituted  by the fact that the many disk   models, such as the majority of thin disks,  are  essentially  not geometric in   the sense explained in  Sec.\il(\ref{Sec:models}). An important example of the ``non-geometrical nature'' of these models  is the determination of the rotational law for each torus and  the specific angular momentum  $\ell$ here assumed a constant parameter or otherwise known function of $ r $ and other variables.
It is clear that, to carry out the analysis given here, we should provide  an appropriate set of parameters to determine and characterize the decomposition.
Finally,  we focused on  the analysis of the  critical configurations, here considered  in the two modes   of $C_x$ or $O_x$. As  proved in Sec.\il(\ref{App:opem}), these configurations are generally the inner ones  and must be interpreted by saying that the Boyer theory  is only able to
describe the dynamical situations at $t$ fixed. The discussion of these cases is expected to be  carried out in a future work.

We also stressed the possibility that, during its unstable  phases, an  interaction with the attractor can occur with the consequent  change of the geometrical characteristics of the spacetime. The mutation of the attractor properties will  determine  in turn a change of the dynamical properties of the orbiting structure   resulting in an iterative process to be analyzed, which could  lead also to  a form of runaway instability\citep{Abr-Nat-Run,Abramowicz:1997sg,Font:2002bi,Lot2013,Hamersky:2013cza,ergon}.

As a first attempt to characterize these structures, a part of this article was necessarily  descriptive: a number of definitions has been introduced, and the formalism has been  extensively explained.
\section{Summary and Conclusions}\label{Sec:s-E}
The main steps of this analysis are as follows:
\begin{description}
\item[I-Part: Introduction of the model] This part was developed  in Secs.\il(\ref{Sec:models}) and (\ref{Sec:intro}). We  provided  a precise  definition of ringed accretion disks  by clarifying  the fundamental concepts of ring, decomposition of the internal structure and sub-sequences. We  emphasized the role of the$\ell$counterrotating sub-sequences. We have shown that the decomposition can be mixed or isolated according to definition of Sec.\il(\ref{Sec:models}). We characterized the  unstable configurations showing that there are essentially two types of unstable systems or   a combination of these. Depending on whether the instability is due to   a  P-W mechanism for one or more elements of the ring decomposition    or to  some phenomena of   collisions between material belonging to consecutive rings.
We provided the definition of some morphological properties. Having naturally identified  the inner (outer) edge of the ringed disk as that of the first ring of its ordered decomposition, we defined the elongation and the spacing and, discussing   the minimum number of quantities to be specified to characterize the entire decomposition, we showed that in general the macro-structure can be  a geometrically thin disk.
\item[II part: characterization  of the ringed disk decomposition] The second part of the article, developed in  Sec.\il(\ref{Sec:caracter}), was dedicated to the analysis of the macro-structure and its characterization. In general, we focused on   the macro-structures in equilibrium.
 We then characterized the decomposition showing that there are only two types of configurations, and accretion disks  can exist in one   or in a combination of these.
The role of the model parameters has  been extensively characterized.
Rings are not possible with equal specific angular momentum, with the exclusion   of   the  $\ell$corotating couple $(C, O_x)$, where   condition in  (\ref{Condizioni}) could be relaxed. We show  that the  configurations determined by special relationships between $K$ parameters,  have a maximum number of rings  (order of decomposition) equal to $2$ or $4$.
An important point of this work was the definition of the effective potential of the whole macro-structure addressed in Sec.\il(\ref{Sec:effective}). The discussion of this aspect has naturally emphasized the nature and value of the boundary conditions. We  introduced the definition of differential rotation of the disk and different notions of the specific angular momentum in Sec.\il(\ref{Sec:rolel}).
 We found a number of results  related to the  configurations with instability points,  especially for the $\ell$counterrotating case.
The  elongation  of the equilibrium disk is smaller than the elongation of its accretion closed configuration
 and increases  with the magnitude  of the specific angular momentum.
The marginally stable orbit  is included in the accretion disk. Some key considerations have driven this analysis.
The specific angular momentum of each $\ell$corotating sub-sequence increases in magnitude with the configuration index.
Given a ringed disk  of the  order $n$, each  $\ell$corotating sub-sequence
allows \emph{only} one  $C_x$ configuration, that is only  the  inner  ring of the disk can accrete onto the Kerr attractor  and  the  inner one has the  lower specific angular momentum in magnitude.
Thus the outer configuration must be  in  $ C$ topology and the rank $\mathfrak{r_x}_{Max}=1$.
It follows then that the rings of the  $\ell$corotating sub-sequences  satisfy, whether or not consecutive,  the property $\bar{\mathfrak{C}}_{1b}$ as in (\ref{C1b}).
Increasing the spacing between the rings centers, or by adding a ring  to a $\ell$corotating ringed configuration,  it is necessary  to supply additional specific angular momentum to the outer ring such that
$\ell_{i/i+1}|\in]0,1[$, or
the distance  between the two minima  $(r_{min}^a,r_{min}^b)$ increases
with increasing difference between the two specific angular momenta in magnitude. The derivative $\partial_r(\mp \ell^{\pm})$  has a maximum point at $r^{\pm}_{\mathcal{M}}$, respectively, different for each attractor, as  we detailed in  Sec.\il(\ref{Sec:app-maxmin}).
Increasing of  the specific angular momentum  with the radial distance  from the attractor  magnitude  is not constant but it rises up to a limiting radius,
  different for the $\ell$counterrotating sequences and for any spacetime, being the  $r^{\pm}_{\mathcal{M}}$ function of the attractor spin-mass ratio. This is a relativistic effect that disappears in the Newtonian limit  when the orbital distance is large enough  with respect to    $r^{\pm}_{\mathcal{M}}$,  these  effects are mainly due to the presence of the  intrinsic spin of the attractor.

The special behavior of the specific angular momenta  in the Kerr black hole spacetimes distinguishes the two $\ell$corotating  sub-sequences   of corotanting and    counterrotating rings respectively  of a  generic decomposition. The Kerr \textbf{BH} spin has a stabilizing effect for the corotating  matter  and  destabilizing for the counterrotating one. As $\partial_{|a|}r_{\mathcal{M}}^\mp\lessgtr0$, the region where less additional specific angular momentum  is due increases with the spin-mass ratio of the spacetime in one case, and decreases in the other,  in Sec.\il(\ref{Sec:app-maxmin}).Many  morphological characteristics of a toroidal  accretion disks show a symmetry under the transformation $\ell\rightarrow-\ell$, then we expect  a similar  symmetry to be  conserved in the macro-structure $\mathbf{C}^n$, and so find  analogies  between the $\ell$corotating  and  $\ell $counterrotating   sub-sequences.
We therefore characterized the configurations with critical points from the point of view of the differential rotation and for the presence of points of accretion  and  jets.
\end{description}
Table\il(\ref{Table:sum-BH})  \begin{table*}
\centering
\resizebox{1.1\textwidth}{!}{%
\begin{tabular}{llll}
 \hline \hline $\bar{\mathfrak{C}}_0$-$\ell$counterrotating;&$n_{Max}(\bar{\mathfrak{C}}_0)=2$&$C_-<C_+$&
 $(a\in[a_{\aleph},M]-C_+\rightarrow O_x^+)$
 \\
$\bar{\mathfrak{C}}_{1a}$-$\ell$counterrotating;&
$n_{Max}(\bar{\mathfrak{C}}_{1a})=2$&$C_-<C_+$-($C_+<C_-$--$a\gtrapprox0$)
&\\
$\bar{\mathfrak{C}}_{1b}$-$\ell$counterrotating-$\ell$corotating;&$n_{Max}(\bar{\mathfrak{C}}_{1b})
=\infty$&$ $
&($\ell$corotating\;-$\mathfrak{r_x}_{Max}=1$; $C^1_x$)
\\
 \hline \hline
$\ell$counterrotating-$\ell_{i/j}=-1$\;$\bar{\mathfrak{C}}_{1b}$;&$n_{Max}=2 $ &$C_+<C_-$
&$\exists\; r_{Max}^-\;a<a_{\aleph}: $-$\breve{\mathrm{C}}_{I}:\; a\in[0,a_{\aleph_1}[ $ $\breve{\mathrm{C}}_{II}:\; a\in[a_{\aleph_1},a_{\aleph}[  $
\\
 &&$ C_-\rightarrow O_x^-\; (a<a_{\aleph})$, $C_-\rightarrow C_x^-\; (a<a_{\aleph_2})$;
&
$C_+\rightarrow(C_x^+,O_x^+)\;a>a_{\aleph_0}\; (\breve{\mathrm{C}}_{\alpha}-\; \breve{\mathrm{C}}_{\beta})$
\\
\hline\hline
$K_{Max}^c=K_{Max}^d=K_{min}^a=K_{min}^b$&$n_{Max}=4$
\\
$K_{Max}^a=K_{min}^b$&$n_{Max}=4$& n=2- $\ell
$counterrotating;$C_b>C_a$& n=2-$\ell
$corotating-$C_a^-\lessgtr C_b^+\quad C_a^+<C_b^-\;(a\in]a,a_{\aleph_1}[-]a_{\aleph_1},M])$
\\
$K_{min}^a=K_{min}^b$& $n_{Max}=2$&$\ell$counterrotating& $C_a^+<C_b^-$
\\
\hline\hline
\end{tabular}}
\caption{Overview of some restrictions on the decomposition order $n$    of the macro-structure $\mathbf{C}^n$. It follows the analysis of Sections (\ref{Sec:procedure},\ref{Sec:i4cases},\ref{Sec:K=K}). The existence of the minimum of the hydrostatic pressure (maximum of the effective potential)  $r_{Max}$  is assumed; therefore there  is  $\ell\in]\ell_{mso},\ell_{\gamma}[$, both for the corotating  and counterrotating configurations. The sequentiality, according  to the location of the maximum points of the  hydrostatic pressure is also briefly summarized. Some considerations on the existence of cusped topologies,  associated with the unstable configurations according to the Paczy\'nski-Wiita mechanism,   are provided $(\rightarrow)$.  $\bar{\mathfrak{C}}_0,\bar{\mathfrak{C}}_{1a},\bar{\mathfrak{C}}_{1b}$ are the principal classes of decompositions.}
\label{Table:sum-BH}
\end{table*}
%
gives an overview of the most  relevant results.
{
This work is mostly  focused on the equilibrium configurations leaving, for  a following  analysis, \citep{coop}, a more careful treatment of the topologies associated with the unstable phases of the macro-structure.
However, through the study of internal structure of the ringed disk as the collection of many parts, the rings, we pointed out two types of  processes    emerging  possibly  with  evidences
 in various astrophysical phenomena.
 Firstly, the existence of any configuration  composed by  several orbiting rings, in which the single ring may reasonably undergo the model introduced  in Sec.\il(\ref{Sec:models}) (which is
 well-known  in the literature and widely used, especially in the analysis of geometrically thick disks)  should be  finally adjusted in the analysis presented here, and subjected to the constraints related to the orbital locations  and specific angular momentum here provided  in details.
  It is clear that the morphological characterization of the  equilibrium is a  first step in the study also for the  gravitational environments  where the   traces  of  these objects would be expected, and  whose effects should be evident.
  In fact, one of the objectives of this work is to propose    precisely the existence of such formations, characterize them and open the way for further analysis. Secondly, it is evident from the thoughtful discussions in this paper, that the ringed disk is  subjected to a very  peculiar  internal dynamics, in which rings have also their own independent evolution that, at some stage, may be unstable: rings could move, collide and even grown one into one another,  giving  rise to  accretion phenomena with a sort of feeding--dry processes between two rings\citep{coop}.
 Effects, usually related to the ``direct'' interaction  of unstable matter in  accretion  with the attractor, may possibly be associated to this type of internal dynamics,
including jets, as it will be shown  in more details in \citep{coop}.
A further, interesting,  possibility would be that some  phenomena should lead traces  of this ringed,``discrete'', structure  of the initial equilibrium state.
The perturbative analysis, only hinted here in Sec.\il(\ref{Sec:pertur}), tries to enter into the argumentation of this issue.
Thirdly, the ringed disk can be considered  actually one orbiting  body: one of the challenges of this work was  the construction and interpretation of this aspect of  the ringed disk model. We realized this through  what we have been defined as constraints on  the decomposition, for example in the definition of the effective potential of the  ringed disk. Therefore,  those phases and phenomena usually attributable to an accretion disk in interaction with the attractor and with its environment could be also analyzed in view of an internal dynamics of a ringed disk model.
 The  dynamics of a ringed disk, as a single body in interaction should be affected by, as shown here, the internal phases and  particularly by the distribution of angular momentum, especially when formed by the $\ell$counterrotating fluids.  We can therefore conclude that the  analysis of the ringed disks around supermassive black hole could give a new insight into the accretion processes  in active galactic nuclei and quasars, and could be helpful especially in relation to the instability of the   ringed disks.}
\acknowledgments
The authors acknowledge the institutional support of the Research Center of Theoretical Physics and Astrophysics at the  Faculty of Philosophy and Science of the Silesian University of Opava.
Z.S.  acknowledges the Albert Einstein Center for Gravitation and Astrophysics supported by the Czech Science Foundation grant No. 14-37086G.
 D. P. also acknowledges useful discussions  with  Prof. J. Miller, Prof. M. A. Abramowicz, Prof. V. Karas and Prof. W. Kluzniak,  in the early stages of this work.

\appendix
\section{Notes of the ringed disk morphology and the effective potential of the ringed disk}\label{Sec:App-mor-p}
We introduce the definition of   ringed disk  \emph{height} and thickness, following the discussion of  Sec.\il(\ref{Sec:principaldef}).
\begin{description}
\item[]
We define the $\mathbf{C}^n$   disk \emph{height} $h_{{\mathbf{C}}^n}$ as:
\be\label{Eq:max-height}
h_{{\mathbf{C}}^n}\equiv\max\{h_i\}_{i=1}= h_h,
\ee
where $h_i$  is the  height associated to  the sub-configuration $C_i$, or the maximum point { $x_{Max}^i$  (\emph{not} $2 x_{Max}^i$)  corresponding to the point $y_{Max}^i\in \Lambda_i$ of the surface $\partial C_i$ in  Fig.\il(\ref{Fig:Quanumd})\footnote{We note that generally $r_{min}^i\neq y_{Max}^i$, when both exist.}}. More generally,  in the set  $ H^n=\{h_i\}_{i=1}^n$,  associated to the elements of the ordered decomposition $\{C_i\}_{i=1}^n$,  the toroidal surface $\partial \mathbf{C}^n$ can show up with several local maxima and relative minima. More precisely let us define $h_{i}$ to be a \emph{local maximum} of the set  $H^n$ if
$h_{i-1}<h_{i}>h_{i+1}$, and we define \emph{local minimum}, if  $h_{i-1}>h_{i}<h_{i+1}$. The set $H^n$ has one and only one maximum if $h_{1}<...<h_{i-1}<h_{i}>h_{i+1}>...>h_{n}$.
Definition  (\ref{Eq:max-height}) assumes implicitly  $h_{Max}$ will be an absolute maximum  correspondent to  one and only one sub-configuration $C_h$, however this can be not the case. The case of several  configurations with   $h_{Max}\equiv\sup{h_i}$, can occur up to $n$-sub-configuration with $h_i=h_h\; \forall i\in\{1,...,n\}$ for a $\mathbf{C}^n$ macro-structure.
  \\
\item[]
We can define
the \emph{thickness} $R_{\mathbf{C}^n}$  and $R_{\mathbf{C}_{\odot}^n}$ of a $\mathbf{C}^n$ ringed disk and of a saturated $\mathbf{C}_{\odot}^n$  ringed disk, respectively, as
\bea\label{Eq:wolne}
R_{\mathbf{C}^n}&\equiv& \frac{2h_{\mathbf{C}^n}}{\lambda_{\mathbf{C}^n}}=
\frac{2h_{Max}}{\sum_i^n \lambda_i+
\bar{\lambda}_{\mathbf{C}^n}}=\frac{R_{h}}{1+\sum_{i=1}^{n\neq h} \lambda_i+
\bar{\lambda}_{\mathbf{C}^n}\lambda^{-1}_{h}}
\\
R_{\mathbf{C}_{\odot}^n}&\equiv&\frac{R_{h}}{1+\sum_{i=1}^{n\neq h} \lambda_i}\quad \mbox{for}\quad \mathfrak{r}=\mathfrak{r}_{Max},
\eea
 where $R_i\equiv 2h_i/\lambda_i\cong1$ is  the thickness of a single torus.
 However from (\ref{Eq:wolne})  it follows that the macro-configuration $\mathbf{C}^n$, made by   thick rings, can be  considered  properly to be  a geometrically thin disk as $R_{\mathbf{C}^n}<1$ or even $R_{\mathbf{C}^n}\ll1$.
 In the  definition  (\ref{Eq:wolne}) we  considered   Eq.\il(\ref{Eq:lambdac1s})
 where $h_{Max}$ is associated to one and only one ring of the decomposition and  $R_{\mathbf{C}}$ is related to the only $R_h$. We can  generalize Eq.\il(\ref{Eq:wolne})   
to the case where there are $m<n$ rings with equal, maximum height $h_{Max}$. In this case  the ringed disk thickness,  defined in   (\ref{Eq:wolne}), can be written as $R_{h}\equiv \max\{R_j\}_{j=1}^{m}$,  in other words  in the evaluation of $R_{h}$, we use the elongation  $\lambda_h\equiv\min\{\lambda_j\}_{j=1}^{m}$.
 \end{description}
\subsection{The effective potential of the  $\mathbf{C}^n$ macro-structure}\label{Sec:effective}
In this Section we
provide   a  definition for an  effective potential of the macro-structure  $\mathbf{C}^n$. We will use  the fact that the decomposition $\{C_i\}_{i=1}^n$,  introduced in Sec.\il(\ref{Sec:intro}),  is uniquely determined by considering a fixed $\mathbf{p}_{\mathbf{C}^n}$ in the appropriate range of values determined by the existence conditions for $\mathbf{C}^n$.
The rings of the decomposition  $\{C_i\}_{i=1}^n$ are not dynamically tied,  as ruled by  $n$ different   Euler equations, but they  are related only by  being rings of $\mathbf{C}^n$. The  sub-configurations  are bound by the boundary conditions  given  on each scalar function of the  effective potential  $V^i_{eff}$ regulating its own stability, morphology and  the  location in the ordered decomposition.  One can ask whether it is possible to associate the macro-configuration  $\mathbf{C}^n$ to a single scalar, say an effective potential  or  $V_{eff}^{\mathbf{C}^n}=V_{eff}^{\mathbf{C}^n}(\mathbf{p},r,a)$, depending  on the   sequences  $\{\ell_i\}_{i=1}^n$ and  $\{K_i\}_{i=1}^n$  and  regulating the macro-structure  and its decomposition, just as  $V_{eff}^i$ is associated  for  each ring $C_i$.
Clearly, each $V^i_{eff}$  is determined by a dynamical  equation for the pressure of the $C_i$ ring and by the constraint of the ring separations, i.e., Eq.\il(\ref{Condizioni}). As we  have  no such an equation for the entire ringed disk  $\mathbf{C}^n$, any attempt to elaborate an effective potential approach for  the  $\mathbf{C}^n$  macro-configuration reduces in this scheme to build  up an    ad hoc  scalar, from  the set $\{V_{eff}^i\}_{i=1}^n$ and the model parameters. However, the potential of the macro-structure should be able to regulate the dynamics of each ring, as not  isolated extended  bodies but as being  parts of one   decomposition  to  describe the situation in the spacing regions of the ringed disk.

\medskip

In general, if $V_{eff}^i$ is the effective potential regulating the $C_i$ ring of a macro-structure  $\mathbf{C}^n$, at fixed  $\{\ell_i\}_{i=1}^n$, or at fixed order  $n$ and  $\{ K_i\}_{i=1}^n$,   we can introduce the following two definitions regulating the ringed disk:
\begin{description}
\item[The effective potential $\left.V_{eff}^{\mathbf{C}^n}\right|_{K_i}$ of the  \emph{decomposed} $\mathbf{C}^n$ macro-structure]
defined as
\be\label{Eq:def-partialeK}
\left.V_{eff}^{\mathbf{C}^n}\right|_{K_i}\equiv\bigcup_{i=1}^n V_{eff}^{i}\Theta(-K_i),
\ee
where $\Theta(-K_i)$ is the Heaviside (step) function  such that $\Theta(-K_i)=1$  for $V_{eff}^{i}<K_i$ and $\Theta(-K_i)=0$ for $V_{eff}^{i}>K_i$, so that the curve $V_{eff}^{\mathbf{C}}(r)$ is the union of each curve $V_{eff}^{i}(r)<K_i$ of its decomposition. This is a piecewise smooth curve\footnote{ {In fact, each potential $ V_{eff}^{i}$ and, therefore, in particular  $V_{eff}^{i}\Theta(-K_i)$   is clearly a smooth function of $r$ in $\Lambda_i$ (certainly $\ell_i$ is a fixed parameter),  with a minimum point $r_{min}^i\in \Lambda_i$ as there is  $K_i\geq K_{min}^i$. Therefore   it is certainly possible to evaluate the minimum points $\{r_{min}^i\}_{i=1}^n$  of the overall potential $\left.V_{eff}^{\mathbf{C}^n}\right|_{K_i}$ by differentiating the single potential $V_{eff}^{i}$.  As stated in the text, the curve $\left.V_{eff}^{\mathbf{C}^n}\right|_{K_i}$  shall be in general discontinuous at the   edges $(y_1^i, y_3^i)$--see for example Fig.\il(\ref{Figs:RightPolstex}) where the $\{K_i\}_{i=1}^n$ are explicitly shown. Finally, we note that the ``cut'' provided by  the Heaviside (step) function  $\Theta(-K_i)$   makes sure that there is in fact no ``overlap'' of curves, in the spacings $\bar{\Lambda}_{i, i+1}$, for the  effective potential $\left.V_{eff}^{\mathbf{C}^n}\right|_{K_i}$ of the  \emph{decomposed}  macro-structure  (a part  the case of double points when there is  $\bar{\Lambda}_{i, i+1}=0$ and also $K_i=K_{i+1}$). Thus, except  some  cases of $\mathbf{C}_{\odot}$ configurations, in any ranges $\Lambda_k$  the potential $\left.V_{eff}^{\mathbf{C}^n}\right|_{K_i}= V_{eff}^{k}$ is well defined, with no overlapping with  the potential $(V_{eff}^{k-1},V_{eff}^{k+1})$ of the consecutive rings,  and  with $\left.V_{eff}^{\mathbf{C}^n}\right|_{K_i}= 0$ in any   range $\bar{\Lambda}_{i, i+1}\neq0$. Finally, we stress that  in Eq.\il(\ref{Eq:def-partialeK}), as in the entire work, we are neglecting to consider the innermost surface (associated with the $y_2^i$ solutions) for any topology but not the $C_x$ one for an accretion torus.}}

The curve $\left.V_{eff}^{\mathbf{C}^n}\right|_{K_i}$ has $n$ minimum points, defining the order of the decomposition and  the centers of each rings, a maximum number $n-1$ of ranges $\bar{\Lambda}_{i+1 1}$ (spacing), reduces to zero for saturated  $\mathbf{C}^n_{\odot}$,  bounded for  the $\Lambda_{\mathbf{C}^n}=[y_3^1, y_1^n]$ with $2n-2$ points of discontinuity,
and eventually up to $n$-maximum points for  a saturated $\mathbf{C}^n_{x}$ macro-structure. 
Thus,  for   fixed   couple $(C_i, K_i)$, the problem to find and characterize the  decomposition  of  $\mathbf{C}^n$ is reduced to the problem to fix  the  sequence  $\{\ell_i\}_{i=1}^n$. The usefulness of this definition is that at fixed ordered sequence $\{K_i\}_{i=1}^n$, it remains to determine the  elongations ranges $\Lambda_{i}$ and the specific angular momentum to determine the decomposition \footnote{{Definition  (\ref{Eq:def-partialeK}) stands not as the simple collection of the single, independent, effective potential of the rings,  but rather it includes  the collection of the constraints (as they are the $\{K_i\}_{i=1}^n$) that  each  single ring has to comply with as a part of the overall structure of the ringed disk. This will be in fact more clear dealing with the perturbation approach as  in Sec.\il(\ref{Sec:pertur}). Therefore definition (\ref{Eq:def-partialeK}) captures a key aspect of the   coupling of  the potentials of each toroidal ring, by the   ``boundary conditions'' provided by  each sub-configuration  determining    the macro-configuration.}}.
In  Fig.\il(\ref{Figs:CrystalPl})  an example of decompositions at fixed $\{\ell_i\}_{i=1}^n$, being   induced by different choices of eligible $\{K_i\}_{i=1}^n$,   is shown, where  each decomposition corresponds to a different  $\left.V_{eff}^{\mathbf{C}^n}\right|_{K_i}$ potential.
\\
\item[ {The effective potential  $V_{eff}^{\mathbf{C}^n}$ of the configuration} ]

More generally one can define {the effective potential of the configuration} ${\mathbf{C}^n}$ of order  $n$ as
\be\label{Eq:Vcomplessibo}
V_{eff}^{\mathbf{C}^n}\equiv\bigcup_{i=1}^n V_{eff}^{i}(\ell_i)\Theta(r_{min}^{i+1}-r)\Theta(r-r_{min}^{i-1}),\quad  r_{min}^{0}\equiv r_{+},\quad r_{min}^{n+1}\equiv+\infty
\ee
The potential $V_{eff}^{\mathbf{C}^n}$  certainly is not a function but has the advantage to describe the situation in the   spacing regions among the rings and the orbital ranges $]r_+,r_{min}^1[$ and $r>r_{min}^n$. For each choice of eligible $\{K_i\}_{i=1}^n$,   different decompositions  will be identified, as  shown in   Fig.\il(\ref{Figs:CrystalPl}) where, at fixed $\{\ell_i\}_{i=1}^n$, the potential  $V_{eff}^{\mathbf{C}^n}$ of the ringed disk is uniquely determined for each decomposition.
The potential given by the definition (\ref{Eq:def-partialeK}) regulates the set of  rings in the configuration  $\mathbf{C}^n$ at fixed $\{K_i\}_{i=1}^n$, but it  is not able to characterize  completely  the matter with proper specific angular momentum $\ell_i$ in any of the spacings of consecutive rings $C_i$. In other words, the ranges 
$\bar{\lambda}_{i\pm k_{\pm},i\pm k_{\pm}-1}$ for  $k_+\geq2$   and $\forall k_-\geq1$,  the cut is imposed by the double,  right and left,  Heaviside function  in (\ref{Eq:def-partialeK}).
Indeed,
the spacings  $(\bar{\lambda}_{i+1,i},\bar{\lambda}_{i,i-1})$ are obviously governed only by the  potential of the consecutive sub-configurations   with respect to   $C_i$, that is the rings $C_{i\pm1}$. This is because the possibility of forming separate tori  is governed only by the potential of the next  closest rings and, therefore, the cuts provided  by  $(K_{i}, K_{i\pm1})$ in definition (\ref{Eq:def-partialeK}). The parameters  $(K_{i}, K_{i\pm1})$  in fact determine the location of the inner and outer edges of the sub-configurations and   $(\Lambda_i,\Lambda_{i\pm1})$, respectively. It is clear then that Eq.\il(\ref{Eq:Vcomplessibo}) \emph{is} the potential associated with each  separate ring \emph{and}  the  spacings, {and therefore it  is not the simple collection of potentials associated with each  individual ring, seen as independent and isolated, but rather the  actual effective potential of the entire structure, formed by the rings, as components of a single macro-configuration, and  related through conditions  at  their boundaries and the spacings.} Thus,  potential  (\ref{Eq:Vcomplessibo}) fully describes the macro-configuration $\mathbf{C}^n$: by setting the ordered sequence  $\{\ell_{i}\}_{i=1}^n$, and then the \emph{ordered sequence of potentials}  $\{V_{eff}^i\}_{i=1}^n$, where the definition of  ordered sequence of potentials  is inferred  straightforwardly from the order of the sequence of minima.
Then we say $V_{eff}^{\mathbf{C}^n}$ is  the potential of the $\mathbf{C}^n$ configuration  and the set $\{V_{eff}^{i}(\ell_i)\Theta(r_{min}^{i+1}-r)\Theta(r-r_{min}^{i-1})\}_{i=1}^n$, its \emph{decomposition}.
The determination of decomposition of the disk potential  is then essential for the determination of the decomposition of the disk itself.
\end{description}

We conclude this Section with some notes on potentials introduced in Eq.\il(\ref{Eq:def-partialeK})  and Eq.\il(\ref{Eq:Vcomplessibo}).
Firstly potential $V_{eff}^{\mathbf{C}^n}$ in Eq.\il(\ref{Eq:Vcomplessibo}) has been defined explicitly having into account $r_{min}^i$ {of each rings}, as parameters, in contrast with the potential of the  individual rings. {In fact, despite the fact that the effective potential in Eq.\il(\ref{Eq:Vcomplessibo}) fails to be a well defined function because of  the discontinuity points due to the ``cuts'' imposed by the Heaviside functions $\Theta(r_{min}^{i+1}-r)\Theta(r-r_{min}^{i-1})$, and for the consequent ``overlapping''   of the curves $(V_{eff}^{i},V_{eff}^{i+1})$ pertaining to the potentials of the consecutive rings, we can still associate to the potential  $V_{eff}^{\mathbf{C}^n}$ the collection of critical points $\{r_{min}^i\}_{i=1}^{n}$ (and possibly $\{r_{Max}^i\}_{i=1}^{n}$) as, for simplicity, the ``minimum (maximum) points'' of the potential  $V_{eff}^{\mathbf{C}^n}$.}
Potential $V_{eff}^{\mathbf{C}^n}$ has, as  $\left.V_{eff}^{\mathbf{C}^n}\right|_{K_i}$ in Eq.\il(\ref{Eq:def-partialeK}),
$n$ minimum-points, but we can have also  $m\leq n$ maximum points, then we are neglecting the possible maximum $r_{Max}^{i+1}<r_{min}^i$. The unstable phase of $C_{i+1}$ would violate  the principle of non-penetration of matter. Perhaps a   more interesting situation would be when   $K_{Max}^{i+1}>0$, that is an outer  $O_x^{i+1}$ sub-configuration occurs. In this case, the assumptions of not-overlapping  of matter could  be relaxed.  This case, related to the matter jet  production will be also addressed for completeness in the next  Sections and particularly in Sec.\il(\ref{App:opem}).  We note that the constraint imposed by the step functions in  Eq.\il(\ref{Eq:Vcomplessibo}) do not exclude  the $\mathbf{C}_{\odot}^n{}^x$ configurations.
Moreover, the curve $V_{eff}^{\mathbf{C}^n}$  can be interwoven and, as $\left.V_{eff}^{\mathbf{C}^n}\right|_{K_i}$ has  $2n-2$ discontinuity points, apart from the interweaving points, it is not properly a function,
but a
multivalued function or,
 in the strict sense a
multi-valued map in the spacing ranges
$\bar{\Lambda}_{i+1,i}$
with double points
(as it should be considered as only consecutive potentials or potentials of consecutive rings).
The spacings are not  necessarily equidistant. As we will see in Sec.\il(\ref{Sec:pertur}) and Sec.\il(\ref{Sec:app-maxmin}) their location and the  precise arrangement is relevant in the treatment of perturbations of the macro-structure as  a perturbation \emph{of } and \emph{inside} its decomposition.
Finally we can  discuss the relation between the potential $\left.V_{eff}^{\mathbf{C}^n}\right|_{K_i}$   of the  decomposed $\mathbf{C}^n$ macro-structure in  (\ref{Eq:def-partialeK}) and the definition $V_{eff}^{\mathbf{C}^n}$   in (\ref{Eq:Vcomplessibo}). Clearly, the last one is  more general and includes any possible potentials $\left.V_{eff}^{\mathbf{C}^n}\right|_{K_i}$ (at $\{\ell_i\}_{i=1}^n$ fixed).
Therefore by fixing  $\{K_i\}_{i=1}^n$  in   Eq.\il(\ref{Eq:Vcomplessibo}),  we  obtain
\be\label{Eq:Vcomplesssprouibo}
\left.V_{eff}^{\mathbf{C}^n}\right|_{K_i}=\bigcup_{i=1}^n
V_{eff}^{i}(\ell_i)\Theta(y_{3}^{i+1}-y)\Theta(y-y_{1}^{i-1})=\bigcup_{i=1}^n
V_{eff}^{i}(\ell_i)\left[\Theta(y-y_{1}^{i-1})-\Theta(y-y_{3}^{i+1})\right].
\ee
{
However, we stress that  if $V_{eff}^{\mathbf{C}^n}$ is known, one can use it to establish its decomposition and therefore the ringed disk decomposition together with other  basic features, as  the envelope surface  mentioned in Sec.\il(\ref{Sec:principaldef}). It  can be used to set up the perturbation analysis according to the discussion of Sec.\il(\ref{Sec:pertur}). In fact, by  means  of the definitions (\ref{Eq:def-partialeK}) and especially  definition  (\ref{Eq:Vcomplessibo}), each ring turns to be regulated by an effective potential  defined by taking into account the constraints imposed due to the fact that the rings actually form part of a whole macro-structure and, as such, they must comply to specific constraints,  captured here by the presence of the Heaviside functions $\Theta(r_{min}^{i+1}-r)\Theta(r-r_{min}^{i-1})$
in Eq.\il(\ref{Eq:Vcomplessibo})
and  $\Theta(-K_i)$ in Eq.\il(\ref{Eq:def-partialeK})}.
\section{Maximum points of the function $\ell'\equiv\mp\partial_r\ell^{\pm}$}\label{Sec:app-maxmin}
In this Section we  discuss in detail the properties of the ring specific angular momentum  as  function of  $r$, and we will prove  some of the  results  mentioned in Section  (\ref{Sec:caracter}) and  particularly   in (\ref{Sec:effective}) and  (\ref{Sec:rolel}). The results are however of   general relevance related to the properties of  rings governed by the Boyer model  and then a generic macro-structure of the  order $n$.
This analysis leads in a natural way to the characterization
of two special macro-configurations $\mathbf{C}^n$, defined by   constraining  the   $\ell$corotating sub-sequences   of its decomposition $\{C_i\}_{i=1}^n$: in particular we assume some conditions on the set of  centers $\{r_{min}^i\}_{i=1}^n$, or the parameters  $\{\ell_i\}_{i=1}^n$. The first disk  made  by  rings with equally spaced   centers, or otherwise spaced with a known general relation, the second disk
with equally spaced specific angular momenta, or  related by a known law. In other words, the elements of each sequence   $\{r_{min}^i\}_{i=1}^n$ (for example this is the case in Fig.\il(\ref{Figs:CrystalPl})) or  $\{\ell_i\}_{i=1}^n$ respectively, are not  independent but related by a fixed relation.
{The presence of this maximum point has an immediate consequence on the formation and dynamics of the ringed disks  through the characterization of the orbital location and specific angular momentum distribution  of its decomposition in particular, in relation to the density of rings and the differential rotation of the ringed disk. Clear distinction can be seen in the presence of  the $\ell$counterrotating subsequences.
In fact,  study of the two particular cases  helps to better enlighten this distinction: firstly, the  maximum points also correspond to points of maximum density of the rings,
located in different orbits depending on the rotation with respect to the  attractor. Secondly, there are six different orbital areas,  respectively, three for the two  $\ell$counterrotating  subsequences, for which a change of the  angular momenta in magnitude  has very different effects on the location of the disk, not changing the ring topology and therefore the stability of the disk: specifically,
the inner orbital area of the maximum points, a neighborhood of the maximum and the outside area. In \citep{coop} this analysis will be extended also to the unstable topologies. }
The simplest case  is when  the difference between the elements  is independent from the index $i$ of the sub-configuration and  are constants, for example  $\epsilon_{i\phantom\ i+1}=\epsilon_{j\phantom\ j+1}:\; \forall\; (i,j)$, where $\epsilon_{ij}$ are the displacement matrices introduced in Eq.\il(\ref{Eq:Marix-ex-spo}). Given the constraint on  $\{r_{min}^i\}_{i=1}^n$ in the first case, and separately $\{\ell_i\}_{i=1}^n$ in the second case, we want to characterize the decomposition $\{C_i\}_{i=1}^n$ (i.e. the two $\ell$corotating sequences) finding how  it is changed the sequence  $\{\ell_i\}_{i=1}^n$ in the first case,   $\{r_{min}^i\}_{i=1}^n$  in the second one. We have noted in  different points in this work that each  $\ell_i$, in the model we are considering, set uniquely $r_{min}^i$ and,  viceversa,  if one considers the  $\ell$corotating  sequences only. However in this Section  we intend to consider a variable  configuration index  as already  done  in  Sec.\il(\ref{Sec:diff-rot}) and  for example in Sec.\il(\ref{App:notes-index}).
 These results are intended to be considered  in the perturbative approaches  with the perturbation of  $\mathbf{C}^n$ as  a perturbation \emph{in} and \emph{of} its decomposition $\{C_i\}_{i=1}^n$, introduced in  Sec.\il(\ref{Sec:pertur}).
 In this Appendix we will focus primarily on the  $\ell$corotating sub-sequences of the  decomposition, some of these results can be extended considering  more general  situations in mixed decompositions \ref{(--)}.
\subsection{Analysis of the  decomposition and  the differential rotation of the disk}\label{Sec:rotdisApp}
The specific angular momentum of the  $\ell $corotating sub-configurations in equilibrium, corresponding to the $C$ topology,   grows in magnitude with the distance from the source, which means that  $|\ell_{i+j}|>|\ell_i|\;\forall(i,j)$ and $\ell_{i+j}\ell_i>0$. However, this increasing is not constant with the radial distance:
 the curves  $\ell'\equiv\mp\partial_r\ell^{\pm}$, have a maximum point, as function of $r/M$, in  $r_{\mathcal{M}}^{\pm}(a)>r_{mso}^{\pm}$ respectively\footnote{The maximum point    $r_{\mathcal{M}}$ is a root  of the  polynomial: $-9 a^6-8 a^4 r+6 a^2 \left(-14+23 a^2\right) r^2+3 \left(48-124 a^2+75 a^4\right) r^3+\left(288-389 a^2\right) r^4-30 \left(-4+a^2\right) r^5-24 r^6+r^7$. For convenience here we consider a dimensionless $a$ and $r$. }, see Figs.\il(\ref{Figs:PlotLKKdisol},\ref{Figs:PlotLKKdisolb}).
From now on, to avoid encumbering the notation and discussion, where there will create no confusion, we  shall always  understand an increase or decrease in magnitude of the specific angular momenta  and accordingly we will avoid to report the  notation $|\phantom\ |$.

This means that, moving the center of the  $\ell$corotating rings (no matter whether consecutive or not)  outwards, it is necessary to increase the magnitude of the ring specific angular momentum, in a way not proportional to the displacement of the ring centers, i.e.,
\be
\delta \ell_{i+j,i}\equiv |\ell_{i+j}-\ell_i|\neq \alpha (r_{cent}^{i+j}-r_{cent}^i)\equiv \alpha \delta r_{cent}^{i+j,i},\quad\alpha=\mbox{constant}, \forall j>0, \quad \ell_i\ell_j>0,
\ee
$\ell_i$, is as usual the specific angular momentum of the $C_i$ ring. For the adopted model of thick accretion tori, there is $\ell_i\equiv\ell^{\pm}(r_{cent}^i)$   if the fluid is counterrotating or corotating.
Or in other words, to move the center of the $C_i$ ring  far  from the attractor, for the  shift $\mathfrak{S}_{i, i+j}:\;C_i\rightarrow C_{i+j}$  preserving the  stability of the  $C_{i+j}$ ring,  $\delta \ell_{i+j,i}$ should be  increasingly large as  $i$ increases, at fixed $j$ (which means physically equidistant center rings),  i.e. $\partial_r\delta \ell_{i+j,i}>0$, at equal $\delta r_{cent}^{i+j,i}$  up to the maximum $r_{\mathcal{M}}(a)$  after which  $\partial_r\delta \ell_{i+j,i}<0$, see also considerations in \ref{NoteIII}. We are considering a macro-configuration $\mathbf{C}^n$ consisting  of sequences of  $\ell$corotating  equidistant rings see--Figs.\il(\ref{Figs:CrystalPl}), along  the entire disk $\mathbf{C}^{n}$  with respect to their center, or $\left.\partial_i\delta r_{cent}^{i+j,i}\right|_{j}=0$, and we  are investigating  how the ordered sequence of the specific angular momentum $\{\ell\}_{i=1}^n$ should be fixed. In the model  we are considering here, the index  $i$ corresponds to one and only one center (and therefore to a single $\ell_i$) then $\partial_r$ also becomes $\partial_i$ at fixed (shift) $j$ and the former relations can then also be written in terms of the variation  on the configuration index  $i$:  i.e. $\partial_i|\ell_{i}|>0$ $\forall i$ but  $\partial^2_i|\ell_{i}|>0$  up to a $i_{\mathcal{M}}$ correspondent  to the ring with the center in $r_{\mathcal{M}}$ where\footnote{In fact, the physical meaning of the  point of maximum $r_{\mathcal{M}}$ in terms of the gap in the specific angular momentum is immediate and it trivially follows the definition of relative maximum of a smooth real function:  $\forall j\approx 0$ there exists a couple of sub-configurations, $C_{i_{\mathcal{M}}\pm j}$
such that $\delta \ell_{i_{\mathcal{M}},i_{\mathcal{M}}+ j}-\delta \ell_{i_{\mathcal{M}},i_{\mathcal{M}}- j}=0$ respectively.}
$\partial^2_i|\ell_{i}|=0$. Then  $\partial^2_i|\ell_{i}|<0$ for higher indexes  $i>i_{\mathcal{M}}$ or also:
\be
\partial_i\left.\delta \ell_{i+j,i}\right|_j\gtreqqless0\quad\mbox{for}\quad i \lesseqqgtr i_{\mathcal{M}}\quad\mbox{and}\quad i\in]i_{mso},+\infty[,
\ee
 where $i_{mso}$ corresponds to the ring with  the {\emph{cusp}}\footnote{So far we have considered the configuration index as attached to the ring center, that is at $r_{min}>r_{mso}$. This is because the critical points, for each $\ell$corotating sequence  at $r>r_{mso}$ are only minimum points of the effective potential. At $r=r_{mso}$, there is not a minimum of the effective potential. } in $r_{mso}$. Clearly   the  commuted relation with  respect to the couple $(i,j)$:  $\partial_j\left.\Delta \ell_{i+j,j}\right|_i>0$ $\forall \; j >0\; i>i_{mso}$ always holds, see also Sec.\il(\ref{Sec:diff-rot}).

The surplus of specific angular momentum $\delta \ell_{i+j,i} $ that should be provided  increases more and more slowly in far away regions  ($r (i)\gg r_{\mathcal{M}}(a) (i_{\mathcal{M}})$), where the Newtonian limit  could be considered and, as asymptotically   it is $\lim_{r\rightarrow\infty}\ell'=0$, with\footnote{Actually, the limit under consideration is related to the dimensionless ratio  $r/M\rightarrow\infty$, equivalently, one can see a similar asymptotic behavior with respect to the dimensionless quantity  $R\equiv r/a\rightarrow\infty$.  The asymptotic limit with respect to $r/M$ goes to zero as $r^{-1/2}$.} $\lim_{r\rightarrow\infty}\ell^{\pm}=\pm\infty$, then $\delta \ell_{i+j,i}$ remains constant.

Equivalently, increasing regularly  the specific angular momentum $\ell_i$ (in  magnitude) of a $C_i$ sub-configuration of a sequence of $\ell$corotating rings for a certain constant amount $\bar{\delta}_{\ell}$, i.e. $\epsilon_{i i+1}=\bar{\delta}_{\ell}\;\forall i$,   the disk center   $r_{cent}^i$  moves outwards, but the distance $\delta r_{cent}^{i,i+\epsilon}$ between  the centers  of the configurations   decreases until  $r=r_{\mathcal{M}}$. Indeed,  $\delta r_{cent}^{i,i+\epsilon}= \delta\ell_{i,i+\epsilon}/ |\ell '|= \bar{\delta}_{\ell}/|\ell'|$
as $\epsilon \gtrapprox 0$. Approaching $r_{\mathcal{M}}$, the distance $\delta r_{cent}^{i,i+\epsilon}$  decreases  (as $\epsilon\approx 0$):
\bea
\left.\partial_{\epsilon} r_{cent}^{i+\epsilon}\right|_{i}>0\quad\mbox{and}\quad\left.\partial_{i} r_{cent}^{i+\epsilon}\right|_{\epsilon}>0, \quad \partial_{|\ell|}r_{cent}^{i}(\ell)>0\quad\forall i>i_{mso}\quad\epsilon>0, \quad
\partial_i\delta r_{cent}^{i,i+\epsilon}\lesseqqgtr0, \quad i \lesseqqgtr i_{\mathcal{M}}
\eea
for $i\in]i_{mso},+\infty[$. Note that in this case, unlike the similar case for the specific angular momentum (where the shift parameter $j$ was kept constant), the shift parameter $\epsilon$ must vary as  $\delta\ell_{i,i+\epsilon}=\bar{\delta}_{\ell}$ is kept constant. Then the analysis of the variation of  $\delta r_{cent}^{i, i+\epsilon}$ actually corresponds to the analysis of the variation of the shift (which is always spatial)  $\epsilon$ required to keep the  shift in the specific angular momentum constant.

In the region $r>r_{\mathcal{M}}$, with constant,  $\bar{\delta}_{\ell}$, i.e. equal  difference between the specific angular moments of the rings, the disk decomposition becomes more and more extended, the separation between the centers of the sub-configurations more and more spaced, so that   spacings among the rings  from  the maximum point  increase  while  they are closer  in the region   $r\in]r_{mso}, r_{\mathcal{M}}]$. In terms of the rings density, \ref{NoteIV},
one could say that
\be\label{Eq:densi}
\partial_r\left.n(r)\right|_{\bar{\delta}_{\ell}}\gtrless0\quad\mbox{for}\quad r\lessgtr r_{\mathcal{M}},
\ee
for any $\ell$counterrotating sub-sequence.
 The ring edges are still to be defined.    To do this we need to  combine the given  analysis with some evaluations of the choice of the $K$  parameter,  investigating   consequences of the assumed conditions on $\{r_{cent}^i\}_{i=1}^n$ or  $\{\ell_i\}_{i=1}^n$  in the sequence $\{K_i\}_{i=1}^n$.
In any cases, these considerations  are relevant  to determine the differential rotation of the disk $\mathbf{C}^n$ as discussed in Sec.\il(\ref{Sec:diff-rot}),  the specific angular momentum of the configuration according to the definitions (\ref{Eq:stand})  and discussed in Sec.\il(\ref{Sec:diff-rot}), and finally to  define the effective potential for the ringed disk according to  Eq.\il(\ref{Eq:def-partialeK}) and Eq.\il(\ref{Eq:Vcomplessibo}) as discussed in Sec.\il(\ref{Sec:effective}).
We can consider  these results with respect to a   mixed or isolated  decomposition of a generic ringed disk  $\mathbf{C}^n$.
Given the two  $\ell$corotating sub-sequences   $\{C_i\}_{i=1}^{n_{\mp}}$ of  $\mathbf{C}^n$  
with $n= n_-+n_+$, there are  two relative maxima of the magnitude of the specific angular momentum  $\ell_{Max}\equiv\{\ell_{n_-},\ell_{n_+}\}$, clearly  related to the outer edge of the disk (still undetermined) of $\mathbf{C}^n$. Thus $\ell_{n_\pm}$ are respectively the specific angular momentum of the outer or inner ring of the two sequences, but do not determine $\ell_n$ as we can have   $C_n=C_{n_-}$ or $C_n=C_{n_+}$, neither the inner edge, as we can have  $C_1=C_{1_-}$ or $C_n=C_{1_+}$, where $\ell_{1_{\pm}}=\min\{|\ell|_{i_{\pm}}\}_{i_{\pm}=1}^{n_{\pm}}$.
Moreover, to identify  the specific angular momenta   $(\bar{\ell}_{\mathbf{C}}^n,\bar{\bar{\ell}}_{\mathbf{C}}^n)$ or  $(\bar{\ell}_{\mathbf{C}^n}^l,\ell_h)$ as defined in Sec.\il(\ref{Sec:diff-rot})
  Eqs\il(\ref{Eq:stand},\ref{Def:leadingl}), we have to  present some  further considerations on the morphological features of the rings $C_i$, as the height $h_i$ and the area $\mathcal{A}_i$:
it is then  $y_3^{n_\pm} \ell_{n_\pm}=\sup\{y_3^{i} \ell^{n^i}\}_{i=1_{\pm}}^{n_{\pm}}$ \emph{whatever}  $K$ is. The greater is $K$, the   greater  is the elongation  $\lambda_{i}$, in other words $\partial_K\lambda_i>0\; \forall K\in]K_{mso},1[$.

Concerning the two $\ell$counterrotating sub-sequences, it is worth noting that
\be
r_{\mathcal{M}}^{+}>r_{\mathcal{M}}^-,\quad
-\ell_{\mathcal{M}}^{+}>\ell_{\mathcal{M}}^-,\quad
\mp\ell_{\mathcal{M}}^{\pm}\in]\mp\ell_{mbo}^{\pm},\mp\ell_{\gamma}^{\pm}[,\quad
r_{\mathcal{M}}^{\pm}> r_{mso}^{\pm},
\ee
while $a>0.93M$ it is $\ell_{\mathcal{M}}^->\ell_{\gamma}^-$, when  the corotating rings have to be  in equilibrium \citep{coop},  see also Sec.\il(\ref{Sec:procedure})\footnote{The function $\ell'=|\ell_{i+\epsilon}-\ell_i|/\delta_{cent}^{i,i+\epsilon}$, obtained for composition of   the specific angular momenta of  two corotating consecutive disks (for  $\epsilon\approx 0$ the rings  $C_{i+\epsilon}$, the outer,  and  $C_i$, the inner, are very close) increases with $r/M$,  as the rings are sufficiently close to the source or  $r\in]r_{mso}^{\pm},r_{\mathcal{M}}]$,  at greater distances  ($r>r_{\mathcal{M}}$) this rise reduces increasingly to become zero.
These results apply in any Kerr spacetime, but depending on the attractor spin, the  maximum $r_{\mathcal{M}}$,  being  a function of $a$,  occurs in a different orbital regions. As shown in Figs\il(\ref{Figs:PlotLKKdisol},\ref{Figs:PlotLKKdisolb})) for the rings  counterrotating with respect to the black hole, the position $r_{\mathcal{M}}^+$ increases with the \textbf{BH}  spin, then the greater the \textbf{BH} spin, the greater  is the specific angular momentum of the  $\ell$corotating  rings made up by counterrotating  matter with respect to the attractor.
Inversely, for the case of corotating rings,  the region of increasing gap of specific angular momentum decreases with the spin, that is, the increasing of the  \textbf{BH} spin  corresponds to a decrease for the additional specific angular momentum to be supplied to locate the ring centers in an exterior region. Thus, one could say in general that if the $\ell$corotating   disks are corotating with respect to the black hole we need less additional specific angular momentum to locate outwardly ($\ell$corotating and corotating) rings with  respect to the other case. In this sense, one could say  also that the corotation with respect to the \textbf{BH} rotation  has a stabilizing effect for the orbital properties  of ringed disks in equilibrium.
}.
Therefore, $r_{\mathcal{M}}^{\pm}$ can be always a center for a counterrotating or corotating disk
 but, as it turns from the analysis of the specific angular momenta, the only instability possible for such a ringed disk is in its open configuration. It cannot be in accretion, but only  jets are possible because of the high  specific angular momenta.  The launching point of the jet would be at
$]r_{\gamma}^{\pm},r_{mo}^{\pm}[$ respectively.
Moreover, as mentioned in Sec.\il(\ref{Sec:rolel}),
\be
\partial_{a_*} r_{\mathcal{M}}^{\pm}\gtrless0,\quad \partial_{a_*} K_{\mathcal{M}}^{\pm}\gtrless0,\quad
\mp\partial_{a_*} \ell_{\mathcal{M}}^{\pm}\gtrless0\quad a_*\equiv a/M,
\ee
see Figs\il(\ref{Figs:PlotLKKdisol},\ref{Figs:PlotLKKdisolb})).
We could say that for high attractor spins the Newtonian limit for the  corotating sub-configuration is  reached, in regions close to the source, and with lower specific angular momenta magnitude with  respect to the counterrotating one. This is in agreement with the discussion in  Sec.\il(\ref{Sec:rolel}). The spin of spacetime clearly distinguishes the two types of configurations
 and the   relativistic effects are essentially determined by the rotation of the attractor.
Indeed, considering  the two mixed sequences and  Figs\il(\ref{Figs:PlotLKKdisol},\ref{Figs:PlotLKKdisolb})),  we note  that for increasing spin:
\be
 r_{\mathcal{M}}^-(a_{\mathcal{M}_1})=r_{mso}^+(a_{\mathcal{M}_1}),\quad
 r_{\mathcal{M}}^-(a_{\mathcal{M}_1})=r_{mbo}^+(a_{\mathcal{M}_2}),\quad
r_{\mathcal{M}}^-(a_{\mathcal{M}_3})=r_{\gamma}^+(a_{\mathcal{M}_3}),\quad a_{\mathcal{M}_1}<a_{\mathcal{M}_2}<a_{\mathcal{M}_3}.
\ee
Then
 we see a clear difference for sources with
$a>a_{\mathcal{M}_1}$
and particularly with $ a>a_{\mathcal{M}_3}$ where the Newtonian limit for the corotating  rings approaches the source. In terms of the ring density, in $]r_{\mathcal{M}}^-,r_{\mathcal{M}}^+[$, the  $\ell$counterrotating sub-configurations  can have  points of  maximum pressure (thus particularly  for  $a>a_{\mathcal{M}_1}$ at $]r_{mso}^+,r_{\mathcal{M}}^+[$). Having in mind the relation  (\ref{Eq:densi}), we find the relation $\partial_r\left.n^{\pm}(r)\right|_{\bar{\delta}_{\ell}}\gtrless0
$.
\subsection{Characterization of the two macro-structures}
In  Sec.\il(\ref{Sec:rotdisApp}) we   considered two  different ringed disks: the  first, say $\mathbf{C}^n_r$   disk made by  equally spaced  rings (according to the location of their centers), and we have  characterized the differential rotation of the ($\ell$corotating) rings.
In the second case,  say   $\mathbf{C}^n_{\ell}$ disk,    we  considered a constant  specific angular momenta  gap between the   $\ell$corotating sub-configurations and we determined the  ring location  in  each sub-sequence $\mathbf{C}^{n_{+}}$ or $\mathbf{C}^{n_{-}}$ respectively.  We have found that the ring density defined  as the density  of the  centers in  a closed orbital region, i.e.  $n=n_{min}/\delta r$ where $n_{min}$ is the number of centers located in the orbital range $\delta r$, is greater  close to the  maximum $r\leq r_{\mathcal{M}}$.  Then the maximum variation of the orbital specific angular momentum corresponds to the maximum of the density of  ring number (not necessarily of the maximum  disk mass, as it depends also on $K_i$, and   on the distance $r_{min}$ from the center) at constant gap of   specific angular momentum (whatever the value is of specific angular momentum)  for every $\ell$corotating sequence belonging to the generic decomposition  $\{C_i\}_{i=1}^n$, see Eq.\il(\ref{Eq:densi}).

There are two  density peaks corresponding to the two $\ell$counterrotating sequences.
Obviously,  to ensure the separation between the rings, the sequences should be  fixed accordingly, so that each value  $K_i$ is small enough to guarantee the separation. Then it follows that   higher number density corresponds to a lower spacing of the disk in the range of fixed $r$ and smaller rings in its decomposition.
We expect that increasing the number of rings in a neighborhood of the $r_{\mathcal{M}}$, the disk height will be lower  with lower spacing $\bar{\lambda}$ and the rings  would be considered constituting a geometrically   thin  disk.
These considerations have some relevance  in the analysis of the perturbations of $\mathbf{C}^n$ rings and  precisely in  the sequence  $\{\ell_i\}_{i=1}^n$ or $\ell$-modes.
Indeed, more generally  one  can use a general   functional relation in the  differential rotation. For example, this is certainly the case in the perturbation theory, where the displacement matrices $\epsilon_{ij}$ could also be not constant, see Sec.\il(\ref{Sec:diff-rot}) and Sec.\il(\ref{Sec:pertur}).

Suppose  that in  the $\mathbf{C}^n_r$ disk  the  distance $\delta r_{min}^{i+j,i}$ among the  centers is not constant, and  for the $\mathbf{C}^n_{\ell}$ disk model the gap between the specific angular momenta is not constant  but there is a (known) relationship such that  $(\clubsuit)_r:\; \delta r_{min}^{i+j,i}=\delta r_{min}^{i+j,i}(r)$ or, using the ring indices   $j=j(i)$ for the first disk and, in the same way for the second  model $(\diamondsuit)_{\ell}:\; \delta\ell_{i+\epsilon,i}=\delta\ell_{i+\epsilon,i}(\ell)$. We aim to find a characterization of a ringed disk subjected to generic  conditions $(\clubsuit)_r$ and   $(\diamondsuit)_{\ell}$ respectively.
First, one  needs to clarify  the choice of the functional dependence in the conditions
 $(\clubsuit)_r,\quad(\diamondsuit)_{\ell}$. Regarding  $(\clubsuit)_r$, the condition on the indices $j=j(i)$ is immediate:  we mean that the distance between the centers, defined by the shift  $j$, depends on the disk $C_i$, and thus on its position as $r_{min}^i$, which is moved  by  the translation $\mathfrak{S}_{i, i+j}:\;  C_i\rightarrow C_{i+j}$ that changes with  $i$. Since this is a  one-dimensional problem, the shift is uniquely determined by $ j $, if  $j=j(i)$ then   in terms of  $r$  it is $\delta r_{min}^{i+j,i}(r_{min}^i)$.

A different consideration must be made for the conditions of   $(\diamondsuit)_{\ell}$;
 in terms of configuration index it should be $(\diamondsuit):\; \epsilon=\epsilon(i)$. For the definition of  $\ell_i$  the parameter  $\epsilon$ follows the movement associated with the centers of the configuration, and this obviously depends on the distance between the specific angular momenta fixed a priori. Even if not constant, the specific angular momentum $\ell_{i+\epsilon}$ then  will be a function of the momentum $\ell_i$.
(The indexes of $\ell$ are always intended to be related to the ring position, but  $\epsilon$ in the definition of  $(\diamondsuit)_{\ell}$ is not constant).
These considerations can be generalized to any mixed decompositions having in mind the  conditions and constraints for the separation of matter, discussed in Sec.\il(\ref{Sec:pertur}) on the perturbative $\ell$-modes and $K$-modes.
\begin{figure}[h!]
\begin{center}
\begin{tabular}{c}
\includegraphics[scale=0.34]{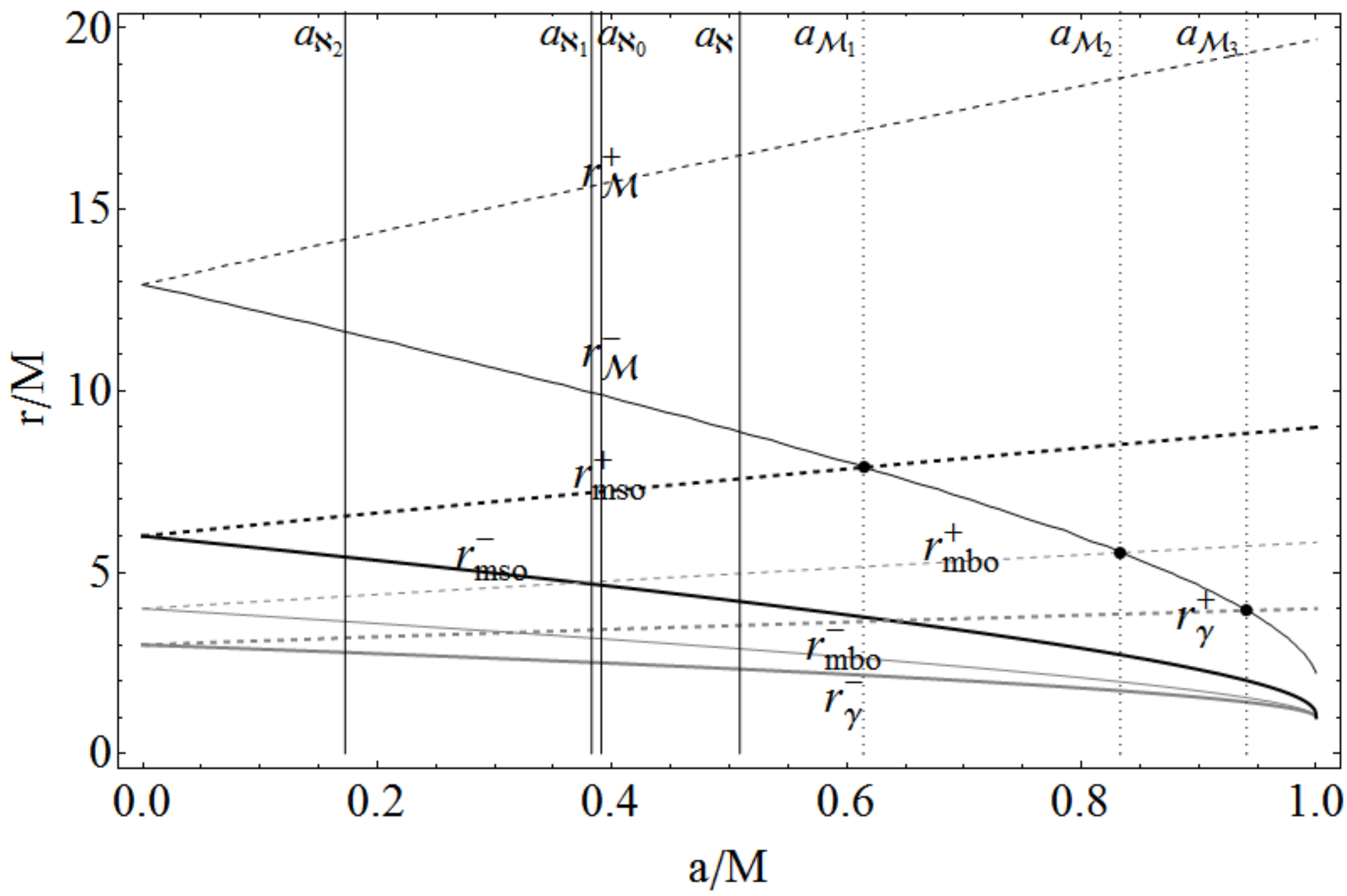}
\end{tabular}
\caption{The marginally stable circular orbits $r_{mso}^{\pm}$, marginally circular orbits $r_{\gamma}^{\pm}$, last bounded orbits $r_b^{\pm}$  functions of the spin-mass ratio of the attractor $a/M$. $r_{\mathcal{M}}^{\pm}$ are the orbits of maximum growing of the specific angular momentum magnitude, for corotating $(\ell_-, K_-)$ and counterrotating  $(\ell_+<0,K_+)$ fluids.  }\label{Figs:PlotLKKdisol}
\end{center}
\end{figure}
\begin{figure}[h!]
\begin{center}
\begin{tabular}{cc}
\includegraphics[scale=0.3]{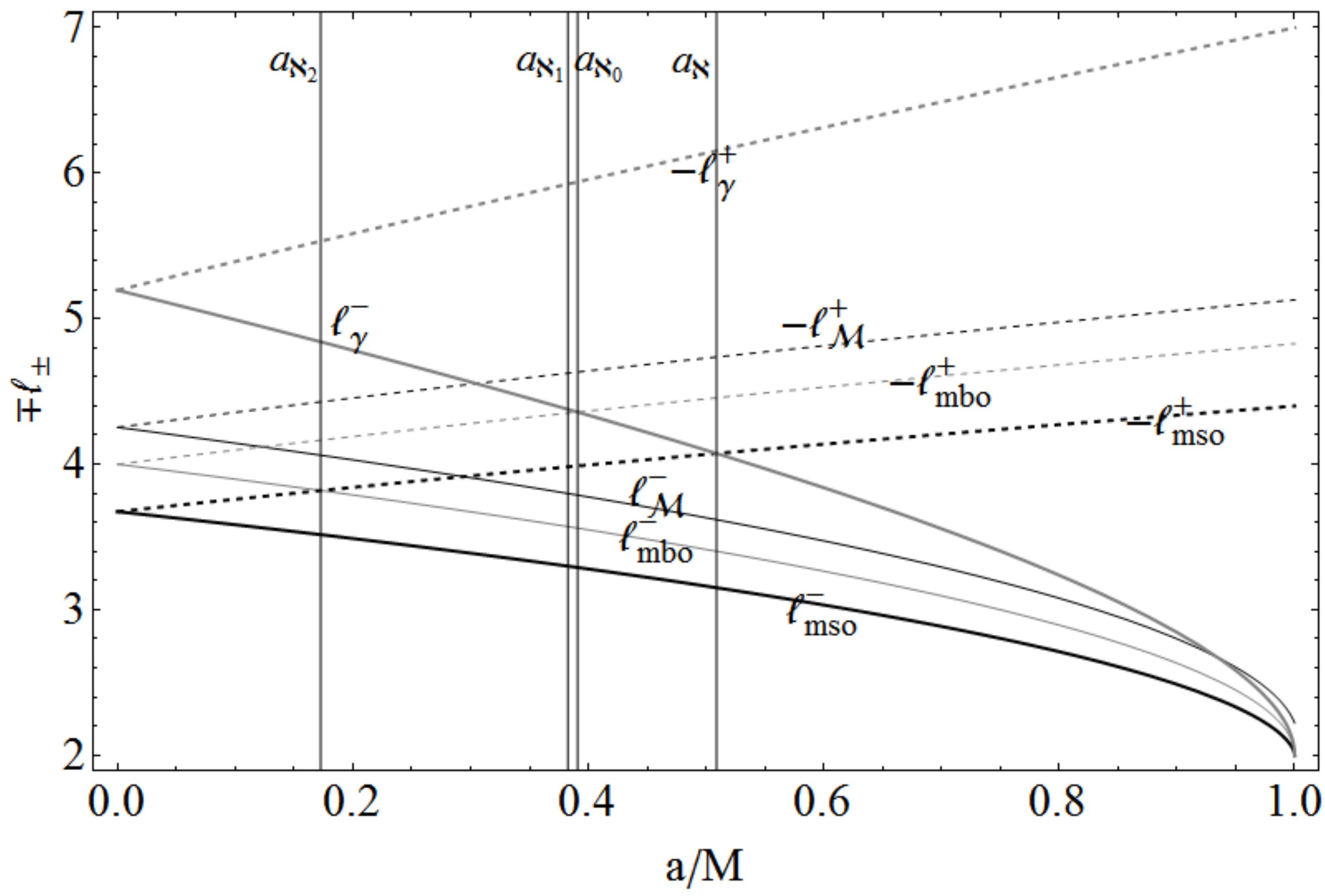}
 \includegraphics[scale=0.3]{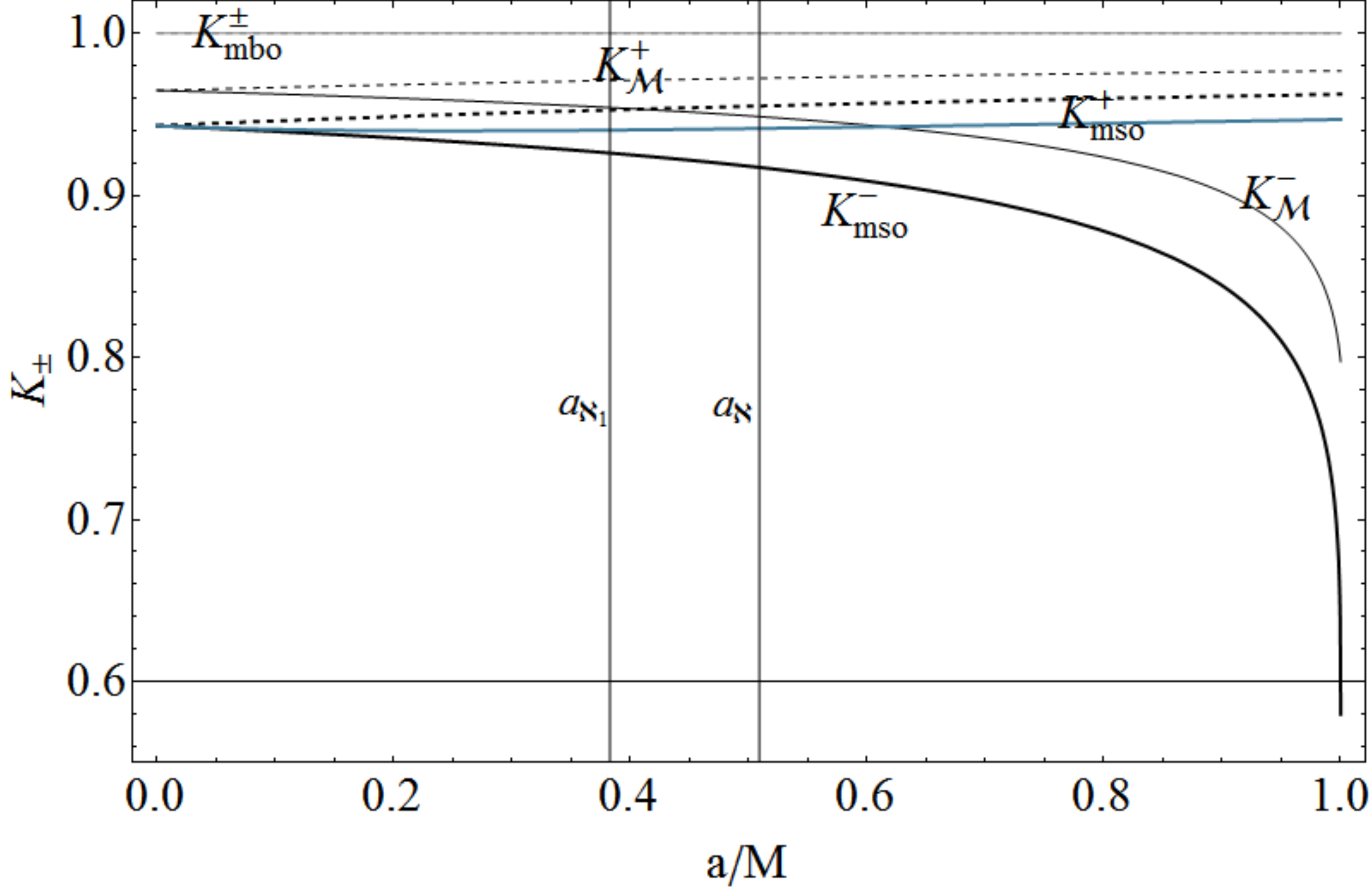}
\end{tabular}
\caption{Angular momenta (left)  and parameter $K$ (right), for corotating $(\ell_-, K_-)$ and counterrotating  $(\ell_+<0,K_+)$ fluids on the marginally stable circular orbits $r_{mso}^{\pm}$, last circular orbits $r_{\gamma}^{\pm}$, last bounded orbits $r_b^{\pm}$  functions of the spin-mass ratio of the attractor $a/M$. 
}\label{Figs:PlotLKKdisolb}
\end{center}
\end{figure}
\section{Some notes on the open configurations}\label{App:opem}
Following   the discussion in Sec.\il(\ref{Sec:procedure}), In this Section we  investigate some aspects of the decompositions with one or more critical P-W points and in particular the decompositions with   $O_x$ sub-configurations,
as shown in  Fig.\il(\ref{Figs:openFig}).
We  avoid to constraint the    $O_x$ surfaces with  the principle of non-overlapping (non-penetration) of matter considering therefore the possibility of a decomposition with more than one launching point of the jet.
For the critical configuration, it is useful to introduce, together with the notation $<$ and $>$ for the ordered sequence of maximum points of the pressure (or $r_{min}$, minimum of the effective potential and the disk centers for the closed sub-configurations), also the
symbols $\succ$ and $\prec$ intended to refer to  sequentiality    ordered location of the minimum points of the pressure (or $r_{Max}$ of the effective potential, or the instability points of accretion or launching of jets), while the terms ``internal'' (equivalently inner) or ``external'' (equivalently outer), in relation to a couple  of rings, will always refer, unless otherwise specified,  to the usual ordinate sequence according to the location of the  centers.

\medskip

\textbf{Critical points in a $\ell$corotating sequence}

We consider first the case of $\ell$corotating sequences, focusing first on a ring couple.

In the case of an \emph{outer  jet}
it can be:  $O_x^i<O_x^o$, $C_x^i<O_x^o$ or $C_i<O_x^o$ where, for definition, it is always
$r_{min}^i<r_{min}^o$, and  $r_{Max}^i>r_{Max}^o$ (according to the existence  conditions) and then if  $()^i<()^o$,  where $\ell_i\ell_o>0$, then $()^i\succ()^o$.

In the case
\be
O_x^i<O_x^o\quad\mbox{it is }\quad O_x^i\succ O_x^o\quad, |\ell_i|<|\ell_o|\in]|\ell_{\gamma}|,|\ell_b|[:
\ee
 there are  two $\ell$corotating   jets starting from the points $(r_{Max}^{o},r_{Max}^{i})$.

In the second case:
\be
C_x^i<O_x^o\quad\mbox{it is }\quad C_x^i\succ O_x^o\quad |\ell_i|<|\ell_o|\in]|\ell_{\gamma}|,|\ell_b|[\quad\mbox{  where}\quad |\ell_i|\in]|\ell_{mso}|,|\ell_{b}|[;
\ee
the launching point of the jet is  internal with respect to the accretion point.
An interesting situation occurs when  the two instability points (location of minimum of the pressure)  are very close or coincide; the last case is possible only for  $\ell$counterrotating couple.

In the third case,  there is only one instability point and it has to be
\be
C_i<O_x^o\quad \ell_i<\ell_o\quad
r_{min}^i<r_{min}^o\quad K_i<1.
\ee
We now consider the case of an \emph{inner  jet}:
 $O_x^i<C_x^o$ or $O_x^i<C_o$.
Regarding the  case, $O_x^i<C_x^o$,  it  is  has to be  $O_x^i\succ C_x^o$, but this is not a possible configuration  for  $\ell$corotating couple--in fact it should be  $|\ell_i|<|\ell_o|\in]|\ell_{mso}|,|\ell_b|[$  and on the other hand it has to be  $|\ell_i|\in]|\ell_b|,|\ell_{\gamma}|[$  that is contradictory. In a  $\ell$corotating couple  with an accretion point and a jet, the launching point must be \emph{internal}, with respect to the accretion point.
The  case   $O_x^i<C_o$ and  $O_x^i\succ C_o$, similarly to the case  $O_x^i<C_x^o$,
 leads  to a contradiction.
\begin{figure}[h!]
\begin{center}
\begin{tabular}{cc}
\includegraphics[scale=0.3]{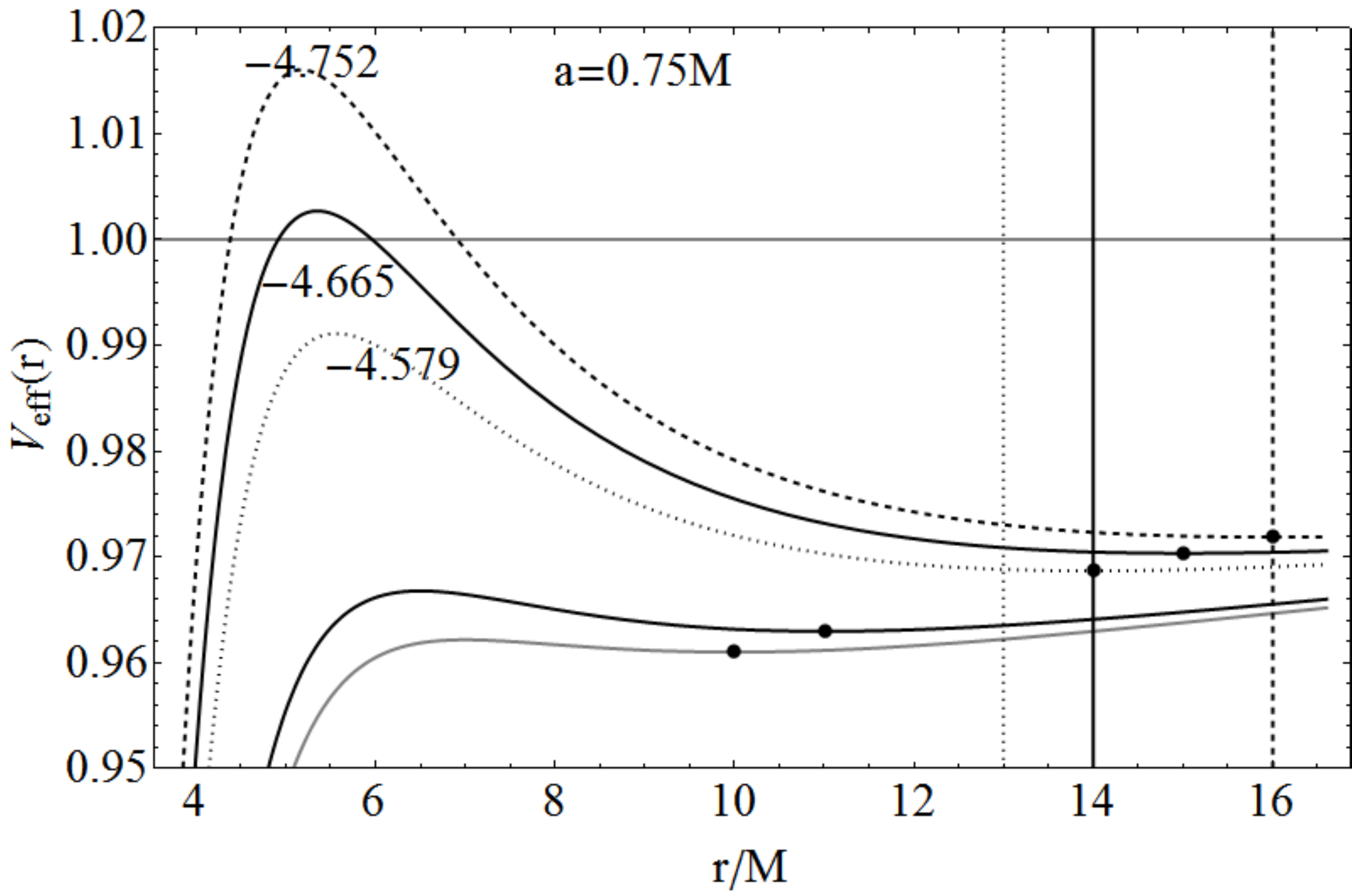}
\includegraphics[scale=0.3]{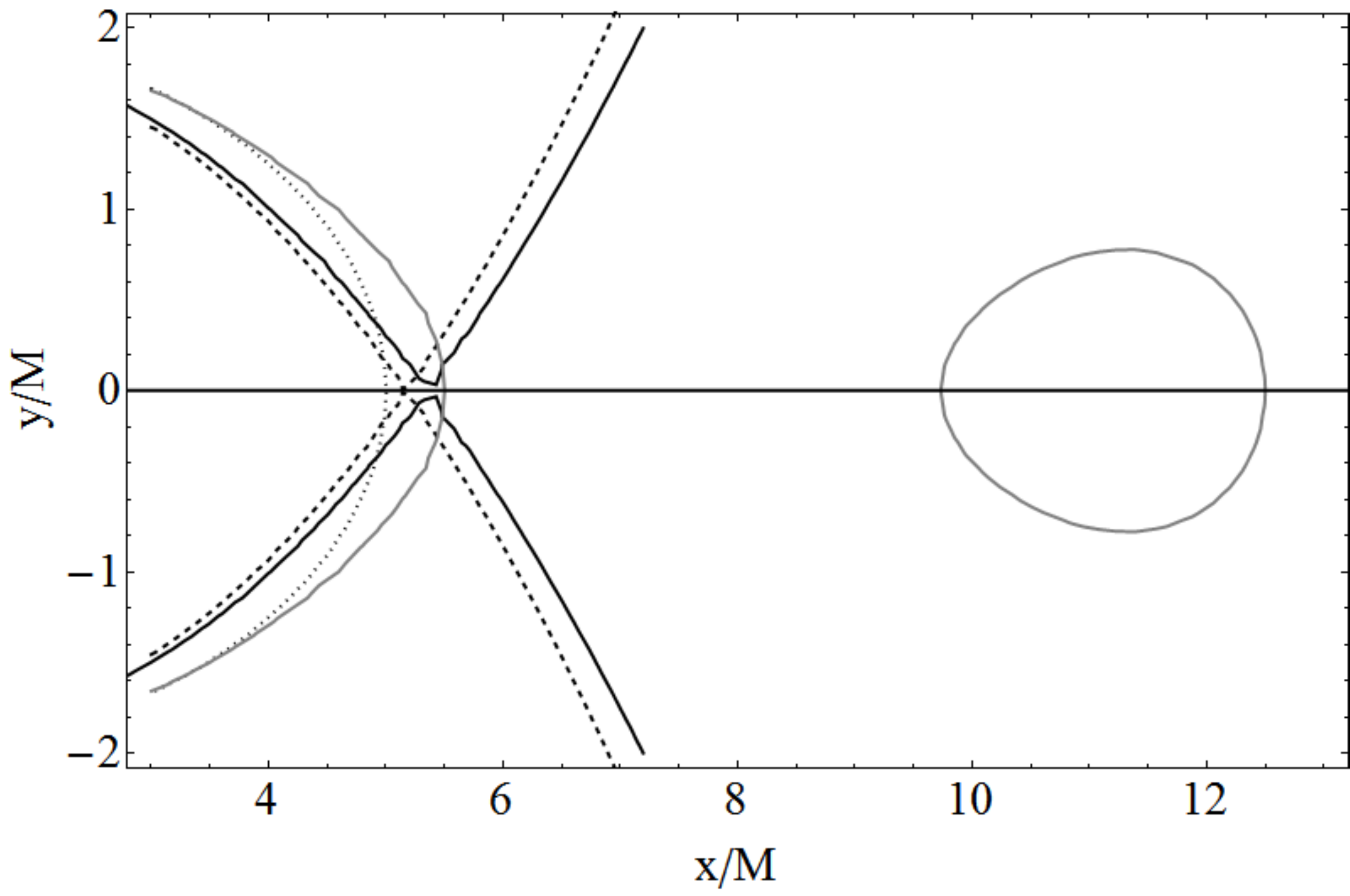}
\end{tabular}
\caption{Spacetime spin $a=0.75M$, $\ell$corotating sequences, $\ell_i\ell_j>0$, of counterrotating disks  $\ell_i a<0$ $\forall i j$. Decompositions including open-cusped sub-configurations $O_x$.  The outer horizon is  at $r_+=1.66144M$.}\label{Figs:openFig}
\end{center}
\end{figure}

\medskip

\textbf{Double critical points in a $\ell$counterrotating sequence}

This note follows the discussion in Sec.\il(\ref{Sec:procedure}), a part of the results shown here are discussed in the section on the $\ell$counterrotating sequences from Eq.\il(\ref{Eq:ban-appl-cho}).
First, we consider the case of  $C_x^+\succ O_x^-$, meaning an outer accretion point  with counterrotating matter followed by an inner instability point where matter, initially corotating with the black hole opens in jets, and
\be
\Delta_J^-\cap\Delta_x^+=\emptyset \quad\mbox{ it is} \quad C_x^+\succ O_x^- \quad\mbox{ and } \quad  C_x^+\nprec O_x^-  \quad\mbox{since } \quad \Delta_J^-<\Delta_x^+
\ee
see also Figs.\il(\ref{Figs:PlotLKKdisol}).

Consider now a couple of $\ell$corotanting  jet configurations.
Then
\bea\label{Eq:lagespicn}
O_x^+\succ O_x^-\quad\mbox{for}\quad a \gtrsim 0.35M; \quad
O_x^+\succ O_x^-\quad\mbox{or}\quad O_x^+\prec O_x^- \quad\mbox{for}\quad a \lesssim 0.35M,
\eea
see Fig.\il(\ref{Figs:PlotLKKdisol}), i.e.  at smaller attractor spin-mass ratio, where $r_{mbo}^->r_{\gamma}^+$, it can be also:
 $O_x^+\prec O_x^-$. But as  Eq.\il(\ref{Eq:lagespicn}) stands, for large  values of spin the inner jet must be corotating.

On the other hand, the couple
$C_x^-\succ O_x^+$, with an inner counterrotating  jet   and  an outer accretion point   from a corotating ring are possible for small enough  spin, i.e. $a\lesssim0.6$,   where $r_{mso}^->r_{\gamma}^+$, while for larger spin,  $C_x^-\prec O_x^+$, or  the accretion point of corotating material  follows the launch of a counterrotating jet.

For the double disk in accretion, it has to be
$C_x^-\prec C_x^+ $ for every spin. We note that  for large spin, $a>a_{\aleph_1}$, this relation could not be inverted to $C_x^+\prec C_x^-$, as clear from Fig.\il(\ref{Figs:PlotLKKdisol})\footnote{A  pointed out in Sec.\il(\ref{Sec:rolel}), it follows from the relation $r_{mso}^{\pm}\in C_x^{\pm}$, respectively, and the considerations of   Eq.\il(\ref{Eq:board-point}): in the case  $C_x^+\prec C_x^-:\; r_{mbo}^+<r_{Max}^+<r_{Max}^-<r_{mso}^-<r_{mso}^+$ and it has to  be in accordance to the Eq.\il(\ref{Eq:board-point}):
$r_{Max}^+<r_{Max}^-<r_{mso}^-<r_{mso}^+<y_{1}^+$  and  $y_1^->r_{mso}^-$,  should lead to a penetration of material between the first lobe of the two configurations, thus  this case is always forbidden.} . For the  couple $C_x^-\prec C_x^+ $ a penetration of matter occurs from the material of  the outer accreting ring to the inner one.
However the case of two disks in accretion clearly results in a penetration of matter that affects the more internal lobe  of the outer configuration  and the outer lobe of the internal one.

Finally, we note that a critical decomposition  $\mathbf{C}_{\odot}^x$ could be formed by a couple of  critical  sub-configurations
when
$y_1^i=y_3^{i+1}=r_{Max}^{i+1}$.  The  condition  in Eq.\il(\ref{Eq:condi-sss})  still holds formally as  here we are not considering  the inner lobe of the ring in accretion, however, within this setup, no more then two accreting points in a ringed disk can exist and they must be $\ell$counterrotating (indeed, $r^{\pm}_{mso}\in C_x^{\pm}$ respectively).
The spacings $\bar{\lambda}_{i+1,i}$ are in general increasing with the attractor spin and decreasing with the magnitude of their specific angular momentum  (corresponding to a decrease of their elongation)  see Eqs\il(\ref{Eq:neut-w},\ref{Eq:lastdon+-}).
As can be seen also from Fig.\il(\ref{Figs:PlotLKKdisol}), this case can be always possible,
 and clearly the outer ring of the $\ell$counterrotating couple has to be counterrotating independently of  the equilibrium state of the inner one, with  $y_1^-=y_3^+=r_{Max}^{+}\in\Delta_{mso}^{\pm}$.
 To simplify the discussion, we consider here  an unstable  inner corotating ring  implying   $\lambda<\lambda_x$, according to  Eq.\il(\ref{Eq:board-point}) and we can refer to Fig.\il(\ref{Figs:PlotLKKdisol}).
The inner ring is then in accretion but it could be $C_x^-\prec C_x^+$  or  $C_x^-\prec O_x^+$, and the considerations made before apply, distinguishing slow and fast attractor according to the former analysis--see also \citep{coop}.

\section{General considerations on the perturbations of the ringed disks and their equilibrium}\label{Sec:pertur}
In this section we consider  possible perturbative approaches to the study of the equilibrium and the   unstable phases of  $\mathbf{C}^n$ as defined in Sec.\il(\ref{Sec:intro}).
The  perturbations arising in the ringed disk structure are meant to be  perturbations of  its decomposition $\{C_i\}_{i=1}^n$, generated by perturbing     the decomposition effective potential in  Eq.\il(\ref{Eq:def-partialeK}) or
Eq.\il(\ref{Eq:Vcomplessibo}), or perturbations of the sequences of parameters $\mathbf{p}_{\mathbf{C}^n}=\{\mathbf{p}_i\}_{i=1}^n$.

We  introduced  in Sec.\il (\ref{Sec:effective}) the effective potential $V_{eff}^{\mathbf{C}^n}$ of  Eq.\il(\ref{Eq:Vcomplessibo}),  governing  the equilibrium structure   of  the  $\mathbf{C}^n$ macro-configuration. By coupling  the potentials of each toroidal ring, we have defined  the  boundary conditions for each sub-configuration  determining   the structural  ``rigidity'' of the macro-configuration.
In this work  we will not perturb the macro-configuration  $\mathbf{C}^n$, but we limit here to  set up the problem    clarifying  some  aspects of the configuration  such as the relationship between the elongation  and the spacings as  discussed in  Sec.\il(\ref{Sec:principaldef}) and Sec.\il(\ref{Sec:effective}).
We can indeed perturb the structure by a perturbation of its effective potential, as defined in Eq.\il(\ref{Eq:def-partialeK}) or \ref{Eq:Vcomplessibo}. The ringed  disk $\mathbf{C}^n$  is determined by the multidimensional ordered parameter  $\mathbf{p}_{\mathbf{C}^n}=\{\mathbf{p}_i=(K_i,
\ell_i)\}_{i=1}^n$.
The perturbations of the decompositions  can be constructed   by perturbing one or both sequences  of the  $\mathbf{p}_{\mathbf{C}^n}$ parameter:  $\{K_i\}_{i=1}^n$, and therefore the boundary conditions, see for example definition (\ref{Eq:def-partialeK}), or $\{\ell_i\}_{i=1}^n$. Keeping the other parameters  fixed, we could then speak  of \emph{$K$-mode} or  \emph{$\ell$-mode} of the perturbation  respectively. The ringed disk can be perturbed in one of these modes or in a combination of  these, and each mode can have some  sub-modes defined   by the restrictions  on the perturbation, bounded by specific relationships on the spacings and elongations.   From  the analysis of the  decomposition in Sec.\il(\ref{Sec:roleofp})  it follows that the
 associated order parameters of the perturbation, the sequences    $\epsilon_K=\{\epsilon_{K_i}\}_{i=1}^n$ and  $
\epsilon_{\ell}=\{\epsilon_{\ell_i}\}$ respectively, are not independent.

The system is unstable, if it endures   one of the unstable topologies  introduced  in  Sec.\il(\ref{Sec:unsta}) that is $\mathbf{C}^n_x$ or $\mathbf{C}_{\odot}^n$  or a combination of these.
The maximum rank  of  $\mathbf{C}^n_x$  is  $\mathfrak{r}=2$
(for $\ell$counterrotating  disks
of the  order $n\geq2$)  and the cusps are located in  the inner tori of the two ordered sequences of corotating and counterrotating rings of the decomposition. No feeding between the $\ell$-corotating disks   is possible, but an exchange of matter can occur as driven by the cusp. However, the instability of the ringed disk occurs due, for this case, to  the combined effect of the violation of the hydrostatic equilibrium by means of the P-W mechanism at the cusps  of one or more sub-configurations, and  the collision or penetration of matter from one ring to the other.
An interesting  situation occurs for the perturbation of a decomposition of  $\mathbf{C}^n$
that preserves the stability of each sub-configuration, i.e., for variation of the
parameters  $\mathbf{p}_{\mathbf{C}^n}$ far  from the  values for a P-W instability for each of its rings.
The breakdown of the equilibrium of the entire macro-configuration  causes  the emergence   of the  $\mathbf{C}_{\odot}^n$ topology, up to a saturated configuration   where the  total spacing is reduced to $\bar{\Lambda}=0$. In this case, as  mentioned in Sec.\il(\ref{Sec:rolel})  for the deformations introduced by the displacement matrix $\epsilon_{ij}$, the perturbation is  strongly constrained  on the initial decomposition. In fact, as mentioned before,
the order parameters  $\epsilon_{\mathbf{p}}\equiv(\epsilon_
{\ell},\epsilon_K)$ are related each other  being  constrained by conditions on
the spacings
 $
\bar{\lambda}_{i+1,i}$ and the distances between the
minima (or equivalently  the elongations). In the following discussion  we always consider   sequences of  ordered parameters and  maintaining a constant  $K_i$ or $\ell_i$ means to consider a  constant  $K$ or $\ell$ for  the  $i$-ring that maintains its identity as fixed in the initial decomposition\footnote{If  $\breve{\mathbf{Q}}$ are the perturbed quantities $\mathbf{Q}$, and $\mathbf{Q}_0$ the initial ones,  keeping $\mathbf{Q}_i$ constant does not mean assuming  $\exists j\neq i:\; \breve{\mathbf{Q}}^i=\mathbf{Q}^j_0$ but instead  $\breve{\mathbf{Q}}^i=\mathbf{Q}^i_0$. The perturbations are not intended to produce a shift in the ordered sequence of the background even if  this could be possible (these are somehow formalized as Eulerian and not Lagrangian perturbations).  The configuration index   has been assigned to ring  center or equivalently to its specific angular momentum  $\ell_i$, as such  in the  $\ell$mode perturbation which induces a change of the $i$  index. We  follow the trajectory of the single center which undergoes a translation along the equatorial axis. Then  in the $\ell$mode  perturbation it is   $\breve{\mathbf{Q}}^i ={\mathbf{Q}}^{\breve{i}}$,  we say that the   $C_i$ ring is now in the position $r_{min}^{\hat{i}}$, with specific angular momentum $\ell_{\hat{i}}$.}
\begin{description}
\item[The $K$-modes]
Keeping the sequence of parameters $\{\ell_i\}_{i=1}^n$   fixed,   the  perturbations will be only on the (variable of ) Heaviside functions in the effective potential (\ref{Eq:def-partialeK}). Then we are dealing with  a  one dimensional problem in each  $K_i\in[K^i_{min},K^i_{Max}]$, the  boundary of each range being fixed by $\ell_i$.  It should be noted that a perturbation of $K_i$ is equivalent to a   (rigid) perturbation of the inner and  outer edges of each ring (or also the elongations $\lambda_i$ at fixed $r_{min}^i$),  shift-parameters of the Heaviside in  (\ref{Eq:Vcomplesssprouibo}); the perturbation of an edge is transmitted rigidly in the second ring margin, being bounded by the  $K_i$ value only. For this to set a perturbation in terms  of  (\ref{Eq:def-partialeK}) would be more convenient, keeping constant  the centers $r_{c}^i$,  and resulting finally  in an expansion or contraction of the sub-configurations,  i.e.  causing  a change in the morphological properties of the rings. In any case  the maximum pressure  points (but not the minimum pressure points) are  fixed as initial data. Thus one can study the sub-$K$-mode considering whether  the spacings  between   consecutive  rings are fixed or not.
  If so, in the  \emph{$K_{\bar{\lambda}}$-modes},  then  the sequence $\{\bar{\lambda_i}\}$ is fixed, and the perturbation is bound to preserve the initial internal topology of the ringed disk. It cannot reach the $\mathbf{C}_{\odot}$,  but it could still be unstable for the P-W mechanics with a $C_x^i$ in its decomposition  or allowing even a $O_x$. 
 Then,  $\bar{\lambda}_{i, i-1}=\mbox{constant}$
 would connect rigidly the two consecutive parameters  $(K_i, K_{i-1})$. In other words, the perturbation of one of the couple  will propagate, through this constraint, rigidly to the other parameters $(K_{i+1}, K_{i-1})$. Proceeding iteratively,   fixing one starting point  of the perturbation, say a $K_*$, attached to any of the rings of the decomposition,  will set the perturbation of the entire decomposition in the $K_{\bar{\ell}}$ modes.
This subcase however clearly preserves the initial condition of stability of the decomposition, as long as  $K_*$ obeys precise constraints. Then we should repeat   similar arguments  to the  Eqs.\il(\ref{Eq:cononK})
 or Eq.\il(\ref{Eq:cononKlcontro}) for the determination of ring spacings.
In the procedure introduced  in Sec.\il(\ref{Sec:procedure}),   it was not necessary that   the starting point of the iterative process was   the inner edge of the first configuration (or $K_1$), but it can be as well any of the margins associated with any  elements of the decomposition $\{C_i\}_{i=1}^n$. The outlined procedure is a  good example to test the effects of the rigidity of this structure on the choice of the parameters.
The reasons for  the choice of the   $K_1$  as initial data  was motivated by the  convenience, as the inner edge of $C_1$, that is also the inner edge of the  $\mathbf{C}^n$ configuration, is  bounded below by the marginally stable orbit  as the outer edge  of $C_n$, and  $\mathbf{C}^n$ can extend outward to infinity. It is sufficient to have only a single initial data on a ring of the  decomposition to constrain the range of variation for the parameter of  the entire initial   decomposition.
Having in mind  also Eqs\il(\ref{Eq:neut-w},\ref{Eq:lastdon+-}
,\ref{Eq:le-no-ji}) we obtain for  generic indices  of the decomposition the relations
\be
\partial_{K_{j}}\lambda_j>0,\quad
\partial_{K_s}\lambda_j\lessgtr0 \quad \mbox{if}\quad j \gtrless s.
\ee
It is worth noting   that the procedure is antisymmetric with  respect to the difference $(j-s)$, i.e. for increasing  configuration index  moving towards  the outer rings, or decreasing indices,  going inwardly towards the attractor.
Applying a {chain rule}, there is  
\be
K_{j-1}=K_{j-1}(K_s),\quad
\partial_{K_s}{K_j}<0,\quad
\partial_{K_{j-1}}{K_j}<0.
\ee
We finally note that fixing the distances $(\lambda_i,\bar{\lambda}^{i, j})$
(or equivalently functions of these distances) does not fix the position of margins that could  translate on the radial  axis $r$.
The spacings $\bar{\lambda}_{i+1 i}$ in this scheme
 must be bound as:
$
 \bar{\lambda}_{i+1,i}=\delta_{min}^{i+1,i}-(\delta_
{min_1}^{i}+\delta_{min_3}^{i+1})$,
where we introduced  the part of elongation
$\delta_{min_3}^i=r_{min}^i-y_{3}^1$ in general different from  $\delta_{min_1}^i=y_{1}^1-r_{min}^i:\; \lambda_i=\delta_{min_3}^i+\delta_{min_1}^i$.
Noting that $\delta_{min}^{i+1,i}  $  does not depend on  $ K$,
 then
\be
\partial_K\delta_{min}=0,\quad
\partial_{K_i}\bar{\lambda}_{i+1,i}=-\partial_{K_i}
(\delta_{min_1}^{i}+\delta_{min_3}^{i+1}),
\ee
the first term in the brackets is always greater then zero and the second  one is negative;  the sum can be positive, negative or zero
 considering the constraint
$\delta_{min_3}^{i+1}\leq\delta_{min_1}^{i}$.
The choice of  $K_1$ is restricted to the values
$K^i\in[K_{min}^i,V_{eff}(\ell_{i},r_{min}^{i+1})] $, in particular this is true for  $i=1$ that is the starting point of  the procedure in  Eq.\il(\ref{Eq:cononK})
 or Eq.\il(\ref{Eq:cononKlcontro}).
But the constraint can be  further
restricted  for any $K^i$ (general condition on the elements of the sequence $\{K_i\}_{i=1}^n$) by
$K^i\in[K_{min}^{i-1},K_{\bullet}^{i+1}]$
 being true in particular for $K^i=K_*$, where:
\bea
K_{\bullet}^{i}=\inf \{V_{eff}(\ell_{i},y_1^{i-1}),V_{eff}(
{\ell_{i},y_{min}^{i+1}})\}\quad \mbox{for}\quad i\neq 1\; i\neq n
\\
\quad\mbox{ with}\quad
K_{\bullet}^{1}=V_{eff}({\ell_{1},y_{min}^{2}}),\quad
K_{\bullet}^{n}=V_{eff}({\ell_{n},y_{min}^{n-1}}).
\eea
These can set the boundary conditions for the  perturbation of our model  depending  on the initial data.
Thus we can write now:
\be
\partial_{K_i}\bar{\lambda}_{i+1,i}=-\partial_{K_i}
(\delta_{min_1}^{i}+\delta_{min_3}^{i+1}).
\ee
Assuming  the spacing set a priori, not necessarily zero, not necessarily constant with respect to the configuration index,   the spacing is rigid but the elongation varies,
and  in this case we obtain
\be
\partial_K^1\delta_{min_1}^{i}=-\partial_K^1\delta_
{min_3}^{i+1}.
\ee
We finally note that
fixing the spacing means ensuring the  hypothesis of rigidity on the ringed disk  for  the spacing would be independent of the perturbations.
\\
\item[The $\ell$-modes]  An increase in  the specific angular momentum magnitude means a  perturbation
outwards, leading to a radial movement
outside.
On the other hand,  varying  $\ell_i$ at fixed $K_i$, the Boyer surfaces   are not rigidly translated (a shift in $r_{min}^i$) on the radial direction, but the ring morphology changes  and in particular  its thickness. This  affects the   morphological characteristics of the ringed  structure  as the thickness and elongation, but not the  total spacing $\bar{\Lambda}_{\mathbf{C^n}}$ if the spacings are kept fixed, as it is in the \emph{$\ell_{\bar{\lambda}}$-modes}.  We  note that  considering fixed the   $\{K_i\}_{i=1}^n$ sequence in the effective potential  of the decomposed structure means keeping  fixed the Heaviside functions in Eq.\il(\ref{Eq:def-partialeK} but not  in   Eq.\il(\ref{Eq:Vcomplesssprouibo})
or in Eq.\il(\ref{Eq:Vcomplessibo}). Indeed, at $K_i$  fixed, varying $\ell_i$  does not imply neither  fixed, ring edges neither fixed  elongation $\lambda_i$, but  one should impose a priori this further restrictions, considering the combined  variation in Eqs\il(\ref{Eq:neut-w},\ref{Eq:lastdon+-},\ref{Eq:le-no-ji}) and \ref{NoteII}.
This is again a one dimensional problem in the radial direction, where different submodes are possible:
if we keep  constant    $\lambda_i$ during the perturbation, we could have \emph{$\ell_{\lambda}$-modes}, and these segments can move forward or backward in the radial direction (increasing and decreasing respectively the specific angular momentum in magnitude, having in mind that one can possibly avoid to consider a ``turning point'' at $\ell=0$), oscillating along $r$,   undergoing  only  radial translations by translation of $r_{min}^i$ preserving  the length (but clearly not the verticality of each ring $h_i$).

However, we have to point out  that formally the $\ell_{\lambda_i}$-modes do not allow   the condition $K_i=$constant (it could be perhaps possible for a $\ell$couuterrotating couple,  but the case $\breve{\ell}_i\ell_0^i<0$ is   excluded here), so that we still consider $\ell$ the perturbation variable,  having $\lambda_i=$constant for the perturbation and therefore let the $K_i$ parameter changing accordingly.

A further submode is due to the  double restriction of $\lambda_i=$constant  \emph{and} $\bar{\lambda}_{i,i+1}=$constant, or \emph{$\ell_{\bar{\lambda}\lambda}$-mode}, for  each pair of consecutive indices of the decomposition.
  In this case, however, the perturbation leads to a change of the verticality of each ring $(h_i)$ and the  shift  (oscillation of) the entire elongation of the ringed disk  $\lambda_{\mathbf{C}_{n}}$, without a fixed point. Optionally, one can relax these conditions assuming a known generic relation between these displacements; for a more general discussion see also  Sec.\il(\ref{Sec:app-maxmin}).

In  the $\ell_{\bar{\lambda}}$-mode,  as in  the  $\ell_{\bar{\lambda}\lambda}$-mode, and  for the $K_{\bar{\lambda}}$-mode,  the initial internal topology of $\mathbf{C}_n$ is preserved,  making  possible a  P-W instability for a ring of the decomposition. Each ring finally changes the thickness.
A perturbation of $ \ell $ (then a
change in velocity in the toroidal direction), at $ K $ fixed, leads in general to a radial perturbation with consequent displacement of the constraint $\lambda_i$.
The perturbations under consideration here  induce a radial
oscillation of the ringed disk  with  general non-periodic or quasi-periodic character depending on the spacing. The frequencies of each one, attached to each spacing is then linked to an equation representing  a  chain of coupled oscillators. The  change of specific angular momentum creates an overall variation of the associated chain of oscillators which may be used for example in models based on
observed QPOs  \citep{KA021,KA02,AKBHRT,SS11,TAKS05,BAF2006} and also \citep{RH05,RH05a,RYMZ,RVW,SM10,Nagar:2006eu}.
The oscillator is then centered in
the center of each sub-configuration $C_i$.
Each oscillator is  associated with a section of the torus in the equatorial plane. The oscillators are connected by the spacings within the constraint that the spacing can be at most zero but not negative (no penetration of matter, this ``rigidity'' of the surfaces should not be broken).
\end{description}
This kind of perturbations  preserves the equatorial  symmetry of the toroidal structure for which the only degree of  freedom is the radial direction. In other words, it represents  a one-dimensional problem. But the shift of the center of each ring in the radial direction, varying the specific angular momentum, or of its elongation in the case of $K$-modes (note that there cannot be a $K_{\lambda}$-mode) the perturbation generally leads  also  to a change in the vertical direction with a change of the height $h_i$ in each sub-configuration.
We stress  that the  constraints on the spacings  or the elongations, for example in the  $()_{\lambda}$ or $()_{\bar{\lambda}}$ modes, address  also the problem  of how many different decompositions are possible in fixed sequence $\{\lambda_i\}_{i=1}^n$ or $\{\bar{\lambda}_{i i+1}\}_{i=1}^n$ respectively.
The perturbation of a sequence of parameters of the decomposition, holding the other  constant $\ell$-mode or $K$-mode respectively\footnote{We note  that the  $K_i$ parameters are related to the specific enthalpy and temperature  of each ring, see for example \citep{Pugtot}.}, is then one of two possible ways to deform the ringed structure.  The  $K$-modes  do not allow a  translation of the centers of the rings, but only a deformation (a kind of breathing, expansion or contraction of the rings).

In the  $\ell$-modes the center of pressure moves \emph{together} with the ring leaving the elongation in the $\ell_{\lambda}$-modes, and  the  spacings in the $\ell_{\bar{\lambda}}$-modes, fixed, but with    vertical deformation of the sub-configuration.
In other words, one  cannot keep the elongation $\lambda_i=$constant and  the shift parameters in the Heaviside potential  Eq.\il(\ref{Eq:Vcomplessibo})  fixed, and let $r_{min}^i$, i.e. the point of maximum pressure in the ring $C_i$, to move inside the ring, with the corresponding change in $h_i$. Such a case would require a combined variation of  $\ell_i$ and  $K_i$, to compensate  the effects of the variation of $K$, therefore a combination of $\ell$- and $K$-modes (however, this case albeit possible may require, at least for some range of variation for the  parameters,  orders of $\epsilon_K$ and $\epsilon_{\ell}$ significantly different, therefore  in some situations one  cannot deal   with  ``small'' perturbations of the sequences of initial parameters anymore). A similar situation was  mentioned   in Sec.\il(\ref{Sec:diff-rot}),   where a compensatory effect between combined perturbations of the same  decomposition through perturbation of its differential rotation  was  considered .

In the case of  $\ell_{\lambda}$-modes  being (as mentioned with the  $K$-modes analysis), $\delta_{min_1}^i$ and  $\delta_{min_3}^i$,  generally nor equal or constant for perturbations  at $K$ or $\ell$ (and therefore $r_{min}$) fixed, one could set $\lambda_i=$constant, but the ring translates (not in  rigid way) along   $r$ and in general  $\breve{\delta}_{min_1}^i$ and  $\breve{\delta}_{min_3}^i$  are  such that $r_{min}$  is moved in the range $ \Lambda_i $, and also  relatively to the inner and outer edge on the ring. In other words,  combining a translational motion of the structure $C_i$, to a translational motion  within the structure, we are considering the variations of the measures  ${\delta}_{min_3}^i$ and $\breve{\delta}_{min_1}^i$ with respect to a shift of $r_{min}$.
 In  this way,  the perturbation induces a variation of the pressure gradient in the radial direction, and therefore a change in its vertical dimension, even if it is not leading  to an instability of the ringed disk.
On the other hand,  the   $\ell$-modes induce  in general changes of the pressure gradient inside the ringed disk, even if they are not leading necessary to an unstable point.

We conclude this Section with the following general considerations on the ringed model and its perturbations:
each sub-configuration of   $\mathbf{C}^n$ is an equilibrium configuration at $\ell=$constant, and  dynamically  not related with the other rings of the decompositions when this is  far from its instability phase or not in contact.

By requiring that the barotropic surfaces uniquely define  the  ring surfaces, we can provide a solution  characterized by the parameter $p_i$, but
there is \emph{no} physical law   in this scheme that relates  $p_i\neq p_j$ for two different rings, but   a relation must be established in advance by an ad hoc considerations,  considering  properly the boundary conditions. We have addressed this aspect in many points in this work and especially in Sec.\il(\ref{Sec:effective}) where we have provided an effective potential for the ringed disk $\mathbf{C}_n$, using the effective potential of each ring and the boundary conditions  trough the Heaviside functions.
When one perturbs $p_i$, it is   $p_i\neq p_j=p_i+\epsilon_i \check{p}_i$ where $\epsilon_i$ sets the perturbation order of the $p_i$  model parameter. For example, we can refer to Eq.\il(\ref{Eq:Marix-ex-spo}) where the displacement matrices $\epsilon_{ij}=\epsilon_i \check{p}_i$ were introduced in this perturbative scheme.

Any perturbation on the potential  (\ref{Eq:def-partialeK}) or (\ref{Eq:Vcomplesssprouibo}), or also (\ref{Eq:Vcomplessibo}), introduces implicitly a dynamical relationship  between the rings through the perturbation of the  boundary data, in each case by the cuts imposed by means of the Heaviside functions. As in the case  of any  evolutionary model built up with  the time-independent Boyer  model,  discusses in  \citep{Pugtot}, any dynamic relation among  the various evolutionary phases  of the ringed disk, has to be explained considering  the time variable in the model,  but also the interaction with the surrounding matter  with  which the configuration may be in interaction.
Analogue arguments can be  done for  the  perturbations that could be  possibly associated to  some  real physical process,   being   not just a study of the structure stability for changes in initial conditions, testing various  eligible decompositions for the  appropriate  model setup.

\section{Some  Notes}\label{App:notes-index}
\begin{description}

\item[For $\ell$corotating  sequences]

\begin{description}

\item[\namedlabel{a.}{a.}]For $\mathbf{|\ell_a|>|\ell_b|}$ the relations  $r_{Max}^a<r_{Max}^b<r_{mso}<r_{min}^b<r_{min}^a $,  $K_{Max}^b<K_{Max}^a$ and $K_{min}^b<K_{min}^a$ hold necessarily. Note that this is not fixing the  relation between the couples $(K_{Max}^b,K_{min}^a)$ and $(K_{Max}^a,K_{min}^b)$, because for $|\ell_a|-|\ell_b|$ sufficiently large, one can also have $K_{min}^a>K_{Max}^b$ and viceversa.
    \\
    \item[\namedlabel{b.}{b.}]For $\mathbf{r_{min}^b<r_{min}^a }$  we should have  ${|\ell_a|>|\ell_b|}$,  $K_{min}^b<K_{min}^a$ and $r_{Max}^a<r_{Max}^b$;   therefore, also  $K_{Max}^b<K_{Max}^a$. One of the four inequalities  is \emph{enough}, within the condition $\ell_a\ell_b>0$, to ensure the remaining three hold, see also \ref{NoteIII}.
\end{description}
\item[
For $\ell$counterrotating sequences]

It follows from the discussion in  Sec.\il(\ref{Sec:procedure}) that

\begin{description}
\item[\namedlabel{(i)}{i)}]
If $\mathbf{\ell_->-\ell_+>\ell_{mso}^+}$,
{then  always }
$r_{Max}^-<r_{Max}^+<r_{mso}^+<r_{min}^+<r_{min}^-$.
 \\
 \item[\namedlabel{(ii)}{ii)}]
If $\mathbf{\ell_-<-\ell_+}$
neither the maximum or the minima are fixed, and it can be either $C_-<C_+$ or $C_+<C_-$, and   $r_{Max}^-<r_{Max}^+$ or $r_{Max}^+<r_{Max}^-$.
\end{description}
From the analysis in Sec.\il(\ref{Sec:procedure}) it follows that:
\begin{description}
\item[\namedlabel{I.}{I.}]
 If $\mathbf{r_{min}^-<r_{min}^+}$,    then  $\ell_-<-\ell_+$, but  the relation between the maximum points has still to be established, see \ref{(ii)}
  \\
\item[\namedlabel{II.}{II.}]
  But if  $\mathbf{\ell_-\in]\ell_{mso}^-,-\ell_+[}$,  then either   $r_{min}^-<r_{min}^+$, \emph{or}  $r_{min}^-\in]r_{min}^+,\bar{r}_-[$ where $\bar{r}_-:\;\ell_-(\bar{r}_-)=-\ell_+$, see \ref{(ii)}.
   \\
\item[\namedlabel{III.}{III.}]
   Therefore, if $\mathbf{r_{min}^->r_{min}^+}$,  then either
$\ell_->-\ell_{+} $ if $r_{min}^->\bar{r}_-$ see (\ref{(i)}), \emph{or} $\ell_-<-\ell_{+} $ if $r_{min}^-<\bar{r}_-$.
\end{description}
\end{description}
\subsubsection{Some notes on $(K_{crit},r_{crit},\ell)$}
Below are  some  general considerations  useful to determine the relationship between $ K_{crit}$, $ r_{crit} $ and $ \ell $ of  two or more sub-configurations of  a generic decomposition.
%
\begin{description}
\item[\namedlabel{NoteI}{Note I}]:
In the  {\large{$\ell$}}corotating case, $\ell_{a}\ell_{b}>0$, at fixed $r$, the effective potential increases with the  magnitude of the specific angular momentum, i.e.,
\be
V_{eff}(\ell_b)>V_{eff}(\ell_a)\quad \mbox{as}\quad |\ell_b|>|\ell_a|\quad\mbox{or}\quad \left.\partial_{|\ell|}V_{eff}\right|_{{r}}>0,
\ee
and  therefore:
\be
r_{Max}^b<r_{Max}^a<r_{mso}<r_{min}^a<r_{min}^b,\quad K_{Max}^b<K_{Max}^a<K_{min}^a<K_{min}^b,
\ee
see Fig.\il(\ref{Fig:bCKGO}). %
\\
\item[\namedlabel{NoteII}{Note II}]:
There is  $\partial_r V_{eff}(\ell)>0$ where $\partial_r|\ell| >0$ and viceversa.
 At fixed $r=\bar{r}$,   $\left.V_{eff}(-\ell)\right|_{\bar{r}}>\left.V_{eff}(\ell)\right|_{\bar{r}}$,
then   $V_{eff}(\ell)$ is increasing with  $|l|$ increasing
 and  $(V_{eff}(\ell_+),V_{eff}(\ell_-))$ and  $(\ell_-,|\ell_+|)$ are mutually decreasing with  $r\in]r_{\gamma}^-,r_{mso}^-]$    or increasing for $r>r_{mso}^+$,
and  viceversa, they are decreasing  with respect to the other  in $r\in]r_{mso}^-,r_{mso}^+[$ where it can be $r_{min}^-=r_{Max}^+$.
\\
\item[\namedlabel{NoteIII}{Note III}]:
Referring  to  Figs\il(\ref{Fig:bCKGO},\ref{Figs:Aslanleph1l}) together, one can say
\bea\nonumber
&&
\left.\partial_{r_{min}}r_{Max}\right|_{ \mathbf{q}}<0\quad  \mathbf{q}\in\{\ell,K_{crit}\},\quad
\partial_{r_{min}}|\ell|>0,\quad \partial_{|\ell|}K_{crit}>0,
\\\label{Eq:partial-easy-p}
&&
\partial_{r_{min}}K_{min}>0,\quad \partial_{r_{Max}}K_{min}<0,\quad \partial_{r_{Max}}K_{Max}<0;
\eea
with $\partial_{\mathbf{B}}\mathbf{Q}>0$,  we intend that the quantity $\mathbf{Q}$ increases where the quantity $\mathbf{B}$ increases and viceversa. Then the first inequality of  Eq.\il(\ref{Eq:partial-easy-p}) means that smaller value of $r_{min}$ etc. corresponds  to larger values of the maximum $r_{Max}$ of a given sub-configuration  if $\mathbf{q}=\ell$, or $\mathbf{q}=K_{crit}=K_{min}=K_{Max}$ at two different configurations, respectively,
\\
\item[\namedlabel{NoteIV}{Note IV}:]
\textbf{On the ring number $n$ and  the ring density}

We consider the  range of specific angular momenta $\Delta\ell$ (difference of  magnitudes), or the range $\Delta K$ for  values of $K\in]K_{min},K_{Max}[$.
Define
$\Delta r_{min}$ as the orbital range whose boundaries are the  minimum points for the  inner and outer rings of a given sub-sequence of the decomposition.
We could consider the order $n$ as function of the quantities $\{\Delta\ell,\Delta K,\Delta r_{min}\}$; one can say that
for  $\Delta\ell\approx0$ there is $K_i\approx K_{min}^i$ and $\Delta K_i-\Delta K_j\approx0$.
This is trivial for the $\ell$corotating configurations, but for the  $\ell$counterrotating ones,   we need to consider separately the isolated  $\overbrace{\mathbf{C}}_{{s}}$, see \ref{(-)}, or mixed $\overbrace{\mathbf{C}}_{{m}}$ decompositions, see \ref{(--)}.
In any case, as  can be seen in  Fig.\il(\ref{Figs:Aslanleph1l}),  $\Delta\ell\approx0$  means $\Delta r_{min}\approx0$ and this happens for $r_{min}$ sufficiently large.
In principle, the number  $n$ of the rings in the range $\Delta r_{min}$, for  given  $K$ as  the sub-configuration parameter and $\delta K$  as the maximum difference of the ring parameters        (or  the outer and inner ring parameters--see discussion in Secs.\il(\ref{Sec:onK}, \ref{Sec:effective}), it has to be:
\be
\left.\partial_{\delta K}n\right|_{\delta r_{min}}<0,\quad\partial_{\delta |\ell_{\pm\pm}|}n>0,\quad\partial_{\delta
|\ell_{+-}|}n>0,
\quad\mbox{and}\quad
\delta |\ell_{+-}|=0 \quad\mbox{with}\quad n=2,
  \ee
where $\ell_{+-}$ is the difference between the \emph{magnitude} of the  specific angular momenta of the outer and inner ring  $\ell$counterrotating  and as $\ell_{\pm \pm}$ for $\ell$corotating ones. The first inequality, depending on the choice of the ring parameters at fixed   $K_{crit}$,  has to be intended following the discussion in  Sec.\il(\ref{Sec:procedure}), especially with references to the  relations in Eqs.\il(\ref{Eq:cononK},\ref{Eq:cononKlcontro}).  
\end{description}

\end{document}